\documentclass[12pt]{article}
\usepackage{latexsym}
\usepackage{verbatim}
\usepackage{setspace}
\usepackage{color}
\usepackage{physics}
\usepackage{tikz}
\usepackage{mathdots}
\usepackage{yhmath}
\usepackage{cancel}
\usepackage{color}
\usepackage{siunitx}
\usepackage{array}
\usepackage{multirow}
\usepackage{gensymb}
\usepackage{tabularx}
\usepackage[normalem]{ulem}
\usepackage{booktabs}
\usetikzlibrary{fadings}
\usetikzlibrary{patterns}
\usetikzlibrary{shadows.blur}
\usetikzlibrary{shapes}
\usepackage{cite}
\usepackage{hyperref}
\usepackage{cleveref}
\usepackage[mathscr]{euscript}
\usepackage{changes}
\usepackage{cite}
\usepackage{xspace}
\usepackage{xparse}
\usepackage{mathtools}
\usepackage{mathrsfs}
\usepackage{fontawesome5}
\usepackage{caption}
\usepackage{subcaption}
\usepackage{tensor}
\usepackage{amsfonts}
\usepackage{amssymb}
\usepackage{amsmath,amssymb}
\usepackage[utf8]{inputenc}
\usepackage{url}

\numberwithin{equation}{section}

\definechangesauthor[name=Vincent,color=blue]{hzc}
\definechangesauthor[name=Ana,color=red]{amr}

\NewDocumentCommand{\modifiedSymbol}{D<>{}D<>{}D<>{}D<>{}O{}O{}mO{}O{}}{%
  \tensor*[^{#5}_{#6}]{#1{#2{#3{#4{#7}}}}}{^{#8}_{#9}}%
}

\newcommand{\cf}{\textit{c.f.\@\xspace}}
\newcommand{\Cf}{\textit{C.f.\@\xspace}}
\newcommand{\eg}{\textit{e.g.\@\xspace}}
\newcommand{\Eg}{\textit{E.g.\@\xspace}}
\newcommand{\ie}{\textit{i.e.\@\xspace}}

\newcommand{\AdS}{\mathrm{AdS}}
\newcommand{\bdy}{\mathrm{bdy}}
\newcommand{\bdyt}{\widetilde{\bdy}}
\newcommand{\blk}{\mathrm{blk}}
\newcommand{\blkt}{\widetilde{\blk}}
\NewDocumentCommand{\CFT}{O{}}{%
  \modifiedSymbol[#1]{\mathrm{CFT}}\relax%
}
\NewDocumentCommand{\cCFT}{O{}}{%
  \modifiedSymbol[#1]{\mathrm{cCFT}}\relax%
}

\newcommand{\cs}{\mathrm{CS}}
\newcommand{\csI}{\mathrm{CSI}}
\newcommand{\edge}{\mathrm{E}}
\newcommand{\edgeI}{\mathrm{EI}}
\newcommand{\Ein}{\mathrm{Ein}}
\newcommand{\ent}{\mathrm{ent}}
\newcommand{\gold}{\mathrm{G}}

\newcommand{\kg}{\mathrm{KG}}

\newcommand{\soft}{\mathrm{soft}}

\newcommand{\Sh}{\mathrm{Sh}}

\newcommand{\vN}{\mathrm{vN}}

\newcommand{\sgn}{\mathrm{sgn}}

\let\Re\relax
\DeclareMathOperator{\Re}{Re}
\let\Im\relax
\DeclareMathOperator{\Im}{Im}
\newcommand{\p}{\partial}
\DeclareMathOperator{\diag}{diag}

\newcommand{\eqDubious}{\stackrel{?}{=}}
\newcommand{\eqReg}{\stackrel{\reg}{=}}
\newcommand{\eqIntegrated}{\stackrel{\reg}{\cong}}
\newcommand{\eqOnShell}{\mathrel{\hat{=}}}
\newcommand{\neqOnShell}{\not\mathrel{\hat{=}}}

\newcommand{\reals}{\mathbb{R}}
\newcommand{\posReals}{\reals_{>0}}

\newcommand{\complexes}{\mathbb{C}}
\newcommand{\integers}{\mathbb{Z}}
\newcommand{\naturals}{\mathbb{N}}

\newcommand{\SL}{\mathrm{SL}}
\newcommand{\SO}{\mathrm{SO}}
\newcommand{\U}{\mathrm{U}}

\newcommand{\kronDelta}{\delta}

\newcommand{\commut}[2]{\left[ #1, #2 \right]}
\NewDocumentCommand{\ketOf}{O{}m}{%
  \modifiedSymbol[#1]{\ket{#2}}
}
\NewDocumentCommand{\braOf}{O{}m}{%
  \modifiedSymbol[#1]{\bra{#2}}
}
\NewDocumentCommand{\hamOf}{O{}}{%
  \modifiedSymbol[#1]{H}
}
\NewDocumentCommand{\softCohKet}{m}{\ket{#1}}
\NewDocumentCommand{\softCohBra}{m}{\bra{#1}}
\NewDocumentCommand{\softCohOp}{O{}}{
  \modifiedSymbol[#1]{W}\relax
}
\NewDocumentCommand{\dnsMat}{O{}}{%
  \modifiedSymbol[#1]{\rho}\relax
}
\newcommand{\prob}{p}
\NewDocumentCommand{\entropy}{O{}}{%
  S_{#1}
}
\NewDocumentCommand{\partFunc}{O{}O{}O{}}{%
  \modifiedSymbol[#1]{Z}[#2][#3]
}

\NewDocumentCommand{\cauchy}{O{}}{%
  \modifiedSymbol[#1]{\Sigma}\relax%
}
\newcommand{\volForm}{\varepsilon} 
\NewDocumentCommand{\deltaFunc}{o}{%
  \IfNoValueTF{#1}
  {\delta}
  {\delta^{#1}}%
}
\NewDocumentCommand{\regDeltaFunc}{sm}{%
  \IfBooleanTF{#1}
  {\tilde{\delta}_{#2}}
  {\delta_{#2}}%
}
\newcommand{\stepFunc}{\Theta}
\newcommand{\regStepFunc}[1]{\stepFunc_{#1}}

\NewDocumentCommand{\diff}{o}{
  \IfNoValueTF{#1}
  {\mathrm{d}}
  {\mathrm{d}_{#1}}%
}
\newcommand{\lie}[1]{\mathcal{L}_{#1}}
\NewDocumentCommand{\covD}{o}{
  \IfNoValueTF{#1}
  {\nabla}
  {\nabla_{#1}}%
}
\NewDocumentCommand{\lap}{o}{
  \IfNoValueTF{#1}
  {\Box}
  {\Box_{#1}}%
}
\NewDocumentCommand{\hodge}{o}{
  \IfNoValueTF{#1}
  {*}
  {*_{#1}}%
}

\NewDocumentCommand{\solnProd}{D<>{}D<>{}D<>{}D<>{}O{}mm}{%
  \modifiedSymbol<#1><#2><#3><#4>[#5]{\left\langle #6, #7 \right\rangle}\relax
}
\newcommand{\kgProd}[2]{\left\langle #1, #2 \right\rangle_{\kg}}
\newcommand{\doubleQuotient}{//}
\newcommand{\entProd}{\times^{\ent}}
\newcommand{\poisson}[2]{\left\{ #1, #2 \right\}}

\NewDocumentCommand{\pullback}{D<>{\pi}o}{
  \IfNoValueTF{#2}
  {#1^*}
  {#1_{#2}^*}%
}

\newcommand{\flatPlane}[1]{\reals^{#1}}
\newcommand{\sphUnit}[1]{S^{#1}}

\newcommand{\ads}[1]{\AdS_{#1}}

\newcommand{\mink}{\flatPlane{1,3}}
\newcommand{\minkProd}[2]{\solnProd{#1}{#2}}
\NewDocumentCommand{\minkSymp}{o}{\symp[#1]}
\newcommand{\minkCommut}[2]{\commut{#1}{#2}}
\newcommand{\minkX}{X}
\newcommand{\minku}{u}
\newcommand{\minkr}{r}
\newcommand{\minkt}{t}
\NewDocumentCommand{\minkz}{o}{%
  \IfNoValueTF{#1}
  {\mathbf{z}}
  {z#1}%
}
\newcommand{\minkzh}{z}
\newcommand{\minkza}{\bar{z}}
\DeclareMathOperator{\lorentz}{\Lambda}
\newcommand{\minkMom}{k}
\newcommand{\spacMom}{\vec{\minkMom}}
\newcommand{\minkMet}{\eta}

\newcommand{\lMilne}{L}
\newcommand{\rMilne}{R}
\newcommand{\milnet}{\tau}
\NewDocumentCommand{\milneCoeff}{m}{%
  \modifiedSymbol{\mathcal{L}}[#1]\relax
}

\newcommand{\hyp}[1]{H_{#1}}
\newcommand{\timeUnit}{\hat{X}}
\newcommand{\lenConst}{\ell}
\newcommand{\hyps}{\sigma} 
\newcommand{\fgr}{\rho} 
\DeclareMathOperator{\hypK}{K} %
\NewDocumentCommand{\blkBdyProp}{D<>{}D<>{}D<>{}D<>{}O{}mO{}}{%
  \modifiedSymbol<#1><#2><#3><#4>[#5]{\mathcal{K}}[#6][#7]
}
\NewDocumentCommand{\blkField}{D<>{}D<>{}D<>{}D<>{}O{}mO{}}{%
  \modifiedSymbol<#1><#2><#3><#4>[#5]{\phi}[#6][#7]
}
\newcommand{\hypVolForm}{\volForm^{(3)}}

\NewDocumentCommand{\nullInfty}{o}{
  \IfNoValueTF{#1}
  {\mathscr{I}}
  {\mathscr{I}^{#1}}%
}
\newcommand{\nullInftyBdy}[2]{\nullInfty[#1]_{#2}}
\newcommand{\spaceInfty}{i_0}

\newcommand{\nullUnit}{\hat{q}}
\newcommand{\refVec}{n}
\newcommand{\polar}{\varepsilon}
\newcommand{\weylFac}{\Omega}
\newcommand{\celPlane}{\flatPlane{2}}
\newcommand{\celMet}{\gamma}
\newcommand{\celVolForm}{\volForm^{(2)}}
\newcommand{\celVol}{\mathrm{vol}^{(2)}}
\newcommand{\celIRReg}{L}
\newcommand{\celRicci}{R^{(2)}}

\NewDocumentCommand{\celCovD}{o}{
  \IfNoValueTF{#1}
  {D}
  {D_{#1}}%
}
\NewDocumentCommand{\celLap}{o}{\lap[#1]^{(2)}} 

\NewDocumentCommand{\celw}{o}{%
  \IfNoValueTF{#1}
  {\mathbf{w}}
  {w#1}%
}
\newcommand{\celwh}{w}
\newcommand{\celwa}{\bar{w}}
\DeclareMathOperator{\celG}{G}
\DeclareMathOperator{\celDeltaFunc}{\delta}
\DeclareMathOperator{\confTrans}{\Lambda}
\newcommand{\celFreq}{\omega}
\NewDocumentCommand{\stereo}{o}{%
  \IfNoValueTF{#1}
  {\vec{\Omega}}
  {\Omega#1}%
}

\newcommand{\celpsi}{\psi}

\newcommand{\celIntVar}{\nu}

\newcommand{\invTemp}{\beta}
\newcommand{\reg}{\epsilon}
\newcommand{\zetaReg}{\mu}
\newcommand{\elemCharge}{q_{\mathrm{e}}}
\NewDocumentCommand{\act}{O{}O{}}{%
  \modifiedSymbol{I}[#1][#2]
}
\newcommand{\anomAct}{\mathcal{A}}
\NewDocumentCommand{\lag}{o}{%
  \IfNoValueTF{#1}
  {L}
  {L^{#1}}%
}
\NewDocumentCommand{\vary}{o}{%
  \IfNoValueTF{#1}
  {\delta}
  {\delta_{#1}}%
}
\newcommand{\eom}{\mathcal{E}}
\NewDocumentCommand{\sympD}{D<>{}D<>{}D<>{}D<>{}O{}O{}}{%
  \modifiedSymbol<#1><#2><#3><#4>[#5]{\omega}[#6]
}
\NewDocumentCommand{\sympPotD}{D<>{}D<>{}D<>{}D<>{}O{}O{}}{%
  \modifiedSymbol<#1><#2><#3><#4>[#5]{\theta}[#6]
}
\NewDocumentCommand{\symp}{om}{%
  \IfNoValueTF{#1}
  {\Omega_{#2}}
  {\Omega^{#1}_{#2}}%
}
\NewDocumentCommand{\sympPot}{om}{%
  \IfNoValueTF{#1}
  {\Theta_{#2}}
  {\Theta^{#1}_{#2}}%
}
\newcommand{\sameInit}[1]{\stackrel{#1}{\sim}}
\newcommand{\unsameInit}[1]{\stackrel{#1}{\nsim}}

\DeclareFontEncoding{LS1}{}{}
\DeclareFontEncoding{LS2}{}{\noaccents@}
\DeclareFontSubstitution{LS1}{stix2}{m}{n}
\DeclareFontSubstitution{LS2}{stix2}{m}{n}
\DeclareSymbolFont{symbols3}      {LS1}{stix2bb}   {m} {n}
\DeclareMathSymbol{\Wedge}                    {\mathbin}{symbols3}{"A3}
\DeclareSymbolFont{arrows3}       {LS2}{stix2tt}   {m} {n}
\DeclareMathSymbol{\Otimes}                   {\mathbin}{arrows3}{"A8}
\newcommand{\fieldTProd}{\Otimes}
\newcommand{\fieldWedge}{\Wedge}
\DeclareMathOperator*{\bigFieldWedge}{\bigwedge\mkern-20mu\bigwedge}
\newcommand{\fieldDiff}{\delta}
\newcommand{\fieldIProd}[1]{I_{#1}}
\newcommand{\fieldLie}[1]{L_{#1}}
\newcommand{\fieldVecGauge}{U}
\NewDocumentCommand{\fieldPullback}{D<>{\Pi}o}{
  \IfNoValueTF{#2}
  {#1^*}
  {#1_{#2}^*}%
}
\newcommand{\fieldHermMetric}{h}
\newcommand{\fieldMetric}{g}
\newcommand{\edgeMetric}{\fieldMetric}
\newcommand{\edgeMetricAll}{\bar{\edgeMetric}}
\newcommand{\scalarFieldMetric}{\hat{\fieldMetric}}
\newcommand{\fieldVolForm}{\mathcal{E}}
\newcommand{\edgeVolForm}{\fieldVolForm}
\newcommand{\edgeVolFormAll}{\bar{\edgeVolForm}}
\DeclareMathOperator{\fieldDeltaFunc}{\delta}
\NewDocumentCommand{\confSpace}{O{}}{%
  \modifiedSymbol[#1]{\mathcal{F}}\relax%
}
\NewDocumentCommand{\solnSpace}{O{}}{%
  \modifiedSymbol[#1]{\bar{\mathcal{F}}}\relax%
}
\NewDocumentCommand{\extPhSpace}{O{}}{%
  \modifiedSymbol[#1]{\Gamma}\relax%
}
\NewDocumentCommand{\cstrSpace}{O{}}{%
  \modifiedSymbol[#1]{\bar{\Gamma}}\relax%
}
\NewDocumentCommand{\phSpace}{O{}}{%
  \modifiedSymbol[#1]{\mathcal{P}}\relax%
}
\NewDocumentCommand{\hilbSpace}{O{}}{%
  \modifiedSymbol[#1]{\mathcal{H}}\relax%
}
\NewDocumentCommand{\cstr}{O{}O{}}{%
  \modifiedSymbol[#1]{Q}[#2]\relax%
}
\NewDocumentCommand{\cstrVal}{O{}O{}}{%
  \modifiedSymbol[#1]{q}[#2]\relax%
}
\NewDocumentCommand{\cstrconj}{O{}O{}}{%
  \modifiedSymbol[#1]{S}[#2]\relax%
}
\NewDocumentCommand{\asCharge}{O{}O{}}{%
  \cstr[#1][#2]%
}
\newcommand{\gaugeParam}{\alpha}
\NewDocumentCommand{\edgeE}{D<>{}D<>{}D<>{}D<>{}O{}O{}O{}}{%
  \modifiedSymbol<#1><#2><#3><#4>[#5]{E}[#6][#7]
}
\NewDocumentCommand{\edgeScalar}{D<>{}D<>{}D<>{}D<>{}O{}O{}O{}}{%
  \modifiedSymbol<#1><#2><#3><#4>[#5]{\varphi}[#6][#7]
}
\newcommand{\blkScalar}{\phi}
\newcommand{\allFields}{\Phi}

\newcommand{\confDim}{\Delta}
\newcommand{\milneFreq}{\lambda}
\NewDocumentCommand{\confPrimA}{D<>{}D<>{}D<>{}D<>{}O{}mO{}}{%
  \modifiedSymbol<#1><#2><#3><#4>[#5]{A}[#6][#7]
}
\NewDocumentCommand{\confTensor}{D<>{}D<>{}D<>{}D<>{}mO{}}{%
  \modifiedSymbol<#1><#2><#3><#4>{m}[#5][#6]
}
\NewDocumentCommand{\confMell}{D<>{}D<>{}D<>{}D<>{}mO{}}{%
  \modifiedSymbol<#1><#2><#3><#4>{V}[#5][#6]
}
\NewDocumentCommand{\confGauge}{D<>{}D<>{}D<>{}D<>{}mO{}}{%
  \modifiedSymbol<#1><#2><#3><#4>{\gaugeParam}[#5][#6]
}
\NewDocumentCommand{\softGaugeg}{D<>{}D<>{}D<>{}D<>{}mO{}}{%
  \modifiedSymbol<#1><#2><#3><#4>{\gamma}[#5][#6]
}
\NewDocumentCommand{\softGaugeb}{D<>{}D<>{}D<>{}D<>{}mO{}}{%
  \modifiedSymbol<#1><#2><#3><#4>{\beta}[#5][#6]
}
\NewDocumentCommand{\mellCoeff}{m}{%
  \modifiedSymbol{\mathcal{K}}[#1]\relax
}
\NewDocumentCommand{\confPrimF}{D<>{}D<>{}D<>{}D<>{}O{}mO{}}{%
  \modifiedSymbol<#1><#2><#3><#4>[#5]{F}[#6][#7]
}

\newcommand{\opA}{A}
\NewDocumentCommand{\operator}{D<>{}D<>{}D<>{}D<>{}O{}O{}O{}}{%
  \modifiedSymbol<#1><#2><#3><#4>[#5]{\mathcal{O}}[#6][#7]
}
\NewDocumentCommand{\confPrimOp}{D<>{}D<>{}D<>{}D<>{}O{}mO{}}{%
  \operator<#1><#2><#3><#4>[#5][#6][#7]
}
\NewDocumentCommand{\goldOp}{D<>{}D<>{}D<>{}D<>{}O{}O{}}{%
  \modifiedSymbol<#1><#2><#3><#4>[#5]{\mathcal{S}}[][#6]
}
\NewDocumentCommand{\goldOpI}{D<>{}D<>{}D<>{}D<>{}O{}O{}}{%
  \modifiedSymbol<#1><#2><#3><#4>[#5]{\mathcal{S}}[\mathrm{I}][#6]
}
\NewDocumentCommand{\csOp}{D<>{}D<>{}D<>{}D<>{}O{}O{}}{%
  \modifiedSymbol<#1><#2><#3><#4>[#5]{\mathcal{Q}}[][#6]
}
\NewDocumentCommand{\csOpI}{D<>{}D<>{}D<>{}D<>{}O{}O{}}{%
  \modifiedSymbol<#1><#2><#3><#4>[#5]{\mathcal{Q}}[\mathrm{I}][#6]
}
\NewDocumentCommand{\entOp}{D<>{}D<>{}D<>{}D<>{}O{}O{}t'}{%
  \modifiedSymbol<#1><#2><#3><#4>[#5]{\mathcal{Q}}[%
  \IfBooleanTF{#7}{\ent'}{\ent}%
  ][#6]
}
\NewDocumentCommand{\confAnhl}{D<>{}D<>{}D<>{}D<>{}O{}mO{}}{%
  \modifiedSymbol<#1><#2><#3><#4>[#5]{a}[#6][#7]
}
\NewDocumentCommand{\milneAnhl}{D<>{}D<>{}D<>{}D<>{}O{}mO{}}{%
  \modifiedSymbol<#1><#2><#3><#4>[#5]{b}[#6][#7]
}

\newcommand{\link}{L}

\newcommand{\shad}{\tilde}

\newcommand{\inv}{\underline}
\newcommand{\invRot}{I}

\newcommand{\bit}{\begin{itemize}}
\newcommand{\eit}{\end{itemize}}
\newcommand{\bd}{\begin{description}}
\newcommand{\ed}{\end{description}}
\newcommand{\bc}{\begin{center}}
\newcommand{\ec}{\end{center}}

\def\bz{\bar{z} }

\begin{document}

\begin{titlepage}

\unitlength = 1mm
\ \\
\vskip 2cm
\begin{center}
{\LARGE{\textsc{Entanglement, Soft Modes, and \\
\vspace{10pt} Celestial CFT
\vspace{10pt}}}}

\vspace{0.8cm}
Hong Zhe Chen${}^{1}$, Robert Myers${}^{2}$ and Ana-Maria Raclariu${}^{3}$\\
\vspace{0.3cm}

${}^1$ \small{\textit{Department of Physics, University of California, Santa Barbara, CA 93106}}\\
${}^{2}$\small{\textit{Perimeter Institute for Theoretical Physics, Waterloo, Canada}}\\
${}^{3}$\small{\textit{King's College London, Strand, London WC2R 2LS, United Kingdom}} \\
\vspace{10pt}

\begin{abstract}
We revisit the calculation of vacuum entanglement entropy in free Maxwell theory in four-dimensional Minkowski spacetime. Weyl invariance allows for this theory to be embedded as a patch inside the Einstein static universe. We use conformal inversions to extend the conformal primary solutions of the equations of motion labelled by a boost-weight $\Delta = 1 + i\lambda$ to an inverted Minkowski patch centered at spacelike infinity of the original patch. For $\lambda \neq 0$ these solutions admit an expansion in terms of wavefunctions supported in the (future) Milne wedges of the original and inverted Minkowski patches, that diagonalize the respective Milne times. The Minkowski vacuum can then be described as a thermofield double state on these two Milne wedges. 
We characterize the soft sectors of $\lambda = 0$ modes supported in the two Milne wedges. Upon reinterpreting the non-pure gauge $\lambda = 0$ wavefunctions as sourced by image charges in the inverse Minkowski patch, we construct an appropriate entangling constraint in the Minkowski theory, thereby characterizing the physical state space. We show that the edge mode contribution to the vacuum entanglement entropy is due to correlations between the soft charges of the two Milne patches, or equivalently non-trivial conformally soft mode configurations at the entangling surface.

\end{abstract}
\vspace{0.5cm}
\end{center}
\vspace{0.8cm}

\end{titlepage}


\tableofcontents

\section{Introduction and summary}



The emergence of near-boundary degrees of freedom is a phenomenon present in a wide range of physical systems.  In gravity, their appearance is accompanied by the promotion of gauge redundancies to physical symmetries. Three-dimensional asymptotically anti-de-Sitter (AdS) spacetimes provide a sharp example where the SL$(2, \mathbb{C})$ isometries get promoted to an infinite set of large diffeomorphisms obeying two copies of the Virasoro algebra \cite{Brown-Henneaux,Balasubramanian:1999re}. These transformations act on the two-dimensional boundary as conformal symmetries, hinting at the existence of a dual conformal field theory (CFT). Boundaries of negatively curved spacetimes are particularly special, as they are conjectured to host local conformal field theories \cite{tHooft:1993dmi,Susskind:1994vu}. While string theory provided a concrete realization of these ideas through the AdS/CFT correspondence \cite{Maldacena:1997re}, coincidences continuing to arise in bottom up approaches to quantum gravity lead one to hope that at least certain universal properties of generic gravitational theories could be captured by conformal field theories in a lower number of dimensions \cite{Strominger:2001pn, deBoer:2003vf,Guica:2008mu, Anninos:2011af,Saad:2019lba}.

Over the past decade it has been realized that the near-boundary structure is not only non trivial  \cite{Bondi:1960jsa, Bondi62}, but also physically meaningful even in (3+1)-dimensional asymptotically flat spacetimes (AFS) \cite{Barnich:2009se,Barnich:2010ojg,Barnich:2010eb}. In a conformal compactification of Minkowski space, the boundary consists of future and past null cones, subtly matched near spacelike infinity \cite{Strominger:2013jfa}. The matching condition is particularly important as it renders the S-matrix well defined and implies the conservation of charges associated with large gauge symmetries \cite{Strominger:2013lka,He:2014cra,Kapec:2015ena}. Moreover, it reveals further symmetries of the gauge and gravity phase space reflected in tree-level scattering amplitudes through a tower of soft theorems \cite{He:2014laa,Cachazo:2014fwa,Kapec:2014opa,Lysov:2014csa,Campiglia:2014yka,Campiglia:2015qka,He:2015zea,Guevara:2021abz,Strominger:2021mtt,Freidel:2021ytz,Freidel:2023gue}. Interestingly, these symmetries are imprinted on physical observables through memory effects \cite{Strominger:2014pwa,Pasterski:2015tva,Strominger:2016wns,Nichols:2017rqr,Compere:2019odm,Grant:2021hga}. Celestial conformal field theories (CCFTs) provide a convenient way of organizing asymptotic scattering states in highest weight representations of the Lorentz SL$(2, \mathbb{C})$. Scattering amplitudes are projected on the two-dimensional celestial space through an integral transform which implements a change a basis from momentum to boost eigenstates \cite{Pasterski:2016qvg,Pasterski:2017kqt}. In the new basis, the Ward identities associated with a universal tower of soft theorems led to the identification of subleading soft particles with generators of infinite dimensional symmetries \cite{Fan:2019emx,Pate:2019mfs,Pasterski:2021fjn,Pasterski:2021dqe,Guevara:2021abz,Himwich:2021dau,  Jiang:2021ovh,Strominger:2021mtt,Donnay:2022sdg}. While this is all suggestive of a holographic correspondence in this context, basic entries in the AFS/CCFT dictionary are still missing.  

\begin{figure}
\centering
 \begin{subfigure}[t]{0.48\textwidth}
 \centering
    \includegraphics[scale=1]{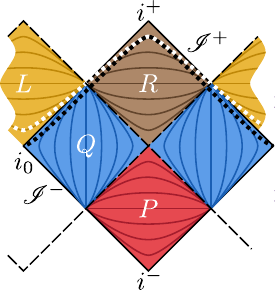}
    \caption{}
    \label{fig:bulkSubregions}
    \end{subfigure}
   \begin{subfigure}[t]{0.48\textwidth}
   \centering
    \includegraphics[scale=1]{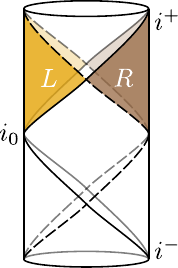}
    \caption{}
	\label{fig:cauchySlices}
 \end{subfigure}
 \caption{(\subref{fig:bulkSubregions}) Penrose diagram of Minkowski space and the
   corresponding inverted Minkowski space (delineated by solid and dashed
   diagonal lines, respectively). They overlap in the patch $Q$, the future
   Milne patches of the two geometries are denoted $R$ and $L$, and $P$ is the
   past Milne patch of the original spacetime. Minkowski (black) and Milne
   (white) Cauchy slices near $\mathscr{I}^+$ are drawn as dashed curves.
   (\subref{fig:cauchySlices}) Conformal mapping of $\rMilne$ and $\lMilne$
   Milne patches to the Einstein static universe.}
    \label{fig:embedding}
\end{figure}

This paper takes a step towards explaining how bulk subregions and their entanglement properties are encoded in the CCFT. To study these questions, we restrict our attention to pure Maxwell theory in (3+1)-dimensional Minkowski spacetime. In addition to admitting an infinity of asymptotic symmetries \cite{He:2014cra}, this theory is Weyl invariant. Hence upon embedding the Minkowski geometry as a patch inside the Einstein static universe $\mathbb{R} \times S^3$, we exploit conformal invariance to extend the Minkowski theory to the full Einstein cylinder. Cauchy slices of the Minkowski theory can then be identified with Cauchy slices of the cylinder (up to subtleties concerning spacelike infinity), allowing for states in the Minkowski theory to be prepared via an Euclidean path integral over $S^4$. The same path integral prepares an entangled thermal state from the perspective of an observer confined to a subregion.  For simplicity we choose the subregion to be the future Milne patch $(R)$ of the Minkowski theory -- the causal future of the origin.  This case is particularly interesting as the associated entanglement can be reinterpreted as entanglement across future null infinity $\mathscr{I}^+$ (see Figure \ref{fig:embedding}).
We show that the Minkowski vacuum state is a thermofield double state with respect to the $R$ patch and the future Milne patch $(L)$ of a Minkowski geometry related to the original one via a conformal inversion -- see Figure \ref{fig:bulkSubregions}.  This decomposition follows directly from the decomposition of conformal primary wavefunctions labelled by an eigenvalue $\lambda$ with respect to the Minkowski Milne time in terms of conformal primary modes supported respectively in the $L$ and $R$ patches. 

In the limit of vanishing Milne energy $\lambda = 0$, these decompositions become degenerate and hence the zero modes (also known as conformal primary and Goldstone modes \cite{Donnay:2018neh}) have to be treated separately. To this end, we  analyze the constraints \cite{Donnelly:2016auv} related to the conservation of large gauge charge in both the Minkowski and the Einstein static theories. 
While the constraints in the two theories differ for generic states, we show that in the case of the common global vacuum state, they imply that the large gauge charges of the $L$ and $R$ Milne patches are correlated. In particular, for the vacuum state defined by the vanishing of the large gauge charge, we show that the correlated fluctuations in the $L$ and $R$ charges lead to a non-trivial contribution to the entanglement entropy across $\mathscr{I}^+$. We demonstrate that, after a judicious choice of cutoff, this contribution coincides with the edge mode contribution to the entanglement entropy identified by Donnelly and Wall (DW) in \cite{Donnelly:2014fua,Donnelly:2015hxa} using different methods. 

This paper is a companion paper to \cite{Chen:2023tvj}. The results herein rely on a series of novel technical points which to the extent of our knowledge have not been discussed in the literature before. We now summarize the main results with an emphasis on the subtleties we encountered in the asymptotic expansions, analytic continuations and regularization of conformal primary wavefunctions, as well as their (conformally) soft limits: 

\begin{enumerate}
\item {\bf Asymptotics of conformal primary wavefunctions.}
In preparation for the analysis the soft contribution to the entanglement entropy at $\mathscr{I}^+$, it was important to revisit the asymptotic, large-radius expansions of the Goldstone and conformally soft wavefunctions. The leading terms in their large-$r$ expansions were shown in \cite{Donnay:2018neh, Donnay:2022sdg} to include  subleading contributions  involving higher and higher powers of $r^{-1}$. We point out that these expansions contain also \textit{logarithmic} terms at large $r$. We show this by direct computation in appendix \ref{sec:asymptotics}, treating the expansions as distributions on the celestial plane. That such terms necessarily appear can also be seen by recalling that the conformal primary wavefunctions are constructed in terms of solutions to the wave-equation on three-dimensional hyperbolic slices inside the Milne patch \cite{Pasterski:2017kqt}, whose asymptotics are known to contain logarithmic terms \cite{FG00,Fefferman:2007rka,Anderson:2004wj}.  Similar terms were recently discovered in gravity from a canonical analysis near spatial infinity and go under the name of logarithmic supertranslations \cite{Fuentealba:2022xsz}. Their Maxwell theory counterparts in a conformal primary basis should be related to conformally soft wavefunctions (see also \cite{Fuentealba:2023rvf} for an analysis of logarithmic terms in electromagnetism). We also work out the asymptotics of various other functions appearing in the expansion of conformally soft wavefunctions in terms of plane waves and conformal primary modes which we evaluate in section \ref{sec:CS-decompositions}. We point out that, due to the presence of delta function contributions, the large-$r$ expansion does not commute with multiplication. Thereby, one needs to be careful not to miss logarithmic contributions in the large-$r$ expansions of conformal primary wavefunctions of spinning particles obtained by dressing the scalar wavefunction with polarization tensors. This is particularly relevant for the expansion of soft photon and graviton conformal primary wavefunctions \cite{Donnay:2018neh}, as well as the tower of subleading counterparts \cite{Donnay:2022sdg}. 


\item {\bf Analytic continuation of conformally soft wavefunctions.} Conformally soft wavefunctions are defined as \cite{Donnay:2018neh}:
\begin{equation}
  \label{eq:CS0}
  \confPrimA{\cs}[a] \equiv \frac{\confPrimA{\log,+}[a] - \confPrimA{\log,-}[a]}{2\pi i}, \quad  \confPrimA{\log,\pm}[a] = -\log(-\minkX_\pm^2)  \confPrimA{1,\pm}[a].
\end{equation}
where $X_{\pm}$ is defined in eq.~\eqref{eq:frame}, $A_a^{1,\pm}$ is the Goldstone wavefunction \eqref{eq:coffee} and the $i\epsilon$ in the definition of $X_{\pm}$ provides a prescription to analytically continue this wavefunction outside the future null cone. It has been previously argued \cite{Donnay:2018neh} that these conformal primary wavefunctions are vanishing inside the future and past light-cones and non-vanishing outside. We point out in section \ref{sec:minkSoftModes} that the $i\epsilon$ prescription in fact implies that $A^{\rm CS}$ is non-vanishing and proportional to the Goldstone wavefunction $A^{\rm G}$ inside the past light-cone. Overall, we find that the pull-back of $A^{\rm CS}$ at $\mathscr{I}^-$ differs from previous expressions in the literature \cite{Donnay:2018neh,Arkani-Hamed:2020gyp} by the addition of $A^{\gold}$. Of course, this does not affect the associated field strengths, nor the inner product between the conformally soft and Goldstone modes. Since the inner product between pure gauge Goldstone modes vanishes, the condition that the conformally soft mode be canonically conjugate only determines it up to a pure gauge term. 
\item {\bf Matching conditions.} Given an outgoing conformally soft wavefunction, the incoming one is completely determined by requiring that it obeys the matching condition across $i^0$ \cite{Strominger:2017zoo}.  The pure gauge contribution to $A^{\rm CS}$ on $\mathscr{I}^-$ discussed above ensures that the matching condition is obeyed. A pure gauge term is also present in the expansion of $A^{\rm CS}$ in terms of plane waves -- see eq.~\eqref{eq:csToV}. This term is again necessary in order to ensure that the matching condition is obeyed. To the best of our knowledge, this pure gauge contribution has not featured in previous constructions of Goldstone operators (defined by the inner product of a field with $A^{\rm CS}$) and the associated dressings \cite{Kapec:2017tkm,Arkani-Hamed:2020gyp,Pasterski:2021dqe}. We expect this subtlety to be relevant in the computation of celestial correlators involving multiple insertions of Goldstone and conformally soft modes. Similarly, by studying the decomposition of $A^{\cs}$ in conformal primary modes, we show that the shadow contribution in the definition of $A^{\rm CS}$ (see eqs.~\eqref{eq:LSoftModeDubiousAsym} and \eqref{eq:shadow-cs}) is necessary to ensure that the matching condition is obeyed. 
\item {\bf Milne conformal primary wavefunctions and charges.} Conformal primary wavefunctions in the original and inverted Minkowski patches are related by an inversion. In particular, we note that in the overlap regions, inversions reduce to diffeomorphisms and use the $i\epsilon$ prescription to analytically continue the wavefunctions to the entire inverted patch. We find the following relation between inverted and shadow conformal primary wavefunctions
\begin{equation} 
\label{eq:shadow-inv}
\underline{A}^{\Delta, \pm}_{a;\mu}(\underline{X}) \eqReg  e^{\pm i \pi (\Delta - 1)} \tilde{A}^{\Delta, \pm}(\underline{X}),
\end{equation}
where $\tilde{A}^{\Delta}$ is a shadow mode with boost-weight $\Delta$ and $\eqReg$ denotes equality in the $\epsilon \rightarrow 0$ limit. These wavefunctions are related to conformal primary wavefunctions of dimension $2 - \Delta$ by a shadow transform. This establishes a spacetime interpretation of the shadow transform in conformally invariant bulk theories. Shadow transforms have been studied in CCFT context before \cite{Fan:2021isc,Kapec:2021eug,Chang:2022jut,Banerjee:2022wht, Jorstad:2023ajr,Chen:2023tvj}, but their physical relevance remained unclear. 

Eq.~\eqref{eq:shadow-inv} allows us to decompose a conformal primary wavefunction as a linear combination of $L$ and $R$ conformal primary wavefunctions supported respectively in the $L$ and $R$ Milne patches -- see eq.~\eqref{eq:LR-dec}. This decomposition together with the completeness of the conformal primary wavefunctions for $\Delta = 1 + i\lambda$ with $\lambda \in \mathbb{R}$ allows us to expand the Maxwell field in terms of $L$ and $R$ modes. This expansion parallels the Rindler decomposition and allows for the global Minkowski vacuum state to be identified with the thermofield double state between the $L$ and $R$ Milne theories. This is natural from an Einstein static universe perspective, where the Rindler and Milne decompositions are simply related by a global time translation. A similar construction applies to de Sitter spacetimes (see for example \cite{Cotler:2023xku}), where global wavefunctions were shown to admit decompositions in terms of modes confined to the north and south static patches (see also \cite{Andy-talk} where related ideas are discussed for a scalar field in flat space). 

In the limit $\lambda \rightarrow 0$, the Milne decompositions become degenerate, hence the soft modes have to be treated separately. Using the transformation properties of the conformally soft and Goldstone wavefunctions under inversion, we propose a construction of associated modes supported in the $L$ and $R$ patches. We show that these are canonically conjugate with respect to inner products inside the respective patches. 

\item {\bf Conformally soft wavefunctions from sources in the inverted patch.} In conformally invariant theories, inversions map solutions of the equations of motion to solutions. The resulting extensions of the conformally soft solutions to the inverted Minkowski patch are therefore sourceless. We show that the same conformally soft field profile in the original patch can also be generated with sources in the inverted patch. This sourced extension differs from the source-less one by the addition of shockwaves along the boundaries $\mathscr{I}^{\pm}$ of the original patch. Equivalently, these shockwaves propagate along the lightcone through the origin of the inverted patch. Both the sourced and the source-less configurations give rise to the same large-gauge charges in the original patch and are therefore indistinguishable from the perspective of a Minkowski observer. From the perspective of the inverse Minkowski patch, the same Coulomb field can be sourced by oppositely charged particles or by radiative modes, as illustrated in Figures \ref{fig:bread} and \ref{fig:avocado}. 
\item {\bf Soft constraints.} Physical states in gauge and gravity theories are subject to constraints \cite{Donnelly:2016auv}. In the pure Maxwell theory, these constraints can be formulated as relations between the $L$ and $R$ soft operators. In particular, we argue in section \ref{sec:entanglement} that the physical state space of the Minkowski theory can be embedded into the product space of the $L$ and $R$ Milne theory by imposing that it is annihilated by the operator
\begin{align}
  \entOp[][a](\celw)
  &=
    \csOp[\lMilne][a](\celw)
    +\csOp[\rMilne][a](\celw)
    +2\left[
    \goldOp[\lMilne][a](\celw)
    -\goldOp[\rMilne][a](\celw)
    \right]
    \;
    \label{eq:entOpMink-intro}
\end{align}
in the limit as the $\epsilon$ regulator is removed.
Here $\csOp[\lMilne, \rMilne][a](\celw)$ and $\goldOp[\lMilne,\rMilne][a](\celw)$ are obtained from the Maxwell fields and the inner product with the corresponding wavefunctions -- see eqs.~\eqref{eq:csOpR} and \eqref{eq:goldOpR}. 
This particular linear combination is selected by requiring that it commutes with the admissible linear combinations of $L$ and $R$ modes involved in the construction of conformally soft and Goldstone modes \eqref{eq:minkSoftOpAsLR}. We demonstrate that this constraint can be interpreted as measuring a violation to Gauss's law due to sources in the inverted patch. On the other hand, states of the pure Maxwell theory in the Einstein static universe ought to have vanishing large gauge charges 
  $\csOp[\lMilne][a](\celw) +\csOp[\rMilne][a](\celw)$. 
  We argue that the only common state to the Minkowski and Einstein static pure Maxwell theories is the vacuum state: generic states of non-vanishing large gauge charge in the Minkowski theory will include sources in the Einstein static universe, while generic states in the Einstein static theory will violate the Minkowski boundary conditions and hence the constraint imposed by  eq.~\eqref{eq:entOpMink-intro}. 
\item {\bf Weyl rescalings of the celestial space.}
For simplicity, we present most of our formulas for a planar section of null infinity (or a celestial plane). In section \ref{sec:entanglementSoftDOFs} we explain how our analysis generalizes to a family of celestial spaces related to the plane by a Weyl rescaling. Given an arbitary conformal frame for the celestial CFT, we establish a relationship between the Fefferman-Graham expansion of the metric on hyperbolic (Euclidean AdS$_{3}$) slices and the Cartesian Minkowski coordinates. The celestial sphere is obtained as a special case. 
\item {\bf DW edge modes versus conformally soft modes.} Donnelly and Wall identified in \cite{Donnelly:2014fua, Donnelly:2015hxa} a non-trivial contribution to the vacuum entanglement entropy due to classical, on-shell field configurations (or edge modes) obeying the Gauss constraint at the entangling surface. After gauge fixing, they identified the edge modes with static modes inside the causal development of the entangling subregion.
In section \ref{sec:entanglementSoftDOFs}, we show that the static field configuration of DW are related to the logarithmic constituents of the conformally soft wavefunctions \eqref{eq:CS0} by a gauge transformation: 
\begin{align}
  \confPrimA{\log,\pm}
  &= -2 \milnet_\pm \, \diff_{\minkX} \confGauge{1,\pm}
  = 2 \left(
    - \diff_{\minkX}(\milnet_\pm \confGauge{1,\pm})
    + \diff\milnet_\pm \, \confGauge{1,\pm}
  \right)
    \;.
    \label{eq:AlogGaugeTransform-intro}
   \end{align}
   Here $d\tau_{\pm} \alpha^{1,\pm}$ is the static gauge field configuration considered in \cite{Donnelly:2014fua, Donnelly:2015hxa}.
Consequently, the DW contribution to the entanglement entropy can be reinterpreted as due to correlated fluctuations in the large gauge charges of the $L$ and $R$ Milne patches. This identification requires the selection of an appropriate relation between the spacetime cutoff at the entangling surface and the $\epsilon$ regulator characterizing the conformally soft wavefunctions, which we motivate by examining the regulated field configurations near the entangling surface.  
\item {\bf On-shell action versus soft effective action.} As a by-product of our analysis, we identify a precise relation between the on-shell action of a subregion of free Maxwell theory in Minkowski spacetime and the soft effective action proposed in \cite{Kapec:2021eug} to capture the infrared dynamics of QED. We demonstrate that the on-shell action coincides with the soft S-matrix computed by Weinberg in \cite{Weinberg:1965nx} after identifying the edge mode profile in the vacuum with the field profile sourced by hard charges (of net vanishing global charge) propagating in spacetime. This identification holds provided that the asymptotic charges associated with the respective field configurations coincide and can be understood from the Einstein static universe point of view, where the edge modes in the original Minkowski patch are sourced by physical charges propagating in the inverted patch. As shown in \cite{Kapec:2021eug}, the same result is obtained from the soft effective action after coupling the soft effective  action to sources and integrating out the Goldstone degrees of freedom. 
\end{enumerate}

This paper is organized as follows. In section \ref{sec:maxwell}, we review the symplectic structure of Maxwell theory, as well as its conformal primary solutions to the free equations of motion. 
In section \ref{sec:minkSoftModes}, we review the zero Milne energy sector of Maxwell theory consisting of conformally soft and Goldstone modes. We discuss their analytic continuations past the null cone through the origin, the matching conditions, as well as the decomposition of the conformally soft solutions in terms of the plane wave and conformal primary bases. We review the computation of inner products among confromal primary wavefunctions and the definition of conformal primary operators.  In section \ref{sec:beyond-Minkowski},  we study various aspects of the embedding of the Minkowski theory as a patch inside  the Einstein static universe, in preparation for the entanglement analysis in section \ref{sec:entanglement}. In section \ref{sec:invtrans}, we explain how conformal primary solutions can be extended to the entire Einstein static cylinder and identify a relationship between inversions of spin-one conformal primary wavefunctions and shadow wavefunctions. We further describe a new physical interpretation of the conformal primary field configurations as sourced by particles propagating outside the original Minkowski patch. In section \ref{sec:crepe}, we characterize the conformal primary sectors of Milne patches $\rMilne$ and $\lMilne$ of respectively the original Minkowski patch  and another Minkowski patch related to it by a conformal inversion on the cylinder. We then characterize the conformally soft and Goldstone wavefunctions associated with these subregions, and describe their transformation under inversions.  Section \ref{sec:entanglement} focuses on entanglement of edge modes across $\mathscr{I}^+.$ In section \ref{sec:pathIntegrals}, we show using the path integral that the Minkowski and Einstein static theories share the same vacuum state. In section \ref{sec:tfd}, we recast the Minkowski vacuum as a thermofield double state with respect to the $\rMilne$ and $\lMilne$ Milne patches. In section \ref{sec:apple}, we analyze the constraints associated with the Minkowski and Einstein static theories. We show that the edge modes identified by Donnelly and Wall in \cite{Donnelly:2014fua, Donnelly:2015hxa} are related to the logarithmic constituents of conformally soft modes by a gauge transformation. We finally reinterpret the edge mode entanglement entropy as due to fluctuations in the large gauge charges associated with the $\lMilne$ and $\rMilne$ Milne patches. 
We conclude with comments about CFT renormalization, discuss the emergence of spacetime entanglement from CCFT and explicitly relate the on-shell action of a subregion in Minkowski space to the soft effective action of \cite{Kapec:2021eug}.


\section{Maxwell theory and Minkowski spacetime $\mink$}
\label{sec:maxwell}


This preliminary section reviews some of the necessary background for Maxwell theory, including its phase space and associated large gauge charges. 
First, in \cref{sec:sympIntro}, we review the symplectic structure for Maxwell theory \cite{Lee:1990nz} in arbitrary Lorentzian spacetimes. Then, in the following subsections, we will review and
establish conventions for discussing Maxwell theory in four-dimensional
Minkowski spacetime $\mink$. In particular, we will review the construction of
conformal primary solutions for the Maxwell equations, that is, wavefunctions of
spin 1 and definite boost-weight $\confDim$ \cite{Pasterski:2017kqt}.
Distinguished among these are the pair of leading soft modes at $\confDim = 1$
which are canonically conjugate
\cite{Donnay:2018neh}.

\subsection{Symplectic structure of Maxwell theory}
\label{sec:sympIntro}

In this paper, we will examine the free Maxwell theory in four-dimensional\footnote{The discussion in this section is valid in any number of dimensions.} Lorentzian
spacetimes for which the bulk action is
\begin{align}
  \act[\blk]\relax
  &= \int \lag[\blk] \;,
  &
    \lag[\blk]
  &= -\frac{1}{2}F \wedge *F
    = -\frac{1}{4} \volForm^{(4)} F^{\mu\nu} F_{\mu\nu}
    \;,
    \label{eq:actAndLag}
\end{align}
where $\volForm^{(4)}$ is the spacetime volume form and $F=\frac12
F_{\mu\nu}\,dx^\mu\wedge dx^\nu$ is the field strength, which as usual is
related to the gauge potential $A$ by $ F = \diff A$.
If the spacetime has a boundary, one may need to supplement this action with
boundary terms.\footnote{For example, see \cite{Marolf:2006nd} for a discussion
  in asymptotically AdS spacetimes.} Let us use $\confSpace$ to denote the space
of field configurations and we will use $\fieldDiff$ to indicate the exterior
derivative on $\confSpace$. Then, applying
the differential $\fieldDiff$ to the bulk Lagrangian, we obtain the equations of
motion and the spacetime differential of the bulk pre-symplectic potential
density \cite{Lee:1990nz,Wald:1993nt,Iyer:1994ys}:
\begin{align}
  \fieldDiff \lag[\blk]
  &= \eom \cdot \fieldDiff A + \diff \sympPotD[][\blk] \;,
    \label{eq:fieldDiffLag}
  \\
  \eom \cdot \fieldDiff A
  &= -\fieldDiff A \wedge \diff * F \;, \qquad
  \sympPotD[][\blk]
  = -\fieldDiff A \wedge * F.
    \label{eq:sympPotDBlk}
\end{align}
In particular, the Maxwell equations of motion can be expressed as
\begin{align}
  & \eom=0
  &
  &\iff
  &
  & \diff * F=0
  &
  &\iff
  &
  & \covD \cdot F=0 \;.
    \label{eq:vacuum-ME}
\end{align}

Gauge symmetries of the theory are shifts of the gauge potential by an exact
differential,
\begin{align}
  A &\mapsto A + \diff \gaugeParam \;,
      \label{eq:AGaugeTrans}
\end{align}
for arbitrary gauge parameters $\gaugeParam$. If Cauchy surfaces have a
nontrivial boundary, those $\gaugeParam$ which vanish at the boundary
parametrize redundancies of the theory. The treatment of gauge transformations
which do not vanish on the boundary, \eg\ at spatial infinity $\spaceInfty$,
however, is more subtle and will be discussed below in
\cref{sec:asymSym}.

Although Maxwell theory will be our primary focus in this paper, we will review
its symplectic structure and the construction of its physical phase space in a
general manner that is easily applied to other field theories
\cite{Lee:1990nz,Wald:1993nt,Iyer:1994ys}. For instance, through integration by
parts, note that the variation of a field theory Lagrangian can generally be
expressed in the form of \cref{eq:fieldDiffLag}, from which one can read off the
theory's bulk pre-symplectic density $\sympPotD[][\blk]$. From
$\sympPotD[][\blk]$, one can construct the bulk symplectic density by taking a
field space differential:
\begin{align}
  \sympD[][\blk]
  &= \fieldDiff \sympPotD[][\blk]
    =
    \fieldDiff A \wedge \hodge \fieldDiff F \;.
    \label{eq:sympD}
\end{align}
Here and elsewhere in this paper, the wedge product of field space one-forms
resulting from the application of the exterior derivative $\delta$ is left
implicit.\footnote{To be precise, given two $\confSpace$ one-forms $a$ and $b$,
  we write $a b = a\fieldWedge b = a \fieldTProd b - b \fieldTProd a$ where
  $\fieldTProd$ and $\fieldWedge$ are respectively the tensor and wedge products
  in $\confSpace$. (This is assuming a bosonic theory or that at least one of
  $a$ and $b$ is non-Grassmann. For Grassmann $a$ and $b$, we instead have
  $a\fieldWedge b = a \fieldTProd b + b \fieldTProd a$.) \label{foot:wedge}}
Integrating $\sympPotD[][\blk]$ and $\sympD[][\blk]$ on a Cauchy surface
$\cauchy$, one obtains the bulk contributions to the pre-symplectic potential
and the corresponding pre-symplectic form:
\begin{align}
  \sympPot[\blk]{\confSpace}
  &= \int_{\cauchy} \sympPotD[][\blk]
    \;,
  &
    \symp[\blk]{\confSpace}
  &= \int_{\cauchy}\sympD[][\blk].
\end{align}
If $\partial\cauchy$ is nontrivial, one may need to supplement these with
boundary terms \cite{Donnelly:2016auv}
\begin{align}
  \sympPot{\confSpace} &= \sympPot[\blk]{\confSpace} + \sympPot[\bdy]{\confSpace} \;,
  &
    \symp{\confSpace} &= \symp[\blk]{\confSpace} + \symp[\bdy]{\confSpace} \;.
\end{align}
We will motivate why one may want to consider such boundary contributions below
in \cref{sec:asymSym} and describe these contributions in detail for the Milne
context in \cref{sec:apple}. As explained in \cref{sec:asymSym}, for Minkowski
spacetime, however, we will treat certain gauge transformations which are
nonvanishing at $\partial\cauchy=\spaceInfty$ not as redundancies but rather as
asymptotic symmetry transformations which are physical.
In this approach, which is standard in the literature of celestial
holography, the boundary contributions to $\sympPot{\confSpace}$ and
$\symp{\confSpace}$ vanish and they are left with only bulk contributions in
Minkowski spacetime.

\begin{figure}
  \centering
  \includegraphics[scale=1.2]{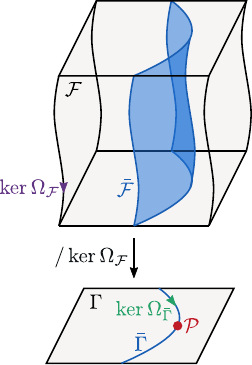}
  \caption{The relationships between the space of field configurations
    \(\confSpace\), the space of solutions \(\solnSpace\), the extended phase
    space \(\extPhSpace\), the constraint surface \(\cstrSpace\), and the
    physical phase space \(\phSpace\). Our construction here mirrors that of
    \cite{Lee:1990nz} --- \cf{} figure 1 therein.}
  \label{fig:fieldSpace}
\end{figure}

In any event, $\symp{\confSpace}$ defines the space $\extPhSpace$ of initial
data through a quotient of $\confSpace$ by the kernel of $\symp{\confSpace}$:
\begin{align}
  \extPhSpace &= \confSpace / \ker(\symp{\confSpace})\;.
                \label{eq:extPhSpace}
\end{align}
That is, an $\confSpace$-space vector $U$ preserves the initial data on
$\cauchy$ if and only if the $\confSpace$-space interior product
$\fieldIProd{U}\symp{\confSpace}$ vanishes. Let us use $\sameInit{\cauchy}$ to
relate two configurations which share the same initial data on the Cauchy
surface $\cauchy$. Then $\extPhSpace$ is the space of equivalence classes under
$\sameInit{\cauchy}$. In particular, we see from \cref{eq:sympD} that, for two
configurations $A$ and $A'$ of the gauge potential,
\begin{align}
  A &\sameInit{\cauchy} A'
  &
  &\implies
  &
    \pullback[\cauchy] A
    = \pullback[\cauchy] A'
  &\text{ and }
    \pullback[\cauchy] * F
    = \pullback[\cauchy] *F' \;,
    \label{eq:sameInit}
\end{align}
where $\pullback$ denotes spacetime pullback. If $\symp[\bdy]{\confSpace}$ is
trivial, then the converse is also true. In \cref{fig:fieldSpace}, we illustrate
the relationships between \(\confSpace\), \(\extPhSpace\), and other
spaces to be discussed below.

The symplectic form $\symp{\extPhSpace}$ on $\extPhSpace$ naturally descends
from $\symp{\confSpace}$ on $\confSpace$ --- to be precise, it is the form
$\symp{\extPhSpace}$ on $\extPhSpace$ such that
\begin{align}
  \symp{\confSpace}
  &= \fieldPullback[\confSpace] \symp{\extPhSpace}
    \;,
    \label{eq:sympConfSpace}
\end{align}
where $\fieldPullback$ denotes field space pullback under the
$\confSpace\to\extPhSpace$ projection \labelcref{eq:extPhSpace}. If we begin
with Cartan's formula
\begin{align}
  \fieldLie{\bullet}
  &= \fieldDiff \fieldIProd{\bullet}
    + \fieldIProd{\bullet} \fieldDiff \;,
    \label{eq:fieldCartan}
\end{align}
where $\fieldLie{\bullet}$ and $\fieldIProd{\bullet}$ are the Lie derivative and
interior product acting on forms on the field space $\confSpace$, then we have
\begin{align}
  \fieldLie{U} \symp{\confSpace}
  &= \fieldDiff \fieldIProd{U} \symp{\confSpace} \;.
\end{align}
Hence $\symp{\confSpace}$ is stationary along its null directions, \ie the
directions noted above which satisfy $\fieldIProd{U} \symp{\confSpace} =0$.
Thus, projecting out $\ker(\symp{\confSpace})$ leads to a well-defined
$\symp{\extPhSpace}$. Moreover, the stationarity of $\symp{\confSpace}$ further
implies that one can construct a symplectic potential $\sympPot{\extPhSpace}$ on
$\extPhSpace$ for which
\begin{align}
  \symp{\extPhSpace}
  &=
    \fieldDiff \sympPot{\extPhSpace} \;,
    \label{eq:extPhSpaceSympSympPot}
\end{align}
because the difference between $\sympPot{\confSpace}$ and its pullback by any
flow produced by null vectors of $\symp{\confSpace}$ is closed. In fact, in
Maxwell theory, $\fieldLie{U}\sympPot{\confSpace}=0$ for any
$U\in\ker(\symp{\confSpace})$ so one may simply define $\sympPot{\extPhSpace}$
by $\sympPot{\confSpace}=\fieldPullback[\confSpace]\sympPot{\extPhSpace}$. By
the construction of $\extPhSpace$, the corresponding symplectic form
$\symp{\extPhSpace}$ is nondegenerate and may be inverted to define a Poisson
bracket. For functionals $a$ and $b$ which are $\extPhSpace$-space
scalars\footnote{Note that we have taken the opposite but more standard sign
  convention relative to (4.14) in \cite{Lee:1990nz} --- our choice ensures the
  expected positive sign for the Poisson bracket between a field and its
  conjugate momentum.},
\begin{align}
  \symp{\extPhSpace}^{-1}(\fieldDiff a,\fieldDiff b)
  &= \poisson{a}{b}
    = -i\, \commut{a}{b} \;.
    \label{eq:poissonBracket}
\end{align}
As indicated in the last equality, the theory is quantized by promoting the
Poisson bracket to commutation relations\footnote{Of course, fermionic fields
  are Grassmann. Note that $\symp{\extPhSpace}(\bullet,\bullet)$ is symmetric
  when both of its arguments are Grassmann. Likewise,
  $\symp{\extPhSpace}^{-1}(\fieldDiff a,\fieldDiff b)$ is symmetric for
  Grassmann variables $a$ and $b$; in this case, one must replace the commutator
  in \cref{eq:poissonBracket} with an anti-commutator.}.

Of course, not all field configurations satisfy the equations of motion
\labelcref{eq:vacuum-ME} and not all initial data are physically valid. Let us
denote by $\solnSpace$ the submanifold of $\confSpace$ corresponding to
solutions of the equations of motion. The pullback
$\fieldPullback[\solnSpace]\sympD[][\blk]$ of $\sympD[][\blk]$ to $\solnSpace$
is spacetime closed, so that, given appropriate boundary conditions at the
boundaries of Cauchy slices, the pre-symplectic form $\symp{\solnSpace}$ on
$\solnSpace$,
\begin{align}
  \symp{\solnSpace}
  &= \fieldPullback[\solnSpace] \symp{\confSpace} \;,
    \label{eq:horse}
\end{align}
is in fact independent of the choice of Cauchy surface $\cauchy$.

Analogously, let us denote by $\cstrSpace$ the space of physically realizable
initial data, given by the image of $\solnSpace$ under the
$\confSpace\to\extPhSpace$ projection \labelcref{eq:extPhSpace}. To specify
$\cstrSpace$ as a submanifold of $\extPhSpace$, one may consider constraint
functionals $\cstr$ on $\extPhSpace$ that vanish on $\cstrSpace$:
\begin{align}
  \fieldPullback[\cstrSpace]
  \cstr
  &= 0.
    \label{eq:cstrSpace}
\end{align}
For gauge theories, these include the charges which generate gauge
transformations that are redundancies of the theory.\footnote{To be specific,
  these are the first class constraints, which generate flows that preserve
  $\cstrSpace$ and are null directions of $\symp{\cstrSpace}$ given below in
  \cref{eq:sympCstrSpace}. As described around there, $\ker\symp{\cstrSpace}$
  are precisely the gauge redundant directions. There may also be second class
  constraints which generate flows that do not preserve $\cstrSpace$; imposing
  these does not result in new null directions for $\symp{\cstrSpace}$.} To be
precise, these charges $\cstr\relax[\gaugeParam]$ are the Hamiltonian generators
for vector fields $\fieldVecGauge[\gaugeParam]$ associated to infinitesimal
transformations parametrized by $\gaugeParam$ along gauge redundancies:
\begin{align}
  \fieldIProd{\fieldVecGauge[\gaugeParam]} \symp{\extPhSpace}
  &= \fieldDiff \cstr\relax[\gaugeParam] \;.
\end{align}

Let us now take a moment to discuss the Hamiltonian generators for arbitrary
symmetries, gauge or not. To begin, we note that a flow with vector field $U$ is
Hamiltonian (\ie{} there exists some scalar functional $H$ on $\extPhSpace$
which generates a flow along $U$:\footnote{Here, we are taking the same sign
  convention as \cite{Donnelly:2016auv} for the vector field generated by a
  Hamiltonian. For a classical particle, the energy $E$ and momentum $p$
  respectively generate vector fields in phase space which act on the position
  variable $x$ as $U_E(x)=-\dot{x}$ and $U_p(x)=-1$. While these may appear to shift
  $x$ backwards in time or space, note that it corresponds to shifting a
  prepared state to a later time or larger $x$.} $\fieldIProd{\fieldVecGauge}
\symp{\extPhSpace} = \fieldDiff H$) if and only if
\begin{align}
  \fieldLie{U} \symp{\extPhSpace}
  &= 0 \;,
    \label{eq:hamVecField}
\end{align}
by Cartan's formula \labelcref{eq:fieldCartan}. \Cref{eq:hamVecField} is a
natural requirement for a symmetry and is obviously crucial if the flow $U$ is
meant to be a gauge redundancy of the theory. Given \cref{eq:hamVecField}, one
can also make $\sympPot{\extPhSpace}$ stationary:
\begin{align}
  \fieldLie{U} (
  \sympPot{\extPhSpace}
  + \Xi_U
  )
  &= 0 \;,
    \label{eq:gaugeInvariantSympPot}
\end{align}
where we have included the possibility of shifting $\sympPot{\extPhSpace}$ by
closed one-form $\Xi_U$ on $\extPhSpace$. It then follows by
\cref{eq:fieldCartan} that
\begin{align}
  \fieldIProd{U} \symp{\extPhSpace}
  &= -\fieldDiff \fieldIProd{U} (\sympPot{\extPhSpace} + \Xi_U) \;,
\end{align}
and so we may take
\begin{align}
  H
  &= -\fieldIProd{U} ( \sympPot{\extPhSpace} + \Xi_U )\;.
    \label{eq:constructingHamGenerator}
\end{align}
This general discussion doesn't tell us how to construct $\Xi_U$. We note
however that given the standard construction of Noether currents and charges
\cite{Lee:1990nz,Iyer:1994ys}, one finds that $\fieldIProd{U}\Xi_U$ includes the
total derivative by which the bulk Lagrangian varies under the symmetry $U$. Of
course, this is integrated over $\cauchy$ to produce a boundary term but there
may be other contributions at $\partial\cauchy$ as well.

For the proper gauge transformations of Maxwell theory though,
$\sympPot{\extPhSpace}$ by itself is already stationary\footnote{For
  field-independent gauge transformations in Maxwell theory, as considered in
  this paper, the $\sympPot[\blk]{\extPhSpace}$ naturally descending from
  $\sympPot[\blk]{\confSpace}$ is stationary
  $\fieldLie{\fieldVecGauge[\gaugeParam]}\sympPot[\blk]{\extPhSpace}=0$ even for
  $\gaugeParam$ nonvanishing at the boundary $\partial\cauchy$. As shown in
  \cite{Donnelly:2016auv}, for field-dependent gauge transformations, it is
  necessary to introduce boundary degrees of freedom as well as a boundary
  contribution $\sympPot[\bdy]{\confSpace}$ for \cref{eq:gaugeInvariantSympPot}
  to be satisfied.} under $\fieldVecGauge[\gaugeParam]$ and so
$\Xi_{\fieldVecGauge[\gaugeParam]}=0$. Considering for now just
$\sympPot[\blk]{\extPhSpace}$ descending from the bulk part
$\sympPot[\blk]{\confSpace}$, we then find
\begin{align}
  \cstr[][\blk]\relax[\gaugeParam]
  &= \int_{\cauchy} \diff\gaugeParam \wedge \hodge F \;
    \label{eq:cstrBlk}
\end{align}
for the gauge transformations \labelcref{eq:AGaugeTrans}. For $\gaugeParam$ which
vanish at $\partial\cauchy$, the above clearly vanishes on-shell upon applying
Stokes' theorem. We will discuss how to treat $\gaugeParam$ that do not vanish
at $\partial\cauchy$ below in \cref{sec:asymSym} and further in the Milne
context in \cref{sec:apple}.

While one can pull back the symplectic potential $\sympPot{\extPhSpace}$ and
symplectic form $\symp{\extPhSpace}$ to obtain
\begin{align}
  \sympPot{\cstrSpace}
  &=\fieldPullback[\cstrSpace] \sympPot{\extPhSpace}
    \;,
  &
    \symp{\cstrSpace}
  &=\fieldPullback[\cstrSpace] \symp{\extPhSpace}
    \label{eq:sympCstrSpace}
\end{align}
on the constraint surface $\cstrSpace$, the resulting form $\symp{\cstrSpace}$
will be degenerate in gauge theories like Maxwell theory. Indeed, these
degeneracies $\ker\symp{\cstrSpace}$ are precisely the gauge redundancies
generated by $\cstr\relax[\gaugeParam]$. To obtain the physical phase space
$\phSpace$, we must divide out by these redundancies:
\begin{align}
  \phSpace
  &= \cstrSpace / \ker{\symp{\cstrSpace}}
    \;.
    \label{eq:phSpace}
\end{align}
For the theory to have a good initial value formulation, the
physical phase space $\phSpace$ must be bijective to the space of solutions
$\solnSpace$ modulo gauge redundancies.
The overall operation going from $\extPhSpace$ to $\phSpace$, \ie{} setting
constraints $\cstr$ to zero and dividing out by any resulting degeneracies in the
pullback of the symplectic form, is known as the double quotient
\cite{Donnelly:2016auv}:
\begin{align}
  \phSpace
  &= \extPhSpace\doubleQuotient \{\cstr\}\;.
    \label{eq:doubleQuotient}
\end{align}
Just as $\sympPot{\extPhSpace}$ and $\symp{\extPhSpace}$ descend from
$\sympPot{\confSpace}$ and $\symp{\confSpace}$, a symplectic potential
$\sympPot{\phSpace}$ and non-degenerate symplectic form $\symp{\phSpace}$ on the
physical phase space $\phSpace$ descend from $\sympPot{\cstrSpace}$ and
$\symp{\cstrSpace}$:
\begin{align}
  \symp{\cstrSpace}
  &= \fieldPullback[\cstrSpace] \symp{\phSpace}
    \;,
  &
    \symp{\phSpace}
  &=
    \fieldDiff \sympPot{\phSpace}
    \;.
\end{align}
In fact, analogous to the comment made below \cref{eq:extPhSpaceSympSympPot},
because $\sympPot{\extPhSpace}$ and hence $\sympPot{\cstrSpace}$ are stationary
under gauge transformations, one may simply define $\sympPot{\phSpace}$ by
$\sympPot{\cstrSpace} = \fieldPullback[\cstrSpace]\sympPot{\phSpace}$.
Altogether, the double quotient \cref{eq:doubleQuotient} therefore takes a
symplectic space $\extPhSpace$ to a symplectic space $\phSpace$.

As a final remark, we note that, upon quantization, the double quotient
operation on Hilbert spaces is actually quite simple to describe \cite{Donnelly:2016auv}. In particular,
let us consider a Hilbert space $\hilbSpace(\extPhSpace)$ corresponding to a
classical symplectic space $\extPhSpace$, and constraints $\cstr$ that are
promoted to operators acting on $\hilbSpace(\extPhSpace)$. Then, the Hilbert
space $\hilbSpace(\phSpace)$ corresponding to $\phSpace$ given by the double
quotient \labelcref{eq:doubleQuotient} is simply the kernel
\begin{align}
  \hilbSpace(\phSpace)
  &= \ker \cstr
    \label{eq:hilbSpaceCstr}
\end{align}
of the operator $\cstr$. Classically, apart from setting the constraints $\cstr$ to zero, one
must also divide out by any flow they generate in $\cstrSpace$. However, this
second step is automatically accomplished at the quantum level, because any state which is a zero eigenstate of a
constraint $\cstr$ is also invariant along the flow generated by $\cstr$.

\subsubsection{Asymptotic symmetries}
\label{sec:asymSym}

There remains a question of how one should treat gauge transformations for which
the gauge parameter $\gaugeParam$ does not vanish at the boundary
$\partial\cauchy$. For Maxwell theory in particular, the quantity in
\cref{eq:cstrBlk} no longer vanishes on-shell:
\begin{align}
  \cstr[][\blk]\relax[\gaugeParam]
  &\eqOnShell \int_{\partial\cauchy} \gaugeParam \hodge F \;,
    \label{eq:cstrBlkOnShell}
\end{align}
where we have introduced $\eqOnShell$ to mean equality on-shell.

One approach \cite{Donnelly:2016auv} is to treat gauge transformations that are
nonvanishing at $\partial\cauchy$ as redundancies, in which case we must have
$\cstr\relax[\gaugeParam]\eqOnShell 0$, in contrast to
\cref{eq:cstrBlkOnShell}.\footnote{Here, we are considering the classical theory
  and this is a constraint on the classical solutions. Of course, this
  discussion of constraints can also be lifted to the corresponding quantum
  theory.} One remedy would then be to introduce extra boundary degrees of
freedom and a new boundary contribution $\sympPot[\bdy]{\confSpace}$ to the
potential, 
so that the corresponding boundary contribution $\cstr[][\bdy]$ cancels against
$\cstr[][\blk]$ for on-shell configurations in the sum
\begin{align}
  \cstr\relax[\gaugeParam]
  &= \cstr[][\blk]\relax[\gaugeParam] + \cstr[][\bdy]\relax[\gaugeParam]
    \eqOnShell 0\;.
\end{align}
Although the gauge transformations which are nonvanishing at $\partial\cauchy$
are still viewed as redundancies, the introduction of boundary degrees of
freedom leads to the appearance of new `boundary symmetries' which are physical.
That is, the action of $\cstr[][\blk]\relax[\gaugeParam]$ alone becomes a
physical transformation of the gauge fields. 
In
\cref{sec:apple}, we review this approach of \cite{Donnelly:2016auv} for the
Milne patches that appear in Minkowski space.

Another approach is to treat gauge transformations that are nonvanishing at
$\partial\Sigma$ not as redundancies, but physical symmetries, \eg{} see
\cite{Strominger:2017zoo}. More precisely, one defines these asymptotic
symmetries as
\begin{align}
  \text{asymptotic symmetries}
  &= \frac{\text{gauge symmetries}}{\text{trivial gauge}}\;,
    \label{eq:asymSymStrominger}
\end{align}
where trivial gauge transformations are those satisfying
$\cstr\relax[\gaugeParam]\eqOnShell 0$ and where $\sympPot{\confSpace}$ and
$\cstr$ are simply given by the expected bulk contributions
$\sympPot[\blk]{\confSpace}$ and $\cstr[][\blk]$ for Maxwell theory in Minkowski
spacetime. While trivial gauge transformations are viewed as redundancies,
asymptotic symmetries are viewed as physical symmetries, so it is acceptable for
their $\cstr\relax[\gaugeParam]\neqOnShell 0$. (Here, we continue to use the
same expression to define $\cstr\relax[\gaugeParam]$ even though these are not
constraints in the case of asymptotic symmetries.) Indeed,
$\cstr\relax[\gaugeParam]$ are precisely the charges associated to these
asymptotic symmetries.

Note that, to have a good initial value formulation, it is necessary to have
certain boundary conditions in place near spatial infinity to prevent these
symmetries from being `local' in time, \eg{}, their action should be the same
for different Cauchy slices. For this, it suffices for the boundary conditions
to constrain field configuration space such that $\symp{\solnSpace}$ on the
space $\solnSpace$ of on-shell solutions is independent of the choice of Cauchy
slice $\cauchy$, as indicated below \eqref{eq:horse}. In the Minkowski case,
where we will often be interested in $\cauchy=\nullInfty[\pm]$, these boundary
conditions give rise to certain antipodal matching conditions
\cite{Kapec:2015ena,He:2014cra} near $\spaceInfty$, as we review in
\cref{sec:minkSoftModes}.

Since this second approach of viewing nontrivial gauge transformations at
$\partial\cauchy$ as physical asymptotic symmetries is more common in the celestial
holography literature, we will follow this approach throughout the following in
considering Minkowski space. However, it should be noted that the two approaches
described above are equivalent. In particular, the boundary symmetries of
\cite{Donnelly:2016auv}, upon fixing the gauge redundancies at the boundary,
become the asymptotic symmetries of the second approach.

\subsubsection{Free fields, vacua, and inner products}
\label{sec:introFreeFields}

In the remainder of this subsection, we provide a review of the quantization of
free field theories and their vacua \cite{Ashtekar:1975zn}. This will also
motivate the definition of an `inner' product between modes, which is used
frequently in the celestial holography literature
\cite{Pasterski:2017kqt,Donnay:2018neh}.

For fields whose target space is some general vector space (as is the case for
Maxwell theory), the field configuration space $\confSpace$ itself
inherits a vector space structure from the target space in the obvious
pointwise-manner. In particular, $\confSpace$ is its own tangent space. A key
property of free theories is the further flatness of the solution space
$\solnSpace$ and thus of phase space $\phSpace$. In particular, any solution to
the equations of motion can be obtained by taking a linear combination of some
basis of wavefunctions which are themselves solutions and so, $\solnSpace$ also
is its own tangent space. It follows that the same must also be true of
$\phSpace$ (which may be identified with $\solnSpace$ modulo gauge redundancies,
as remarked around \cref{eq:phSpace}), provided the gauge redundancies generate
a vector subspace in $\solnSpace$.

It is helpful to introduce some notation to describe free theories further. Let
us use the symbol $\allFields$ to denote all fields collectively in the free
field theory. We may establish coordinates on the phase space $\phSpace$ by
writing
\begin{align}
  \allFields &=  b^A\,V_A
               \label{eq:generalModeExpansion}
\end{align}
where $V_A$ are a basis wavefunctions and the coefficients $b^A$ can be viewed
as coordinates on $\phSpace$. Of course, the $V_A$ also have a second meaning as
the field space vectors producing variations, \eg{}
$\fieldIProd{V_A}\fieldDiff\allFields=\partial_A\allFields=V_A$, concretely
realizing the identification between $\phSpace$ and its tangent space.

While we have been referring only to the classical phase space and classical
fields, the above comments readily lift to the quantum theory.
Eq.~(\ref{eq:generalModeExpansion}) then describes the full set of operators for
the quantum Hilbert space. The wavefunctions $V_A$ remain c-numbers, \ie{}
solutions of the classical equations, while the coefficients $b^A$ become the
annihilation and creation operators or soft operators on the Hilbert space of
the free theory.

Keeping in mind the notion that the $V_A$ are tangent vectors on the phase
space, we may consider objects such as $\symp{\phSpace}(V_A,\allFields)$ which
provide an alternative set of operators in the quantum theory. The commutators
of these (bosonic) operators are readily evaluated using
\cref{eq:poissonBracket,eq:generalModeExpansion}:
\begin{align}
  \commut{\allFields}{\symp{\phSpace}(V_A,\allFields)}
  &= V_B \commut{b^B}{b^C} \symp{\phSpace}(V_A,V_C)
    = -i V_A
    \label{eq:fieldShiftByStarAlgebra}
  \\
  \commut{\symp{\phSpace}(V_A,\allFields)}{\symp{\phSpace}(V_B,\allFields)}
  &= -i\symp{\phSpace}(V_A,V_B) \;.
    \label{eq:starAlgebra}
\end{align}
These $\symp{\phSpace}(V_A,\allFields)$ operators form the so-called $*$-algebra
appearing in \cite{Ashtekar:1975zn},\footnote{Our conventions include an extra
  sign in \cref{eq:starAlgebra} relative to eq.~(3) in \cite{Ashtekar:1975zn}.}
whose commutators are directly read off from the symplectic form. 

It remains to construct the Hilbert space of states which give a representation
to the operators $b^A$ or alternatively $\symp{\phSpace}(V_A,\allFields)$. In
general, $V_A$ might include so-called `soft modes', which
typically include zero-energy modes such as those corresponding to asymptotic
symmetries.\footnote{However, there may also be `soft' modes which are not soft
  in the conventional sense. An example are the conformally soft modes
  $\confPrimA{\cs}$ defined in \cref{eq:CS} which, as we show in
  \cref{eq:csToV}, have a decomposition including plane waves of arbitrarily
  high frequencies. However, as seen in \cref{eq:minkProdAcsNonsoft}, in the
  $\reg\to 0$ limit, its symplectic products give a vanishing distribution over
  the non-soft conformal primary modes on the principal series. Their only
  nontrivial symplectic product is \cref{eq:softMinkProd} with the Goldstone
  modes $\confPrimA{\gold}$ describing asymptotic symmetries.} However, the
representation for these modes does not follow the standard Fock space
construction, so let us set these to the side for the moment.  We will assume that
$\symp{\phSpace}$ has a block decomposition between the soft modes and the
remaining (\ie{} non-soft) modes. The Fock space construction for the non-soft modes then follows upon equipping the phase space with a complex structure which, upon quantization, allows for the Hilbert space to be decomposed into positive and negative energy components \cite{Ashtekar:1975zn}. This is reviewed in appendix \ref{app:Fock}.

Finally, it is helpful to define the product
\begin{align}
  \solnProd{V}{V'}
  &= i\, \symp{\confSpace}(V,V'^*) \;,
    \label{eq:sympProd}
\end{align}
between field configurations. This product is manifestly linear and anti-linear
respectively in its first and second arguments, and moreover satisfies the
identities
\begin{align}
  \solnProd{V'}{V}
  &= \solnProd{V}{V'}^*
    = -\solnProd{V^*}{V'^*} \;.
\end{align}
The product
\cref{eq:sympProd} is colloquially called an `inner' product between modes,
\eg{} in celestial holography literature
\cite{Pasterski:2017kqt,Donnay:2018neh}, though really it is only positive
definite between the holomorphic modes $U_M$, as we review in appendix \ref{app:Fock}. In the absence of boundary
contributions, it gives, for the free scalar, the standard Klein-Gordon inner product
\begin{align}
  \kgProd{\blkScalar}{\blkScalar'}
  &= i \int_{\cauchy} \left(
    \blkScalar \hodge \diff \blkScalar'^*
    - \blkScalar'^* \hodge \diff \blkScalar
    \right)
    \;,
\end{align}
while, for free Maxwell theory, one finds
\begin{align}
  \solnProd{A}{A'}
  &= i \int_{\cauchy} \left(
    A \wedge \hodge F'^*
    -
    A'^* \wedge \hodge F
    \right) \;.
\end{align}

\subsection{Minkowski spacetime, celestial space, and null momenta}
\label{sec:minkConventions}
Of primary interest in this paper is four-dimensional Minkowski spacetime
$\mink$, points in which we will denote by $\minkX$. We will denote Minkowski
Cartesian coordinates $\minkX^\mu$ with indices $\mu,\nu,\ldots$ ranging from
$0,\ldots,3$. Also of interest to us is a celestial space. While this is commonly referred to
as the celestial sphere in the celestial holography literature, we will find
it most convenient to consider a conformal frame where the celestial space is a
plane $\celPlane$ --- we discuss the generalization to other conformal frames in section \ref{sec:entanglementSoftDOFs}.

\begin{figure}
  \begin{subfigure}{\textwidth}
      \centering \includegraphics[scale=1.2]{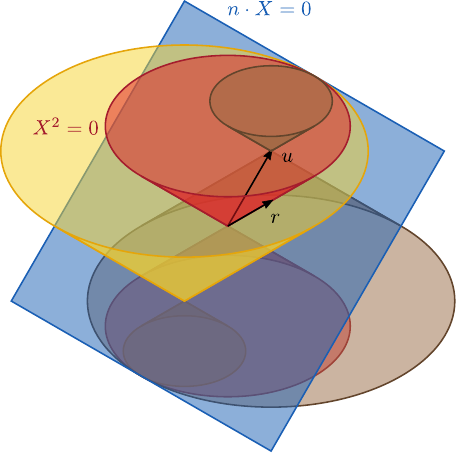}
      \caption{Foliation of Minkowski spacetime by lightcones. One spherical direction is suppressed.}
      \label{fig:planarRetardedCoordinates3D}
  \end{subfigure}
  \par\bigskip
  \begin{subfigure}{\textwidth}
      \centering \includegraphics[scale=1.2]{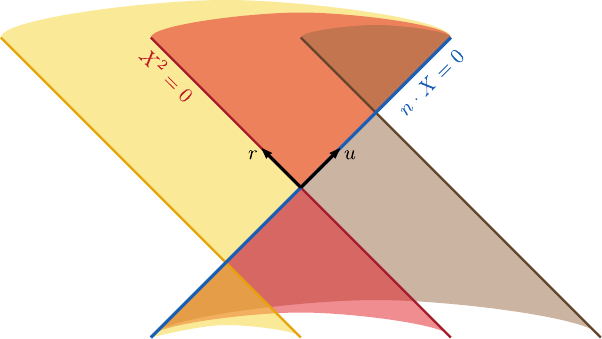}
      \caption{A side view of (\subref{fig:planarRetardedCoordinates3D}) cut in half.}
  \end{subfigure}
  \caption{The planar retarded coordinates defined in \cref{eq:cupcake-intro} describe a
    foliation of
  Minkowski spacetime by lightcones. The shifts $\minku \refVec$ of the lightcone
  centers from the origin $\minkX=0$ are parameterized by $\minku$;
  on a given lightcone,
  the radial direction $\minkr \nullUnit(\minkz)$ is parameterized by $\minkr$
  and the spherical directions are parametrized by $\minkz$. (The future and past components of a
  lightcone are described by $r>0$ and $r<0$ respectively.) The $\minkX^2=0$,
  \ie{} $\minku=0$, lightcone is shown in red. The blue
  $\minkr=-\refVec\cdot\minkX=0$ plane may be viewed as the
  $\minku\to -\infty$ ($\minku\to+\infty$) limit of a future (past) lightcone.}
  \label{fig:planarRetardedCoordinates}
\end{figure}

The celestial space can be viewed as living on the $\minkX^2=0$ section of
$\nullInfty[\pm]$. To make this precise, we introduce a set of planar retarded
coordinates $(\minku,\minkr,\minkzh,\minkza)$ for Minkowski spacetime, by
writing
\begin{align}
  \minkX^{\mu} &= \minku\, \refVec^\mu + \minkr \nullUnit^{\mu}(\minkz) \;,
                 \label{eq:cupcake-intro}
\end{align}
where $u = t - r$ is the retarded time,  $\nullUnit$ is a null vector\cite{Pasterski:2017kqt}\footnote{Our
  definition \labelcref{eq:null-vector} for $\nullUnit$ contains an extra factor
  of $1/2$ relative to many other literature --- see \eg{} (2.9) in
  \cite{Pasterski:2017kqt}, (2.4) in \cite{Donnay:2018neh}, and (2.1) in
  \cite{Pasterski:2020pdk}. We have made this choice to obtain the standard
  normalization \labelcref{eq:yogurt} for the planar metric in complex
  coordinates $(z,\bar{z})$. \label{foot:normq} } towards the point $\minkz$ on the celestial space at
$\nullInfty[+]$
\begin{align}
  \nullUnit^{\mu}(\minkz)
  &= \frac{1}{2}\left(
    1 + \minkzh\minkza, \minkzh + \minkza, -i(\minkzh - \minkza), 1 - \minkzh\minkza
    \right) \;,
    \label{eq:null-vector}
\end{align}
and $\refVec$ is given by
\begin{align}
  \refVec^\mu
  &= \frac{1}{2}\celLap \nullUnit^\mu
    = (1,0,0,-1)\;.
    \label{eq:apple}
\end{align}
(We treat $\nullUnit$ as a spacetime vector and celestial scalar.) Here $\Box^{(2)} = D\cdot D = 2 \p_z \p_{\bz}$ is the celestial Laplacian. Minkowski
spacetime is covered by retarded times $\minku\in\reals$, radii
$\minkr\in\reals$, and complex coordinates $(\minkzh,\minkza)$ parametrizing
celestial space $\celPlane$. Surfaces of constant $\minku$, for $\minkr>0$
($\minkr<0$) are future (past) lightcones emanating from the points $\minkX^\mu
= \minku \refVec^\mu$ --- see \cref{fig:planarRetardedCoordinates}; surfaces of
constant $\minkr = -\refVec\cdot\minkX$ are null planes. The Minkowski spacetime
regions covered by various signs of $\minku$ and $\minkr$ are summarized in
\cref{tab:rice}. In planar retarded coordinates \eqref{eq:cupcake-intro}, the Minkowski metric becomes
\begin{equation}
  \label{eq:planar-Mink-intro} \diff s_{\mink}^2 = -2 \diff\minku \diff\minkr +
  \minkr^2 \celMet_{\alpha \beta} \diff \minkz[^\alpha] \diff \minkz[^\beta] \,,
\end{equation}
where $\celMet_{z\bz} = \frac{1}{2}$ is the planar celestial
metric.\footnote{One may consider extending the calculations of this paper to
  more general metrics obtained by Weyl-rescaling $\celMet \to \Omega^2 \celMet$
  with an arbitrary function $\Omega(\celw)>0$. The spacetime description of
  this will be a Penrose-Brown-Henneaux (PBH) diffeomorphism
  \cite{Imbimbo:1999bj} along (A)dS slices foliating the Minkowski spacetime; in
  particular, on the lightcone $\minkX^2=0$, this reduces simply to rescaling
  $\nullUnit\to\Omega\nullUnit$ with $u=0$ fixed in \cref{eq:cupcake-intro}. In
  general, an arbitrary PBH transformation will render the analogues of
  \cref{eq:cupcake-intro,eq:planar-Mink-intro} rather complicated, but a simpler
  transformation, \eg{} to obtain the celestial \emph{sphere}, may be tractable.
  Some new considerations that appear include: the fact that, under Weyl
  rescalings, the conformal primary wavefunctions ought to renormalize as
  $\confPrimA{\confDim}[a] \to \Omega^{-\confDim+1} \confPrimA{\confDim}[a]$
  (more precisely, given a conformal primary operator $\confPrimOp{\confDim}[a]
  = i\minkProd{A}{\confPrimA{\confDim^*}[\bar{a}]}$ in a celestial CFT with
  metric $\celMet$, the analogous operator in the celestial CFT with metric
  $\weylFac\celMet$ should be $\weylFac^{-\confDim+1}\confPrimOp{\confDim}[a]$);
  and certain calculations will be corrected by celestial curvature
  terms. \label{foot:celestialHoloConfTrans}} Null infinity $\nullInfty[\pm]$ is
reached with $\minkr\to\pm\infty$ and parametrized by the full range of
$(\minku,\minkzh,\minkza)$.

\begin{table}
  \centering
  \begin{tabularx}{\textwidth}{|c|X|X|}
    \cline{2-3}
    \multicolumn{1}{c|}{}
    & \multicolumn{1}{c|}{$\minkr>0$}
    & \multicolumn{1}{c|}{$\minkr<0$}
    \\ \hline
    $\minku>0$
    & Inside the future $\minkX^2=0$ lightcone.
    & Outside the $\minkX^2=0$ lightcone and below the $\refVec\cdot\minkX=0$ null plane.
    \\ \hline
    $\minku<0$
    & Outside the $X^2 = 0$ lightcone and above the $\refVec\cdot\minkX=0$ null plane.
    & Inside the past $\minkX^2=0$ lightcone.
    \\ \hline
  \end{tabularx}
  \caption{Minkowski spacetime regions covered by various signs of $\minku$ and
    $\minkr$.}
  \label{tab:rice}
\end{table}

An advantage of the planar retarded coordinates is the ability to easily
describe both past and future null infinities. A disadvantage is that, the
$\minkr=0$ plane is a rather singular surface where $\minku\to -\infty$ as
approached from above and $\minku\to +\infty$ as approached from below. Indeed,
Minkowski spacetime is covered by two coordinate patches separated by the
$\minkr=0$ plane. Making a consistent choice for spacetime orientation, we take
\begin{align}
  \volForm^{(4)}
  &= \diff\minkX^0 \wedge \diff\minkX^1 \wedge \diff\minkX^2 \wedge \diff\minkX^3
    = \sgn(\minkr)\, \minkr^2\, \diff\minku \wedge \diff\minkr \wedge \celVolForm
    \label{eq:minkOrientation}
\end{align}
which clearly exhibits the coordinate singularity at $\minkr=0$.

It will also be convenient to introduce the polarization
tensor\cite{Pasterski:2017kqt}\footnote{Our polarization vector $\polar$ is
  $1/2$ times that of \cite{Pasterski:2017kqt}'s (2.11) and $1/\sqrt{2}$ times
  that of many other literature --- \cf{}, for example, \cite{Donnay:2018neh}'s
  (2.8), \cite{Arkani-Hamed:2020gyp}'s (110), and \cite{Pasterski:2020pdk}'s
  (2.5). (Recall that our normalization for $\nullUnit$ is also somewhat
  nonstandard, as described in footnote \ref{foot:normq}.) Our choice ensures
  \cref{eq:chocolate} is satisfied without extra factors, where $\celMet$ is the
  standard metric \labelcref{eq:yogurt} on the plane.}
\begin{equation} \polar(\celw) = \diff[\celw] \nullUnit(\celw)\;,
  \label{eq:icecream}
\end{equation}
where $\diff[\celw]$ is the differential on celestial space. Having
one spacetime index and one celestial index, the polarization tensor obeys the
orthogonality and completeness relations
\begin{align} \polar_a \cdot \polar_b &= \celMet_{ab} \;,
                                        \label{eq:chocolate} \\ \polar^\mu \cdot \polar^\nu &= \eta^{\mu\nu} +
                                                                                              \nullUnit^{(\mu} \celLap \nullUnit^{\nu)} \;.
\end{align}
From these, we can derive a few identities which will be useful later:
\begin{align}
  \celLap \log(\nullUnit\cdot \minkX)
  &= - \frac{\minkX^2}{(\nullUnit\cdot \minkX)^2} \;,
    \label{eq:celLaplogqX}
  \\
  \frac{2\minkX^{\mu}}{\minkX^2} \celLap \log(\nullUnit\cdot \minkX)
  &= - \frac{2\minkX^{\mu}}{(\nullUnit\cdot \minkX)^2}
    = 2 \celCovD\cdot \frac{\polar^{\mu}}{\nullUnit\cdot \minkX}
    - \diff[\minkX_{\mu}] \frac{\celLap \nullUnit \cdot \minkX}{\nullUnit\cdot \minkX} \;.
    \label{eq:cheesecake}
\end{align}
Here, $\diff[\minkX]$ is the spacetime differential. (Where it is obvious
whether differentials are acting on spacetime or celestial space and the points
at which they are acting are clear, we may sometimes omit subscripts on
differentials in this paper.) We refer the reader to appendix \ref{sec:minkConventions} for further details on notation and conventions.

\subsection{Maxwell theory and conformal primary wavefunctions}
\label{sec:steak}
In four-dimensional Minkowski spacetime the free Maxwell's equations are solved
by conformal primary solutions of the form
\cite{Pasterski:2017kqt,Pasterski:2020pdk}\footnote{In this paper (as is common
  in literature on celestial holography), we are implicitly rescaling
  $\minkX^\mu$, $\minkr$, and $\minku$ by some dimensionful unit of length
  $\lenConst$ which we notationally suppress. Reinstating this constant would be
  required to give all conformal primary wavefunctions a consistent mass
  dimension, \eg~ there will be an additional factor of $\ell^{\Delta - 1}$ on the RHS of \cref{eq:maxwell-cpw}.} \label{foot:noodles}
\begin{align}
  \label{eq:maxwell-cpw}
  \confPrimA{\confDim, \pm}[a; \mu](\celw; \minkX)
  &=  \frac{\confTensor{\pm}[a;\mu](\celw; \minkX)}{(-\nullUnit\cdot \minkX_{\pm})^{\confDim}},
\end{align}
where we have introduced
\begin{align}
  \label{eq:frame}
  \confTensor{\pm}[a;\mu](\celw; \minkX)
  &= \polar_{a; \mu} + \frac{\polar_{a} \cdot \minkX_{\pm}}{-\nullUnit\cdot \minkX_{\pm}} \nullUnit_{\mu}\;,
  &
    \minkX_{\pm}^{\mu}
  &= \minkX^{\mu} \mp i\reg n^{\mu}\;.
\end{align}
The $\pm i\reg$ regulators provide instructions for passing through the
$\nullUnit\cdot \minkX=0$ surface in \cref{eq:maxwell-cpw}, with the sign
bearing a physical interpretation explained shortly. In particular, we shall
take the branch cut of the logarithm of $-\nullUnit\cdot\minkX_\pm$ to be over
the negative reals:
\begin{align}
  \arg(-\nullUnit\cdot\minkX_\pm)
  &\eqReg \mp\pi \stepFunc(\nullUnit\cdot\minkX) \,.
    \label{eq:argqX}
\end{align}
We have introduced here the symbol $\eqReg$ meaning an equality that holds in
the $\reg\to 0$ limit. In the following we will omit the spacetime and celestial indices on the conformal primary wavefunctions to reduce the clutter. 

Under spacetime Lorentz transformations paired with corresponding conformal
transformations \labelcref{eq:confTrans} of the celestial space, the conformal
primary wavefunctions transform as their name suggests (apart from an adjustment
to the regulator $\reg$):
\begin{align}
  \confPrimA{\confDim,\pm} (\confTrans(\celw); \lorentz \cdot \minkX)
  &\eqReg (cw + d)^{2h} (\bar{c}\bar{w}+\bar{d})^{2\bar{h}}
    \lorentz \cdot
    \confPrimA{\confDim,\pm} (\celw; \minkX)\;,
    \label{eq:confPrimATrans}
\end{align}
where the holomorphic and antiholomorphic weights $(h,\bar{h})$ depend on
$\confDim$ and the helicity $J$:
\begin{align}
  h &= \frac{\confDim+J}{2} \;,
  &
    \bar{h} &= \frac{\confDim-J}{2} \;,
  &
    J &= \begin{cases}
           1 & \confPrimA{\confDim}[w] \\
           -1 & \confPrimA{\confDim}[\bar{w}]
         \end{cases}
        \;.
\end{align}
In this paper, the $\reg\to 0$ limit will always be the last to be taken. This
gives $\eqReg$ a richer distributional meaning than merely a naive equality of
point-wise limits. For example, \cref{eq:confPrimATrans} does not merely say
that the $\reg\to 0$ point-wise limit of the LHS equals the point-wise limit of
the RHS away from $\nullUnit\cdot \minkX=0$. At  $\nullUnit\cdot \minkX=0$ both point-wise limits are
undefined. Instead, \cref{eq:confPrimATrans} says that when the LHS and RHS of are integrated against any smooth
spacetime function --- even those nonvanishing on $\nullUnit\cdot\minkX=0$ ---
and then the $\reg\to 0$ is taken, the integrals become equal. (Later in
\cref{sec:butter}, we will define a weaker relation $\eqIntegrated$ which
further restricts consideration to smooth smearing in the conformal dimension.)

It is helpful to note that the conformal primary wavefunctions can be decomposed
as \cite{Donnay:2018neh}
\begin{align}
  \confPrimA{\confDim,\pm}(\celw; \minkX)
  &= \mellCoeff{\confDim,\pm} \, \confMell{\confDim,\pm}(\celw;\minkX)
    + \diff[\minkX] \confGauge{\confDim, \pm}(\celw; \minkX) \;,
  &
    \mellCoeff{\confDim,\pm}
  &= \frac{(\pm i)^\confDim(\confDim-1)}{\Gamma(\confDim+1)} \;,
    \label{eq:cake}
\end{align}
splitting the wavefunction into a piece which is the Mellin transform of plane
waves
\begin{align}
  \confMell{\confDim,\pm}[a;\mu](\celw;\minkX)
  &=
    (\mp i)^\confDim \Gamma(\confDim) \frac{\polar_{a;\mu}}{(- \nullUnit \cdot \minkX_{\pm})^{\confDim}}
  =\polar_{a;\mu}
    \int_0^{\infty} \diff\celFreq\, \celFreq^{\confDim - 1} e^{\pm i\celFreq \nullUnit\cdot \minkX_{\pm}} \;,
    \label{eq:Mellin1}
\end{align}
and a pure gauge piece
\begin{align}
  \confGauge{\confDim,\pm}[a](\celw; \minkX)
 &= \frac{1}{\confDim} \frac{\polar_a \cdot \minkX_{\pm}}{(-\nullUnit \cdot \minkX_{\pm})^{\confDim}}.
    \label{eq:cpw}
\end{align}
From \cref{eq:Mellin1}, we see that the interpretation of the sign $\pm$ in
conformal primary wavefunctions is to distinguish solutions constructed from
positive- and negative-frequency modes. Note that \cref{eq:Mellin1} relates the remaining part of the conformal primary
wavefunction to more familiar plane waves. Moreover, the decomposition
\labelcref{eq:cake} will be useful, because as will be described later, symplectic
products involving the pure gauge piece
\begin{align}
  \diff_{\minkX}\confGauge{\confDim,\pm}
  &= \frac{1}{\confDim} \diff[\celw] \frac{\nullUnit}{(-\nullUnit\cdot\minkX_\pm)^\confDim}
    = \frac{1}{(\mp i)^\confDim \Gamma(\confDim+1)} \diff[\celw]\left(
    \nullUnit \int_0^{\infty} \diff\celFreq\, \celFreq^{\confDim - 1} e^{\pm i\celFreq \nullUnit\cdot \minkX_{\pm}}
    \right)
    \label{eq:pureGaugeMellin}
\end{align}
are often vanishing.

\section{Soft modes and their physical interpretation}
\label{sec:minkSoftModes}

In this section we review the conformally soft and Goldstone solutions to the free Maxwell equations \cite{Donnay:2018neh}. We discuss their large-$r$ expansions and matching conditions, as well as their analytic continuation across the future and past lightcones through the origin. In section \ref{sec:CS-decompositions} we decompose the conformally soft solutions in terms of plane waves and (non-soft) conformal primary wavefunctions. In section \ref{sec:butter} we review their inner products, while in section \ref{sec:sandwich} we define the associated celestial operators.


An important role will be played by the pure gauge, \ie{} Goldstone,
wavefunctions obtained by evaluating \cref{eq:maxwell-cpw} at $\confDim = 1$
\begin{align}
  \confPrimA{\gold}[a](\celw,\minkX)
  &\equiv
    \frac{\confPrimA{1,+}[a](\celw; \minkX) + \confPrimA{1,-}[a](\celw; \minkX)}{2}
    = \diff[\minkX] \confGauge{\gold}[a](\celw,\minkX)
    \;,
    \label{eq:G-wf}
  \\
  \confGauge{\gold}[a](\celw,\minkX)
  &\equiv \frac{\confGauge{1,+}[a](\celw,\minkX) + \confGauge{1,-}[a](\celw,\minkX)}{2}
    \;,
\end{align}
where
\begin{align}
  A^{1,\pm}_a(\celw;\minkX)
  &= \diff[\minkX] \confGauge{1,\pm}[a](\celw;\minkX)
    \;,
    \label{eq:coffee}
  \\
  \confGauge{1,\pm}(\celw;\minkX)
  &= -\diff_{\celw} \log(-\nullUnit\cdot \minkX_\pm)
    = - 4\pi \diff[\celw] \celG(\minkz,\celw) \left[
    1 + \frac{2(\minku\mp i\reg)}{\minkr|\celwh-\minkzh|^2}
    \right]^{-1}
    \;.
    \label{eq:cappuccino}
\end{align}
Here the 2D Green's function $\celG(\minkz,\celw)$ was defined in appendix \ref{sec:minkConventions} and the last equality follows from:
\begin{align}
  -\nullUnit(\celw)\cdot \minkX_\pm
  &= \minku \mp i\reg + \frac{\minkr\abs{\celwh-\minkzh}^2}{2}.
    \label{eq:qX1}
\end{align}

Physically, turning on $\confPrimA{\gold}$ corresponds to applying an asymptotic
symmetry transformation meaning that, while $\confPrimA{\gold}$ is pure gauge, the gauge
parameter $\confGauge{\gold}$ is nontrivial near $\spaceInfty$. We shall treat
these symmetries as physical in Minkowski spacetime. A consistency requirement
to do so requires the symmetry transformation alter the initial data on all
Cauchy slices. For example, $\confGauge{\gold}$ is nontrivial on both past and
future null infinities $\nullInfty^{\pm}$ (reached, as we recall, by $\minkr\to
-\infty$ and $\minkr\to \infty$ respectively).
From \cref{eq:cappuccino}, the leading terms in the asymptotic
expansions\footnote{While the logarithmic corrections in \eqref{eq:tea} may seem
  surprising, they capture the divergences that come from viewing the naively
  linear order term of the Taylor expansion of \cref{eq:cappuccino} in $\fgr$ as
  a distribution. This is derived and explained in \cref{sec:asymptotics}.
  Though we only require the leading order term in \cref{eq:tea} for the present
  discussion, the next order term will play an important role in
  \cref{sec:entanglementSoftDOFs}.} may be obtained
\begin{align}
  \confGauge{1,\pm}(\celw;\minkX)
  &= - 4\pi \diff[\celw] \celG(\celw,\minkz)
    + \order{\minkr^{-1}\log\minkr}
    =
    \confGauge{\gold}(\celw;\minkX).
    \label{eq:tea}
\end{align}
 It is standard to demand antipodal matching conditions for
gauge transformations of Maxwell theory in Minkowski spacetime, that is for the
gauge parameter $\gaugeParam$ on the past boundary $\nullInftyBdy{+}{-}$ of
$\nullInfty[+]$ and the future boundary $\nullInftyBdy{-}{+}$ of $\nullInfty[-]$
to agree antipodally \cite{He:2014cra,Kapec:2015ena}:
\begin{align}
  \pullback[\nullInftyBdy{+}{-}]\gaugeParam
  &= \pullback[\nullInftyBdy{-}{+}]\gaugeParam
    \;,
    \label{eq:antipodalGaugeParam}
\end{align}
where evaluation at the same point $\minkz$ in the celestial space parametrizing
$\nullInftyBdy{\pm}{\mp}$ is understood on both sides of
\cref{eq:antipodalGaugeParam}. (Note that a given point $\minkz$ of the celestial
plane points in antipodal directions in Minkowski spacetime for $\minkr$ of
opposite signs.) This
matching condition ensures that the action of an asymptotic symmetry on the
early Cauchy slice $\nullInfty[-]$ will evolve to a corresponding transformation
on the later Cauchy slice $\nullInfty[+]$ --- as expected for a physical
symmetry transformation. The antipodal nature of the matching condition is a
consequence of Lorentz invariance. A boundary condition on field configurations
which enforces \cref{eq:antipodalGaugeParam} is
\begin{align}
  \pullback[\nullInftyBdy{+}{-}] A
  &= \pullback[\nullInftyBdy{-}{+}] A \;.
    \label{eq:antipodalA}
\end{align}
Clearly from \cref{eq:tea}, we see that $\confGauge{1,\pm}$,
$\confGauge{\gold}$, $\confPrimA{1,\pm}$, and $\confPrimA{\gold}$ satisfy
\cref{eq:antipodalGaugeParam,eq:antipodalA}.

As can also be seen from \cref{eq:tea}, the difference
$\confGauge{1,+}-\confGauge{1,-}$, in contrast to $\confGauge{\gold}$, vanishes
at both null infinities.\footnote{In fact, from results derived in
  \cref{sec:logqXAsymptotics}, it can be shown that the
  $\sim\minkr^{-1}\log\minkr$ term in $\confGauge{1,+}-\confGauge{1,-}$ vanishes
  in the $\reg\to 0$ limit, in which case $\confGauge{1,+}-\confGauge{1,-}$ has
  an even faster $\order{\minkr^{-1}}$ falloff at large
  $\minkr$. \label{foot:fasterFalloff}} Any gauge transformation (such as
$\confPrimA{1,+}-\confPrimA{1,-}$) permissible by boundary conditions, for which
the gauge parameter on a full Cauchy slice (such as past or future null
infinity) is zero, will fail to alter the initial data on that Cauchy slice and
must therefore be a redundancy of the theory, \ie{} trivial gauge.

Asymptotic symmetries are generated by asymptotic charges
\cite{He:2014cra,Kapec:2015ena}. In pure Maxwell theory, they are given by
\cref{eq:cstrBlk} and on-shell by \cref{eq:cstrBlkOnShell}, which at
$\nullInfty^{\pm}$ become\footnote{More
  generally, when coupled to charged matter, the RHS of \cref{eq:cstrBlk} gives
  a `soft charge', which combines with a `hard charge' contribution from charged
  matter to give the asymptotic charge. The asymptotic charge is always given
  on-shell by the RHS of
  \cref{eq:cstrBlkOnShell}. \label{foot:softVsAsymCharge}}
\begin{align}
  \asCharge[][\pm]\relax[\gaugeParam]
  &= \int_{\nullInfty[\pm]} \diff\gaugeParam \wedge * F
    \eqOnShell
    \int_{\nullInftyBdy{\pm}{\mp}} \gaugeParam \hodge F
    \;.
    \label{eq:sundae}
\end{align}
Conservation $\asCharge[][+] = \asCharge[][-]$ of these charges is equivalent to
the antipodal matching
\begin{align}
  \pullback[\nullInftyBdy{+}{-}] \hodge F
  &= \pullback[\nullInftyBdy{-}{+}] \hodge F
    \label{eq:antipodalF}
\end{align}
of Coulomb fields near $\spaceInfty$.

As shown in \cite{Donnay:2018neh} and we rederive later, the Goldstone modes
\eqref{eq:G-wf} are the canonical partners of conformally soft, \ie{} memory,
wavefunctions
\begin{equation}
  \label{eq:CS}
  \confPrimA{\cs}[a] \equiv \frac{\confPrimA{\log,+}[a] - \confPrimA{\log,-}[a]}{2\pi i},
\end{equation}
where
\begin{align}
  \confPrimA{\log,\pm}[a]
  &\equiv \lim_{\confDim \rightarrow 1} \partial_{\confDim}\left[
    \confPrimA{\confDim,\pm}[a] + \confPrimA<\shad>{2 - \confDim, \pm}[a]
    \right]
    = -\log(-\minkX_\pm^2)  \confPrimA{1,\pm}[a]
    \;.
    \label{eq:log}
\end{align}
Here $\tilde{A}_a^{2-\Delta, \pm}$ are the shadow transforms of the spin-1
conformal primary wavefunctions \cite{Pasterski:2017kqt}
\begin{align}
  \confPrimA<\shad>{2 - \confDim, \pm}[a]
  = (-X_{\pm}^2)^{1-\Delta}  \confPrimA{2-\confDim,\pm}[a]
  \;.
  \label{eq:confPrimAShad}
\end{align}

We shall choose the branch of $\log(-\minkX_{\pm}^2)$ so that
\begin{align}
  \arg(-\minkX_{\pm}^2)
  &\eqReg
    \begin{cases}
      0 & \text{$\minkX^2<0$ and $\minkX^0>0$} \\
      \mp \pi & \text{$\minkX^2>0$} \\
      \mp 2\pi & \text{$\minkX^2<0$ and $\minkX^0<0$}
    \end{cases} \;.
    \label{eq:argX2}
\end{align}
Note that we are making the standard choice of $\arg(-\minkX_{\pm}^2)\eqReg 0$
in the Milne patch to the future of the origin; the $i\reg$ prescription then
determines $\arg(-\minkX_{\pm}^2)$ everywhere else in the Minkowski spacetime.
In particular, it forces us to choose $\arg(-\minkX_{\pm}^2)\eqReg \mp 2\pi$ in
the Milne patch to the past of the origin even though
$\minkX^2 < 0$ here too.

Next, using
\begin{align}
  -\minkX_{\pm}^2
  &= 2\minkr (\minku\mp i \reg) \;,
\end{align}
we can further deduce the asymptotic behaviour of $\confPrimA{\cs}$ by relating
$\confPrimA{\cs}$ to $\confPrimA{\gold}$ and $\confPrimA{1,+}-\confPrimA{1,-}$:
\begin{align}
  \confPrimA{\cs}
  &= \left[ 
    \regStepFunc{\reg}(-\minku) +
    \stepFunc(-\minkr)
    \right]
    \confPrimA{\gold}
    -\frac{
    \log(
    -\minkX_+^2
    )
    + \log(
    -\minkX_-^2
    )
    }{
    2\pi i
    }
    \,
    \frac{\confPrimA{1,+}-\confPrimA{1,-}}{2}
    \;,
    \label{eq:csToPm}
\end{align}
where the regulated step function is defined by\footnote{For the logarithms in
  \cref{eq:regStepFunc}, we take the standard choice $\arg \posReals=0$. Recall
  however that consistency with the $i\reg$ prescription over spacetime led us
  to choose \cref{eq:argX2} for $\log(-\minkX_{\pm}^2) = \log[2\minkr (\minku\mp
  i \reg)]$. Some care is required to reconcile these conventions to derive the
  first term of the RHS in \cref{eq:csToPm}. This nuance seems to have been
  missed in previous works, leading to incorrect expressions for
  $\confPrimA{\cs}$ --- for instance, compare our \cref{eq:csToPm} with
  \cite{Donnay:2018neh}'s (3.21). \label{foot:AcsLogSubtlety}}
\begin{align}
  \regStepFunc{\reg}(-\minku)
  &\equiv \frac{\log(\minku + i \reg) - \log(\minku - i \reg)}{2\pi i}
    \;.
    \label{eq:regStepFunc}
\end{align}
In particular, we have the equality
\begin{align}
  \pullback[\nullInfty[\pm]] \confPrimA{\cs}
  &= 
    [1\mp \regStepFunc{\reg}(\pm \minku)] \,
    \pullback[\nullInfty[\pm]]\confPrimA{\gold}
    = 
    4\pi 
    [-1\pm \regStepFunc{\reg}(\pm \minku)] \,
    \diff[\celw] \diff[\minkz] \celG(\celw,\minkz)
    \label{eq:AcsAsym}
\end{align}
of pullbacks to $\nullInfty[\pm]$.

However, $\confPrimA{\cs}\unsameInit{\nullInfty[\pm]}{ [1\mp
  \regStepFunc{\reg}(\pm \minku)]}\confPrimA{\gold}$ --- recall
$\sameInit{\nullInfty[\pm]}$ means a comparison of initial data, as defined
around \cref{eq:sameInit} --- because $\confPrimA{\cs}$ has nontrivial Coulomb
fields near $\spaceInfty$. For instance, from \cref{eq:AcsAsym} and the constraint equations at
$\nullInfty[\pm]$, one finds
\begin{align}
  \pullback[\nullInfty^{ \pm}] \hodge[\minkX] \confPrimF{\cs}
  &= 4\pi \left\{ 
    \mp \regDeltaFunc{\reg}(\minku) \,
    \diff\minku \wedge \hodge[\minkz]\diff[\minkz]
    \left[ \diff[\celw] \celG(\celw,\minkz) \right]
    + \regStepFunc{\reg}(\mp\minku) \,
    \celVolForm(\minkz) \,
    \diff[\celw]\celDeltaFunc(\celw,\minkz)
    \right\}
    \;,
    \label{eq:FcsAsym}
\end{align}
where
\begin{align}
  \regDeltaFunc{\reg}(\minku)
  &= \partial_\minku \regStepFunc{\reg}(\minku)
    \;
\end{align}
and $\hodge[\minkz]$ is the Hodge star with respect to the celestial metric.
We see, in particular that \cref{eq:FcsAsym} satisfies the conservation
\eqref{eq:antipodalF} of asymptotic charge; indeed, it does so nontrivially, as
we describe shortly.

\begin{figure}
  \centering \includegraphics[scale=1.2]{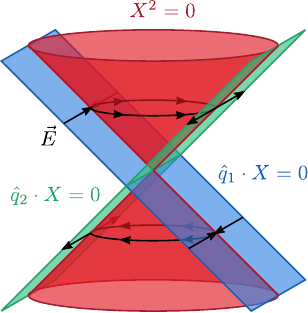}
  \caption{The field strength of $\confPrimA{\csI}(\celw_1,\celw_2)$, defined in
    \cref{eq:lettuce} by integrating $\confPrimA{\cs}(\celw)$ from $\celw_1$ to
    $\celw_2$, is localized to spherical and planar shockwaves on $\minkX^2=0$,
    $\nullUnit_1\cdot \minkX =0$, and $\nullUnit_2\cdot \minkX=0$. The electric
    field lines along time slices of these shockwaves are also shown. }
  \label{fig:bacon}
\end{figure}

More generally, while $\confPrimA{\gold}$ clearly has a vanishing field
strength, the conformally soft mode $\confPrimA{\cs}$ has a field strength which
is localized to spherical and planar shockwaves on the $\minkX^2=0$
and $\nullUnit\cdot \minkX=0$ surfaces respectively, as described in \cite{Donnay:2018neh}.
Actually, the electric fields on the spherical shockwave resemble those produced
by a dipole placed at $\nullUnit$ on a sphere; moreover, the planar shockwave is
best thought of as two infinitesimally nearby planes. For ease of illustration,
it is helpful to separate the two singular points of the dipole on the spherical
shockwave and the two planes by integrating
\begin{align}
  \confPrimA{\csI}(\celw_1,\celw_2;\minkX)
  &\equiv \int_{\celw_1}^{\celw_2} \confPrimA{\cs}(\celw;\minkX)
    \label{eq:lettuce}
  \\
  &= \frac{1}{2\pi i}\left\{ 
    \log(-\minkX_+^2) \diff[\minkX] [
    \log(-\nullUnit_2\cdot \minkX_+)
    -\log(-\nullUnit_1\cdot \minkX_+)
    ]
    - (\minkX_+\to \minkX_-)
    \right\}
    \;.
\end{align}
Here $\confPrimA{\cs}$ is viewed as a one-form in the celestial space which can
be integrated along a path connecting points $\celw_1$ and $\celw_2$.\footnote{From
\cref{eq:coffee,eq:cappuccino,eq:CS,eq:log}, it is evident that
$\confPrimA{\cs}$ is an exact form in the celestial space, so that the above
integral is path-independent.} The field strength of
$\confPrimA{\csI}(\celw_1,\celw_2)$ is localized to spherical and planar
shockwaves on $\minkX^2=0$, $\nullUnit_1\cdot \minkX =0$, and $\nullUnit_2\cdot
\minkX=0$ --- this is illustrated in \cref{fig:bacon}. As shown, the
$\nullUnit_1\cdot \minkX=0$ shockwave brings electric field lines from infinity
to the intersection line with the $\minkX^2=0$ spherical shockwave, providing
the positive source of the now separated dipole on spherical time slices here.
The electric field lines on the spherical shockwave then run to the sink at the
intersection with $\nullUnit_2\cdot \minkX=0$. Finally, the $\nullUnit_2\cdot
\minkX=0$ planar shockwave diverts the electric field lines back out to
infinity. 

Because $\confPrimA{\csI}$ and $\confPrimA{\cs}$ modify the electric field at
infinity, particularly near $\spaceInfty$, turning on these modes corresponds to
changing the value of asymptotic charges defined by \cref{eq:sundae}. Indeed, we
may calculate from \cref{eq:FcsAsym} or pictorially  from
\cref{fig:bacon} that \footnote{Note that the $\nullUnit_1\cdot\minkX=0$ and
  $\nullUnit_2\cdot\minkX=0$ planes in \cref{fig:bacon} graze $\nullInfty^{\pm}$
  respectively along $\celw_1$ and $\celw_2$ for $\mp\minku>0$. Along these
  lines of $\nullInfty^{\pm}$, the blue (green) shockwave carries away (deposits)
  electric field lines from (to) $\nullInftyBdy{\pm}{\mp}$.}
\begin{align}
  \asCharge[][\pm]\relax[\gaugeParam, \confPrimA{\cs}]
  &= 4\pi\, \diff \gaugeParam \;,
  &
    \asCharge[][\pm]\relax[\gaugeParam, \confPrimA{\csI}]
  &= 4\pi [\gaugeParam(\celw_2) -\gaugeParam(\celw_1)] \;.
    \label{eq:asymChargeCSI}
\end{align}
These charges will be given an intuitive meaning in \cref{sec:honey} when we
discuss sources that appear in the Einstein static universe outside the
Minkowski patch. Since $\confPrimA{\cs}$ turns on a nontrivial asymptotic
charge, it is not surprising that it is canonically paired with asymptotic
symmetries $\confPrimA{\gold}$, as we demonstrate in \cref{sec:butter}.

\subsection{Decomposing the conformally soft mode $\confPrimA{\cs}$}
\label{sec:CS-decompositions}
Before moving on, it will be helpful for later discussions to derive
decompositions of the conformally soft mode $\confPrimA{\cs}$ in terms of plane
waves and in terms of conformal primary wavefunctions $\confPrimA{\confDim}$
without their shadow counterpart.

\paragraph{In terms of plane waves:} Since plane waves offer a complete basis of solutions up to gauge symmetries,
we expect to be able to decompose $\confPrimA{\log}$ and $\confPrimA{\cs}$ as linear combinations of
plane waves plus pure gauge pieces. Indeed, applying \cref{eq:grape} to
\cref{eq:coffee,eq:cappuccino,eq:log}, we obtain
\begin{align}
  \begin{split}
    \MoveEqLeft[1]\confPrimA{\log,\pm}(\celw)
    \\
    &= \log[-\minkX_\pm^2] \int \celVolForm(\celw') \,
      \diff[\celw] \celG(\celw,\celw')  \celLap \diff[\minkX]\log(-\nullUnit'\cdot \minkX_\pm)
  \end{split}
  \\
    &= \int \celVolForm(\celw')\,
      \diff[\celw] \celG(\celw,\celw')
      \celLap \left\{
      \diff[\minkX] \left[ \log(-\minkX_\pm^2) \log(-\nullUnit'\cdot \minkX_\pm)\right]
      -\frac{2 \minkX_\pm}{\minkX_\pm^2} \log(-\nullUnit'\cdot \minkX_\pm)
      \right\}
  \\
    &= \pm 2i \int \celVolForm(\celw') \,
      \diff[\celw] \celG(\celw,\celw')
      \celCovD\cdot \confMell{1,\pm}(\celw')
      + \diff[\minkX] \softGaugeg{\log,\pm}.
      \label{eq:syrup}
\end{align}
Here we have defined
\begin{align}
  \softGaugeg{\log,\pm}
  &\equiv \log(-\minkX_\pm^2) \diff[\celw] \log(-\nullUnit\cdot \minkX_\pm)
    + \int \celVolForm(\celw')\,
    \diff[\celw] \celG(\celw,\celw')
    \frac{\celLap \nullUnit'\cdot \minkX_\pm}{\nullUnit'\cdot \minkX_\pm}
  \\
  &= -\log(-\minkX_\pm^2) \confGauge{1,\pm}
    + \int \celVolForm(\celw')\,
    \diff[\celw] \celG(\celw,\celw')
    \left\{
    -\celCovD\cdot\confGauge{1,\pm}(\celw')
    + [\confGauge{1,\pm}(\celw')]^2
    \right\}
    \;,
    \label{eq:gammaLog}
\end{align}
and we have substituted \cref{eq:cheesecake,eq:cappuccino} in
\cref{eq:syrup,eq:gammaLog}. Recall $\confMell{\confDim,\pm}$ is given by
the Mellin transform \labelcref{eq:Mellin1} of plane waves. Similarly, we can
decompose $\confPrimA{\cs}$ defined in terms of $\confPrimA{\log}$ by
\cref{eq:CS}:
\begin{align}
  \confPrimA{\cs}(\celw)
  &=  \frac{1}{\pi} \int \celVolForm(\celw') \,
    \diff[\celw] \celG(\celw,\celw')
    \celCovD\cdot \left[
    \confMell{1,+}(\celw')
    + \confMell{1,-}(\celw')
    \right]
    + \diff[\minkX] \softGaugeg{\cs}
    \;,
    \label{eq:csToV}
  \\
    \softGaugeg\cs
  &= \frac{\softGaugeg{\log,+}-\softGaugeg{\log,-}}{2\pi i} \;.
  \label{eq:gammacs}
\end{align}
 As we will see in \cref{sec:butter}, this decomposition is consistent with the
inner product between $\confPrimA{\cs}$ and plane waves
\cite{Arkani-Hamed:2020gyp}. 

Naively, one might expect the first term to be sufficient in the construction of Goldstone operators \cite{Donnay:2018neh}. However, as we now show, the two terms separately violate the matching condition \eqref{eq:sundae}. This can be seen by examining the large-$r$ behavior of the gauge parameter $\softGaugeg{\cs}$.
Oddly, $\softGaugeg{\cs}$
vanishes on $\nullInfty[+]$ but not on $\nullInfty[-]$, which can be seen as
follows. Firstly, following nearly the same steps
leading to \cref{eq:AcsAsym}, we find
\begin{align}
  \frac{
  -\log(-\minkX_+^2)\confGauge{1,+}
  +\log(-\minkX_-^2)\confGauge{1,-}
  }{2\pi i}
  &= -4\pi [\regStepFunc{\reg}(-\minku) + \stepFunc(-\minkr)] \diff[\celw] \celG(\celw,\minkz)
    + \order{\minkr^{-1}\log^2\minkr}
    \;.
    \label{eq:logX2confGaugeAsym}
\end{align}
Secondly, from \cref{eq:tea}, the first term in the integrand of
\cref{eq:gammaLog} cancels in $\softGaugeg{\cs}$, at least up to
$\order{\minkr^{-1}\log\minkr}$ corrections. Finally, using the asymptotic expansion of
$(\confGauge{1,+})^2 - (\confGauge{1,-})^2$ derived in
\cref{eq:differOfconfGauge2Asym}, we find\footnote{Similar to the comment in \cref{foot:fasterFalloff}, note that the $\sim
  \minkr^{-1} \log^2\minkr$ terms in the corrections to
  \cref{eq:logX2confGaugeAsym,eq:gammaCSAsym} vanish in the $\reg\to 0$
  limit, in which case the correction would have even faster
  $\order{\minkr^{-1}\log\minkr}$ falloffs at large $\minkr$. The same is true
  of \cref{eq:softGaugebAsym,eq:confGaugecsAsym} further below. In fact, in the
  $\reg\to 0$ limit, the correction to \cref{eq:softGaugebAsym} has no
  $\sim\minkr^{-1}\log\minkr$ term either and becomes $\order{\minkr^{-1}}$.
}
\begin{align}
  \softGaugeg{\cs}
  &= -8\pi \,
    \stepFunc(-\minkr) \,
    \diff[\celw] \celG(\celw,\minkz)
    + \order{\minkr^{-1}\log^2\minkr}
    \;.
    \label{eq:gammaCSAsym}
\end{align}
This violates the antipodal matching condition
\cref{eq:antipodalGaugeParam}, and consequently neither $\diff[\minkX]\softGaugeg{\cs}$ nor
\begin{align}
  \pullback[\nullInfty[\pm]](\confPrimA{\cs}-\diff[\minkX]\softGaugeg{\cs})
  &= \mp 4\pi \regStepFunc{\reg}(\mp \minku) \diff[\celw]\diff[\minkz]\celG(\celw,\minkz)
\end{align}
can be independently turned on without violating
\cref{eq:antipodalA}. (But, note that $\confPrimA{\cs}$ \textit{does satisfy}
\cref{eq:antipodalA}, as can be seen from \cref{eq:AcsAsym}.)
The fact that the
linear combination of plane waves given by the integral of \cref{eq:csToV}
violates boundary conditions near $\spaceInfty$ is an illustration of the
dangers of including arbitrarily soft plane waves.\footnote{
  We should also caution against another trap related to arbitrarily soft
  plane waves. Usually, one expects plane waves to localize in a particular
  celestial direction at $\nullInfty[\pm]$, by applying a stationary phase
  analysis to $\exp(\pm i \celFreq\nullUnit\cdot\minkX) = \exp[\mp i \celFreq
  \left(u+ \frac{r\abs{\celwh-\minkzh}^2}{2}\right)]$ at large $\minkr$.
  However, the convergence of the stationary phase approximation worsens with
  small $\celFreq$. This failure is demonstrated, for example by considering the
  profiles of \cref{eq:Mellin1,eq:pureGaugeMellin} at $\nullInfty[\pm]$ which are
  obviously not purely contact terms at $\celw=\minkz$ --- see \eg{}
  \cref{eq:tea}. In the present calculation, if a naive
  stationary phase analysis is applied to \cref{eq:csToV} one would have
  \textit{incorrectly} concluded that
  \begin{align}
    \pullback[\nullInfty[\pm]](\confPrimA{\cs}-\diff[\minkX]\softGaugeg{\cs})
    \eqDubious
    4\pi \left[
    \frac{1}{2} -\regStepFunc{\reg}(- \minku)
    \right] \diff[\celw]\diff[\minkz]\celG(\celw,\minkz)
    \;.
  \end{align}
} In particular, if
one takes a linear combination of plane waves $e^{\pm i\celFreq\nullUnit\cdot
  \minkX}$ with coefficients that are $\order{\celFreq^0}$ near $\celFreq=0$, then the resulting
profiles at $\nullInfty[\pm]$ may not die off at large $\minku$, so one
must be careful to engineer the required matching \cref{eq:antipodalA}
between $\nullInftyBdy{\pm}{\mp}$.

\paragraph{In terms of conformal primary wavefunctions $\confPrimA{\confDim}$:}
Also useful for later discussions is a decomposition of $\confPrimA{\cs}$ in
terms of the conformal primary wavefunctions $\confPrimA{\confDim}$
\emph{without} their shadows. To derive this, we use
\begin{align}
  (\partial_\confDim \confPrimA{\confDim,\pm})_{\confDim\to 1}
  &= \pm i \,\confMell{1,\pm} - \frac{1}{2} \diff[\minkX] \beta^{\log,\pm},
    \label{eq:waffle}
  \\
  \beta^{\log,\pm}
    &\equiv -2 (\partial_\confDim \confGauge{\confDim,\pm})_{\confDim\to 1}
      = 2[1+\log(-\nullUnit\cdot\minkX_\pm)] \confGauge{1,\pm}
  \\
  &= -\diff[\celw]\left\{
    [2+\log(-\nullUnit\cdot\minkX_\pm)] \log(-\nullUnit\cdot\minkX_\pm)
    \right\}
    \label{eq:betaLog}
\end{align}
to relate $\confMell{1}$ to
$(\partial_\confDim \confPrimA{\confDim})_{\confDim\to 1}$ in \cref{eq:syrup}.
This gives
\begin{align}
  \confPrimA{\log,+}(\celw)
  &=
    2\int \celVolForm(\celw') \,
    \diff[\celw] \celG(\celw,\celw')
    \celCovD\cdot \left[
    \partial_\confDim \confPrimA{\confDim,\pm}(\celw')
    \right]_{\confDim = 1}
    + \diff[\minkX] \confGauge{\log,\pm}(\celw),
  \\
  \confGauge{\log,\pm}[a]
  &\equiv \gamma^{\log,\pm}_a + \beta^{\log,\pm}_a
\end{align}
and therefore 
\begin{align}
  \confPrimA{\cs}(\celw)
  &= \frac{1}{\pi i}
    \int \celVolForm(\celw') \,
    \diff[\celw] \celG(\celw,\celw')
    \celCovD\cdot
    \partial_\confDim
    \left[
    \confPrimA{\confDim,+}(\celw') - \confPrimA{\confDim,-}(\celw')
    \right]_{\confDim=1}
    + \diff[\minkX] \confGauge{\cs}(\celw)
    \;,
    \label{eq:csToConfPrimA}
  \\
  \confGauge{\cs}[a]
  &\equiv \frac{\confGauge{\log,+}[a] - \confGauge{\log,-}[a]}{2\pi i}
    = \softGaugeg{\cs}[a] + \softGaugeb{\cs}[a]
    \;,
    \qquad\qquad
    \softGaugeb{\cs}[a]
    \equiv \frac{\softGaugeb{\log,+}[a] - \softGaugeb{\log,-}[a]}{2\pi i}
    \;,
\end{align}
where $\softGaugeg{\cs}[a]$ was defined in \cref{eq:gammacs}.
Note that upon integration by parts and expanding at large-$r$, the integration kernel in the first term reduces to a linear combination of the identity and a 2D shadow transform. Taking the inner product with the Maxwell field, we see that these are the same operations involved in the construction of celestial Goldstone operators \cite{Arkani-Hamed:2020gyp}. Therefore we expect the second term to vanish in the large-$r$ limit.  

Indeed, the second term $\diff[\minkX]\confGauge{\cs}$ on the RHS of \cref{eq:csToConfPrimA}
is trivial gauge, as we will now show by demonstrating that $\confGauge{\cs}$
vanishes at $\nullInfty$. Again from \cref{eq:tea}, the first term on the RHS of \cref{eq:betaLog} will
cancel in the difference $\softGaugeb{\cs}$, at least up to
$\order{\minkr^{-1}\log\minkr}$ corrections. To treat the last term of
\cref{eq:betaLog}, we apply the identities
\begin{align}
\diff[\celw]\log^2(-\nullUnit\cdot\minkX_\pm)
    &= \int \celVolForm(\celw')\,
    \diff[\celw] \celG(\celw,\celw') \celLap \log^2(-\nullUnit'\cdot\minkX_\pm)
    \;,
  \\
  \celLap \log^2(-\nullUnit\cdot\minkX_\pm)
  &= 2 \log(-\nullUnit\cdot\minkX_\pm)
    \celLap\log(-\nullUnit\cdot\minkX_\pm)
    + 2(\confGauge{1,\pm})^2
    \;,
    \label{eq:laplog2qX}
\end{align}
with the asymptotic expansion of \cref{eq:laplog2qX} given by \cref{eq:differOfconfGauge2Asym,eq:logqXlaplogqXAsym}.
Altogether, we find 
\begin{align}
  \softGaugeb{\cs}
  &= 8\pi \,
    \stepFunc(-\minkr) \,
    \diff[\celw] \celG(\celw,\minkz)
    + \order{\minkr^{-1}\log^2\minkr}
    \label{eq:softGaugebAsym}
\end{align}
and thus
\begin{align}
  \confGauge{\cs}
  &= \order{\minkr^{-1}\log^2\minkr}
    \label{eq:confGaugecsAsym}
\end{align}
vanishes at $\nullInfty$ as claimed.

Since the combination
\begin{align}
  \frac{
  \partial_\confDim
  \left(
  \confPrimA{\confDim,+} - \confPrimA{\confDim,-}
  \right)_{\confDim=1}
  }{2\pi i}
  &= \frac{
    -\log(-\nullUnit\cdot\minkX_+) \confPrimA{1,+}
    +\log(-\nullUnit\cdot\minkX_-) \confPrimA{1,-}
    }{2\pi i}
\end{align}
that shows up in \cref{eq:csToConfPrimA} will make a reappearance in
\cref{sec:confPrimMilne}, let us also consider its pullback to $\nullInfty$
here. Using the asymptotic expansions
\cref{eq:logqXpzalphaDifferAsym,eq:logqXpualphaAsym}, we see that
\begin{align}
  \pullback[\nullInfty]\frac{
  \partial_\confDim
  \left(
  \confPrimA{\confDim,+} - \confPrimA{\confDim,-}
  \right)_{\confDim=1}
  }{2\pi i}
  &= \begin{cases}
    -2\pi \regStepFunc{\reg}(-\minku)
       \celMet \celDeltaFunc(\celw,\minkz)
       & \text{at $\nullInfty[+]$}
    \\
       2\pi [1-\regStepFunc{\reg}(-\minku)]
       \celMet \celDeltaFunc(\celw,\minkz)
       +4\pi \diff[\minkz] \diff[\celw]\celG(\celw,\minkz)
       & \text{at $\nullInfty[-]$}
     \end{cases}
    \;,
    \label{eq:LSoftModeDubiousAsym}
\end{align}
which clearly violates the matching condition \labelcref{eq:antipodalA}. Note,
however that the celestial divergence of the above,
\begin{align}
\label{eq:shadow-cs}
  \pullback[\nullInfty[\pm]]\frac{
  \celCovD[\celw] \cdot \partial_\confDim
  \left(
  \confPrimA{\confDim,+} - \confPrimA{\confDim,-}
  \right)_{\confDim=1}
  }{2\pi i}
  &=
    2\pi [1\mp\regStepFunc{\reg}(\pm\minku)]
    \diff[\minkz] \celDeltaFunc(\celw,\minkz)
    \;,
\end{align}
does satisfy \cref{eq:antipodalA} and, when integrated in
\cref{eq:csToConfPrimA}, is consistent with the pullback \labelcref{eq:AcsAsym}
of $\confPrimA{\cs}$ to $\nullInfty[\pm]$.

\subsection{Inner products of conformal primary wavefunctions}
\label{sec:butter}
The inner products of conformal primary wavefunctions can be derived by starting
from the Klein-Gordon product of plane waves:
\begin{align}
  \kgProd{
  e^{\pm i \celFreq \nullUnit\cdot \minkX}
  }{
  e^{\pm i \celFreq' \nullUnit' \cdot \minkX}
  }
  &= \pm 2 (2\pi)^3 \celFreq\nullUnit^0  \deltaFunc[3](\celFreq \nullUnit - \celFreq' \nullUnit') \\
  &= \pm 2 (2\pi)^3  \celFreq^{-1}
    \deltaFunc(\celFreq - \celFreq') \celDeltaFunc(\celw,\celw')
    \;.
    \label{eq:biscuit}
\end{align}
Of course, the inner products of plane waves with oppositely signed frequencies
vanish. Thus, we find agreement with \cref{eq:solnProdModeProperties}, where the
metric $g$ between modes is indeed positive-definite. One can show, using the
polarization orthogonality condition \labelcref{eq:chocolate} and
\cref{eq:pureGaugeMellin,eq:biscuit}, that
\begin{equation}
  \minkProd{ \diff[\minkX] \confGauge{\confDim, \pm}[a]}{ \polar_b e^{\pm i \celFreq \nullUnit' \cdot \minkX_{\pm}} } = 0
  \;,
  \label{eq:pureGaugePlaneWaveMinkProd}
\end{equation}
which further implies, by \cref{eq:cake,eq:Mellin1} that
\begin{align}
  \minkProd{
  \diff[\minkX] \confGauge{\confDim, \pm}[a]
  }{
  \confMell{\confDim',\pm}[b]
  } &= 0
      \;,
  &
    \minkProd{
    \diff[\minkX] \confGauge{\confDim, \pm}[a]
    }{
    \confPrimA{\confDim',\pm}[b]
    } &= 0
        \;.
        \label{eq:pureGaugeMinkProd}
\end{align}
Actually, some conditions are attached to
\cref{eq:pureGaugePlaneWaveMinkProd,eq:pureGaugeMinkProd}. In particular, if
$\Re\confDim<1$ and $\celFreq=0$, then the evaluation of
\cref{eq:pureGaugePlaneWaveMinkProd} using \cref{eq:biscuit} gives an indefinite
result. Similarly, the first equality of \cref{eq:pureGaugeMinkProd} is not
necessarily true if $\Re(\confDim + \confDim'^*) \le 2$. For example, from the
results we will derive further below in a separate analysis of soft modes, it
will be clear that the first inner product in \cref{eq:pureGaugeMinkProd} is
nonzero for $\confDim=\confDim'=1$. Note, however, that the second inner product
of \cref{eq:pureGaugeMinkProd} obviously vanishes in this case because
$\confPrimA{1,\pm}$ is pure gauge.

Now, to evaluate the inner product between the conformal primaries
$\confPrimA{\confDim}$, it suffices to consider the $\confMell{\confDim}$ part
given by \labelcref{eq:Mellin1} in terms of plane waves. Applying a Mellin
transform to \cref{eq:biscuit}, one finds \cite{Donnay:2018neh}
\begin{align}
  \minkProd{
  \confMell{1+i\milneFreq,\pm}[a](\celw)
  }{
  \confMell{1+i\milneFreq',\pm}[\bar{b}](\celw')
  }
  &= \pm 2 (2\pi)^4 \regDeltaFunc*{2\reg}(\milneFreq - \milneFreq'^*) \,
    \celMet_{a b} \celDeltaFunc(\celw,\celw') \;,
    \label{eq:peach}
  \\
  &= \frac{1}{\mellCoeff{1+i\milneFreq,\pm} \mellCoeff{1-i\milneFreq'^*,\mp}}
    \minkProd{
    \confPrimA{1+i\milneFreq,\pm}[a](\celw)
    }{
    \confPrimA{1+i\milneFreq',\pm}[\bar{b}](\celw')
    }
    \;,
    \label{eq:pear}
\end{align}
where $\mathcal{K}^{\Delta,\pm}$ were defined in \eqref{eq:cake}. As with the plane waves, the inner products between $+$ and $-$ modes vanish and
again we have agreement with \cref{eq:solnProdModeProperties}. In
\cref{eq:peach}, we have defined \cite{Donnay:2020guq}
\begin{align}
  2\pi \regDeltaFunc*{\reg}(\milneFreq)
  &= \int_0^\infty \diff\celFreq\, \celFreq^{i\milneFreq -1} e^{-\reg \celFreq}
    = \reg^{-i\milneFreq} \Gamma(i\milneFreq)
  &
    (\Im \milneFreq <0),
    \label{eq:ham}
\end{align}
where the constraint $\Im\milneFreq <0$ is required for convergence of the integral. This regularized delta function has also appeared in \cite{Donnay:2020guq} in relation to conformal primary wavefunctions with $\Delta \in \mathbb{Z}$. Here, the $\epsilon$ regulator is directly inherited from the plane waves.

Therefore, let us now consider conformal primary wavefunctions
$\confPrimA{1+i\milneFreq}$ with an infinitesimally negative imaginary part in
$\milneFreq$. For such $\confPrimA{1+i\milneFreq}$, the regulated
$\regDeltaFunc*{2\reg}$-function in \cref{eq:peach} behaves similar to a true
$\deltaFunc$-distribution when integrated against functions of $\milneFreq$ (or
$\milneFreq'^*$) which are smooth near $\milneFreq\in\reals$
($\milneFreq'\in\reals$). This is because the rapid oscillations of the
$\reg^{-i(\milneFreq-\milneFreq'^*)}$ factor in
$\regDeltaFunc*{2\reg}(\milneFreq-\milneFreq'^*)$ dampens the integral
everywhere except near the $\milneFreq=\milneFreq'^*$ pole of
$\Gamma[i(\milneFreq-\milneFreq'^*)]$; moreover, the
$\reg^{-i(\milneFreq-\milneFreq'^*)}$ allows us to close the $\milneFreq$
($\milneFreq'^*$) integration contour around the upper half $\milneFreq$ plane
(lower half $\milneFreq'^*$ plane), thereby picking up the residue at the
$\milneFreq=\milneFreq'^*$ pole. In this sense, \cref{eq:peach} gives, in the
$\reg\to 0$ and $\Im\milneFreq,\Im\milneFreq'\to 0$ limit,
\begin{align}
  \label{eq:cpw-ip}
  \minkProd{
  \confPrimA{1+i\milneFreq,\pm}[a](\celw)
  }{
  \confPrimA{1+i\milneFreq',\pm}[\bar{b}](\celw')
  }
  &\eqIntegrated \pm 2(2\pi)^4
    \frac{\milneFreq \sinh(\pi \milneFreq) e^{\mp \pi \milneFreq}}{\pi(1 + \milneFreq^2)}
    \deltaFunc(\milneFreq - \milneFreq') \,
    \celMet_{a b}\celDeltaFunc(\celw,\celw')\;,
\end{align}
where we have used
\begin{align}
  \mellCoeff{1+i\milneFreq,\pm} \mellCoeff{1-i\milneFreq,\mp}
  &=
    \frac{\milneFreq \sinh(\pi\milneFreq) e^{\mp\pi\milneFreq}}{\pi(1+\milneFreq^2) } \;.
    \label{eq:raisin}
\end{align}
In this paper, we shall use the symbol $\eqIntegrated$ to mean equality in the
sense described above: when integrated, along the deformed principal series (or
its conjugate), against functions of $\milneFreq$ (or $\milneFreq'^*$) which are
smooth near $\milneFreq\in\reals$ (or $\milneFreq'^*\in\reals$), and the
$\reg\to 0$ and $\Im\milneFreq,\Im\milneFreq'\to 0$ limits are taken. Here we refer to the set of conformal dimensions $\Delta \in 1 + i \mathbb{R}$ as the principal series. We define the deformed principal series as the principal series deformed infinitesimally to the right, $1^+ + i\reals$.  Depending
on context, at $\lambda=0$, the smearing functions may be permitted to have
certain poles or required to have zeros; for example, in going from
\cref{eq:pear} to \cref{eq:cpw-ip}, it is clear that the smearing functions of
the latter can have $1/\milneFreq$ or $1/\milneFreq'^*$ simple poles, like
$1/\mellCoeff{1+i\milneFreq,\pm}$ or $1/\mellCoeff{1-i\milneFreq'^*,\mp}$.

The physical measurements of non-soft degrees of freedom should not be able to
distinguish infinitesimally separated $\milneFreq$. Thus, for these
considerations, it seems reasonable to smear modes
$\confPrimA{1+i\milneFreq}/\mellCoeff{1+i\milneFreq,\pm}$ against smooth
distributions over $\milneFreq$. In these cases, it suffices to consider
\cref{eq:cpw-ip}, which says that the conformal primary wavefunctions
$\confPrimA{1+i\milneFreq}$ form an orthogonal basis of solutions provided that $\lambda \in \reals$. It can be
further shown that this basis is complete, spanning the space of all
non-constant plane waves \cite{Pasterski:2017kqt} (up to gauge). To be precise, we have argued that non-soft degrees of
freedom are described by conformal primary wavefunctions
$\confPrimA{\confDim}/\mellCoeff{\confDim}$ smoothly smeared over $\confDim$ in
the deformed principal series.

However, as described in \cref{sec:sandwich}, (the operators associated to) the
soft modes $\confPrimA{\gold}$ and $\confPrimA{\cs}$ by themselves have special
physical significance. As a special case of \cref{eq:pureGaugeMinkProd}, we
already know that the inner product of $\confPrimA{\gold}$ with
$\confPrimA{\confDim}$ vanishes for $\confDim \in 1^+ + i\reals$, so it only
remains to study the inner products involving $\confPrimA{\cs}$. It is simplest
to do so by considering the Cauchy surfaces $\nullInfty[\pm]$, on which the
initial data for $\confPrimA{\cs}$ are given by \cref{eq:AcsAsym,eq:FcsAsym}. In
particular, one finds the following inner product with plane waves of nonzero
frequency \cite{Arkani-Hamed:2020gyp}\footnote{As remarked in footnote 17
  therein, \cite{Arkani-Hamed:2020gyp} drops the latter term of our
  \cref{eq:csToPm} in their definition (79) for ``$\confPrimA{\cs}$'', but that
  supposedly does not affect their calculation of inner products. Our
  \cref{foot:AcsLogSubtlety} further explains a difference between the first
  term of our \cref{eq:csToPm} and \cite{Arkani-Hamed:2020gyp}'s (79). Beyond
  this, the ``$\confPrimA{\gold}$'' and ``$\confPrimA{\cs}$'' written
  respectively in (78) and (79) of \cite{Arkani-Hamed:2020gyp} also contain
  extra factors of $\frac{1}{e^2}$ and $-\frac{1}{4\pi}$ relative to ours.
  Additionally, the ``$\nullUnit$'' and ``$\polar$'' defined in their (107) and
  (110) are $2$ and $\sqrt{2}$ times ours. Finally, we have kept a factor of
  $e^{-\celFreq \reg}$ in \cref{eq:pineapple} which is dropped in
  \cite{Arkani-Hamed:2020gyp} as they take $\reg\to 0$.}:
\begin{align}
  \minkProd{ \polar e^{\pm i\celFreq \nullUnit\cdot \minkX} }{ \confPrimA{\cs}(\celw')}
  &= \mp \frac{(4\pi)^2}{\celFreq} e^{-\celFreq \reg}
    \diff[\celw] \diff[\celw'] \celG(\celw,\celw')
    \;.
    \label{eq:pineapple}
\end{align}
Note that the consistency of \cref{eq:Mellin1,eq:biscuit,eq:syrup} with
\cref{eq:pineapple} is equivalent to the vanishing of inner products between
plane waves and the pure gauge piece $\diff[\minkX]\gamma^{\cs}$. Now, using
\cref{eq:cake} and by either Mellin transforming \cref{eq:pineapple} or applying
\cref{eq:peach} to \cref{eq:syrup}, one finds, for conformal primary
wavefunctions on the deformed principal series,
\begin{align}
  \minkProd{ \confPrimA{1+i\milneFreq,\pm}(\celw) }{ \confPrimA{\cs}(\celw') }
  &= -\frac{(4\pi)^2 i}{(\mp 2 i\reg)^{i\milneFreq} (1+i\milneFreq)}
    \diff[\celw] \diff[\celw'] \celG(\celw,\celw')
    \label{eq:apricot}
  \\
  &\eqIntegrated 0
    \;.
    \label{eq:minkProdAcsNonsoft}
\end{align}
Though it is not clear why this derivation of \cref{eq:apricot} should survive
the $\milneFreq\to 0$ limit, the inner product with $\confPrimA{\gold}$ can be
easily calculated at $\nullInfty[\pm]$ using the profile \labelcref{eq:tea} of
$\confGauge{\gold}$ and initial data \labelcref{eq:AcsAsym,eq:FcsAsym} of
$\confPrimA{\cs}$ there \cite{Donnay:2018neh}:
\begin{equation}
  \minkProd{ \confPrimA{\gold}(\celw) }{ \confPrimA{\cs}(\celw')}
  = -(4\pi)^2 i \,\diff[\celw] \diff[\celw'] \celG(\celw,\celw')
  \;.
  \label{eq:softMinkProd}
\end{equation}
Evidently, this does turn out to be the $\milneFreq\to 0$ limit of
\cref{eq:apricot}.

\subsection{Conformal primary operators}
\label{sec:sandwich}
The gauge potential admits the following decompositions in terms of conformal
primary wavefunctions
\begin{align}
  \begin{split}
    \opA
    &= \frac{1}{\sqrt{2} (2\pi)^2}
      \int \celVolForm(\celw)
      \int_{\reals - i 0^+} \diff\milneFreq \,
\left[(\mellCoeff{1+i\milneFreq,+})^{-1}\confPrimA{1+i\milneFreq,+}[a](\celw)
      a^a_{\milneFreq}(\celw)
      +
      (\cdots)^\dagger
       \right]
      \\
    &\phantom{{}={}}
      + \frac{1}{4\pi}
      \int\celVolForm(\celw)
      \int\celVolForm(\celw')
      \celG(\celw,\celw')
      \left[
      \cstrconj(\celw)
      \celCovD\cdot\confPrimA{\gold}(\celw')
      - \cstr(\celw)
      \celCovD\cdot\confPrimA{\cs}(\celw')
      \right]
\;.
  \end{split}
  \label{eq:confPrimModeExpansion}
\end{align}
We have separated the decomposition of non-soft and soft modes between the first
and second lines of \cref{eq:confPrimModeExpansion} respectively. Based on the
discussion following \cref{eq:cpw-ip}, it is reasonable to require the
coefficients $a_\milneFreq$ and $a_\milneFreq^\dagger$ of the non-soft modes be
smooth functions over $\milneFreq$ and $\milneFreq^*$ respectively for
$\milneFreq$ near $\reals$. For the soft modes, the celestial divergences and
the smearing against the celestial Green's function $\celG$ ensure that there is
not an over-complete description of the soft degrees of freedom --- despite
being celestial one-forms, recall that $\confPrimA{\gold}$ and $\confPrimA{\cs}$
are both exact in celestial space, so they each correspond to one celestial
scalar degree of freedom.

To quantize the theory, as described in \cref{sec:introFreeFields}, the
coefficients $a$, $a^\dagger$, $\cstrconj$, and $\cstr$ in the mode expansion
\eqref{eq:confPrimModeExpansion} are promoted to operators. In particular, the
$a$ are annihilation operators which annihilate the vacuum.
From the field operator $\opA$, one can also construct conformal primary operators
through symplectic products with conformal primary wavefunctions
\cite{Donnay:2020guq}, as described in \cref{sec:introFreeFields}:
\begin{align}
  \confPrimOp{\confDim,\pm}[a](\celw)
  &=
    \symp{\phSpace}(
    \confPrimA{\confDim,\pm}[a](\celw),
    \opA
    )
    = -i \minkProd{\confPrimA{\confDim,\pm}[a](\celw)}{\opA}
    =\confPrimOp{\confDim^*,\mp}[\bar{a}](\celw)^\dagger
    \;,
    \label{eq:confPrimOp}
  \\
  \csOp[][a](\celw)
  &=
    \symp{\phSpace}(
    \confPrimA{\gold}[a](\celw),
    \opA
    )
    = -i \minkProd{\confPrimA{\gold}[a](\celw)}{\opA}
    \;,
    \label{eq:csOp}
  \\
  \goldOp[][a](\celw)
  &=
    \symp{\phSpace}(
    \confPrimA{\cs}[a](\celw),
    \opA
    )
    = -i \minkProd{\confPrimA{\cs}[a](\celw)}{\opA}
    \;.
    \label{eq:goldOp}
\end{align}
As in \cref{eq:starAlgebra}, the commutators for these operators can be read off
from the inner products \labelcref{eq:pear,eq:apricot,eq:softMinkProd}:
\begin{align}
  \minkCommut{
  \confPrimOp{1+i\milneFreq,+}[a](\celw)
  }{
  \confPrimOp{1-i\milneFreq'^*,-}[b](\celw')
  }
  &= -2(2\pi)^4\mellCoeff{1+i\milneFreq,+} \mellCoeff{1-i\milneFreq'^*,-} \,
    \regDeltaFunc*{2\reg}(\milneFreq - \milneFreq'^*) \,
    \celMet_{a b} \celDeltaFunc(\celw,\celw')
  \\
  &\eqIntegrated
    -\frac{4(2\pi)^3\milneFreq \sinh(\pi\milneFreq) e^{\mp\pi\milneFreq}}{1+\milneFreq^2}
    \deltaFunc(\milneFreq - \milneFreq') \,
    \celMet_{a b} \celDeltaFunc(\celw,\celw')
    \;,
    \label{eq:minkCommutConfPrimOp}
  \\
  \minkCommut{
  \confPrimOp{1+i\milneFreq,\pm}(\celw)
  }{
  \goldOp(\celw)
  }
  &= \frac{(4\pi)^2 i}{(\mp 2 i\reg)^{i\milneFreq} (1+i\milneFreq)}
    \diff[\celw] \diff[\celw'] \celG(\celw,\celw')
    \eqIntegrated 0
    \;,
  \\
  \minkCommut{
  \csOp(\celw)
  }{
  \goldOp(\celw')
  }
  &= (4\pi)^2 i \,\diff[\celw] \diff[\celw'] \celG(\celw,\celw')
    \;,
    \label{eq:minkCommutSoft}
\end{align}
with other commutators vanishing exactly.
In terms of the operators appearing in
the mode expansion \labelcref{eq:confPrimModeExpansion}, the conformal primary
operators are given by
\begin{align}
  \begin{split}
    \confPrimOp{1+i\milneFreq,+}(\celw)
    &= - i \sqrt{2} (2\pi)^2 \mellCoeff{1+i\milneFreq,+}
      \int_{\reals + i 0^+} \diff\milneFreq'^* \,
      \regDeltaFunc*{2\reg}(\milneFreq - \milneFreq'^*) \,
      a_{\milneFreq'}(\celw)^\dagger
    \\
    &\phantom{{}={}}
      -\frac{4\pi}{(\mp 2 i\reg)^{i\milneFreq} (1+i\milneFreq)}
      \int \celVolForm(\celw')
      \diff[\celw]\celG(\celw,\celw') \cstr(\celw')
      \label{confPrimO}
  \end{split}
  \\
    &\eqIntegrated
      - i \sqrt{2} (2\pi)^2 \mellCoeff{1+i\milneFreq,+}
      a_{\milneFreq}(\celw)^\dagger\;,
  \\
  \csOp(\celw)
    &= -4\pi
      \int \celVolForm(\celw')
      \diff[\celw]\celG(\celw,\celw') \cstr(\celw')\;,
      \label{eq:csOpAsCstr1}
  \\
  \begin{split}
    \goldOp(\celw)
    &=
      \frac{4}{\sqrt{2}}
      \int \celVolForm(\celw')
      \left\{
      \int_{\reals - i 0^+} \diff\milneFreq\,
      \frac{
      a_{\milneFreq}(\celw') \cdot
      \diff[\celw'] [\diff[\celw] \celG(\celw,\celw')]
      }{
      \mellCoeff{1+i\milneFreq,+}
      (\mp 2 i\reg)^{i\milneFreq} (1+i\milneFreq)
      }
      + (\cdots)^\dagger
      \right\}
    \\
    &\phantom{{}={}} - 4\pi
      \int\celVolForm(\celw')
      \diff[\celw]\celG(\celw,\celw')
      \cstrconj(\celw')\;,
  \end{split}
  \\
    &\eqReg - 4\pi
      \int\celVolForm(\celw')
      \diff[\celw]\celG(\celw,\celw')
      \cstrconj(\celw')\;.
\end{align}
The commutation relations between the operators on the RHSs can then be deduced
by referring to \cref{eq:minkCommutConfPrimOp,eq:minkCommutSoft}. In particular,
the creation and annihilation operators satisfy the commutation
relations
\begin{align}
  [a_{a,\milneFreq}(\celw), a_{b,\milneFreq}(\celw')^{\dagger}]
  \eqIntegrated
  \deltaFunc(\milneFreq - \milneFreq')
  \celMet_{a\bar{b}}
  \celDeltaFunc(\celw,\celw') \;.
\end{align}

To understand the physical meaning of the soft operators $\csOp$ and $\goldOp$,
we recall the expression \labelcref{eq:cstrBlk} for the charges associated to
asymptotic symmetries (as discussed in \cref{sec:asymSym}). We see that $\csOp$
is precisely the asymptotic charge for a gauge transformation with parameter
$\confGauge{\gold}$:
\begin{align}
  \csOp[][a](\celw)
  &= \cstr\relax[\confGauge{\gold}[a](\celw)] \;.
    \label{eq:csOpAsCstr2}
\end{align}
Conversely, evaluating the symplectic product \labelcref{eq:csOp} at
$\nullInfty[\pm]$, where $\confGauge{\gold}$ has the asymptotic profile
\labelcref{eq:tea}, one can use the Green's function identity
\labelcref{eq:grape} to express $\cstr\relax[\gaugeParam]$ for arbitrary
$\gaugeParam$ in terms of $\csOp$:
\begin{align}
  \cstr\relax[\gaugeParam]
  &= \int \celVolForm(\celw) \,
    \gaugeParam(\celw) \,
    \cstr(\celw) \;,
  &
    \cstr(\celw)
  &= -\frac{1}{4\pi} \celCovD\cdot\csOp(\celw) \;.
    \label{eq:cstrAscsOp}
\end{align}
Above, we have also used the Green's function identity to
invert \cref{eq:csOpAsCstr1}. Altogether, we see that the operators $\csOp$
provide a complete description of the asymptotic charges of Maxwell
theory.\footnote{As mentioned in \cref{foot:softVsAsymCharge},
  \cref{eq:cstrBlk,eq:csOp} only give the soft component of asymptotic charge in
theories coupled to charged matter.} As expressed in \cref{eq:minkCommutSoft},
the $\goldOp$ or, alternatively, the
\begin{align}
  \cstrconj
  &\eqReg
    -\frac{1}{4\pi} \celCovD\cdot \goldOp
\end{align}
appearing in the mode expansion \labelcref{eq:confPrimModeExpansion} are the
canonical partners of the asymptotic charges. Thus, the $\goldOp$ are referred
to as Goldstone operators. For later discussions, it is helpful also to
introduce the integrated Goldstone operators
\begin{align}
  \goldOpI(\celw_1,\celw_2)
  &=
    \symp{\phSpace}(
    \confPrimA{\csI}(\celw_1,\celw_2),
    \opA
    )
    = -i \minkProd{\confPrimA{\csI}(\celw_1,\celw_2)}{\opA}
    =\int_{\celw_1}^{\celw_2} \goldOp(\celw)
    \label{eq:goldOpI}
\end{align}
associated to the integrated conformally soft modes \labelcref{eq:lettuce}.




\section{Beyond the Minkowski patch}
\label{sec:beyond-Minkowski}
\subsection{Inversion transformation}
\label{sec:invtrans}

\begin{figure}
  \centering
  \begin{subfigure}[t]{0.48\textwidth}
    \centering \includegraphics[scale=1.2]{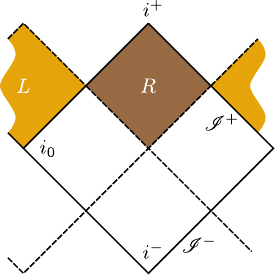}
    \caption{The original Minkowski
geometry and its $\rMilne$ Milne patch (brown), together with the $\lMilne$ patch (yellow) of the inverted Minkowski geometry.}
\label{fig:potato}
  \end{subfigure}
  \hfill
  \begin{subfigure}[t]{0.48\textwidth}
    \centering \includegraphics[scale=1.2]{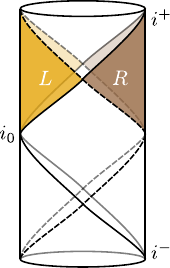}
    \caption{Conformal embedding into the Einstein static universe}
    \label{fig:einsteinStaticUniverse}
  \end{subfigure}
  \caption{}
  \label{fig:conformalDiagrams}
\end{figure}

Conformal invariance of pure Maxwell theory in four dimensions allows one to
conformally map the theory in Minkowski spacetime into a patch in the Einstein
static universe \cite{hawking_ellis_1973, Jorstad:2023ajr}. On the Einstein static universe, one can then extend the theory
beyond the original Minkowski patch. In particular, through another conformal
map, one may consider the theory from the perspective of the `inverse' Minkowski
patch centred on the $\spaceInfty$ of the original Minkowski spacetime. The
overall conformal transformation from the original Minkowski spacetime, with
coordinates $\minkX^\mu$, to the inverse Minkowski spacetime, with coordinates
\begin{align}
  \inv{\minkX}^{\mu} = \frac{\minkX^{\mu}}{\minkX^2} \;,
  \label{eq:soda}
\end{align}
is an inversion transformation, given by the above coordinate transformation and a compensating
Weyl rescaling of the metric by $\inv{\minkX}^{4}=\minkX^{-4}$.
The spacetime region $\minkX^2>0$, \ie{}
$\inv{\minkX}^2>0$, where the two opposing Minkowski patches overlap is mapped
to itself under the diffeomorphism implementing the above change of coordinates.
Since the Maxwell field is Weyl-invariant, a given field configuration $A$ in
the original Minkowski patch will appear here, from the perspective of the
inverse patch, simply as the push-forward $\inv{A}$ of $A$ under this
diffeomorphism. Explicitly,
\begin{align}
  \inv{A}_{\mu}(\inv{\minkX})
  &= \frac{\p \minkX^{\nu}}{\p \inv{\minkX}^{\mu}} A_{\nu}(\minkX) \;,
  &
    (\minkX^2>0).
    \label{eq:burger}
\end{align}
If $A$ is a solution, then $\inv{A}$ will also be a solution that can be evolved
from the Cauchy data in the $\inv{\minkX}^2>0$ region to the rest of the inverse
Minkowski patch (at least up to the possible presence of sources in the regions
$\inv{\minkX}^2 \le 0$ --- we will further discuss this later).

Our goal now will be to see how the conformal primary wavefunctions extend into
the inverse Minkowski patch. We begin by calculating
\begin{align}
  \frac{\p \minkX^{\nu}}{\p \inv{\minkX}^{\mu}}
  &= \minkX^2 \invRot_\mu^\nu(\minkX)
    = \inv{\minkX}^{-2} \invRot_\mu^\nu(\inv{\minkX}) \;,
  &
    \invRot_\mu^\nu(\minkX)
  &= \delta_{\mu}^{\nu} - \frac{2 \minkX_{\mu}\minkX^{\nu}}{\minkX^2}
    = \invRot_\mu^\nu(\inv{\minkX})
    \;.
\end{align}
Since the polarizations $m_{a; \mu}^{\pm}$ (introduced in \cref{eq:frame}) appearing
in the conformal primary wavefunctions \labelcref{eq:maxwell-cpw} are invariant
under the action of the inversion tensor, i.e.
\begin{align}
  \invRot_\mu^\nu(X) \, \confTensor{\pm}[a;\nu](\celw;\minkX)
  &\eqReg \confTensor{\pm}[a;\mu](\celw;\minkX)
    \eqReg \confTensor{\pm}[a;\mu](\celw;\inv{\minkX}) \;,
    \label{eq:invm}
\end{align}
we simply find that
\begin{align}
  \label{eq:inversion}
  \confPrimA<\inv>{\confDim,\pm}[a; \mu](\celw; \inv{\minkX})
  &\eqReg  (\inv{\minkX}_{\pm}^2)^{\confDim - 1}
    \confPrimA{\confDim,\pm}[a; \mu](\celw; \inv{\minkX})
    = e^{\pm i\pi (\confDim -1)}
    \confPrimA<\shad>{\confDim,\pm}[a;\mu](\celw;\inv{\minkX}).
\end{align}
Here $\tilde{A}^{\Delta, \pm}$ are the shadow conformal primary wavefunctions introduced in \cref{eq:confPrimAShad}. 
Because inversion is its own inverse, we also have, from
\cref{eq:inversion},
\begin{align}
  \confPrimA<\inv><\shad>{\confDim,\pm}[a;\mu](\celw; \inv{\minkX})
  &\eqReg e^{\mp i\pi(\confDim -1)} \confPrimA{\confDim,\pm}[a;\mu](\celw; \inv{\minkX}) \;.
    \label{eq:cherry}
\end{align}

Some explanation is needed to accompany this derivation. Firstly, note that the
$\eqReg$ relations are not exact equalities at finite $\reg$. However,
recall the purpose of the $\reg$ regulator in the conformal primary
wavefunction \labelcref{eq:maxwell-cpw} is to tell us how $\confPrimA{\confDim,\pm}$
passes through the $\nullUnit\cdot \minkX=0$ surface. Moreover, we are concerned
with applying the matching \labelcref{eq:burger} only in the spacetime region
$\minkX^2>0$. Within this spacetime region, the expressions of \cref{eq:invm,eq:inversion} related by $\eqReg$ pass through the $\nullUnit\cdot
\minkX=0$ surface in the same way and, in the $\reg\to 0$ limit, become
identical. Secondly, one may then wonder why we specified
$(\inv{\minkX}_\pm^2)^{\confDim-1}$ as the factor in the middle expression of
\cref{eq:inversion}, and not $(\inv{\minkX}^2)^{\confDim-1}$ or
$(\inv{\minkX}_\mp^2)^{\confDim-1}$ --- after all, the three choices are
indistinguishable in the aforementioned $\eqReg$ sense within
$\inv{\minkX}^2>0$. The reason we are forced to select
$(\inv{\minkX}_\pm^2)^{\confDim-1}$ is that, upon extending to
$\inv{\minkX}^2\le 0$, we are seeking a solution in the inverse Minkowski patch,
and Maxwell's equations will force us to select a consistent choice of constant
shift $\mp i \reg n^\mu$ of $\inv{\minkX}^\mu$ in both the prefactor
$(\inv{\minkX}_{\pm}^2)^{\confDim - 1}$ and $\confPrimA{\confDim,\pm}[a; \mu](q;
\inv{\minkX})$. As shown in \cref{eq:inversion}, the way that
$(\inv{\minkX}_\pm^2)^{\confDim-1}=(\inv{\minkX}^2\mp i \reg n\cdot
\inv{\minkX})^{\confDim-1}$ continues through $\inv{\minkX}^2=0$ determines the
numerical factor in the final expression.

On this subject, it will be helpful for later discussions to note that, under
inversion, discontinuities of conformal primary wavefunctions across the surface
$\inv{\minkX}^2=0$ remain sharp at finite $\reg$. That is, the
$\reg$ regulator does \textit{not} smooth out discontinuities in the LHSs of
\cref{eq:inversion,eq:cherry} across $\inv{\minkX}^2=0$, contrary to the RHSs. This
can be seen by viewing the $\reg$-regulator as an imaginary spacetime
diffeomorphism. In particular, note that the $\reg$-regulated conformal primary
wavefunction \labelcref{eq:maxwell-cpw} and its shadow
\labelcref{eq:confPrimAShad} in the original Minkowski spacetime are obtained
from their $\reg\to 0$ limits by shifting $X$ by $\mp i \epsilon n $. In doing so, the
discontinuity across any surface not tangential to $\refVec$ becomes smoothed
out. Similarly, the $\reg$-regulated inverted modes $\confPrimA<\inv>{\confDim,\pm}[a;
\mu](\celw; \inv{\minkX})$ and
$\confPrimA<\inv><\shad>{\confDim,\pm}[a;\mu](\celw; \inv{\minkX})$ should be
obtained from their $\reg\to 0$ limits through a pullback by $\reg$ along the flow generated by
$\mp i$ times
\begin{align}
  \frac{\partial \inv{\minkX}^\mu}{\partial \minkX^\nu}
  \refVec^\mu
  &=
    \inv{\minkX}^2 \refVec^\mu
    - 2 \inv{X}^\mu \inv{X}\cdot\refVec
    \;.
    \label{eq:refVecInversion}
\end{align}
Note that this is a conformal Killing vector of the inverse Minkowski spacetime
(and thus preserves the equations of motion). Since this vector is tangential to
the $\inv{\minkX}^2=0$ surface, the $\reg$-regulator, implemented by pulling
back along the generated flow, does not smooth out any discontinuities across this
surface.

To summarize, we have shown that, up to numerical factors and some
transformations of the $\reg$-regulator, spacetime inversions map spin-1
conformal primary solutions to their shadows. We expect the equivalence between
shadow transforms and inversions to be a general property of theories with
spacetime conformal invariance --- both transformations act on a conformal
primary wavefunction by flipping the sign of its Milne frequency \cite{Brown:2022miw,Chen:2023tvj,Jorstad:2023ajr}.
Note that a boost around the bifurcation surface that evolves time forward in the $R$ Milne patch will be a backward time evolution in $L$, therefore the corresponding modes are positive energy with respect to the boost flows  -- see also section \eqref{sec:tfd}.

\subsubsection{Soft modes}
\label{sec:honey}
From \cref{eq:inversion,eq:cherry}, we find that the $\confDim=1$
and logarithmic modes \labelcref{eq:log} transform under inversion as
\begin{align}
  \confPrimA<\inv>{1,\pm}(\inv{\minkX})
  &\eqReg \confPrimA{1,\pm}(\inv{\minkX}) \;,
  &
    \confPrimA<\inv>{\log,\pm}(\inv{\minkX})
  &\eqReg  - \confPrimA{\log,\pm}(\inv{\minkX}) \pm 2\pi i \confPrimA{1,\pm}(\inv{\minkX})
    \;.
\end{align}
and thus the Goldstone and conformally soft modes transform as 
\begin{align}
  \confPrimA<\inv>{\gold}(\inv{\minkX})
  &\eqReg \confPrimA{\gold}(\inv{\minkX})
    \;,
    \label{eq:cheddar}
  \\
  \confPrimA<\inv>{\cs'}(\inv{\minkX})
  &\eqReg 2\confPrimA{\gold}(\inv{\minkX})
    - \confPrimA{\cs}(\inv{\minkX}) \;.
    \label{eq:cheese}
\end{align}
For illustrative purposes, it is helpful to consider the $\confPrimA{\csI}$ mode
defined in \cref{eq:lettuce}:
\begin{align}
  \begin{split}
  \confPrimA<\inv>{\csI'}(\celw_1,\celw_2;\inv{\minkX})
  &\eqReg 2\diff[\minkX] \left[
   - \log
    (-\nullUnit_2\cdot \inv{\minkX}_+)
    +  \log
    (-\nullUnit_1\cdot \inv{\minkX}_+)
    \right]
    + (\minkX_+\to \minkX_-)
    \\
    &\phantom{{}={}} - \confPrimA{\csI}(\celw_1,\celw_2;\inv{\minkX})
    \;.
  \end{split}
  \label{eq:citrus}
\end{align}
(The reason for the primes on $\confPrimA<\inv>{\cs'}$ and
$\confPrimA<\inv>{\csI'}$ in \cref{eq:cheese,eq:citrus} will be explained
shortly below.)

\begin{figure}
  \centering
  \begin{subfigure}[t]{0.48\textwidth}
    \centering \includegraphics[scale=1.2]{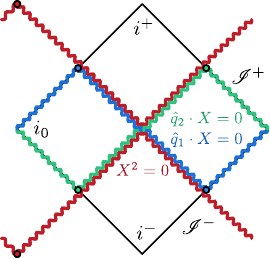}
    \caption{}
    \label{fig:croissant}
  \end{subfigure}
  \hfill
  \begin{subfigure}[t]{0.48\textwidth}
    \centering \includegraphics[scale=1.2]{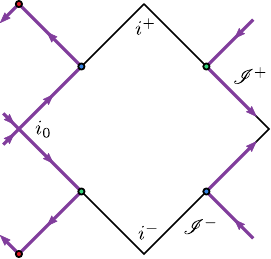}
    \caption{}
    \label{fig:bun}
  \end{subfigure}
  \caption{(a) Redrawing the shockwaves of \cref{fig:bacon} in the Minkowski
      patch and extending the spherical shockwave beyond. The shockwave colors match those in \cref{fig:bacon}. Note that in general the planar shocks may be tangent to arbitrary points on the celestial sphere. Their components along $\mathscr{I}^{\pm}$ are an artifact of the conformal compactification. (b) Sources corresponding to the shockwave configuration in \cref{fig:croissant}. The left and right edges of each figure are identified.}
  \label{fig:bread}
\end{figure}

Turning on the conformally soft mode $\confPrimA{\cs}$ or its integral
$\confPrimA{\csI}$ changes the asymptotic charge \labelcref{eq:sundae} because
it modifies the electric field near $\spaceInfty$, as written in
\cref{eq:FcsAsym,eq:asymChargeCSI}. This becomes obvious when viewing the
conformally soft mode from the perspective of the Einstein static universe. In
particular, redrawing the shockwaves displayed in \cref{fig:bacon} in the
Minkowski patch, one obtains the picture shown in \cref{fig:croissant}. There, we
see that the planar shockwaves can be viewed as emanating from and collapsing to
points at the intersection of $\nullInfty[\pm]$ with the lightcone $\minkX^2=0$;
moreover, the spherical shockwave on this lightcone can be extended beyond the
Minkowski patch where it too is seen to emanate from one point and collapse to
another. By tracing the electric field lines shown in \cref{fig:bacon} to their
start and end points, one finds that sources run along the trajectories shown in
\cref{fig:bun} which has cusps which are precisely responsible for the emission
and absorption of the shockwaves. Due to these sources (more precisely, the blue
and green shockwaves produced by them), the asymptotic charge
\labelcref{eq:sundae} evaluated in this field configuration does not vanish but
rather is given by \cref{eq:asymChargeCSI}.\footnote{In contrast, had there been no
artifacts near $\spaceInfty$, then the asymptotic charge \labelcref{eq:sundae}
will obviously vanish when viewed from the perspective of the inverse Minkowski
patch, where the $\nullInftyBdy{\pm}{\mp}$ of the original Minkowski patch
shrinks to infinitesimal surfaces.} While \cref{fig:bread} illustrates the
integrated conformally soft mode $\confPrimA{\csI}(\celw_1,\celw_2)$, the
conformally soft mode $\confPrimA{\cs}$ is simply obtained in the
$\celw_1\to\celw_2$ limit (and dividing by the separation in celestial space). That is, rather than have finitely separated branches of the source
trajectory with opposite charges in \cref{fig:bun}, the sources of
$\confPrimA{\cs}$ would be a dipole. By default in the absence of primes, we shall
consider extensions of the conformally soft modes $\confPrimA{\cs}$ and
$\confPrimA{\csI}$ to the Einstein static universe which include sources
passing through $\spaceInfty$, as illustrated for $\confPrimA{\csI}$ in
\cref{fig:bread}.

\begin{figure}
  \centering
  \begin{subfigure}[t]{0.48\textwidth}
    \centering \includegraphics[scale=1.2]{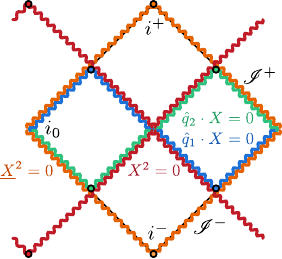}
    \caption{Adding another shockwave (orange) to \cref{fig:croissant} along
      $\nullInfty[\pm]$ of the original Minkowski patch.}
    \label{fig:lemon}
  \end{subfigure}
  \hfill
  \begin{subfigure}[t]{0.48\textwidth}
    \centering \includegraphics[scale=1.2]{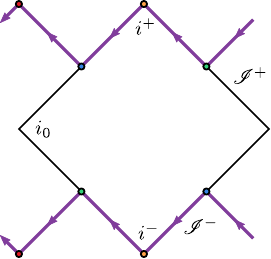}
    \caption{Sources corresponding to \cref{fig:lemon}, which now avoid
      the interiors of both the original and inverse Minkowski patches.}
  \end{subfigure}
  \par\bigskip
  \begin{subfigure}[t]{0.48\textwidth}
    \centering \includegraphics[scale=1.2]{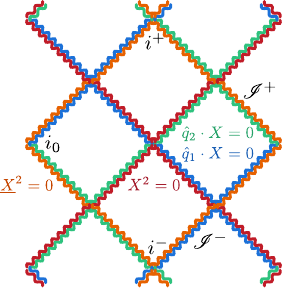}
    \caption{By further extending the shockwaves indefinitely, a completely
      source-free solution on the Einstein static universe is obtained.}
    \label{fig:melon}
  \end{subfigure}
  \caption{}
  \label{fig:avocado}
\end{figure}

This is to be contrasted with the extensions of the conformally soft modes
written in \cref{eq:cheese,eq:citrus} which are source-free in both the original
and inverse Minkowski patches --- recall they were constructed using
\cref{eq:inversion,eq:cherry}, which are required to be solutions to the free
Maxwell's equations. In fact, \cref{eq:citrus} is better described by the
configurations sketched in \cref{fig:avocado}, where the sources, if they exist,
now avoid the interiors of both the original and inverse Minkowski patches. We
will use primes in the symbols $\confPrimA{\cs'}$ and $\confPrimA{\csI'}$ to
indicate these source-free extensions beyond the original Minkowski patch, as
described by \cref{eq:cheese,eq:citrus,fig:avocado}. Compared to $\confPrimA{\cs}$
and $\confPrimA{\csI}$, the source-free extensions $\confPrimA{\cs'}$ and
$\confPrimA{\csI'}$ have extra shockwaves running along $\nullInfty[\pm]$.
\label{textBegin:AcsPrimeSharpShockwave}
From
the discussion around \cref{eq:refVecInversion}, we learn that this shockwave
is sharp, even at finite $\reg$. Thus, $\confPrimA{\cs'}$ and
$\confPrimA{\csI'}$ are actually
infinite energy configurations from the perspective of the Einstein static
universe (but this is not an issue if the Einstein static universe is only a
tool for us to study the Minkowski and Milne theories).
\label{textEnd:AcsPrimeSharpShockwave}

Notice that $\confPrimA{\cs}$ and $\confPrimA{\csI}$ are indistinguishable from
$\confPrimA{\cs'}$ and $\confPrimA{\csI'}$ within the interior of the
original Minkowski patch. For instance, the asymptotic charges are the same,
since the RHS of \cref{eq:sundae} should really be thought of as evaluated on a limiting
surface from within the (original) Minkowski interior. Moreover, it must be that
the Minkowski inner product is indifferent to the various extensions of
conformally soft modes beyond the Minkowski interior:
\begin{align}
  \minkProd{ \confPrimA{\cs}}{\bullet }
  &= \minkProd{ \confPrimA{\cs'}}{\bullet } \;,
  &
    \minkProd{ \confPrimA{\csI}}{\bullet }
  &= \minkProd{ \confPrimA{\csI'}}{\bullet }\;.
    \label{eq:cscsPrimEquivalence}
\end{align}
Thus, from the Minkowski perspective, they merely provide alternative
descriptions of the same physical configurations. The source-free extension of
the conformally soft modes, by construction, more naturally meshes with the
extensions of the non-soft $\confDim\ne 1$ modes. On the other hand, we will see
in \cref{sec:confPrimMilne} that, to relate the Milne and Minkowski symplectic products, it is
helpful to consider $\confPrimA{\cs}$ or $\confPrimA{\csI}$. Moreover, in
\cref{sec:apple}, the
presence of sources in these modes will motivate a modification to the prescription
described by \cite{Donnelly:2016auv} for gluing subregions.

\subsection{Complementary Milne patches}
\label{sec:crepe}

In this section we characterize the conformal primary wavefunctions associated to a subregion in Minkowski spacetime and its complement inside the Einstein static universe. We also describe their conformally soft and Goldstone counterparts, as well as their relation to the conformal primary wavefunctions of Minkowski spacetime. 

\subsubsection{Milne and hyperbolic geometry}
\label{sec:milnegeo}

Let us begin by considering the Milne spacetime $\rMilne$ given by the patch of
the (original) Minkowski spacetime to the future of the origin $\minkX^\mu=0$ --- see
\cref{fig:conformalDiagrams}. As we review now, the geometry of this Milne patch is
conformally $\reals\times \hyp{3}$, where $\hyp{3}$ is the unit
three-dimensional hyperboloid.

We begin by defining the Milne time $\milnet$ and Hyperbolic bulk coordinate
$\hyps$ in terms of the $u$ and $r$ coordinates introduced in
\cref{eq:cupcake-intro}:
\begin{align}
  u &= \frac{e^{\milnet-\hyps}}{2} \;,
  &
    r &=  e^{\milnet+\hyps} \;.
\end{align}
Equivalently, the Cartesian Minkowski coordinates \labelcref{eq:cupcake-intro} are given by
\begin{align}
  \minkX^\mu &= e^{\milnet} \timeUnit^\mu \;,
      \\
  \timeUnit^\mu &= \frac{1}{2}
                  \left(
                    e^{-\hyps} + e^\hyps (1+z\bar{z}) \,,
                    e^\hyps (z+\bar{z}) \,,
                    -i e^\hyps (z-\bar{z}) \,,
                    -e^{-\hyps} + e^\hyps (1-z\bar{z})
                  \right)
                  \;.
\end{align}
For each instant of Milne time
\begin{align}
  \milnet
  &= \frac{1}{2}\log\left(
    2\minkr \minku
    \right)
    = \frac{1}{2}\log(-\minkX^2)
    \;,
    \label{eq:milnet}
\end{align}
this gives the usual Poincar\'e parametrization of a three-dimensional
hyperboloid, \ie{} Euclidean $\ads{3}$, with curvature radius\footnote{As mentioned
  in \cref{foot:noodles}, recall that we are suppressing in this paper a unit of length
  $\lenConst$ which is implicitly rescaling certain dimensionful
  quantities.} given by $e^\milnet$. Moreover,
$e^{-\hyps}$ and
\begin{align}
  \fgr &= e^{-2\hyps}
         \label{eq:fgr}
\end{align}
are the standard Poincar\'e and Fefferman-Graham bulk coordinates.
Indeed, the Minkowski metric becomes
\begin{align}
  \label{eq:Milne-metric}
  \diff s^2
  &= e^{2\milnet}\left(-\diff \milnet^2 + \diff s_{\hyp{3}}^2\right) \;,
\end{align}
with the geometry on $\hyp{3}$ expressed in terms of a
planar foliation\footnote{There is also a spherical foliation
  \begin{align}
    \diff s_{\hyp{3}}^2
    &= \diff \varsigma^2 + \sinh^2\varsigma\, \diff s_{\sphUnit{2}}^2 \;,
    &
      \diff s_{\sphUnit{2}}^2
    &= \frac{4\, \diff \zeta\, \diff\bar{\zeta}}{(1+\zeta\bar{\zeta})^2} \;,
      \label{eq:stereometric}
  \end{align} 
  obtained from coordinates which parametrize the unit hyperboloid
  $\hyp{3}$ through $
    \timeUnit^\mu
    = \left( 
      \cosh\varsigma \,,
      \stereo \sinh\varsigma 
      \right)$
  where 
  \begin{align}
    \stereo[^i]
    &= \left( 
      \frac{\zeta + \bar{\zeta}}{1 + \zeta\bar{\zeta}} \,,
      \frac{-i(\zeta - \bar{\zeta})}{1 + \zeta\bar{\zeta}}\,,
      \frac{1-\zeta\bar{\zeta}}{1 + \zeta\bar{\zeta}}
      \right)
  \end{align}
  gives the stereographic parametrization of the unit normal to the $\sphUnit{2}$.
}
\begin{align}
  \diff s_{\hyp{3}}^2
  &= \diff\hyps^2 + e^{2\hyps} \diff s_{\celPlane}^2
    = \frac{\diff\fgr^2}{4\fgr^2} + \frac{\diff s_{\celPlane}^2}{\fgr}\;,
     \label{eq:Poincare-metric}
\end{align}
where
\begin{align}
  \diff s_{\celPlane}^2
  &= \celMet_{ab} \diff z^a \diff z^b
    = \diff z \, \diff \bar{z} 
\end{align}
is the same metric on celestial space as introduced in
\cref{eq:yogurt}. The Minkowski volume form \cref{eq:minkOrientation} and
$\hyp{3}$ volume form read
\begin{align}
  \volForm
  &= e^{4\milnet} \, \diff\milnet \wedge
    \hypVolForm
    \;,
    &
    \hypVolForm
    &=
      e^{2\hyps} \diff\hyps \wedge \celVolForm
      = -\frac{\diff\fgr \wedge \celVolForm}{2\fgr^2}
      \;.
\end{align}

\subsubsection{Conformal primary wavefunctions and Milne patches}
\label{sec:confPrimMilne}
The conformal primary solutions \labelcref{eq:maxwell-cpw} are particularly
well-adapted to the hyperbolic foliation of the Milne patch described in
\cref{sec:milnegeo} (and more generally the (A)dS foliation of Minkowski
spacetime). In particular, each conformal primary wavefunction
$\confPrimA{1+i\milneFreq}$ has definite Milne energy $\milneFreq$ in the
$\reg\to 0$ limit:
\begin{align}
  \lie{\partial_\milnet}
  \confPrimA{1+i\milneFreq}
  &\eqReg -i\milneFreq \confPrimA{1+i\milneFreq}
  &(\text{Milne interior})\;,
\end{align}
where $\lie{\bullet}$ is the spacetime Lie derivative which in this case reduces to $\partial_{\tau}$. As we will see below and
in \cref{sec:tfd}, this makes the conformal primary wavefunctions an ideal set
of solutions to consider when studying the entanglement of Milne patches. In
\cref{sec:apple}, we will further connect the conformally soft ($\lambda = 0$) modes $\confPrimA{\cs}$ and
$\confPrimA{\gold}$ to edge modes and boundary degrees of
freedom for Milne patches.

To prepare for these interesting discussions, we shall begin here by examining
how the conformal primary wavefunctions restrict to the Milne patch $\rMilne$,
belonging to the original Minkowski spacetime, and a complementary Milne patch
$\lMilne$ analogously placed in the inverse Minkowski spacetime. These two
complementary Milne patches are illustrated in \cref{fig:conformalDiagrams} (see also \cref{fig:cauchySlices}), and are
respectively the Cauchy developments of two hemispheres of a full $\sphUnit{3}$
Cauchy slice of the Einstein static universe.

While the conformal primary wavefunctions $\confPrimA{\confDim}$ were introduced
in \cref{sec:steak} in the original Minkowski spacetime which houses the
$\rMilne$ patch, we also saw in \cref{sec:invtrans} how these solutions can be
extended to the Einstein static universe, and to the inverse
Minkowski spacetime containing the $\lMilne$ patch. Let us view each conformal
primary wavefunction $\confPrimA{\confDim}$ as a linear combination of two
solutions, labelled by $\modifiedSymbol[\lMilne]\bullet$ and
$\modifiedSymbol[\rMilne]\bullet$, which, in the $\reg\to 0$ limit, each vanish
in the interior of the Milne patch complementary to its label. One can view each
of these two pieces as the result of evolving the initial data of
$\confPrimA{\Delta}$ with restricted support to a hemisphere of the
aforementioned $\sphUnit{3}$ Cauchy surface (with some $\reg$-fuzziness at the
equator). In light of \cref{eq:inversion}, we shall write this decomposition for
non-integer $\confDim$ (leaving the soft case for later discussion below) as
\begin{align}
  \label{eq:LR-dec}
  \confPrimA{\confDim,\pm}
  &= e^{\pm i\pi (\confDim -1)}\, \confPrimA<\shad>[\lMilne]{\confDim}
    + \confPrimA[\rMilne]{\confDim} \;.
  &
    (\confDim\not\in\integers)
\end{align}
In the $\reg\to 0$ limit, the solution we call $\confPrimA[\rMilne]{\confDim}$
agrees with $\confPrimA{\confDim,\pm}$ in the interior of $\rMilne$ but vanishes
in the interior of $\lMilne$, while $e^{\pm i\pi (\confDim -1)}\,
\confPrimA<\shad>[\lMilne]{\confDim}$ agrees with $\confPrimA{\confDim,\pm}$ in
the interior of $\lMilne$ but vanishes in the interior of $\rMilne$. Our
notation for the latter is motivated by the perspective of the inverse Minkowski
spacetime, which according to \cref{eq:inversion}, sees
$\confPrimA{\confDim,\pm}$ as a shadow conformal primary wavefunction with a
proportionality factor $e^{\pm i\pi (\confDim -1)}$.

\Cref{eq:LR-dec} is exactly analogous to the Unruh construction a wavefunction
in the span of positive- (or negative-) Minkowski-frequency solutions from
Rindler wavefunctions of definite Rindler frequency
\cite{Unruh:1976db}.\footnote{See also section 4.5 of \cite{Birrell:1982ix} for a
  review.} Indeed, the $\lMilne$ and $\rMilne$ Milne patches are conformally
related to two complementary Rindler patches, with $\milnet$ being mapped to
Rindler time. The Minkowski spacetime to which these Rindler patches belong is
shifted on the Einstein static universe with respect to the original Minkowski
spacetime; however, the span of positive-Minkowski-frequency solutions for any
Minkowski patch agrees with the span of positive-Einstein-static-frequency
solutions.\footnote{Here, we are omitting subtleties involving the soft modes,
  which we treat carefully in \cref{sec:apple}.} We will elaborate on these
ideas further in \cref{sec:entanglement} under the guise of discussing vacua and
Bogoliubov transformations.

\label{textBegin:regMinkLR}
For the $\confPrimA<\shad>[\lMilne]{\confDim}$ and
$\confPrimA[\rMilne]{\confDim}$ modes to be viewed as smooth solutions in
Minkowski spacetime, we shall think of them as inheriting the $\reg$ regulator
of the Minkowski modes $\confPrimA{\confDim,\pm}$. This shall be implicit
whenever we relate $\lMilne$ and $\rMilne$ modes to Minkowski modes, as in
\cref{eq:LR-dec}, or when we evaluate the \emph{Minkowski} inner product of the
$\lMilne$ and $\rMilne$ modes in \cref{eq:minkProdRRLL,eq:minkProdLR} below.
However, in later discussions, we shall also  assign the symbols
$\confPrimA<\shad>[\lMilne]{\confDim}$ and $\confPrimA[\rMilne]{\confDim}$
another meaning as simply $\reg$-free modes of definite Milne-frequency, which
shall be implicit when taking the perspective of the Milne theories.\footnote{Recall
$\reg$ regulates the conformal primary wavefunctions
$\confPrimA{\confDim}(\nullUnit,\minkX)$ at the
$\nullUnit\cdot\minkX=\nullUnit\cdot\inv{\minkX}=0$, $\minkX^2=0$, and
$\inv{\minkX}^2=0$ surfaces, which all lie outside the interiors of the
$\lMilne$ and $\rMilne$ patches.}
\label{textEnd:regMinkLR}

For non-integer $\confDim$, we can invert \cref{eq:LR-dec}, to obtain
\begin{align}
  \confPrimA<\shad>[\lMilne]{\confDim}
  &= i\, \frac{
    \confPrimA{\confDim, +} - \confPrimA{\confDim, -}
    }{
    2 \sin \pi \confDim
    }
    \;,
  &
    \confPrimA[\rMilne]{\confDim}
  &= i\, \frac{
    e^{-i\pi\confDim} \confPrimA{\confDim, +} - e^{i\pi\confDim} \confPrimA{\confDim, -}
    }{
    2 \sin \pi\confDim
    }
    \;,
  &
    (\confDim\not\in\integers).
    \label{eq:inv}
\end{align}
Note that \cref{eq:inv} can be taken as the definition of the (non-soft)
$\confPrimA<\shad>[\lMilne]{\confDim}$ and $\confPrimA[\rMilne]{\confDim}$ modes
as Minkowski solutions, including how the $\reg$ regulator is inherited
from the $\confPrimA{\confDim,\pm}$ modes. The Minkowski inner products of the
$\lMilne$ and $\rMilne$ modes can then be evaluated using \cref{eq:pear}:
\begin{align}
  -\minkProd{
  \confPrimA<\shad>[\lMilne]{1+i\milneFreq}[a](\celw)
  }{
  \confPrimA<\shad>[\lMilne]{1+i\milneFreq'}[\bar{b}](\celw')
  }
  &= \minkProd{
    \confPrimA[\rMilne]{1+i\milneFreq}[a](\celw)
    }{
    \confPrimA[\rMilne]{1+i\milneFreq'}[\bar{b}](\celw')
    }
  \\
  &= (2\pi)^4 \milneCoeff{1+i\milneFreq} \milneCoeff{1-i\milneFreq'^*}
    \sinh\left[ \frac{\pi(\milneFreq+\milneFreq'^*)}{2} \right]
    \regDeltaFunc*{2\reg}(\milneFreq - \milneFreq'^*) \,
    \celMet_{a b} \celDeltaFunc(\celw,\celw')
    \label{eq:minkProdRRLL}
  \\
  &\eqIntegrated (2\pi)^4 \frac{\lambda}{\pi(1+\lambda^2)}
    \deltaFunc(\milneFreq - \milneFreq') \,
    \celMet_{a b} \celDeltaFunc(\celw,\celw')\;,
    \label{eq:limitMinkProdRRLL}
\end{align}
while
\begin{align}
  \minkProd{
  \confPrimA<\shad>[\lMilne]{1+i\milneFreq}[a](\celw)
  }{
  \confPrimA[\rMilne]{1+i\milneFreq'}[\bar{b}](\celw')
  }
  &= (2\pi)^4 \milneCoeff{1+i\milneFreq} \milneCoeff{1-i\milneFreq'^*}
    \sinh\left[ \frac{\pi(\milneFreq-\milneFreq'^*)}{2} \right]
    \regDeltaFunc*{2\reg}(\milneFreq - \milneFreq'^*) \,
    \celMet_{a b} \celDeltaFunc(\celw,\celw')
    \label{eq:minkProdLR}
  \\
  &\eqIntegrated 0
    \label{eq:limitMinkProdLR}
    \;.
\end{align}
Here we have defined
\begin{align}
\label{eq:Lnorm}
  \milneCoeff{\confDim}
  &= \frac{1-\confDim}{\sin(\pi\confDim) \,\Gamma(1+\confDim)} \;,
\end{align}
which satisfies
\begin{align}
  \milneCoeff{1+i\milneFreq} \milneCoeff{1-i\milneFreq}
  &= \frac{\lambda}{\pi(1+\lambda^2)\sinh(\pi\lambda)}
    \;.
\end{align}
The meaning of the $\eqIntegrated$ relations is the same as introduced below
\cref{eq:raisin} --- namely, equality when the previous line is smeared against
functions of $\milneFreq$ or $\milneFreq'^*$ that are smooth\footnote{For
  \cref{eq:limitMinkProdRRLL,eq:limitMinkProdLR}, the smearing functions are
  permitted to have simple poles respectively in $\milneFreq+\milneFreq'^*$ and
  $\milneFreq-\milneFreq'^*$.} near $\reals$ and the $\reg\to 0$ and
$\Im\milneFreq,\Im\milneFreq'\to 0$ limits are taken. Again, such a smearing
seems physically justified when considering non-soft degrees of freedom.

\begin{figure}
  \centering
  \begin{subfigure}[t]{0.48\textwidth}
    \centering \includegraphics[scale=1.2]{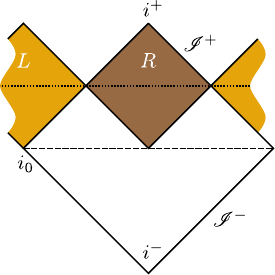}
    \caption{Two Cauchy slices related by Einstein static time translation. }
    \label{fig:cauchySlicesEinsteinStatic}
  \end{subfigure}
  \hfill
  \begin{subfigure}[t]{0.48\textwidth}
    \centering \includegraphics[scale=1.2]{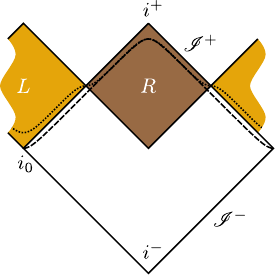}
    \caption{Two nearby Cauchy slices.}
    \label{fig:cauchySlicesNull}
  \end{subfigure}
  \caption{Dashed and dotted curves respectively mark Cauchy slices for the
    original Minkowski patch and the $\lMilne$ and $\rMilne$ Milne patches. The dashed surface may be regarded as the union of Cauchy slices associated with Rindler patches inside the original Minkowski spacetime.}
  \label{fig:cauchySlices}
\end{figure}

Though we have arrived at \cref{eq:limitMinkProdRRLL,eq:limitMinkProdLR} above
using \emph{Minkowski} inner products, we note that these expressions may also
be interpreted as giving inner products in the \emph{Milne} theory. This is
because, as illustrated in \cref{fig:cauchySlices}, a Minkowski Cauchy surface
(plus $\spaceInfty$) serves also as a Cauchy surface for the Einstein static
universe; similarly, the sum of the Cauchy surfaces for the $\lMilne$ and
$\rMilne$ Milne patches (plus the bifurcation surface between them) is also a
Cauchy surface for the Einstein static universe. Moreover, one can more directly
see an agreement between certain limits of Cauchy surfaces of Minkowski
spacetime and the Milne patches --- this is illustrated in
\cref{fig:cauchySlicesNull}. Thus, we find
\begin{align}
    \minkProd{\bullet}{\bullet}
  &\eqIntegrated \solnProd[\lMilne]{\bullet}{\bullet}
    +\solnProd[\rMilne]{\bullet}{\bullet} \;,
    \label{eq:minkAsSumOfLR}
\end{align}
where 
$\minkProd{\bullet}{\bullet}$, $\solnProd[\lMilne]{\bullet}{\bullet}$, and
$\solnProd[\rMilne]{\bullet}{\bullet}$ are the Minkowski, $\lMilne$ and
$\rMilne$ inner products. In \cref{sec:apple}, we will more carefully treat (and
eliminate by gauge fixing and constraints) possible boundary contributions to
the Milne symplectic form --- let us set these issues aside for now so that
inner products and symplectic forms are given purely by integrating the
symplectic potential density $\sympD$. Still, there is an underlying assumption
in \cref{eq:minkAsSumOfLR} that the modes under consideration continue across
the past boundary of $\lMilne$ in such a way that these integrals on the
different Cauchy surfaces drawn in \cref{fig:cauchySlices} give the same
answers. In fact, the modes \labelcref{eq:LR-dec,eq:inv} are not smooth in the
conventional sense across this surface. Nonetheless, for $\Re \milneFreq\ne 0$,
\cref{eq:LR-dec} has rapid oscillations $\sim(\inv{\minkX}^2)^{i\milneFreq}$
near this surface which prevent the discontinuities here from contributing to
the integrals, as the two Cauchy surfaces of
\cref{fig:cauchySlicesEinsteinStatic} are evolved to each other. Thus,
\cref{eq:minkAsSumOfLR} applies at least for the non-soft modes.

Let us now turn to the $\confDim=1$ soft modes. Naively inserting integer
$\confDim$ into \cref{eq:LR-dec}, one finds that the RHS cannot resolve the
difference $\confPrimA{\confDim,+}-\confPrimA{\confDim,-}$ which, in the
$\reg\to 0$ limit, has no support in the interior of either Milne patch.
Correspondingly, if one tries to take the integer $\confDim$ limit of
\cref{eq:inv}, one finds divergent contributions proportional to
$\confPrimA{\confDim,+}-\confPrimA{\confDim,-}$. Happily, at least at
$\confDim=1$, this difference is trivial gauge as described in
\cref{sec:minkSoftModes}. So, one may be tempted to define
$\confPrimA<\shad>[\lMilne]{\confDim,1}$ and $\confPrimA[\rMilne]{\confDim,1}$
through the limit of \cref{eq:inv} with this divergent trivial gauge piece
subtracted off:
\begin{align}
  \confPrimA<\shad>[\lMilne]{1}
  &\eqDubious \lim_{\confDim\to 1}\left(
    \confPrimA<\shad>[\lMilne]{\confDim}
    -i\, \frac{
    \confPrimA{1, +} - \confPrimA{1, -}
    }{
    2 \sin \pi \confDim
    }
    \right)
    =
    \frac{
    \partial_\confDim (\confPrimA{\confDim,+} - \confPrimA{\confDim,-})_{\confDim=1}
    }{
    2 \pi i
    }
    \;,
    \label{eq:LSoftModeDubious}
  \\
  \confPrimA[\rMilne]{1}
  &\eqDubious \lim_{\confDim\to 1}\left(
    \confPrimA[\rMilne]{\confDim}
    +i\, \frac{
    \confPrimA{1, +} - \confPrimA{1, -}
    }{
    2 \sin \pi \confDim
    }
    \right)
    = \confPrimA{\gold} - \confPrimA<\shad>[\lMilne]{1}
    \label{eq:RSoftModeDubious}
    \;.
\end{align}
Indeed, by their construction from \cref{eq:LR-dec}, these have in the $\reg\to 0$
limit the desired property
\begin{align}
  \confPrimA[\lMilne]{\gold}
  &\eqReg \begin{cases}
        \confPrimA{\gold} & \text{in $\lMilne$}
        \\
        0 & \text{in $\rMilne$}
      \end{cases}
    \;,
  &
    \confPrimA[\rMilne]{\gold}
  &\eqReg \begin{cases}
       0 & \text{in $\lMilne$}
       \\
       \confPrimA{\gold} & \text{in $\rMilne$}
     \end{cases}
    \;.
    \label{eq:ALRGoldInsidePatches}
\end{align}
Unfortunately, the modes \labelcref{eq:LSoftModeDubious,eq:RSoftModeDubious} are
unsatisfactory from the Minkowski perspective, where the matching condition
\labelcref{eq:antipodalA} is violated --- in fact, \cref{eq:LSoftModeDubious} is
precisely the solution whose pullback to $\nullInfty[\pm]$ is written in
\cref{eq:LSoftModeDubiousAsym}.

Instead, we propose that the appropriate decomposition of Minkowski soft modes
into Milne soft modes is obtained from examining the inversion transformations
\labelcref{eq:cheddar,eq:cheese}. After all, as we've already seen in
\cref{sec:minkSoftModes}, the Minkowski soft modes $\confPrimA{\gold}$ and
$\confPrimA{\cs}$ do satisfy the matching condition \labelcref{eq:antipodalA}
and so must any of their linear combinations (\eg{}
\labelcref{eq:milneSoftModesDressed,eq:milneSoftModesDressedPrime} below). The
transformations \labelcref{eq:cheddar,eq:cheese} motivate the
decompositions\footnote{\Cref{eq:cheddar,eq:cheese} more directly lead to
  \cref{eq:AcsPrimeAsLR}. The reason for the extra factor of $2$ multiplying
  $\confPrimA[\lMilne]{\gold}$ in \cref{eq:AcsPrimeAsLR} relative to
  \cref{eq:AcsAsLR} is due to the $\inv{\minkX}^2=0$ shockwave present in
  $\confPrimA{\cs'}$ which is replaced by image sources in $\confPrimA{\cs}$ ---
  compare \cref{fig:bread,fig:avocado}.}
\begin{align}
  \confPrimA{\gold}
  &= \confPrimA[\lMilne]{\gold} + \confPrimA[\rMilne]{\gold}
    \label{eq:AGoldAsLR}
\end{align}
and
\begin{align}
  \confPrimA{\cs}
  &= \confPrimA[\lMilne]{\gold}
    + \confPrimA[\lMilne]{\edge}
    + \confPrimA[\rMilne]{\edge}
    \label{eq:AcsAsLR}
\end{align}
when considering the extension $\confPrimA{\cs}$ of the Minkowski conformally soft mode with image
sources. Let us now describe the Milne soft modes appearing on
the RHSs. Firstly, we have the Goldstone modes $\confPrimA[\lMilne]{\gold}$ and
$\confPrimA[\rMilne]{\gold}$ defined by
\begin{align}
  \confPrimA[\lMilne]{\gold}
  &= \lim_{\reg\to 0} \confPrimA{\cs}
    \;,
    &
    \confPrimA[\rMilne]{\gold}
  &= \lim_{\reg\to 0} (\confPrimA{\gold}-\confPrimA{\cs})
    \label{eq:ALRGoldAcs}
\end{align}
in terms of $\confPrimA{\cs}$.

In the case of the source-free extension $A^{\cs'}$, \eqref{eq:AcsAsLR} and \eqref{eq:ALRGoldAcs} are respectively replaced by
\begin{align}
  \confPrimA{\cs'}
  &= 2 \, \confPrimA[\lMilne]{\gold}
    + \confPrimA[\lMilne]{\edge}
    + \confPrimA[\rMilne]{\edge}
    \label{eq:AcsPrimeAsLR}
\end{align}
and
\begin{align}
  \confPrimA[\lMilne]{\gold}
  &= \lim_{\reg\to 0} \frac{\confPrimA{\cs'}}{2}
    \;,
    &
    \confPrimA[\rMilne]{\gold}
  &= \lim_{\reg\to 0} \left(\confPrimA{\gold}-\frac{\confPrimA{\cs'}}{2}\right)
    \label{eq:ALRGoldAcsPrime}.
\end{align}
Note that
  \cref{eq:ALRGoldAcs,eq:ALRGoldAcsPrime} do not agree in the original Minkowski spacetime.
They define the Milne Goldstone modes in the alternative but distinct
cases where the Minkowski conformally soft mode is extended with image sources
($\confPrimA{\cs}$) or without sources ($\confPrimA{\cs'}$) to the inverse Minkowski patch.

 Like
\cref{eq:LSoftModeDubious,eq:RSoftModeDubious}, these also satisfy
\cref{eq:ALRGoldInsidePatches}; unlike
\cref{eq:LSoftModeDubious,eq:RSoftModeDubious}, the $\reg$ regulator has been
removed by definition, for reasons which will be explained shortly. As described
on \cpageref{textBegin:regMinkLR,textEnd:regMinkLR}, the $\reg$ regulator in
modes can generally be removed while maintaining smoothness within the interiors
of the Milne patches.

\label{textBegin:AEdgeDefinition}
When stitching together modes from the Milne patches to obtain Minkowski modes
however, the $\reg$ regulator is needed to smooth the interpolation between
$\lMilne$ and $\rMilne$ modes within the (original) Minkowski
spacetime.\footnote{Recall that we previously, on
  \cpageref{textBegin:regMinkLR,textEnd:regMinkLR}, swept this subtlety into
  what is implicitly meant by the symbols $\confPrimA<\shad>[\lMilne]{\confDim}$
  and $\confPrimA[\rMilne]{\confDim}$ depending on whether we are taking the
  Minkowski or Milne perspective.} For the conformally soft mode
\labelcref{eq:AcsAsLR}, this interpolation will play an important physical role.
Thus, we have explicitly introduced the Milne soft modes
$\confPrimA[\lMilne]{\edge}$ and $\confPrimA[\rMilne]{\edge}$ which are
localized to the $\reg$-regulated shockwaves of \cref{fig:croissant} and are
responsible for rendering the RHS of \labelcref{eq:AcsAsLR}\footnote{In
  contrast, \cref{eq:AcsPrimeAsLR} further has a sharp (even at finite $\reg$)
  shockwave running along $\nullInfty[\pm]$ --- see \cref{fig:avocado} and the
  discussions around \cref{eq:refVecInversion} and on
  \cpageref{textBegin:AcsPrimeSharpShockwave,textEnd:AcsPrimeSharpShockwave}.}
smooth (apart from the image sources). The important physical point is that
these modes represent the $\lMilne$ and $\rMilne$ halves of the $\reg$-width
$\minkX^2=0$ shockwave of $\confPrimA{\cs}$; by considering the electric fields
carried by this shockwave, we shall argue in \cref{sec:milneSoft} that
$\confPrimA[\lMilne]{\edge}$ and $\confPrimA[\rMilne]{\edge}$ are in fact the
edge modes \cite{Donnelly:2015hxa,Donnelly:2014fua} of their respective Milne
spacetimes. We should note though that, while $\confPrimA[\lMilne]{\edge}$ and
$\confPrimA[\rMilne]{\edge}$ in \cref{eq:AcsPrimeAsLR,eq:AcsAsLR} inherit their
$\reg$ regulator from the Minkowski mode, it will later be more convenient to
consider an $\reg$ regulator which respects the Milne time translation
(conformal) symmetry. Thus, our precise definitions for
$\confPrimA[\lMilne]{\edge}$ and $\confPrimA[\rMilne]{\edge}$ will also
implicitly depend on whether we are building the previously defined Minkowski
modes, as in \cref{eq:AcsPrimeAsLR,eq:AcsAsLR}, or starting from a Milne
perspective, as in \cref{sec:apple}.
\label{textEnd:AEdgeDefinition}

As a precursor for later discussions in \cref{sec:apple}, we note that although
there are four separate Milne modes $\confPrimA[\lMilne]{\gold}$,
$\confPrimA[\rMilne]{\gold}$, $\confPrimA[\lMilne]{\edge}$, and
$\confPrimA[\rMilne]{\edge}$, the Minkowski theory only allows two combinations
of these modes:
\begin{align}
  \confPrimA[\lMilne]{\gold}
  + \confPrimA[\lMilne]{\edge}
  + \confPrimA[\rMilne]{\edge}
  &=
    \confPrimA{\cs}
    \;,
  &
    \confPrimA[\rMilne]{\gold}
    - \confPrimA[\lMilne]{\edge}
    - \confPrimA[\rMilne]{\edge}
  &=
    \confPrimA{\gold}
    -\confPrimA{\cs},
    \label{eq:milneSoftModesDressed}
\end{align}
where the Milne Goldstone modes are accompanied by edge
modes that ensure smoothness within the Minkowski spacetime. For the alternative source-free extension $\confPrimA{\cs'}$ of the Minkowski
conformally soft mode we instead have
\begin{align}
  \confPrimA[\lMilne]{\gold}
  + \frac{
  \confPrimA[\lMilne]{\edge}
  + \confPrimA[\rMilne]{\edge}
  }{2}
  &=
    \frac{\confPrimA{\cs'}}{2}
    \;,
  &
    \confPrimA[\rMilne]{\gold}
    - \frac{
    \confPrimA[\lMilne]{\edge}
    + \confPrimA[\rMilne]{\edge}
    }{2}
  &=
    \confPrimA{\gold}-\frac{\confPrimA{\cs'}}{2}.
    \label{eq:milneSoftModesDressedPrime}
\end{align}
 We will later
interpret this reduction in degrees of freedom as resulting from the constraint
used to glue together the $\lMilne$ and $\rMilne$ theories.

\begin{figure}
  \centering
  \begin{subfigure}[t]{0.48\textwidth}
    \centering \includegraphics[scale=1.2]{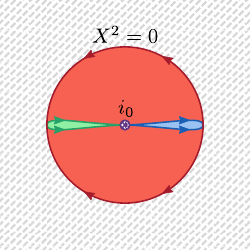}
    \caption{The $\confPrimA{\csI}$ wavefunction on the dashed Cauchy slice of
      \cref{fig:cauchySlicesNull}.}
    \label{fig:squash}
  \end{subfigure}
  \hfill
  \begin{subfigure}[t]{0.48\textwidth}
    \centering \includegraphics[scale=1.2]{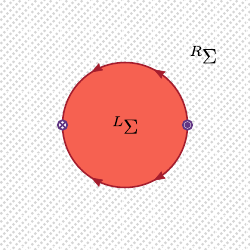}
    \caption{The $\confPrimA{\csI}$ wavefunction on the dotted Cauchy slice of
      \cref{fig:cauchySlicesNull}.}
    \label{fig:pumpkin}
  \end{subfigure}
  \caption{The $\confPrimA{\csI}$ mode on the Cauchy slices shown in
    \cref{fig:cauchySlicesNull} for the Minkowski and $\lMilne,\rMilne$ Milne
    patches. The subfigures are centred on $\inv{X}^i=0$ in the
    inverse Minkowski patch. In each subfigure, $\confPrimA{\csI}$ vanishes in
    the uncoloured regions, varies smoothly within each coloured region, and
    jumps rapidly (but smoothly for finite $\reg$) across the boundaries of
    these regions. These jumps give rise to ($\reg$-width) shockwaves to which
    the field strength $\confPrimF{\csI}$ is localized; the directions of
    electric field lines on these shockwaves are shown. These electric field
    lines connect positive (circled dot) and negative (circled cross) image
    sources, marked in purple --- see \cref{fig:bun}. In the limit where the two
    Cauchy surfaces illustrated in \cref{fig:cauchySlicesNull} approach each
    other, $\confPrimA{\csI}$ tends to the same profile on \cref{fig:squash} as
    on \cref{fig:pumpkin}, apart from the green and blue slivers which become
    infinitesimally thin in \cref{fig:squash} but are always absent in
    \cref{fig:pumpkin}.}
  \label{fig:AcsOnCauchySlices}
\end{figure}

Using \cref{eq:softMinkProd,eq:minkAsSumOfLR}, we may evaluate the inner
product between the soft modes appearing in
\cref{eq:milneSoftModesDressed,eq:milneSoftModesDressedPrime}:
\begin{align}
  \solnProd[\lMilne]{\confPrimA[\lMilne]{\gold}}{\confPrimA[\lMilne]{\edge}}
  &= \solnProd[\rMilne]{\confPrimA[\rMilne]{\gold}}{\confPrimA[\rMilne]{\edge}}
    = \frac{1}{2}
    \minkProd{\confPrimA{\gold}}{\confPrimA{\cs}}
    = -8\pi^2 i \,\diff[\celw] \diff[\celw'] \celG(\celw,\celw')
    \;,
    \label{eq:LRSoftMinkProd}
\end{align}
with all other soft mode inner products vanishing. The applicability of \cref{eq:minkAsSumOfLR} here requires some explanation,
since, as described around that equation, it assumes that
modes continue across the past boundary of $\lMilne$ in a controlled way. It is
simplest to consider the extension $\confPrimA{\cs}$ of the Minkowski
conformally soft mode which includes image sources. Eq. \eqref{eq:cscsPrimEquivalence} then implies $\confPrimA{\cs}$ may be freely replaced by
$\confPrimA{\cs'}$ in \cref{eq:LRSoftMinkProd}.
As illustrated in \cref{fig:AcsOnCauchySlices}, $\confPrimA{\csI}$
(and hence $\confPrimA{\cs}$) extends continuously across the past boundary of $\lMilne$, away
from the image sources. The only difference between
$\minkProd{A}{\confPrimA{\csI}}$ and
$\solnProd[\lMilne]{A}{\confPrimA{\csI}}+\solnProd[\rMilne]{A}{\confPrimA{\csI}}$
evaluated respectively on the dashed and dotted Cauchy surfaces of
\cref{fig:cauchySlicesNull} are Wilson lines $\int A$ running along the image
source trajectories on the past boundary of $\lMilne$. As easily seen from
\cref{eq:tea}, the $u$ component of $\pullback[\nullInfty[+]] \confPrimA{\gold}$
clearly vanishes (and, in fact, it is quite standard to use
$(\pullback[\nullInfty[\pm]]A)_u=0$ to fix the gauge redundancy on $\nullInfty[\pm]$). Thus, we may
safely use \cref{eq:minkAsSumOfLR} for soft modes as well. We will reproduce
\cref{eq:LRSoftMinkProd} in \cref{sec:apple} by starting from the Milne
perspective which will serve as further support for the relations
\labelcref{eq:AGoldAsLR,eq:AcsAsLR} between Minkowski and Milne soft modes.


\section{Entanglement of Milne patches}
\label{sec:entanglement}

In this section, we at last turn to the entanglement between the $\lMilne$ and
$\rMilne$ Milne patches. The states of interest are those of the Minkowski
theory, so it will be pertinent to understand how such states appear from the
Milne perspective. Focusing on the Minkowski vacuum $\ket{0}$, in
\cref{sec:pathIntegrals}, we first show by a path integral argument that the
same vacuum state is shared by free Maxwell theory on the Einstein static
universe. We further show that $\ket{0}$ is a thermofield double state of the
$\lMilne$ and $\rMilne$ Milne theories. This is confirmed in \cref{sec:tfd} by
using the earlier derived $\lMilne$ and $\rMilne$ decomposition \cref{eq:LR-dec}
of Minkowski modes to derive Bogoliubov transformations between the annihilation
operators of their respective theories.

Again, special treatment is required for the soft degrees of freedom associated
to the $\confDim=1$ modes. To study the entanglement of these degrees of
freedom, in \cref{sec:apple}, we consider the entangling product
\cite{Donnelly:2016auv} governed by constraints which relate the Minkowski,
Einstein static, and Milne Hilbert spaces. We will see how the sources
illustrated in \cref{fig:bun} correct the naive Gauss's law that acts as a
constraint selecting out the Minkowski Hilbert space. We will find that the
constraints of the Minkowski and Einstein static theories dictate the
entanglement of soft degrees of freedom, which results from large correlated
fluctuations in $\lMilne$ and $\rMilne$ asymptotic charges, \ie{} edge mode
fluctuations \cite{Donnelly:2015hxa,Donnelly:2014fua}.

\subsection{Path integrals and conformally related vacua}
\label{sec:pathIntegrals}

Taking advantage of the conformal invariance of free Maxwell theory, we will now
demonstrate, using a path integral argument, that the same vacuum state is
shared by the Minkowski and Einstein static spacetimes, and is a thermofield
double state of the $\lMilne$ and $\rMilne$ Milne patches.

Let us recall that the vacuum state of a (conformal) quantum field theory in a
given (conformally) static Lorentzian spacetime with time $\minkt$ can be
constructed by a path integral over a Euclidean section given by
$-\infty<i\minkt<0$. The Euclidean Lagrangian here is given by $-i$ times the
Lorentzian Lagrangian. \Eg{} for Maxwell theory, the Euclidean action and
Lagrangian are
\begin{align}
  \act_{i\minkt\in[-\infty,0]}
  &= \int_{i\minkt\in[-\infty,0]}
    (-i \lag[\blk])
    \;,
  &
    -i \lag[\blk]
  &=
    \frac{1}{2} F \wedge (i*F)
    =
    \frac{1}{4} \, (i\volForm) \,F^{\mu\nu} F_{\mu\nu}
    \;.
    \label{eq:actEuclidean}
\end{align}
Staying consistent, we shall always use $\lag[\blk]$ and $\volForm$ to refer to
the Lorentzian Lagrangian and volume form, while $-i\lag[\blk]$ and $i\volForm$
refer to their Euclidean counterparts. (Note that the analytic continuation to
Euclidean time naturally imparts a factor of $i$ to the volume form.) Of course,
the path integral is weighted in the usual way by $e^{-\act_{i\minkt\in
    [-\infty,0]}}$ and has a measure $\fieldVolForm[A]$ where gauge redundancies
have been implicitly divided out.

\begin{figure}
  \centering
  \begin{subfigure}[t]{0.48\textwidth}
    \centering \includegraphics[scale=1.2]{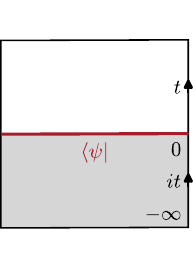}
    \caption{$\braket{\psi}{0}$ in the Minkowski theory with Lorentzian time $t$.}
    \label{fig:pathIntegralMinkowski}
  \end{subfigure}
  \hfill
  \begin{subfigure}[t]{0.48\textwidth}
    \centering \includegraphics[scale=1.2]{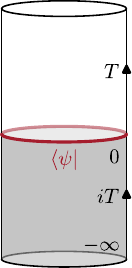}
    \caption{$\braket{\psi}{0}$ in the Einstein static theory with Lorentzian
      time $T$.}
    \label{fig:pathIntegralEinsteinStatic}
  \end{subfigure}
  \par\bigskip
  \begin{subfigure}[t]{0.48\textwidth}
    \centering \includegraphics[scale=1.2]{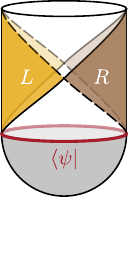}
    \caption{The Euclidean path integral evaluating $\braket{\psi}{0}$ mapped to
      an $S^4$ hemisphere.}
    \label{fig:pathIntegralSphere}
  \end{subfigure}
  \hfill
  \begin{subfigure}[t]{0.48\textwidth}
    \centering \includegraphics[scale=1.2]{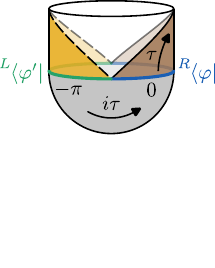}
    \caption{The overlap
      $\braOf[\lMilne]{\varphi'}\braOf[\rMilne]{\varphi}\ket{0}$ with states in
      the Milne theory, with Lorentzian time $\milnet$.}
    \label{fig:pathIntegralMilne}
  \end{subfigure}
  \caption{The path integral on Euclidean geometries (grey) representing the
    shared vacuum $\ket{0}$ of the Minkowski and Einstein static theories. More
    precisely, each illustrated path integral evaluates the overlap of $\ket{0}$
    with a state $\bra{\psi}$ or
    $\braOf[\lMilne]{\varphi'}\braOf[\rMilne]{\varphi}$, which sets the boundary
    condition on a (conformally) static time slice. Conformal maps relate the
    Euclidean geometries in
    (\subref{fig:pathIntegralMinkowski})-(\subref{fig:pathIntegralSphere}). We
    may cut out a portion of Lorentzian Einstein static time evolution, under
    which $\ket{0}$ is invariant, to obtain (\subref{fig:pathIntegralMilne}).}
  \label{fig:pathIntegral}
\end{figure}

Let us consider the Euclidean geometries over which one prepares the vacuum
state by path integration. For Minkowski spacetime, $t$ is the Minkowski
time and the relevant Euclidean geometry is the lower half of
$\flatPlane{4}$, as illustrated in \cref{fig:pathIntegralMinkowski}; for the
Einstein static universe with time $T$, this is the lower half of the
cylinder $\reals\times \sphUnit{3}$, as illustrated in
\cref{fig:pathIntegralEinsteinStatic}. These Euclidean geometries are
conformally related to each other and, moreover, to a hemisphere of
$\sphUnit{4}$ --- see \cref{fig:pathIntegralSphere}. It follows that free
Maxwell theory has the same vacuum state $\ket{0}$ in Minkowski spacetime and
the Einstein static universe.

Moreover, there are rotation symmetries on $\sphUnit{4}$ which suggest thermal
interpretations of $\ket{0}$. Indeed, the thermality of $\ket{0}$ with respect
to Rindler wedges follows immediately upon identifying one such rotation as
Euclidean Rindler time evolution. Below, we will make the analogous argument for
Milne patches more explicitly. (Note that Rindler wedges and Milne patches are
related by a conformal transformation which can be understood as a translation
on the Einstein static universe. Thus, the following argument is essentially
identical to the Rindler case, up to such a translation.)

To understand the vacuum $\ket{0}$ from the perspective of the Milne patches,
let us consider its overlap with arbitrary Milne states
$\braOf[\lMilne]{\varphi'}\braOf[\rMilne]{\varphi}\ket{0}$. This overlap is
evaluated using the path integral illustrated in \cref{fig:pathIntegralMilne}
where the states $\braOf[\lMilne]{\varphi'}$ and $\braOf[\rMilne]{\varphi}$ set
boundary conditions for the Euclidean path integral on Cauchy surfaces of the
Milne patches. Relative to \cref{fig:pathIntegralSphere}, a section of
Lorentzian time evolution has been removed --- this is possible because
$\ket{0}$ is invariant under this evolution.

As indicated in \cref{fig:pathIntegralMilne}, evolution in Milne time $\milnet$
analytically continues to a rotation on $\sphUnit{4}$. Thus, the same path
integral can be viewed as evaluating a transition amplitude
over a period $i\milnet \in [-\pi,0]$ of
Euclidean Milne time:
\begin{align}
  \braOf[\lMilne]{\varphi'}
  \braOf[\rMilne]{\varphi}\ket{0}
  &\propto
    \braOf[\rMilne]{\varphi}
    e^{-\pi \hamOf[\rMilne]}
    \ketOf[\rMilne]{\shad{\varphi}'}
    \;,
    \label{eq:vacuumMatElement}
\end{align}
where $\hamOf[\rMilne]$ is the $\rMilne$ Hamiltonian and a CPT transformation
reinterprets the path integral boundary condition set by the $\lMilne$ state
$\tensor[^\lMilne]{\bra{\varphi'}}{}$ as an $\rMilne$ state
$\tensor[^\rMilne]{\ket{\shad{\varphi}'}}{}$. For the Maxwell case under
consideration, we shall see in \cref{sec:tfd} that this coincides with the
shadow transformation. From \cref{eq:vacuumMatElement}, it follows that
$\ket{0}$ is a thermofield double state on $\lMilne$ and $\rMilne$:
\begin{align}
  \ket{0}
  &\propto \sum_i e^{-\pi E_i}
    \tensor[^\lMilne]{|\tilde{E}_i\rangle}{}
    \tensor[^\rMilne]{\ket{E_i}}{}
    \;,
    \label{eq:tfd}
\end{align}
where the Milne energy eigenstates $\tensor[^\rMilne]{\ket{E_i}}{}$ have
eigenvalues $E_i$. Tracing over either Milne patch leaves a reduced thermal
state in the other Milne patch with temperature $1/2\pi$.

As previously hinted, the above is essentially identical to the perhaps more
familiar path integral argument for the thermality of Rindler wedges. Just as in
the Rindler case, we shall see in \cref{sec:tfd} how \cref{eq:tfd} can
alternatively be deduced using Bogoliubov transformations induced by the
Unruh-like decomposition \labelcref{eq:LR-dec} of Minkowski modes into $\lMilne$
and $\rMilne$.

However, the entanglement of soft degrees of freedom will require separate
special treatment in \cref{sec:apple}. Already in \cref{eq:tfd}, we see some
limitations in the current analysis with regards to these degrees of freedom.
Within any truly soft sector, the thermofield double \labelcref{eq:tfd} is
maximally entangled since the Boltzmann factor $e^{-\pi E_i}$ disappears. However,
\cref{eq:tfd} does not give us the Schmidt decomposition of $\ket{0}$ in this
soft sector. Equivalently, we do not yet know how the soft operators such as
Milne asymptotic charges act on \cref{eq:tfd}. To learn about the soft sector of
$\ket{0}$, in \cref{sec:apple}, we will study constraints at the entangling
surface.

\subsection{The thermofield double of non-soft modes}
\label{sec:tfd}

Equipped with the results of section \ref{sec:beyond-Minkowski} we can expand
the gauge potential $A(X)$ in terms of the $\lMilne$ and $\rMilne$ conformal
primary wavefunctions
\begin{equation}
  \begin{split}
    A(X) =  \int \epsilon^{(2)}({\bf w}) \int_{\mathbb{R} - i0^+} d\lambda \frac{i\gamma^{ab}}{(2\pi)^4 |\mathcal{L}^{1 + i\lambda}|^2 \sinh \pi \lambda} &\Big[ {}^{\lMilne}\tilde{A}^{1+i\lambda}_a(X;{\bf w}) \left({}^L \tilde{\mathcal{O}}^{1 + i\lambda}_{\bar{b}}({\bf w})\right)^{\dagger}\\&-{}^{\rMilne}A^{1+i\lambda}_a(X;{\bf w}) \left({}^R \mathcal{O}^{1 + i\lambda}_{\bar{b}}({\bf w}) \right)^{\dagger}
    \Big] + \cdots,
  \end{split}
\end{equation}
where $\cdots$ denote the soft ($\lambda = 0$) modes that will be discussed in
detail in the following section. Here $\gamma_{ab}$ is the metric on the celestial space defined in \eqref{eq:chocolate} and $\mathcal{L}^{\Delta}$ is the normalization factor in \eqref{eq:Lnorm}. The two sets of operators
${}^{\rMilne}\mathcal{O}^{1+i\lambda}$ and
${}^{\lMilne}\tilde{\mathcal{O}}^{1+i\lambda}$ admit a further decomposition
into positive and negative energy $(\lambda)$ components with respect to the
Milne time $\tau$. Recalling that the Milne time flows in opposite directions in
the $\lMilne$ and $\rMilne$ patches, we can write
\begin{equation}
  \begin{split}
    A =  \int \epsilon^{(2)}({\bf w}) \int_{\lambda > 0} d\lambda &\Big[\mathcal{M}^{1+i\lambda, +}{}^{\rMilne}A^{1+i\lambda}_a({\bf w}) {}^R a^a_{\lambda}({\bf w}) + \mathcal{M}^{1-i\lambda, -}{}^{\rMilne}A^{1-i\lambda}_a({\bf w}) {}^R a^{\bar{a}}_{\lambda}({\bf w})^{\dagger} \\
                                                                  &+ \mathcal{M}^{1+i\lambda,+}{}^{\lMilne}\tilde{A}^{1-i\lambda}_{a}({\bf w}) {}^L \tilde{a}^{a}_{\lambda}({\bf w}) + \mathcal{M}^{1-i\lambda,-}{}^{\lMilne}\tilde{A}^{1+i\lambda}_{a}({\bf w}) {}^L \tilde{a}^{\bar{a}}_{\lambda}({\bf w})^{\dagger}
                                                                    \Big] + \cdots ,
  \end{split}
\end{equation}
where we suppressed the dependence on the spacetime point $X$ and
$\mathcal{M}^{1\pm i\lambda,\pm}$ are normalization constants defined in \eqref{eq:M} below, chosen such that
${}^{\rMilne}a, {}^{\rMilne}a^{\dagger}$ and ${}^{\lMilne}a,
{}^{\lMilne}a^{\dagger}$ obey the canonical commutation relations
\eqref{eq:ccLR}. 

Following the discussion in section \ref{sec:sandwich} and using the relation
\eqref{eq:inv} between global and $\lMilne/\rMilne$ conformal primary
wavefunctions, we find the decompositions of
${}^L\tilde{\mathcal{O}}^{\Delta}_a, {}^R\mathcal{O}^{\Delta}_a$ in terms of
conformal primary operators
\begin{equation}
  \label{LR-cpw-decomposition}
  \begin{split}
    {}^L\tilde{\mathcal{O}}^{\Delta}_a &= -i \langle  \confPrimA<\shad>[\lMilne]{\confDim},A \rangle = \frac{i}{2\sin\pi \Delta} \left(\mathcal{O}^{\Delta,+}_a - \mathcal{O}^{\Delta,-}_a \right), \\
    {}^R\mathcal{O}^{\Delta}_a &= -i \langle  \confPrimA[\rMilne]{\confDim},A \rangle =\frac{i}{2\sin\pi \Delta} \left(e^{-i\pi \Delta} \mathcal{O}^{\Delta,+}_a - e^{i\pi \Delta}\mathcal{O}^{\Delta,-}_a \right),
  \end{split}
\end{equation}
where the operators on the RHS were defined in \cref{eq:confPrimOp}. From
\cref{eq:limitMinkProdRRLL}, we have
\begin{equation} 
\begin{split}
    [{}^L\tilde{\mathcal{O}}^{1+i\lambda}_a(\celw),
  {}^L\tilde{\mathcal{O}}^{1+ i\lambda'}_{b}(\celw')] &=
  -[{}^R\mathcal{O}^{1+i\lambda}_a(\celw), {}^R\mathcal{O}^{1+
    i\lambda'}_{b}(\celw')] \\
    &\eqIntegrated (2\pi)^4
  \frac{\lambda}{\pi(1+\lambda^2)} \deltaFunc(\milneFreq + \milneFreq') \,
  \celMet_{a b} \celDeltaFunc^{}(\celw,\celw')\,.
  \end{split}
\end{equation}
These operators define creation and annihilation operators 
\begin{equation} {}^L\tilde{\mathcal{O}}^{1+i\lambda}_a(\celw) =
  \frac{(2\pi)^2\sqrt{\lambda}}{\sqrt{\pi(1+\lambda^2)}}
  {}^L\tilde{a}_{a,\milneFreq}(\celw), \quad
  {}^R\mathcal{O}^{1+i\lambda}_a(\celw) =
  \frac{(2\pi)^2\sqrt{\lambda}}{\sqrt{\pi(1+\lambda^2)}}
  {}^Ra_{\bar{a},\milneFreq}(\celw)^{\dagger}, \quad \lambda >0
\end{equation}
normalized such that
\begin{align}
  \label{eq:ccLR}
  [{}^L\tilde{a}_{a,\milneFreq}(\celw), {}^L\tilde{a}_{\bar{b},\milneFreq'}(\celw')^{\dagger}] = 
  [{}^Ra_{a,\milneFreq}(\celw), {}^Ra_{\bar{b},\milneFreq'}(\celw')^{\dagger}] =
  \celMet_{a b} \deltaFunc(\milneFreq - \milneFreq')
  \celDeltaFunc^{}(\celw,\celw')\,.
\end{align}
From these relations we can read off
\begin{equation}
\label{eq:M}
\mathcal{M}^{1+i\lambda, \pm } = \frac{\sqrt{\pm \pi(1+\lambda^2)}}{(2\pi)^2 \sqrt{\lambda}}.
\end{equation}
We indeed see that ${}^R a_{\lambda}^{\dagger}$ (${}^L
\tilde{a}_{\lambda}^{\dagger}$) create positive Milne energy states in the right
(left) Milne patches. Further using the expressions for
$\mathcal{O}^{\Delta,\pm}$ in \cref{confPrimO}, we find that
\cref{LR-cpw-decomposition} imply the Bogoliubov transformations
\begin{equation}
  \begin{cases}
    {}^L \tilde{a}_{a,\milneFreq} &=  \dfrac{1}{\sqrt{2\sinh \pi \milneFreq}} \left(
                                    e^{\pi \milneFreq/2} a_{a, -\milneFreq} + e^{-\pi \milneFreq/2}a^{\dagger}_{\bar{a}, \milneFreq}
                                    \right) \vspace{10pt}\\
    {}^Ra_{\bar{a},\lambda}^{\dagger} &= \dfrac{1}{\sqrt{2\sinh \pi \milneFreq}} \left(
                                        e^{-\pi \milneFreq/2} a_{a,-\milneFreq} + e^{\pi \milneFreq/2}a^{\dagger}_{\bar{a}, \milneFreq}
                                        \right) 
  \end{cases}, \quad \milneFreq > 0\,.
\end{equation}
The inverse transformations are given by
\begin{align}
~~  \begin{cases}
    a_{a,-\milneFreq} &= \dfrac{1}{\sqrt{2\sinh \pi \milneFreq}}\left({}^L \tilde{a}_{a,\milneFreq} e^{\pi\lambda/2} -{}^Ra_{\bar{a},\lambda}^{\dagger} e^{-\pi\lambda/2}\right) \vspace{10pt}\\
    a_{\bar{a},\milneFreq}^{\dagger} &=  \dfrac{1}{\sqrt{2\sinh \pi \milneFreq}}\left({}^Ra_{\bar{a},\lambda}^{\dagger} e^{\pi\lambda/2}-{}^L \tilde{a}_{a,\milneFreq} e^{-\pi\lambda/2} \right)
  \end{cases}, \quad \milneFreq > 0\,.
\end{align}
These expansions are completely analogous to the decomposition of global
Minkowski modes in terms of Rindler modes.

It follows that the Minkowski vacuum defined by
\begin{equation}
  a_{a,\milneFreq}|0\rangle = 0
\end{equation}
is a thermal state with respect to the left and right Milne vacua
\begin{equation}
  \tilde{a}^L_{a, \milneFreq}|0\rangle_L = 0, \quad a^R_{a,\milneFreq} |0\rangle_R = 0\,.
\end{equation}
That is, we have
\begin{equation}
  \label{eq:TFD}
  |0\rangle = \exp\left\{ \int \epsilon^{(2)} \int_{\lambda >0} d\milneFreq \left[\log(1-e^{-2\pi \lambda}) + e^{-\pi \milneFreq} \celMet^{\bar{a}\bar{b}} {}^{L}\tilde{a}^{\dagger}_{a, \milneFreq} {}^{R\dagger}a^{\dagger}_{b, \milneFreq} \right]\right\}| 0 \rangle_L |0\rangle_R\,,
\end{equation}
where the logarithmic term in the exponent provides a normalization constant fixed by
$\langle 0| 0\rangle = 1$. 

\subsection{Entangling product and soft modes}
\label{sec:apple}

In this \namecref{sec:apple}, our goal will be to relate the physical phase
spaces and corresponding Hilbert spaces of the $\lMilne$ and $\rMilne$ patches
to those of Minkowski spacetime. In particular, we shall find that a constraint
must be imposed on the $\lMilne$ and $\rMilne$ spaces which results in the
entanglement of $\lMilne$ and $\rMilne$ soft degrees of freedom in Minkowski
states.
Similar ideas have been explored by
\cite{Donnelly:2015hxa,Donnelly:2014fua,Donnelly:2016auv} in settings where the
entangling surface resides within the spacetime interior. It is helpful to review
their framework in our Milne context. First, we must better understand the phase
space and Hilbert space of a single Milne patch, say the $\rMilne$ patch.

\subsubsection{Symplectic structure in Milne spacetime}
\label{sec:sympMilne}

Because Milne spacetime has a boundary at spacial infinity, the symplectic
potential is supplemented with a boundary term for reasons explained in
\cref{sec:asymSym}:
\begin{align}
  \sympPot{\confSpace[\rMilne]}
  &= \sympPot[\blk]{\confSpace[\rMilne]} + \sympPot[\bdy]{\confSpace[\rMilne]}
    \;,
  &
    \sympPot[\blk]{\confSpace[\rMilne]}
  &= \int_{\cauchy[\rMilne]} \sympPotD[][\blk]
    \;,
  &
    \sympPot[\bdy]{\confSpace[\rMilne]}
  &= \int_{\partial\cauchy[\rMilne]} \sympPotD[\rMilne][\bdy]
    \;,
    \label{eq:sympPotR}
  \\
  & & & &
          \sympPotD[\rMilne][\bdy]
  &= \fieldDiff\edgeScalar[\rMilne]~ \, \edgeE[\rMilne] \;.
\end{align}
While $\sympPotD[][\blk]$ is the same symplectic potential density as in
\cref{eq:sympPotDBlk}, the boundary contribution $\sympPotD[\rMilne][\bdy]$
introduces new boundary degrees of freedom, $\edgeScalar[\rMilne]$ and
$\edgeE[\rMilne]$, which are both scalars\footnote{ Note that the ``$\edgeScalar$''  in \cite{Donnelly:2016auv} is the exponential of ours. } on
the boundary $\partial\cauchy[\rMilne]$ of the Cauchy slice $\cauchy[\rMilne]$
of the $\rMilne$ spacetime. Under a gauge transformation with parameter
$\gaugeParam$, the gauge potential continues to transform in the usual way
written in \cref{eq:AGaugeTrans} but $\edgeScalar[\rMilne]$ also transforms
nontrivially if $\gaugeParam$ does not vanish on $\partial\cauchy[\rMilne]$:
\begin{align}
  \edgeScalar[\rMilne] &\mapsto \edgeScalar[\rMilne] + \gaugeParam \;.
                         \label{eq:edgeScalarGaugeTrans}
\end{align}
Thus locally, $\edgeScalar$ may be thought of as giving a trivialization of the gauge
bundle over $\partial\cauchy[\rMilne]$. Since the total pre-symplectic form
$\symp{\confSpace[\rMilne]}$ contains the boundary contribution
$\symp[\bdy]{\confSpace[\rMilne]}=\fieldDiff\sympPot[\bdy]{\confSpace[\rMilne]}$,
the space $\extPhSpace[\rMilne]$ of initial data on $\cauchy[\rMilne]$, as
defined in \cref{eq:extPhSpace}, now includes the new variables
$\edgeScalar[\rMilne]$ and $\edgeE[\rMilne]$. Therefore, $\extPhSpace[\rMilne]$ is
referred to as the extended phase space.

Note that $\sympPot{\extPhSpace[\rMilne]}$ is stationary under gauge
transformations. We may then construct the Hamiltonian generators of gauge
transformations following the procedure described around
\cref{eq:constructingHamGenerator}. We find, analogous to \cref{eq:cstrBlk},
that the generators are
\begin{align}
  \cstr[\rMilne]\relax[\gaugeParam]
  &= \cstr[\rMilne][\blk]\relax[\gaugeParam]
    + \cstr[\rMilne][\bdy]\relax[\gaugeParam]
    \label{eq:cstrR1}
  \\
  &= \cstr[\rMilne][\blkt]\relax[\gaugeParam]
    + \cstr[\rMilne][\bdyt]\relax[\gaugeParam]
    \label{eq:cstrR2}
    \;,
\end{align}
where
\begin{align}
  \cstr[\rMilne][\blk]\relax[\gaugeParam]
  &= \int_{\cauchy[\rMilne]} \diff\gaugeParam \wedge \hodge F
    \;,
  &
    \cstr[\rMilne][\bdy]\relax[\gaugeParam]
  &= -\int_{\partial\cauchy[\rMilne]} \gaugeParam \, \edgeE[\rMilne]
    \label{eq:cstrRBlkBdy}
\end{align}
and
\begin{align}
  \cstr[\rMilne][\blkt]\relax[\gaugeParam]
  &=
    -\int_{\cauchy[\rMilne]} \gaugeParam \, \diff\hodge F
    \;,
  &
    \cstr[\rMilne][{\bdyt}]\relax[\gaugeParam]
  &= \int_{\partial\cauchy[\rMilne]}
    \gaugeParam \, (\hodge F - \edgeE[\rMilne])
\end{align}
offer two ways of splitting $\cstr[\rMilne]$ into bulk and boundary pieces. The
former split \labelcref{eq:cstrR1} is the one naturally following from that of
\cref{eq:sympPotR}; on the other hand, the terms in the latter split
\labelcref{eq:cstrR2} are more easily understood as bulk and boundary
constraints, as we now discuss.

Recall, as explained around \cref{eq:cstrSpace}, the generators $\cstr[\rMilne]$
of gauge transformations also serve as a constraints that pick out a constraint
surface $\cstrSpace[\rMilne]\subset\extPhSpace[\rMilne]$. In particular,
$\cstrSpace[\rMilne]$ is given by the condition
\begin{align}
  \cstr[\rMilne]
  &= 0\;,
  &
  &\text{\ie}
  &
    \cstr[\rMilne][\blk]
  &= -
    \cstr[\rMilne][\bdy] \;,
  &
  &(\text{on $\cstrSpace[\rMilne]$})
    \label{eq:cstrRToZero}
\end{align}
or alternatively from \cref{eq:cstrR2},
\begin{align}
  \pullback[\cauchy[\rMilne]]\, \diff\hodge F
  &=0\;,
  &
    \pullback[\partial \cauchy[\rMilne]]
    \hodge F
  &=
    \edgeE[\rMilne]
    \;
  &
  &(\text{on $\cstrSpace[\rMilne]$})\;.
    \label{eq:localCstrR}
\end{align}
The first equation is simply the component of the equations of motion
\eqref{eq:vacuum-ME} giving Gauss's law, while the latter instructs us to
identify $\edgeE[\rMilne]$ as the perpendicular component of the electric field
at the boundary $\partial\cauchy[\rMilne]$. As described in \cref{sec:asymSym},
having $\cstr[\rMilne]$ vanish on-shell is the motivation for introducing the
boundary degrees of freedom $\edgeScalar[\rMilne]$ and $\edgeE[\rMilne]$ and
adding $\sympPot[\bdy]{\confSpace[\rMilne]}$ to the pre-symplectic potential in
the first place. In this approach, all gauge transformations are treated as
redundancies, even if the gauge parameter $\gaugeParam$ is nontrivial on the
boundary $\partial\cauchy[\rMilne]$.

To obtain the physical phase space $\phSpace[\rMilne]$, we must divide out by
these redundancies, as in \cref{eq:phSpace}. Equivalently, we may fix the
redundancies by a choice of gauge. For example, we may use the gauge freedom on
the Milne spacial infinity to set
\begin{align}
  \edgeScalar[\rMilne]
  &= 0 
    \qquad (\text{on $\phSpace[\rMilne]$})\;,
    \label{eq:edgeScalarGaugeCondition}
\end{align}
while the gauge freedom elsewhere in the spacetime is fixed by other conditions,
\eg{} fixing the null component of the gauge potential at null infinity and
imposing Lorenz gauge.

Separate from gauge redundancies, however, is another set of transformations
that are physical, called surface symmetries. In \cite{Donnelly:2016auv}, these
surface symmetries are defined to act only on the boundary degrees of freedom
while the gauge potential is fixed; for the Maxwell case we are considering
which is Abelian, it is the negative of the transformation written in
\cref{eq:edgeScalarGaugeTrans}. However, since the combination of
eqs. \eqref{eq:AGaugeTrans} and \eqref{eq:edgeScalarGaugeTrans} is a redundancy, one
may equivalently define surface symmetries to be the transformation
\labelcref{eq:AGaugeTrans} of the gauge potential while $\edgeScalar[\rMilne]$
remains fixed. We shall take this latter perspective, since it preserves the
gauge condition \eqref{eq:edgeScalarGaugeCondition}. The Hamiltonian generator
for this surface symmetry is simply $\cstr[\rMilne][\blk]$.

At this point, we have in fact recovered the second approach described in
\cref{sec:asymSym} for treating gauge symmetries reaching spacial infinity. In
particular, the boundary degrees of freedom $\edgeScalar[\rMilne]$ and
$\edgeE[\rMilne]$ have respectively been gauge-fixed in
\cref{eq:edgeScalarGaugeCondition} and identified with the perpendicular
component of the field strength in \cref{eq:localCstrR}. Consequently, the
symplectic potential $\sympPot{\phSpace[\rMilne]}$ on the physical phase space
is the same as though starting with only the bulk contribution
$\sympPot[\blk]{\confSpace[\rMilne]}$. Moreover, we are left with a set of
physical surface symmetries which appear like gauge symmetries that reach the
Milne spacial infinity, had we not introduced $\edgeScalar[\rMilne]$ and $\edgeE[\rMilne]$ in the
first place. These are precisely the asymptotic symmetries described around
\cref{eq:asymSymStrominger}, and as claimed there, they are generated by
$\cstr[\rMilne][\blk]$ arising from $\sympPot[\blk]{\confSpace[\rMilne]}$.

\subsubsection{Soft modes and degrees of freedom in Milne spacetime}
\label{sec:milneSoft}

Before describing the entangling product required to relate the theories in Milne
and Minkowski spacetimes, it is helpful first to discuss the Milne soft modes
and degrees of freedom which will play important roles. We have already seen
these modes appear when expressing Minkowski soft modes in terms of Milne modes
in \cref{eq:AGoldAsLR,eq:AcsPrimeAsLR,eq:AcsAsLR}. In particular, the soft modes
of a Milne spacetime $\rMilne$ are the Goldstone modes
$\confPrimA[\rMilne]{\gold}$ and the edge modes $\confPrimA[\rMilne]{\edge}$.

As expressed in \cref{eq:ALRGoldInsidePatches}, the former are the analogues of
the Minkowski Goldstone modes $\confPrimA{\gold}$ and describe asymptotic
symmetry transformations of the $\rMilne$ patch with gauge parameter
$\confGauge{\gold}$. As in \cref{eq:csOp} for the Minkowski case, we may
similarly define the Milne charges associated to these symmetries:
\begin{align}
  \csOp[\rMilne][a](\celw)
  =
  \symp{\phSpace[\rMilne]}(
  \confPrimA[\rMilne]{\gold}[a](\celw),
  \opA
  )
  = -i \,\solnProd[\rMilne]{\confPrimA{\gold}[a](\celw)}{\opA}
  &= \cstr[\rMilne][\blk]\relax[
    \confGauge{\gold}[a](\celw)
    ].
    \label{eq:csOpR}
\end{align}
The last equality 
is the analogue of \cref{eq:csOpAsCstr2} and equates $\csOp[\rMilne]$ to the
generator $\cstr[\rMilne][\blk]\relax[\confGauge{\gold}]$ of the asymptotic
symmetry with parameter $\confGauge{\gold}$, as described in
\cref{sec:sympMilne}. Conversely, as in \cref{eq:cstrAscsOp}, the generator for
any asymptotic symmetry may be written in terms of $\csOp[\rMilne]$:
\begin{align}
  \cstr[\rMilne][\blk]\relax[\gaugeParam]
  &= -
    \cstr[\rMilne][\bdy]\relax[\gaugeParam]
    =
    -\frac{1}{4\pi}\int_{\partial \cauchy[\rMilne]} \celVolForm
    \gaugeParam \,
    \celCovD\cdot \csOp[\rMilne]
    \qquad (\text{on $\cstrSpace[\rMilne]$}).
    \label{eq:cstrRAscsOpR}
\end{align}
Here, we have made use of the constraint \eqref{eq:cstrRToZero} which holds
on-shell. For example, we may consider the quantum field theory where the field
operator $\opA$ in \cref{eq:csOpR} is on-shell as it decomposes into modes
satisfying the equations of motion. From
\cref{eq:cstrRBlkBdy,eq:cstrRToZero,eq:localCstrR}, we see that
$\cstr[\rMilne][\blk]$ and $-\cstr[\rMilne][\bdy]$ are determined on-shell
purely by the perpendicular component of the electric field at
$\partial\cauchy[\rMilne]$.

\begin{figure}
  \centering
  \begin{subfigure}[t]{0.48\textwidth}
    \centering \includegraphics[scale=1.1]{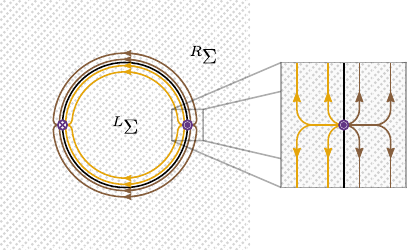}
    \caption{The electric field lines in the configuration $\confPrimA{\csI}$.
      The image sources are shown in purple. (\Cf{}
      \cref{fig:bread,fig:pumpkin}.)  The black line represents the boundary between $\cauchy[\lMilne]$ and $\cauchy[\rMilne]$.}
    \label{fig:milneEdgeE}
  \end{subfigure}
  \hfill
  \begin{subfigure}[t]{0.48\textwidth}
    \centering \includegraphics[scale=1.1]{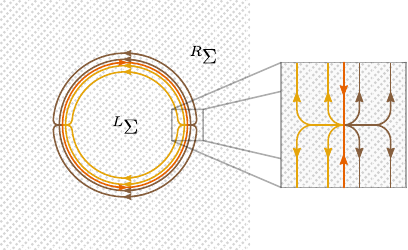}
    \caption{The electric field lines in the configuration $\confPrimA{\csI'}$.
      Shown in orange is the $\inv{\minkX}^2=0$ shockwave, which is sharp even
      at finite $\reg$,
      lies exactly between $\lMilne$ and $\rMilne$, and does not belong to either
      Milne patch. (\Cf{} \cref{fig:avocado}.)}
    \label{fig:milneEdgeEPrime}
  \end{subfigure}
  \caption{The electric fields of Milne edge modes
    $\confPrimA[\rMilne]{\edgeI}$ (brown) and $\confPrimA[\lMilne]{\edgeI}$
    (yellow). These electric fields are part of the $\reg$-regulated $\minkX^2=0$ shockwave
    of the Minkowski conformally soft mode $\confPrimA{\csI}$ (or
    $\confPrimA{\csI'}$), as shown on a Cauchy surface
    $\cauchy=\cauchy[\lMilne]+\cauchy[\rMilne]$ in the main panels. (See dotted surfaces of
    \cref{fig:cauchySlices}.) The magnified side panels show how this shockwave
    contributes to the perpendicular component of the electric field at
    $\partial\cauchy[\lMilne]$ and $\partial\cauchy[\rMilne]$, namely there are $\delta$-functions in the perpendicular component of the electric fields at the two points on the celestial space.}
  \label{fig:finiteRegShockwave}
\end{figure}

We now show that these electric fields are turned on precisely by the edge
modes $\confPrimA[\rMilne]{\edge}$ described on
\cpageref{textBegin:AEdgeDefinition,textEnd:AEdgeDefinition}. We draw the $\reg$-regulated
$\minkX^2=0$ shockwave of $\confPrimA{\csI}$ in \cref{fig:milneEdgeE} and of
$\confPrimA{\csI'}$ in \cref{fig:milneEdgeEPrime} on
a Cauchy surface for the $\lMilne$ and $\rMilne$ Milne patches (\eg{} a dotted surface in \cref{fig:cauchySlicesNull}). We see that the
integrated edge mode
\begin{align}
  \confPrimA[\rMilne]{\edgeI}(\celw_1,\celw_2)
  &= \int_{\celw_1}^{\celw_2}\confPrimA[\rMilne]{\edge}(\celw)
    \;
\end{align}
defined analogously to \cref{eq:lettuce}, turns on $\celDeltaFunc$-functions in
the perpendicular component of the electric field at $\partial\cauchy[\rMilne]$
localized to the celestial positions $\celw_1$ and $\celw_2$ of the image
sources. (In the case of $\confPrimA{\csI'}$, these would-be sources are
replaced by the $\inv{\minkX}^2=0$ shockwave, which is sharp --- recall the
discussions around \cref{eq:refVecInversion} and on
\cpageref{textBegin:AcsPrimeSharpShockwave,textEnd:AcsPrimeSharpShockwave} --- and
sits exactly between the $\lMilne$ and $\rMilne$ patches, and is therefore
hidden from the theories of either patch.) Thus, evaluated on this
configuration, the generator \labelcref{eq:cstrRAscsOpR} for an arbitrary
asymptotic symmetry becomes
\begin{align}
  \cstr[\rMilne][\blk][\gaugeParam, \confPrimA[\rMilne]{\edgeI}]
  &= -\cstr[\rMilne][\bdy][\gaugeParam, \confPrimA[\rMilne]{\edgeI}]
    = 2\pi [\gaugeParam(\celw_2) -\gaugeParam(\celw_1)]
    \;.
    \label{eq:cstrRBlkEdgeI}
\end{align}
Taking the points $\celw_1$ and $\celw_2$ infinitesimally close together, we
further find
\begin{align}
  \cstr[\rMilne][\blk][\gaugeParam, \confPrimA[\rMilne]{\edge}]
  &= -\cstr[\rMilne][\bdy][\gaugeParam, \confPrimA[\rMilne]{\edge}]
    = 2\pi\, \diff \gaugeParam
    \;.
    \label{eq:cstrRBlkEdge}
\end{align}
\Cref{eq:cstrRBlkEdgeI,eq:cstrRBlkEdge} are the Milne equivalents of
\cref{eq:asymChargeCSI}. The only difference is a factor of $1/2$ from splitting
the $\minkX^2=0$ shockwave evenly between $\lMilne$ and $\rMilne$. Just as $\confPrimA{\cs}$
is paired with $\confPrimA{\gold}$ in the Minkowski theory, the edge mode
$\confPrimA[\rMilne]{\edge}$ is the canonical partner of
$\confPrimA[\rMilne]{\gold}$. In particular, from
\cref{eq:csOpR,eq:cstrRBlkEdge}, we recover precisely the inner product
\labelcref{eq:LRSoftMinkProd} between the Milne soft modes, previously derived
there by relation to Minkowski modes.

We are therefore led to consider the Goldstone operator
\begin{align}
  \goldOp[\rMilne][a](\celw)
  &=
    \symp{\phSpace[\rMilne]}(
    \confPrimA[\rMilne]{\edge}[a](\celw),
    \opA
    )
    = -i \,\solnProd[\rMilne]{\confPrimA[\rMilne]{\edge}[a](\celw)}{\opA}
    \label{eq:goldOpR}
\end{align}
(\cf{} \cref{eq:goldOp}), which is the canonical partner of the charges
$\csOp[\rMilne]$. Their commutator, determined by \cref{eq:LRSoftMinkProd}, is
precisely half of the Minkowski analogue \labelcref{eq:minkCommutSoft}:
\begin{align}
  \commut{
  \csOp[\rMilne](\celw)
  }{
  \goldOp[\rMilne](\celw')
  }
  &= 8\pi^2 i \,\diff[\celw] \diff[\celw'] \celG(\celw,\celw')
    \;.
\end{align}
Analogous to
\cref{eq:goldOpI}, it will be useful for later discussions to introduce the
integrated soft operators
\begin{align}
  \goldOpI[\rMilne](\celw_1,\celw_2)
  &=
    \symp{\phSpace[\rMilne]}(
    \confPrimA[\rMilne]{\edgeI}(\celw_1,\celw_2),
    \opA
    )
    =\int_{\celw_1}^{\celw_2} \goldOp[\rMilne](\celw)
    \;.
    \label{eq:goldOpIR}
\end{align}

The relations between the Minkowski and Milne soft degrees of freedom are
straightforwardly read off from the corresponding relations
\cref{eq:AGoldAsLR,eq:AcsAsLR} between soft modes:
\begin{align}
  \csOp(\celw)
  &= \csOp[\lMilne](\celw) + \csOp[\rMilne](\celw) \;,
  &
    \goldOp(\celw)
  &= \csOp[\lMilne](\celw)
    + \goldOp[\lMilne](\celw)
    + \goldOp[\rMilne](\celw)
    \;,
   \label{eq:minkSoftOpAsLR}
\end{align}
and
\begin{align}
\goldOpI(\celw_1,\celw_2)
&=
\cstr[\lMilne][\blk][
\log(-\nullUnit_1\cdot\minkX)
-\log(-\nullUnit_2\cdot\minkX)
]
+ \goldOpI[\lMilne](\celw_1,\celw_2)
+ \goldOpI[\rMilne](\celw_1,\celw_2).
\label{eq:goldOpIAsLR}
\end{align}
Alternatively, with the source-free shocks we have
\begin{align}
  \goldOp'(\celw)
  &= 2\csOp[\lMilne](\celw)
    + \goldOp[\lMilne](\celw)
    + \goldOp[\rMilne](\celw)
  \;,
  \\
  \goldOpI'(\celw_1,\celw_2)
  &= 2
    \cstr[\lMilne][\blk][
    \log(-\nullUnit_1\cdot\minkX)
    -\log(-\nullUnit_2\cdot\minkX)
    ]
    + \goldOpI[\lMilne](\celw_1,\celw_2)
    + \goldOpI[\rMilne](\celw_1,\celw_2)
  \label{eq:goldOpIPrimeAsLR}
\end{align}
considering the source-free extension $\confPrimA{\cs'}$.

\subsubsection{Entangling product of two Milne patches}

So far, in \cref{sec:sympMilne}, we have described the symplectic structure of
Maxwell theory in a single Milne patch $\rMilne$. Let us now consider how the
theories in two complementary Milne patches $\lMilne$ and $\rMilne$, as
illustrated in \cref{fig:potato}, can be glued together to describe a theory on
the Einstein static universe and on Minkowski spacetime. While the former case
follows the same prescription as introduced in \cite{Donnelly:2016auv}, in the
latter, we shall find that the presence of image sources modifies this
prescription for the Minkowski theory. In either case, the soft operators described
\cref{sec:milneSoft} will play important roles.

Let us begin with a very general review of the gluing procedure for two
spacetime patches $\lMilne$ and $\rMilne$ whose (partial) Cauchy slices
$\cauchy[\lMilne]$ and $\cauchy[\rMilne]$ join to cover a full Cauchy surface
$\cauchy$ in a larger spacetime. One must impose additional
constraints relating degrees of freedom on $\partial\cauchy[\lMilne]$ and
$\partial\cauchy[\rMilne]$. To be precise, the constraint space of the theory in
the larger spacetime can be identified as a submanifold $ \cstrSpace
\subset\cstrSpace[\lMilne]\times\cstrSpace[\rMilne] $ of the $\lMilne$ and
$\rMilne$ constraint spaces; this submanifold is specified by the vanishing
\begin{align}
  \cstr[][\ent] &= 0
                  \qquad (\text{on $\cstrSpace$})
\end{align}
of constraints $\cstr[][\ent]$. These new constraints $\cstr[\ent]$ are on top
of the constraints $\cstr[\lMilne]$ and $\cstr[\rMilne]$ used to pick out
$\cstrSpace[\lMilne]\subset\extPhSpace[\lMilne]$ and
$\cstrSpace[\rMilne]\subset\extPhSpace[\rMilne]$. In order for $\cstr[][\ent]$
to be independent of gauge redundancies from the $\lMilne$ and $\rMilne$
perspectives, we demand
\begin{align}
  0
  &= \poisson{\cstr[][\ent]}{\cstr[\lMilne]}
    = -i\commut{\cstr[][\ent]}{\cstr[\lMilne]}
    \;
  &
    0
  &= \poisson{\cstr[][\ent]}{\cstr[\rMilne]}
    = -i\commut{\cstr[][\ent]}{\cstr[\rMilne]}
    \;.
    \label{eq:cstrEntGaugeInvariance}
\end{align}
The presymplectic potential $\sympPot{\cstrSpace}$ and presymplectic form
$\symp{\cstrSpace}$ on $\cstrSpace$ are simply given by summing the
corresponding $\lMilne$ and $\rMilne$ objects:
\begin{align}
  \sympPot{\cstrSpace}
  &= \pullback[\cstrSpace](
    \sympPot{\cstrSpace[\lMilne]}
    +\sympPot{\cstrSpace[\rMilne]}
    )
    \;,
  &
    \symp{\cstrSpace}
  &= \pullback[\cstrSpace](
    \symp{\cstrSpace[\lMilne]}
    +\symp{\cstrSpace[\rMilne]}
    ) \;.
\end{align}

Finally, the physical phase space $\phSpace$ is obtained from $\cstrSpace$ by
dividing out the null directions of $\symp{\cstrSpace}$, as in
\cref{eq:phSpace}. The overall relation between the physical phase spaces
constructed from $\cstrSpace$, $\cstrSpace[\lMilne]$, and $\cstrSpace[\rMilne]$
is known as the fusion or entangling product \cite{Donnelly:2016auv}:
\begin{align}
  \phSpace
  &= \phSpace[\lMilne] \entProd \phSpace[\rMilne]
    = (\phSpace[\lMilne]\times\phSpace[\rMilne])
    \doubleQuotient \{\cstr[][\ent]\}
    \;.
\end{align}
As expressed in the last equality, one may construct $\phSpace$ more directly
from $\phSpace[\lMilne]\times\phSpace[\rMilne]$ through the double quotient
introduced in \cref{eq:doubleQuotient}. This is possible because the invariance
\labelcref{eq:cstrEntGaugeInvariance} of $\cstr[][\ent]$ with respect to the
gauge redundancies of the $\lMilne$ and $\rMilne$ theories allows
$\cstr[][\ent]$, originally introduced on
$\cstrSpace[\lMilne]\times\cstrSpace[\rMilne]$, to be pulled back to a
well-defined function on $\phSpace[\lMilne]\times\phSpace[\rMilne]$. As
described around \cref{eq:hilbSpaceCstr}, the double quotient has a
straightforward interpretation at the level of Hilbert spaces of the quantum
theory as simply restricting to the kernel of the constraints. Denoting the
Hilbert spaces corresponding to $\phSpace$, $\phSpace[\lMilne]$, and
$\phSpace[\rMilne]$ respectively as $\hilbSpace \equiv \hilbSpace(\phSpace)$,
$\hilbSpace[\lMilne]\equiv\hilbSpace(\phSpace[\lMilne])$, and
$\hilbSpace[\rMilne]\equiv\hilbSpace(\phSpace[\rMilne])$, we have
\begin{align}
  \hilbSpace
  &= \ker \cstr[][\ent]
    \;,
\end{align}
where $\cstr[][\ent]$ is viewed as an operator on the larger Hilbert space
$\hilbSpace[\lMilne]\otimes\hilbSpace[\rMilne]=\hilbSpace(\phSpace[\lMilne]\times\phSpace[\rMilne])$.
Because $\cstr[][\ent]$ will involve degrees of freedom from both $\lMilne$ and
$\rMilne$, the restriction to its kernel will produce entanglement between these
degrees of freedom, as we shall see in \cref{sec:entanglementSoftDOFs}.

\begin{figure}
  \centering
  \includegraphics[scale=1.2]{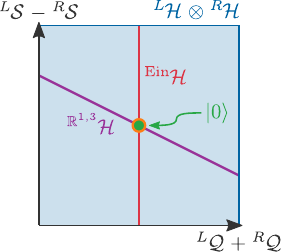}
  \caption{The relation between various Hilbert spaces. Each point in the plane
    represents a particular eigenspace of
    \(\csOp[\lMilne]+\csOp[\rMilne]\) and
    \(\goldOp[\lMilne]-\goldOp[\rMilne]\).}
  \label{fig:hilbSpaces}
\end{figure}

It remains to specify the entangling constraints $\cstr[][\ent]$. Let us first
consider the case where the entangling surface
$\partial\cauchy[\rMilne]=-\partial\cauchy[\lMilne]$ runs through the interior
of the larger spacetime under consideration, \eg{} when gluing the Hilbert
spaces of the $\lMilne$ and $\rMilne$ Milne patches in \cref{fig:potato} to
obtain the Hilbert space of the Einstein static universe. Here, it is natural in
Maxwell theory to require a matching of $\edgeE[\lMilne]$ and $\edgeE[\rMilne]$
such that Gauss's law is satisfied on the entangling surface. Recall from
\cref{eq:localCstrR} that $\edgeE[\rMilne]$ is identified with the perpendicular
component of the electric field at $\partial\cauchy[\rMilne]$, and similarly for
$\lMilne$. Thus, \cite{Donnelly:2016auv} takes as the entangling constraints $
-\cstr[\lMilne][\bdy]\relax[\gaugeParam]
-\cstr[\rMilne][\bdy]\relax[\gaugeParam]$; alternatively, by
\cref{eq:cstrRAscsOpR}, we take
\begin{align}
  \cstr[\Ein][\ent]
  \relax[\gaugeParam]
  &=
    \cstr[\lMilne][\blk]\relax[\gaugeParam]
    +\cstr[\rMilne][\blk]\relax[\gaugeParam]
    \;,
    \label{eq:cstrEntEin}
\end{align}
or equivalently
\begin{align}
  \entOp[\Ein][a](\celw)
  &= \csOp[\lMilne][a](\celw)
    +\csOp[\rMilne][a](\celw)
    \;.
    \label{eq:entOpEin}
\end{align}
The constraint space $\cstrSpace[\Ein]$ of the Einstein static universe thus satisfies
\begin{align}
  \cstr[\Ein][\ent]
  &= 0
    \;,
  &
  &\text{\ie{}}
  &
    \entOp[\Ein]
  &= 0
  & (\text{on $\cstrSpace[\Ein]$}),
\end{align}
which specifies a subspace of the $\lMilne$ and $\rMilne$ constraint space
$\cstrSpace[\lMilne]\times\cstrSpace[\rMilne]$. At the quantum level, states
$\ketOf[\Ein]{\psi}$ in the Hilbert space
$\hilbSpace[\Ein]\subset\hilbSpace[\lMilne]\otimes\hilbSpace[\rMilne]$ of the
Einstein static universe are annihilated by \cref{eq:cstrEntEin,eq:entOpEin}:
\begin{align}
  \cstr[\Ein][\ent]\,
  \ketOf[\Ein]{\psi}
  &= 0
    \;,
  &
  &\text{\ie{}}
  &
    \entOp[\Ein]\,
    \ketOf[\Ein]{\psi}
  &= 0 .
\label{eq:einStateCstr}
\end{align}
In \cref{fig:hilbSpaces}, we illustrate the embedding of the Einstein static
Hilbert space into the product of Milne Hilbert spaces. We also illustrate the
embedding of the Minkowski Hilbert space, to which we now turn.

From the Minkowski perspective, we note from \cref{eq:minkSoftOpAsLR} that 
\cref{eq:cstrEntEin,eq:entOpEin} have the simple interpretation of charges
associated to asymptotic symmetries:
\begin{align}
  \cstr[\Ein][\ent][\gaugeParam]
  &= \cstr[][\blk][\gaugeParam]
    \;,
    &
    \entOp[\Ein]
  &= \csOp
    \;.
    \label{eq:cstrEinEntAsMinkCharge}
\end{align}
Recall from \cref{sec:asymSym} that, when previously considering the Minkowski
theory, we referred to $\cstr[][\blk]$ as simply $\cstr$. Thus, if a state is
common to both the Einstein static and Minkowski Hilbert spaces, then
\cref{eq:einStateCstr} requires the state to have vanishing Minkowski asymptotic
charges. One such state of particular relevance to our discussion will be the
vacuum state $\ket{0}$ shared by the Einstein static and Minkowski theories.

Obviously, other Minkowski states generically do not have vanishing asymptotic
charges. This is already evident at the classical level, where turning on the
conformally soft mode $\confPrimA{\cs}$ leads to a nonvanishing value for
\begin{align}
  \cstr[][\blk][\gaugeParam, \confPrimA{\cs}]
  &= 4\pi\, \diff \gaugeParam
\;,
\end{align}
as can be seen from
\cref{eq:softMinkProd}. This can be
understood as a violation to Gauss's law due to the presence of the image
sources illustrated in \cref{fig:milneEdgeE}. One may also replace
$\confPrimA{\cs}$ above with the sourceless $\confPrimA{\cs'}$, as can be seen from
\cref{eq:AcsPrimeAsLR} and \cref{eq:cscsPrimEquivalence}. Then the apparent
`violation' to Gauss's law arises because neither the $\lMilne$ nor $\rMilne$
theories can see the $\inv{\minkX}^2=0$ shockwave which redistributes electric
field lines --- see \cref{fig:milneEdgeEPrime}.

However, as hinted
around \cref{eq:milneSoftModesDressed}, there is
certainly a restriction in the $\lMilne$ and $\rMilne$ soft modes that one is
allowed to turn on in Minkowski theory. To select an appropriate entangling
constraint, we observe that the two permitted
combinations \labelcref{eq:milneSoftModesDressed} of Milne soft modes both
have vanishing Milne inner product with
$\confPrimA[\lMilne]{\gold}+\confPrimA[\rMilne]{\gold}+2(\confPrimA[\lMilne]{\edge}-\confPrimA[\rMilne]{\edge})$.
Thus, we are led to the notion that permissible Minkowski
configurations are built from Milne configurations which have vanishing inner
product with
$\confPrimA[\lMilne]{\gold}+\confPrimA[\rMilne]{\gold}+2(\confPrimA[\lMilne]{\edge}-\confPrimA[\rMilne]{\edge})$.
Hence the appropriate entangling constraint is 
\begin{align}
  \entOp[][a](\celw)
  &=
    \csOp[\lMilne][a](\celw)
    +\csOp[\rMilne][a](\celw)
    +2\left[
    \goldOp[\lMilne][a](\celw)
    -\goldOp[\rMilne][a](\celw)
    \right]
    \;,
    \label{eq:entOpMink}
\end{align}
instead of \cref{eq:entOpEin}. Equivalently, instead of \cref{eq:cstrEntEin}, we
may consider
\begin{align}
  \cstr[][\ent]
  \relax[\gaugeParam]
  &=
    \cstr[\lMilne][\blk]
    \relax[\gaugeParam]+
    \cstr[\rMilne][\blk]
    \relax[\gaugeParam]
    +2\left(
    \cstrconj[\lMilne]
    \relax[\gaugeParam]
    -\cstrconj[\rMilne]
    \relax[\gaugeParam]
    \right)
    \;,
    \label{eq:cstrEntMink}
\end{align}
where we have defined
\begin{align}
  \cstrconj[\rMilne]\relax[\gaugeParam]
  &= { -\frac{1}{4\pi}}
    \int_{\partial \cauchy[\rMilne]} \celVolForm
    \gaugeParam \,
    \celCovD\cdot \goldOp[\rMilne]
    \qquad(\text{same for $\lMilne$})
\end{align}
in analogy with \cref{eq:cstrRAscsOpR}.
Note that, essentially by construction, 
\begin{align}
  \commut{\cstr[][\ent]}{\goldOp}
  &= 0
    \;,
  &
  &\text{\ie{}}
  &
    \commut{\entOp}{\goldOp}
  &= 0 \;,
    \label{eq:commutEntGoldOp}
\end{align}
where $\goldOp$ is the Minkowski Goldstone operator \labelcref{eq:goldOp}.
Of course, a nearly identical analysis
may be done starting from \cref{eq:milneSoftModesDressedPrime}, which leads to
\begin{align}
  \entOp[][a]'(\celw)
  &=
    \csOp[\lMilne][a](\celw)
    +\csOp[\rMilne][a](\celw)
    +
    \goldOp[\lMilne][a](\celw)
    -\goldOp[\rMilne][a](\celw)
    \label{eq:entOpMinkPrime}
  \\
  \cstr[][\ent']
  \relax[\gaugeParam]
  &=
    \cstr[\lMilne][\blk]
    \relax[\gaugeParam]
    +\cstr[\rMilne][\blk]
    \relax[\gaugeParam]
    +
    \cstrconj[\lMilne]
    \relax[\gaugeParam]
    -\cstrconj[\rMilne]
    \relax[\gaugeParam]
    \label{eq:cstrEntMinkPrime}
\end{align}
when considering the source-free extension $\confPrimA{\cs'}$ of the Minkowski
conformally soft mode.

In contrast to the Einstein static case, however, the constraints as
written in
\cref{eq:entOpMink,eq:cstrEntMink,eq:entOpMinkPrime,eq:cstrEntMinkPrime} do
not vanish identically in the Minkowski constraint space $\cstrSpace$ and do not
identically annihilate Minkowski states. This is because the Milne edge modes
$\confPrimA[\lMilne]{\edge}$ and $\confPrimA[\rMilne]{\edge}$ carry $\reg$
regulators which are then inherited by the Goldstone operators
$\goldOp[\lMilne]$ and $\goldOp[\rMilne]$ defined in \cref{eq:goldOpR}. At
finite $\reg$, the Milne edge modes have nonvanishing inner product with
non-soft Milne modes \labelcref{eq:inv} and so the Goldstone operators in fact
carry some dependence on the non-soft degrees of freedom considered in
\cref{sec:tfd}. It is only in the $\reg\to 0$ limit that we have
\begin{align}
  \cstr[][\ent]
  &\eqReg 0
    \;,
  &
  &\text{\ie{}}
  &
    \entOp
  &\eqReg 0
  & (\text{on $\cstrSpace$})
    \label{eq:minkCstrSpace}
\end{align}
and, for states $\ket{\psi}$ in
the Minkowski Hilbert space $\hilbSpace$,
\begin{align}
  \cstr[][\ent]\,
  \ket{\psi}
  &\eqReg 0
    \;,
  &
  &\text{\ie{}}
  &
    \entOp[]\,
    \ket{\psi}
  &\eqReg 0
    \label{eq:minkStateCstr}
\end{align}
(or similarly for $\cstr[][\ent']$ or $\entOp'$). As we will see
in \cref{sec:entanglementSoftDOFs}, the $\reg$ regulator will play an important
role in the entanglement between $\lMilne$ and $\rMilne$ soft degrees of freedom.

Still, we may construct the Minkowski Hilbert space $\hilbSpace$ as follows. As
stated previously, there is a vacuum state
\begin{align}
  \ket{0}\in\hilbSpace\cap\hilbSpace[\Ein]
\end{align}
common to both the Einstein static and Minkowski theories. Following
\cite{Arkani-Hamed:2020gyp}, we can construct other states $\softCohKet{\cstrVal}$ in
the Minkowski Hilbert space $\hilbSpace$ with arbitrary Minkowski asymptotic
charge as coherent states built from $\ket{0}$:
\begin{align}
  \softCohKet{\cstrVal}
  &= e^{i \goldOp\relax[\cstrVal]} \ket{0}
    \;,
    \label{eq:softStateMink}
    \\
  \goldOp\relax[\cstrVal]
  &=
\symp{\phSpace}(
    \confPrimA{\cs}\relax[\cstrVal],
    \opA
    )
    =
    \frac{1}{4\pi} \int \celVolForm(\celw)
    \cstrVal(\celw)
    \goldOpI(\celw,\infty)
    \;,
    \label{eq:goldOpSmeared}
  \\
  \confPrimA{\cs}\relax[\cstrVal]
  &=
    \frac{1}{4\pi} \int \celVolForm(\celw)
    \cstrVal(\celw)
    \confPrimA{\csI}
    (\celw,\infty)\;.
    \label{eq:AcsSmeared}
\end{align}
Here, $\cstrVal$ is an arbitrary scalar function over celestial space (not to be
confused with $\nullUnit$ introduced in \cref{eq:null-vector} which is also a
spacetime vector) with zero mean and the smeared mode
$\confPrimA{\cs}\relax[\cstrVal]$ and operator $\goldOp\relax[\cstrVal]$ have
been defined in terms of the integrated conformally soft mode
\labelcref{eq:lettuce} and integrated Goldstone operator \labelcref{eq:goldOpI}.
By design, $\confPrimA{\cs}\relax[\cstrVal]$ has asymptotic charge
\begin{align}
  \cstr[][\blk][\gaugeParam, \confPrimA{\cs}\relax[\cstrVal]]
  &=
    i\commut{
    \cstr[][\blk][\gaugeParam]
    }{
    \goldOp\relax[\cstrVal]
    }
    =
    \int \celVolForm \gaugeParam\, \cstrVal
    \label{eq:cstrValOfAcsSmeared}
\end{align}
and so the unitary operator and state in
\cref{eq:softStateMink} satisfy
\begin{align}
  e^{-i \goldOp\relax[\cstrVal]} \cstr[][\blk][\gaugeParam]e^{i \goldOp\relax[\cstrVal]}
  &= \cstr[][\blk][\gaugeParam]
    + \int \celVolForm \gaugeParam\, \cstrVal\;,
    \label{eq:unitaryShiftingCstr}
  \\
  \cstr[][\blk][\gaugeParam]\softCohKet{\cstrVal}
  &= \left(
    \int \celVolForm \gaugeParam\, \cstrVal
    \right)\softCohKet{\cstrVal} \;.
    \label{eq:asymChargeEigenstate}
\end{align}
Thus, $\softCohKet{\cstrVal}$ are Minkowski asymptotic charge eigenstates with
eigenvalues parametrized by $\cstrVal$. The requirement for $\cstrVal$ to have
vanishing mean is closely related to the fact that pure Maxwell theory should
have vanishing global charge, as measured by $\cstr[][\blk][\gaugeParam]$ where
$\gaugeParam$ takes a constant value (in both the interior of $\celPlane$ and at
$\infty$). If $\cstrVal$ has nonvanishing mean in the interior of celestial
space, then there will be localized Coulomb fields on $\nullInfty[\pm]$ at the
angular position corresponding to the $\infty$ point of celestial space. In
fact, such a configuration is quite singular: note that the $\nullUnit_2\cdot\minkX=0$
shockwave of $\confPrimA{\csI}(\celw_1,\celw_2)$ becomes sharp in the
$\celw_2\to\infty$ limit, giving $\confPrimA{\cs}\relax[\cstrVal]$ an infinite
energy unless the smearing function $\cstrVal$ has zero mean. Indeed, from \eqref{eq:qX1} we see that the $\hat{q}_2\cdot X = 0$ shockwave with $\celw_2 \rightarrow \infty$ corresponds to $u \rightarrow -\infty$, or equivalently the blue plane depicted in figure \ref{fig:planarRetardedCoordinates3D}. Consequently the regulator, which  would otherwise give the shockwaves a small width has no effect in this case. Smearing against a function of zero mean on the sphere as in \eqref{eq:AcsSmeared} then removes the associated singularity.

From \cref{eq:commutEntGoldOp}, we see that $\goldOp\relax[\cstrVal]$ commutes with
the Minkowski entangling constraints and so $\softCohKet{\cstrVal}$ satisfies
\cref{eq:minkStateCstr}. Moreover, even at finite $\reg$, we expect
$\softCohKet{\cstrVal}\in\hilbSpace$ since $\goldOp\relax[\cstrVal]$ is a Minkowski operator
which ought to preserve $\hilbSpace$. By further acting on these
$\softCohKet{\cstrVal}$ with non-soft conformal primary operators
$\confPrimOp{1+i\milneFreq,+}$ introduced in \cref{eq:confPrimOp} for the
Minkowski theory, the full Hilbert space $\hilbSpace$ is generated.

Previously, in \cref{sec:tfd}, we have already seen how the Bogoliubov
transformation relating the Minkowski and Milne non-soft operators leads to a
thermal relation between the Minkowski and Milne vacua.\footnote{Recall, however,
that our analysis in \cref{sec:tfd} mostly relies upon $\reg\to 0$ limits. As we
briefly mentioned above \cref{eq:minkCstrSpace} and we will
discuss further in \cref{sec:entanglementSoftDOFs,sec:discussion}, there is some
mixing between the `soft' and non-soft degrees of freedom at finite $\reg$. We will leave a more
careful analysis of this issue for future work.}
In \cref{sec:milneSoft},
we also saw how the Minkowski and Milne soft operators are related. In
particular, using \cref{eq:goldOpIAsLR},
we may rewrite \cref{eq:goldOpSmeared} as
\begin{align}
  \goldOp\relax[\cstrVal]
  &=
    \int \celVolForm(\celw)
    \cstrVal(\celw)
    \cstr[\lMilne][\blk][
    \log(-\nullUnit(\celw)\cdot\minkX)
    ]
    +\frac{
    \goldOp[\lMilne]\relax[\cstrVal]
    +\goldOp[\rMilne]\relax[\cstrVal]
    }{2}
    \label{eq:goldOpSmearedSplitting}
\end{align}
or
\begin{align}
  \goldOp'[\cstrVal]
  &=
    2\int \celVolForm(\celw)
    \cstrVal(\celw)
    \cstr[\lMilne][\blk][
    \log(-\nullUnit(\celw)\cdot\minkX)
    ]
    +\frac{
    \goldOp[\lMilne]\relax[\cstrVal]
    +\goldOp[\rMilne]\relax[\cstrVal]
    }{2},
    \label{eq:goldOpSmearedPrimeSplitting}
\end{align}
by \cref{eq:goldOpIPrimeAsLR}. We have defined, in analogy to \cref{eq:goldOpSmeared,eq:AcsSmeared},
\begin{align}
\goldOp[\rMilne]\relax[\cstrVal]
&=
\symp{\phSpace[\rMilne]}(
\confPrimA{\edge}[\rMilne][\cstrVal],
\opA
)
=
\frac{1}{2\pi} \int \celVolForm(\celw)
\cstrVal(\celw)
\goldOpI[\rMilne](\celw,\infty)
\;,
\\
\confPrimA[\rMilne]{\edge}\relax[\cstrVal]
&=
\frac{1}{2\pi} \int \celVolForm(\celw)
\cstrVal(\celw)
\confPrimA[\rMilne]{\edgeI}
  (\celw,\infty)
  \label{eq:AERSmeared}
\end{align}
and similarly for $\lMilne$. Note the factor of $2$ relative to
\cref{eq:goldOpSmeared,eq:AcsSmeared}: this compensates for the relative factor
of $1/2$ between \cref{eq:cstrRBlkEdgeI,eq:asymChargeCSI} so that
\begin{align}
\cstr[\rMilne][\blk][\gaugeParam, \confPrimA[\rMilne]{\edge}\relax[\cstrVal]]
&=
i\commut{
  \cstr[\rMilne][\blk][\gaugeParam]
}{
  \goldOp[\rMilne]\relax[\cstrVal]
}
=
  \int \celVolForm \gaugeParam\, \cstrVal
  \;,
  \label{eq:edgeModeCharge}
  \\
  e^{-i \goldOp[\rMilne]\relax[\cstrVal]}
  \cstr[\rMilne][\blk][\gaugeParam]
  e^{i \goldOp[\rMilne]\relax[\cstrVal]}
  &= \cstr[\rMilne][\blk][\gaugeParam]
    + \int \celVolForm \gaugeParam\, \cstrVal
    \label{eq:unitaryShiftingCstrR}
\end{align}
similar to \cref{eq:cstrValOfAcsSmeared,eq:unitaryShiftingCstr}.
The fact that $\goldOp\relax[\cstrVal]$ can be decomposed into $\lMilne$ and
$\rMilne$ parts will be useful to our discussion of entanglement in \cref{sec:entanglementSoftDOFs}.

\subsubsection{Entanglement of soft degrees of freedom}
\label{sec:entanglementSoftDOFs}

The focus of this section has been to understand the entanglement of a Milne
patch $\rMilne$ in the vacuum $\ket{0}$ state shared by the Einstein static and
Minkowski theories; moreover, in this subsection we would like to understand the
entanglement in the soft sector of this state. Much of our discussion here
parallels that of \cite{Donnelly:2015hxa,Donnelly:2014fua}, though we will aim
to be more explicit in relating the Euclidean action of edge modes to an energy in
a Boltzmann probability distribution over Milne asymptotic charges.

\paragraph{Weyl transformation of the celestial space.}
So far, we considered the case of a planar celestial space $\mathbb{R}^2$. Since
the celestial conformal frame is naturally inherited by a rescaled geometry of
the entangling surface, it will prove useful to generalize our analysis to
surfaces related to the plane by a Weyl rescaling. The goal of this subsection
is to explain how this can be achieved.

To discuss an arbitrary conformal frame for the celestial CFT, it is helpful to
recall the related discussion in AdS/CFT. Here, Weyl transformations of the CFT
may be related to diffeomorphisms of AdS which are best understood in terms of a
Fefferman-Graham expansion. In particular, within the Minkowski metric
\labelcref{eq:Milne-metric}, the metric on hyperbolic (\ie{} Euclidean AdS$_3$)
slices can take an arbitrary Fefferman-Graham form\footnote{Note that the coefficient of the would-be leading logarithmic term is determined in terms of $\gamma^{(0)}$. For two-dimensional celestial spaces, this is proportional to the metric variation of the Euler character and hence vanishes \cite{deHaro:2000vlm, Skenderis:2000in}.}
\begin{align}
  \diff s_{\hyp{3}}^2
  &= \frac{\diff\fgr^2}{4\fgr^2}
    + \frac{
    \celMet_{a b}(\fgr,\minkz)
    \diff \minkz[^a]
    \diff \minkz[^b]
    }{\fgr}
    \;,
  &
    \celMet(\fgr,\minkz)
  &= \celMet^{(0)}(\minkz)
    + \fgr \celMet^{(2)}(\minkz)
    + \fgr^2 \celMet^{(4)}(\minkz)
    \label{eq:fgExpansion}
\end{align}
for which \cref{eq:Poincare-metric} is just one choice. For an arbitrary CFT
metric $\celMet^{(0)}$ \cite{Skenderis:1999nb},
\begin{align}
  \celMet^{(2)}
  &= \frac{1}{2} \left(
    - R^{(0)} \celMet^{(0)}
    + T
    \right)
    \;,
  &
    \celMet^{(4)}
    &= \frac{1}{4} \celMet^{(2)} (\celMet^{(0)})^{-1} \celMet^{(2)}
\end{align}
where $R^{(0)}$ is the curvature of $\celMet^{(0)}$ and $T$ satisfies
\begin{align}
  (\celCovD^{(0)})^a T_{ab}
  &= 0 \;,
  &
    \tr[(\celMet^{(0)})^{-1} T]
    &= 0 \;.
\end{align}
Thus, $T$ may be identified as the stress tensor of a scalar field $\phi$
satisfying Liouville equations of motion. Concretely, in conformal gauge,
\begin{align}
  (\celMet_\Omega^{(0)})_{ab} \diff \celw[^a] \diff \celw[^b]
  &= \Omega^2  \diff s_{\celPlane}^2
    \;,
      \label{eq:celConfGauge}
\end{align}
where $\Omega$ is an arbitary function on the celestial space,
\begin{align}
  \phi
  &= 2 \log\Omega
    \;.
\end{align}

Let us now tie this discussion back to Cartesian Minkowski coordinates $\minkX^\mu$ and a
generalization of the retarded coordinates $(u,r,\minkz)$.
We take, as in \cref{eq:cupcake-intro},
\begin{align}
  \minkX^{\mu}
  &= \minku\, \refVec_\Omega^\mu(\minkz)
    + \minkr \nullUnit_\Omega^{\mu}(\minkz) \;,
\end{align}
but now, in place of \cref{{eq:null-vector},eq:apple},
\begin{align}
  \hat{q}_{\Omega}^{\mu}
  &= \Omega \hat{q}^{\mu}
    \;,
  &
  \refVec_\Omega^\mu(\minkz)
  &= \frac{1}{2}\left[
    (\celCovD_\Omega^{(0)})^2
    + \frac{\alpha}{2} R_\Omega^{(0)}
    \right] \nullUnit_\Omega^\mu
    \;.
    \label{eq:minkXOmega}
\end{align}
Here, $\alpha$ is a constant which may be chosen arbitrarily at our convenience and   $(\celCovD_\Omega^{(0)})^2 \equiv \gamma_{\Omega}^{ab}(\celCovD_\Omega^{(0)})_a(\celCovD_\Omega^{(0)})_b$ is the Laplacian with respect to \eqref{eq:celConfGauge}.
The Minkowski metric then becomes
\begin{align}
\label{eq:Weyl-bulk-metric}
\diff s^2
&=
- \frac{\alpha-1}{2} R_\Omega^{(0)} \diff u^2
- 2 \diff u \, \diff r
+ \frac{\alpha-1}{2} u \, \diff[\minkz] R_\Omega^{(0)} \, \diff u
  + \frac{\minkX^2}{\fgr}
  (\celMet_\Omega)_{a b}(\fgr,\minkz)
  \,
  \diff \minkz[^a]
  \diff \minkz[^b]
  \;,
\end{align}
where
\begin{align}
  \rho
  &= \frac{4u^2}{\minkX^2}
    \;,
    &
    \minkX^2
  &= 2u \left( r + \frac{\alpha-1}{4} u R_\Omega^{(0)} \right)
    \;.
\end{align}
Of course, this agrees with
\cref{eq:Milne-metric,eq:fgExpansion} upon identifying the Milne time $\milnet$
again by the last expression of \cref{eq:milnet},
\begin{align}
  \tau
  &= \frac{1}{2}\log(-\minkX^2)
    \;.
\end{align}
Clearly, many of the above expressions simplify if one chooses $\alpha=1$ so
that all explicit dependences on $R_\Omega^{(0)}$ disappear.

Taking $\Omega=1$ (with $\alpha$ then dropping out), we recover the planar
celestial space considered thus far in this paper. On the other hand, the
celestial sphere corresponds to
\begin{align}
  \Omega_{S^2}
  &= \frac{2}{1+z\bz}
  \;,
  &
  R^{(0)}_{S^2}
  &= 2
  \;,
  &
    \celMet_{S^2}(\fgr,\minkz)
  &= \left( 1 - \frac{\fgr}{2} + \frac{\fgr^2}{16} \right)\celMet_{S^2}^{(0)}(\minkz)
    \;.
\end{align}
Further taking $\alpha=2$,
\begin{align}
  \hat{n}_{S^2}^\mu &= (1,0,0,0)
  \;,
  &
  \frac{\minkX^2}{\fgr}
    \celMet_{S^2}(\fgr,\minkz)
  &= \minkr^2 \celMet^{(0)}_{S^2}(\minkz)
    \;,
\end{align}
we recover the standard spherical Bondi coordinates $(u,r,\minkz)$ and metric. For simplicity, we will continue working with the planar celestial space. We will consider a spherical entangling surface when computing the edge mode partition function and associated entanglement entropy after \cref{eq:onShellAction}.

\paragraph{Density matrices and soft statistics.}
First, we shall discuss the reduced density matrix $\dnsMat[\rMilne]$ for the
$\rMilne$ Milne patch and glean as much information as possible purely from what
we have learned previously about Milne soft degrees of freedom.

We can begin more generally, by also considering the soft states $\softCohKet{\cstrVal}$
built unitarily from the vacuum $\ket{0}$ in \cref{eq:softStateMink}. Tracing
out the $\lMilne$ system, we are left with a density matrix
$\dnsMat[\rMilne][\cstrVal]$ on $\rMilne$:
\begin{align}
  \dnsMat[\rMilne][\cstrVal]
  &= \tr_{\hilbSpace[\lMilne]} \softCohKet{\cstrVal}\softCohBra{\cstrVal}
    = e^{i \goldOp[\rMilne]\relax[\cstrVal]/2}\,
    \dnsMat[\rMilne][0]\,
    e^{-i \goldOp[\rMilne]\relax[\cstrVal]/2}.
\end{align}
As noted in the last equality, these density matrices for various $\cstrVal$ are
unitarily related as a consequence of the splitting
\labelcref{eq:goldOpSmearedSplitting} (or
\labelcref{eq:goldOpSmearedPrimeSplitting}) of $\goldOp\relax[\cstrVal]$ into
$\lMilne$ and $\rMilne$ pieces. This implies that the same spectrum is shared by
the density matrices $\dnsMat[\rMilne][\cstrVal]$. Consequently, the von
Neumann and R\'enyi entropies
\begin{align}
  \entropy[\vN](\dnsMat)
  &= -\tr(\dnsMat\log\dnsMat)
    = \lim_{n\to 1} \entropy[n](\dnsMat)
    \;,
    \label{eq:entropyvN}
    \\
      \entropy[n](\dnsMat)
  &= -\frac{\log\tr(\dnsMat^n)}{n-1}
    \label{eq:entropyRenyi}
\end{align}
which characterize entanglement, are also independent of $\cstrVal$ when
evaluated for $\dnsMat[\rMilne][\cstrVal]$:
\begin{align}
  \entropy[\vN](\dnsMat[\rMilne][\cstrVal])
  &= \entropy[\vN](\dnsMat[\rMilne][0]) \;,
  &
    \entropy[n](\dnsMat[\rMilne][\cstrVal])
  &= \entropy[n](\dnsMat[\rMilne][0]).
\end{align}
Without loss of generality, in the following, we shall therefore restrict
attention to the vacuum state $\ket{0}$ with reduced density matrix
$\dnsMat[\rMilne][0]$.

As noted below \cref{eq:cstrEinEntAsMinkCharge}, because this vacuum is also
shared by the Einstein static theory, it is annihilated by the sum
\labelcref{eq:cstrEntEin} of $\lMilne$ and $\rMilne$ asymptotic charges. From
this fact, it is easy to show using the cyclicity of the trace that the
$\rMilne$ asymptotic charge
$\cstr[\rMilne][\blk][\gaugeParam]$ commutes with $\dnsMat[\rMilne][0]$:
\begin{align}
  \commut{
  \cstr[\rMilne][\blk][\gaugeParam]
  }{
  \dnsMat[\rMilne][0]
  }
  &= \tr_{\hilbSpace[\lMilne]}
    \commut{
    \cstr[\rMilne][\blk][\gaugeParam]
    }{
    \ket{0}\bra{0}
    }
    =-\tr_{\hilbSpace[\lMilne]}
    \commut{
    \cstr[\lMilne][\blk][\gaugeParam]
    }{
    \ket{0}\bra{0}
    }
    = 0
    \;.
    \label{eq:commutDnsMatCstrR}
\end{align}
Consequently,
\begin{align}
  \dnsMat[\rMilne][0]
  &= \int \edgeVolForm[\cstrVal] \,
    \prob[\cstrVal]\,
    \dnsMat[\rMilne][0,\cstrVal]
    \label{eq:dnsMatBlockDecomposition}
\end{align}
decomposes into blocks
$\dnsMat[\rMilne][0,\cstrVal]$ of definite $\cstr[\rMilne][\blk][\gaugeParam]$:
\begin{align}
  \cstr[\rMilne][\blk][\gaugeParam]\,
  \dnsMat[\rMilne][0,\cstrVal]
  &=
    \dnsMat[\rMilne][0,\cstrVal]\,
    \cstr[\rMilne][\blk][\gaugeParam]
    =
    \left(
    \int \celVolForm \gaugeParam\, \cstrVal
    \right)
    \dnsMat[\rMilne][0,\cstrVal]
    \;.
    \label{eq:dnsMatBlock}
\end{align}
We have introduced in \cref{eq:dnsMatBlockDecomposition} a measure
$\edgeVolForm[\cstrVal]$ over the functions $\cstrVal$ parameterizing
eigenvalues of $\cstr[\rMilne][\blk][\gaugeParam]$ --- such a measure has been
derived in \cite{Donnelly:2015hxa} by taking the continuum limit of lattice
gauge theory.

The density matrices $\dnsMat[\rMilne][0]$ and $\dnsMat[\rMilne][0,\cstrVal]$ are both defined to have unit trace. The normalization of the overall density matrix $\dnsMat[\rMilne][0]$ and the
blocks $\dnsMat[\rMilne][0,\cstrVal]$ implies, using \eqref{eq:dnsMatBlockDecomposition}, that $\prob[\cstrVal]$ is a
probability distribution that is also normalized:
\begin{align}
  1 &= \int \edgeVolForm[\cstrVal]\, \prob[\cstrVal]
      \;.
      \label{eq:dnsMatProbNormalization}
\end{align}
The probability distribution $\prob[\cstrVal]$ is over the functions
$\cstrVal$ determining the $\rMilne$ asymptotic charge, as in \cref{eq:dnsMatBlock}.
In other words, it measures fluctuations in the $\rMilne$ edge modes which turn
on these asymptotic charges.

Associated to this probability distribution are classical Shannon and R\'enyi
entropies
\begin{align}
  \entropy[\Sh][p]
  &= -\int \edgeVolForm[\cstrVal]\,
    p[\cstrVal] \log p[\cstrVal]
  = \lim_{n\to 1} \entropy[n][p]
    \;,
    \label{eq:entropySh}
  \\
  \entropy[n][p]
  &= - \frac{
    \log \left\{
    \int\edgeVolForm[\cstrVal]\, \prob[\cstrVal]^n
    \right\}
    }{n-1}
    \;.
    \label{eq:entropyClassicalRenyi}
\end{align}
By substituting in the block decomposition \labelcref{eq:dnsMatBlockDecomposition},
one finds that the von Neumann entropy \labelcref{eq:entropyvN} of
$\dnsMat[\rMilne][0]$ is given by the sum of the above Shannon entropy and the von
Neumann entropies of $\dnsMat[\rMilne][0,\cstrVal]$ averaged
over $\cstrVal$:
\begin{align}
  \entropy[\vN](\dnsMat[\rMilne][0])
  &= \entropy[\Sh][p]
    + \int \edgeVolForm[\cstrVal]\,
    p[\cstrVal]\,
    \entropy[\vN](\dnsMat[\rMilne][0,\cstrVal])
    \;.
    \label{eq:entropyvNBlockDecomposition}
\end{align}
Meanwhile, the R\'enyi entropies are given by
\begin{align}
  \entropy[n](\dnsMat[\rMilne][0])
  &= - \frac{
    \log \left\{
    \int\edgeVolForm[\cstrVal]\, \prob[\cstrVal]^n
    \tr(\dnsMat[\rMilne][0,\cstrVal]^n)
    \right\}
    }{n-1}
    \;.
    \label{eq:entropyRenyiBlockDecomposition}
\end{align}
Below, we shall see that the blocks $\dnsMat[\rMilne][0,\cstrVal]$ are unitarily
related, so that each of the von Neumann and R\'enyi entropies of
$\dnsMat[\rMilne][0]$ are given the sum of a piece depending only on
$\prob[\cstrVal]$ --- an edge mode contribution --- and a piece depending only
on $\dnsMat[\rMilne][0,0]$ --- a bulk contribution.

We would like to better understand how the density matrix
  description of the state given above can be formulated in terms of path
  integrals. Our starting point
is the observation that the expectation value of arbitrary $\rMilne$ operators
may be expressed in two ways: as a trace against the $\rMilne$ density matrix
\labelcref{eq:dnsMatBlockDecomposition}, or as a path integral:
\begin{align}
  \bra{0}
  \operator[\rMilne] \cdots \operator[\rMilne]
  \ket{0}
  &=
  \tr\left(
    \operator[\rMilne] \cdots \operator[\rMilne]\,
    \dnsMat[\rMilne][0]
    \right)
    = \int \edgeVolForm[\cstrVal] \,
    \prob[\cstrVal]\,
    \tr(
    \operator[\rMilne] \cdots \operator[\rMilne]\,
    \dnsMat[\rMilne][0,\cstrVal]
    )
    \label{eq:operatorRExpValDnsMat}
  \\
  &=\frac{
    \int_{i\milnet=0}^{i\milnet=2\pi} \fieldVolForm[A]\,
    e^{-\act[][2\pi][A]}
    \operator[\rMilne] \cdots \operator[\rMilne]\,
    }{
    \partFunc[\rMilne][][2\pi]
    }
    \;.
    \label{eq:operatorRExpValPathIntegral}
\end{align}
 Note that the path integrals in eqns. \eqref{eq:operatorRExpValDnsMat} and \eqref{eq:operatorRExpValPathIntegral} are over superselection sectors labeled by the $\rMilne$ Milne charge and gauge field configurations in Minkowski spacetime respectively.
For simplicity, we may consider operators $\operator[\rMilne]$ on the
$\milnet=0$ time slice, in which case the path integral need only prepare the
Minkowski (and Einstein static) vacuum bra and ket states and runs over one
thermal period $2\pi$ in $i\milnet$, as described in \cref{sec:pathIntegrals}.
Correspondingly, $\act[][2\pi][A]$ and $\partFunc[\rMilne][][2\pi]$ are the
Euclidean action and partition function over this thermal period. \footnote{For operators
$\operator[\rMilne]$ at other Lorentzian Milne times, one must consider more
general Schwinger-Keldysh path integrals with the time contour deformed to reach
those Lorentzian times.}

The partition function, for an arbitrary thermal period $\invTemp$, is defined
by a path integral
\begin{align}
  \partFunc[\rMilne][][\invTemp]
  &=
    \int_{i\milnet\sim i\milnet+\invTemp} \fieldVolForm[A]\,
    e^{-\act[][\invTemp][A]}
    \label{eq:partFunc}
\end{align}
with no operator insertions. Here, we are considering a Euclidean spacetime
where $i\milnet$ is periodically identified with period $\invTemp$, over which
one has the Euclidean action (\cf{} \cref{eq:actEuclidean})
\begin{align}
  \act[][\invTemp]
  = \int_{i\milnet \in [0,\invTemp]}
  (-i \lag[\blk])
  \;.
  \label{eq:actEuclideanPeriodic}
\end{align}
For $\beta \neq 2\pi$, the Euclidean manifold on which
\eqref{eq:actEuclideanPeriodic} is evaluated has a conical singularity.
As shown by \cite{Donnelly:2014fua,Donnelly:2015hxa}, there are
  two ways to evaluate \cref{eq:partFunc} which produce identical results: by
  regularizing the tip of the conical singularity, or excising the tip with a
  cutoff surface where one fixes then integrates over electric fluxes
  (corresponding to edge mode configurations). While we are agnostic for now, we
  will eventually take the latter approach below.

In \cref{eq:operatorRExpValPathIntegral},
$\partFunc[\rMilne][][2\pi]$ normalizes the expression for consistency with
\cref{eq:dnsMatProbNormalization}. The partition function for $\invTemp$ at
other positive integer multiples of $2\pi$ may be viewed as gluing together
multiple copies of the path integral preparing $\dnsMat[\rMilne][0]$, and can
therefore be related to the trace of higher powers of $\dnsMat[\rMilne][0]$:
\begin{align}
  \tr(\dnsMat[\rMilne][0]^n)
  &=\frac{\partFunc[\rMilne][][2\pi n]}{\partFunc[][][2\pi]^n}
    \;.
\end{align}
Thus, the R\'enyi entropies \cref{eq:entropyRenyi} can alternatively be defined
in terms of the partition functions:
\begin{align}
  \entropy[n](\dnsMat)
  &= \frac{
    n\log\partFunc[][][2\pi]
    -\log\partFunc[][][2\pi n]
    }{n-1}
    \;.
    \label{eq:entropyRenyiPartFunc}
\end{align}
Analytically continuing in $n$, one also obtains from \cref{eq:entropyvN} the
von Neumann entropy
\begin{align}
  \entropy[\vN](\dnsMat)
  &=
    (1-\invTemp\partial_\invTemp)
    \left.
    \log\partFunc[][][\invTemp]
    \right|_{\invTemp=2\pi}
    \;.
    \label{eq:entropyvNPartFunc}
\end{align}
This technique for evaluating entropies is known as the replica trick, \eg  see 
\cite{Calabrese:2004eu,Calabrese:2005zw,Callan:1994py}.

A key observation of \cite{Donnelly:2015hxa,Donnelly:2014fua} is that the
partition function may be decomposed in a manner similar to
\cref{eq:dnsMatBlockDecomposition}. To achieve this, we would like to divide
the partition function
\begin{align}
  \label{eq:partFuncIntegralOverSectors}
\partFunc[\rMilne][][\invTemp]
=
\int \edgeVolForm[\cstrVal]\,
\partFunc[\rMilne][][\invTemp][\cstrVal]
\end{align}
into pieces \(\partFunc[\rMilne][][\invTemp][\cstrVal]\) where the local electric flux
\(\pullback[\partial\cauchy[\rMilne]] (i\hodge F)\) at the entangling surface
\(\partial\cauchy[\rMilne]\) has been fixed to some function \(q\). Very
schematically,
\begin{align}
  \partFunc[\rMilne][][\invTemp][\cstrVal]
  &\eqDubious \int_{i\milnet\sim i\milnet+\invTemp} \fieldVolForm[A]\,
    e^{-\act[][\invTemp][A]}
    \fieldDeltaFunc[
    \cstrVal
    -\hodge[\minkz]\,
    \pullback[\partial\cauchy[\rMilne]]
    (i\hodge F)
    ]
    \;,
\end{align}
 but let us construct \(\partFunc[\rMilne][][\invTemp][\cstrVal]\)
  more carefully in
  terms of boundary conditions on a regulator surface.

  As mentioned below \cref{eq:actEuclideanPeriodic}, the path integral
  \(\partFunc[\rMilne][][\invTemp]\) can be evaluated on a regulated conifold
  with a conical tip that is smoothed over some neighbourhood of the entangling
  surface \(\partial\cauchy[\rMilne]\). Let us call this neighbourhood
  \(\mathcal{S}_1\) and the remainder of the Euclidean spacetime manifold
  \(\mathcal{S}_2\). On \(\partial\mathcal{S}_2 = -\partial\mathcal{S}_1\),
  which we dub the regulator surface, let us use \(\sigma_\milnet\) to refer to
  the sections of constant \(\milnet\). Then, we can split the path integral
  \(\partFunc[\rMilne][][\invTemp]\) into pieces over \(\mathcal{S}_1\) and
  \(\mathcal{S}_2\) by fixing then integrating over boundary conditions at
  \(\partial\mathcal{S}_2 = -\partial\mathcal{S}_1\). (This is equivalent to
  inserting an orthonormal set of states on \(\partial\mathcal{S}_2 =
  -\partial\mathcal{S}_1\).) The consistent set of boundary conditions (\ie{}
  states) we will consider are those specified by \(\pullback[\sigma_\milnet]
  A\) and \(\pullback[\sigma_\milnet] \hodge F\), motivated by
  \cite{Ball:2024tba}.\footnote{The act of integrating over these
    quantities will reproduce the ``dynamical edge mode boundary condition''
    formulated in \cite{Ball:2024tba}. (Note that boundary quantities conjugate
    to those integrated over can be effectively treated as fixed. As a toy
    example, one can consider the path integral representations of a state which
    is read off of the boundary condition at an endpoint. Starting with a fixed
    position boundary condition, one can Fourier transform to momentum basis by
    integrating over position --- see also \cref{foot:boundaryTerms}. The result
    is a state with fixed momentum.) \label{foot:dne}} We thus obtain
\begin{align}
  \begin{split}
    \partFunc[\rMilne][][\invTemp]
    &= \int \edgeVolForm[\pullback[\sigma_\milnet] A]\,
      \edgeVolForm[\pullback[\sigma_\milnet] \hodge F] \Biggl\{
      \left(
      \int_{\substack{\pullback[\sigma_\milnet] A\\ \pullback[\sigma_\milnet] \hodge F}}
    \fieldVolForm[A]
    e^{
    - \int_{\mathcal{S}_1} \frac{1}{2} F \wedge (i\hodge F)
    + \int_{\partial\mathcal{S}_1} A \wedge \pullback[\sigma_\milnet] (i\hodge F)
    }
    \right)
    \\
    &\phantom{{}={}}\left(
      \int_{\substack{\pullback[\sigma_\milnet] A\\ \pullback[\sigma_\milnet] \hodge F}}
    \fieldVolForm[A]
    e^{
    - \int_{\mathcal{S}_2} \frac{1}{2} F \wedge (i\hodge F)
    + \int_{\partial\mathcal{S}_2} A \wedge \pullback[\sigma_\milnet] (i\hodge F)
    }
    \right)
    \Biggr\},
  \end{split}
  \label{eq:partFuncSplit}
\end{align}
where the quantities fixed as the boundary conditions of the two
  parenthesized path integrals over \(\mathcal{S}_1\) and \(\mathcal{S}_2\) are
  identified in the subscripts. As shown, extra boundary terms
  are required in
  the action corresponding to
  the Neumann boundary condition of fixed \(\pullback[\sigma_\milnet] \hodge
  F\).\footnote{ These boundary terms are required to give the action a
    well-defined variational principle \cite{Marolf:2006nd}.
    Moreover, in Lorentz signature and on a spacelike boundary, such a
    boundary term has a very
    simple interpretation in terms of the state read off of the boundary: it implements the Fourier transformation
    between a ``position'' \(A\) basis of states (corresponding to Dirichlet boundary
    conditions) to a conjugate ``momentum'' \(\hodge F\) basis (corresponding to Neumann boundary
    conditions). When sewing together path integrals along an interface by
    path integrating over Neumann boundary conditions, the boundary terms give
    rise to a \(\delta\)-functional which identifies the unfixed components of \(A\)
    across the interface. In Euclidean signature, as in \cref{eq:partFunc}, one
    might be able to tell a similar story after complex deformations of the
    integration contours. \label{foot:boundaryTerms}}Now,
  let us consider the limit where the neighbourhood \(\mathcal{S}_1\) shrinks to
  the entangling surface. In this limit, we may reduce attention to
  \(\pullback[\sigma_\milnet] A\) and \(\pullback[\sigma_\milnet] \hodge F\)
  which are independent of \(\milnet\). Moreover, as we will justify a
  posteriori, we
  claim that the path integral over \(\mathcal{S}_1\) reduces to
\begin{align}
  \label{eq:shrinkingS1}
  \int_{\substack{
  \pullback[\sigma_\milnet] A \\
  \pullback[\sigma_\milnet] (i\hodge F)
  =\hodge[\minkz] \cstrVal
  }}
  \fieldVolForm[A]
  e^{
  -\int_{\mathcal{S}_1} \frac{1}{2} F \wedge (i\hodge F)
  + \int_{\partial\mathcal{S}_1} A \wedge \pullback[\sigma_\milnet] (i\hodge F)
  }
  &= e^{
    \int_{\partial\mathcal{S}_1} \confPrimA[\rMilne]{\edge}\relax[\cstrVal]
    \wedge
    (i\hodge \confPrimF[\rMilne]{\edge}\relax[\cstrVal])
    }
    \;,
\end{align}
  where the edge mode \(\confPrimA[\rMilne]{\edge}\relax[\cstrVal]\) is a solution
  in \(\mathcal{S}_2\)
  satisfying
\begin{align}
  \pullback[\partial\mathcal{S}_2]
  (i\hodge
  \confPrimF[\rMilne]{\edge}\relax[\cstrVal]
  )
  &=
    \hodge[\minkz] \cstrVal
    \;.
\end{align}
  Here, we are slightly overloading the meaning of
  \(\confPrimA[\rMilne]{\edge}\relax[\cstrVal]\), because in \cref{eq:AERSmeared} we
  have already defined it as a soft
  conformal primary wavefunction. But, as we see from \cref{eq:edgeModeCharge},
  this previous definition approximately matches the above description in the limit where
  \(\mathcal{S}_1\) shrinks to the entangling surface. There is nonetheless a
  slight mismatch, which we will discuss further later, due to the difference in regulating with the
  \(\reg\) parameter inherited from conformal primary wavefunctions, versus the neighbourhood
  \(\mathcal{S}_1\). Proceeding with the calculation at hand, the partition function \labelcref{eq:partFuncSplit}
  now reduces to \cref{eq:partFuncIntegralOverSectors} with
\begin{align}
  \partFunc[\rMilne][][\invTemp][\cstrVal]
  &= \int_{\pullback[\partial\mathcal{S}_2] (i\hodge F) = \hodge[\minkz] \cstrVal}
    \fieldVolForm[A]
    e^{
    - \int_{\mathcal{S}_2} \frac{1}{2} F \wedge (i\hodge F)
    + \int_{\partial\mathcal{S}_2} (A-\confPrimA[\rMilne]{\edge}\relax[\cstrVal]) \wedge (i\hodge F)
    }
    \;.
    \label{eq:partFuncWithDelta}
\end{align}

  The dependence on \(\cstrVal\) can be isolated by a change of variables which
  shifts the background by the edge mode solution
  \(\confPrimA[\rMilne]{\edge}\relax[\cstrVal]\). We find that
\begin{align}
\partFunc[\rMilne][][\invTemp][\cstrVal]
  &= e^{
    -\act[][\invTemp]\left[
    \confPrimA[\rMilne]{\edge}\relax[\cstrVal]
    \right]
    }
    \partFunc[\rMilne][][\invTemp][0]
    \;,
\end{align}
  where \(I_\beta\) is the Maxwell action with the standard Lagrangian
  \labelcref{eq:actEuclidean} integrated over \(\mathcal{S}_2\) and without any boundary terms.
The partition function therefore factorizes into an edge mode contribution
$\partFunc[\rMilne][\edge][\invTemp]$ and a bulk contribution $\partFunc[\rMilne][][\invTemp][0]$:
\begin{align}
  \partFunc[\rMilne][][\invTemp]
  &= \partFunc[\rMilne][\edge][\invTemp]
    \, \partFunc[\rMilne][][\invTemp][0]
    \;,
    \label{eq:partFuncDecomposition}
    \\
  \partFunc[\rMilne][\edge][\invTemp]
  &= \int \edgeVolForm[\cstrVal]\,
    e^{
    -\act[][\invTemp]\left[
    \confPrimA[\rMilne]{\edge}\relax[\cstrVal]
    \right]
    }
    \;.
    \label{eq:partFuncRE}
\end{align}
We emphasize that the factorization of \eqref{eq:partFuncDecomposition} into
independent edge mode and bulk contributions is a special property of the free
theory.  Moreover, the cancellation of boundary terms in the action
  was facilitated by the extra factor coming from the \(\mathcal{S}_1\) path
  integral \labelcref{eq:shrinkingS1}. Since \cref{eq:partFuncDecomposition} was
  shown by \cite{Donnelly:2015hxa,Donnelly:2014fua} to agree with the path
  integral on the smoothed conifold \footnote{For \(\beta=2\pi\),
    where conical singularity vanishes, \cite{Ball:2024tba} argues that their
    ``dynamical edge mode boundary condition'' --- see \cref{foot:dne} --- is
    ``shrinkable''. This means that the path integral over \(\mathcal{S}_2\) with
    this boundary condition reproduces the path integral over the full Euclidean
    spacetime (with \(\mathcal{S}_1\) filled in), in the limit where
    \(\mathcal{S}_1\) shrinks to zero size.}, this provides some
  justification for the guess \labelcref{eq:shrinkingS1}, but it would be nice
  to independently verify the latter in the future.

We are now in a position to relate the probability distribution
$\prob[\cstrVal]$ to the action $\act[][\invTemp]\left[
  \confPrimA[\rMilne]{\edge}\relax[\cstrVal] \right]$. For this, we observe that
\cref{eq:operatorRExpValDnsMat,eq:operatorRExpValPathIntegral}, when applied to
$\rMilne$ asymptotic charge operators, both reduce to path integrals over
$\cstrVal$:
\begin{align}
  \bra{0}
  \cstr[\rMilne][\blk][\gaugeParam_1] \cdots \cstr[\rMilne][\blk][\gaugeParam_n]
  \ket{0}
  &=
    \int \edgeVolForm[\cstrVal] \,
    \prob[\cstrVal]
    \left(
    \int \celVolForm \gaugeParam_1 \cstrVal
    \right)
    \cdots
    \left(
    \int \celVolForm \gaugeParam_n \cstrVal
    \right)
  \\
  &=\frac{
    \int_{i\milnet=0}^{i\milnet=2\pi} \edgeVolForm[\cstrVal]\,
    e^{
    - \act[][2\pi]\left[
    \confPrimA[\rMilne]{\edge}\relax[\cstrVal]
    \right]
    }
    \left(
    \int \celVolForm \gaugeParam_1 \cstrVal
    \right)
    \cdots
    \left(
    \int \celVolForm \gaugeParam_n \cstrVal
    \right)
    }{
    \partFunc[\rMilne][\edge][2\pi]\;.
    }
\end{align}
In order for the two lines above to match for all possible $\gaugeParam$, it
must be that the probability distribution is given by
\begin{align}
  \prob[\cstrVal]
  &= \frac{
    e^{
    - \act[][2\pi]\left[
    \confPrimA[\rMilne]{\edge}\relax[\cstrVal]
    \right]
    }
    }{\partFunc[\rMilne][\edge][2\pi]}
    \;.
    \label{eq:probCstrVal}
\end{align}
We note that this probability distribution takes the usual Boltzmann form with inverse
temperature $2\pi$ if we identify the energy of the edge mode
$\confPrimA[\rMilne]{\edge}\relax[\cstrVal]$ as $\act[][2\pi]\left[
  \confPrimA[\rMilne]{\edge}\relax[\cstrVal]
\right]/2\pi$.
Indeed, as we will review later, the edge mode action is generally  proportional to $\invTemp$, \ie, 
\begin{align}
  \act[][\beta]\left[
  \confPrimA[\rMilne]{\edge}\relax[\cstrVal]
  \right]
  &= \frac{\beta}{2\pi}
    \,
    \act[][2\pi]\left[
    \confPrimA[\rMilne]{\edge}\relax[\cstrVal]
    \right]
    \;,
    \label{eq:actBetaDependence}
\end{align}
and so \cref{eq:partFuncRE} is the corresponding Boltzmann partition function at
temperature $1/\invTemp$.
 Of course, this fits consistently within the interpretation of
  the Minkowski vacuum as a thermofield double \labelcref{eq:tfd} which reduces
  to a Milne thermal state. 

One may wonder why the `soft' mode $\confPrimA[\rMilne]{\edge}\relax[\cstrVal]$
has a non-zero energy at all --- as we later discuss, this is a consequence the $\reg$ regulator
inherited by $\confPrimA[\rMilne]{\edge}\relax[\cstrVal]$ from the conformal
primary wavefunctions (or alternatively, a consequence of the
  regulated size of the
  neighbourhood \(\mathcal{S}_1\) introduced around \cref{eq:partFuncSplit}). In
the $\reg\to 0$ limit, the edge mode
has vanishing inner product with the $\milneFreq\ne 0$ modes and becomes truly
soft (in terms of Milne energy); correspondingly, we will find that the Euclidean action
\labelcref{eq:actBetaDependence} vanishes as $\reg\to 0$.
In fact this is also required for the probability
distribution \labelcref{eq:probCstrVal} to be consistent with the vanishing
\labelcref{eq:minkStateCstr} of the Minkowski entangling constraint \labelcref{eq:cstrEntMink}. Together
with the vanishing \labelcref{eq:einStateCstr} of the constraint
\labelcref{eq:entOpEin} in the Einstein static theory, it implies that the shared
vacuum of the Minkowski and Einstein static theories further satisfies:
\begin{align}
  \left(
  \cstrconj[\lMilne]
  \relax[\gaugeParam]
  -\cstrconj[\rMilne]
  \relax[\gaugeParam]
  \right)
  \ket{0}
  &\eqReg 0 \;.
\end{align}
By the same steps leading to \cref{eq:commutDnsMatCstrR}, we find that
\begin{align}
  \commut{
  \cstrconj[\rMilne]\relax[\gaugeParam]
  }{
  \dnsMat[\rMilne][0]
  }
  \eqReg
  0
  \;.
\end{align}
Given the block decomposition \cref{eq:dnsMatBlockDecomposition}, the only way
this can be true is if the probability distribution $\prob[\cstrVal]$ (together
with the measure $\edgeVolForm[\cstrVal]$) is uniform. Indeed, we shall see
below in \cref{eq:dnsMatBlocksUnitaryRelation}
that
$e^{i\goldOp[\rMilne]\relax[\cstrVal]}$ relates the different blocks
$\dnsMat[\rMilne][0,\cstrVal]$ in the density matrix $\dnsMat[\rMilne][0]$.

Using
\cref{eq:probCstrVal,eq:actBetaDependence}, the Shannon entropy
\labelcref{eq:entropySh} and classical R\'enyi entropies
\labelcref{eq:entropyClassicalRenyi} can be expressed in terms of the edge mode
partition function \labelcref{eq:partFuncRE} through relations analogous to
\cref{eq:entropyvNPartFunc,eq:entropyRenyiPartFunc}:
\begin{align}
  \entropy[\Sh][p]
    &= (1-\invTemp\partial_\invTemp)
    \left.
    \partFunc[\rMilne][\edge][\invTemp]
    \right|_{\invTemp=2\pi}
      \;,
      \label{eq:entropyShPartFunc}
    \\
\entropy[n][p]
  &= \frac{
    n\log
    \left(
    \partFunc[\rMilne][\edge][2\pi]
    \right)
    -
    \log\left(
    \partFunc[\rMilne][\edge][2\pi n]
    \right)
    }{n-1}
    \;.
    \label{eq:entropyClassicalRenyiPartFunc}
\end{align}
Moreover, we can show, as claimed below \cref{eq:entropyRenyiBlockDecomposition}, that
the R\'enyi entropy \cref{eq:entropyRenyiBlockDecomposition} decomposes into
edge mode and bulk contributions respectively depending only on $\prob$ and
$\dnsMat[\rMilne][0,0]$. To do so, we apply \cref{eq:partFuncDecomposition} to
the formula \labelcref{eq:entropyRenyiPartFunc} for R\'enyi entropy in terms of
partition functions and compare with \cref{eq:entropyClassicalRenyiPartFunc}:
\begin{align}
  \entropy[n](\dnsMat[\rMilne][0])
  &= \entropy[n][p]
    + \entropy[n](\dnsMat[\rMilne][0,0]).
    \label{eq:entropyRenyiDecomposition}
\end{align}

By comparing this decomposition with \cref{eq:entropyRenyiBlockDecomposition},
we can also argue that $\dnsMat[\rMilne][0,\cstrVal]$ for various $\cstrVal$ are
unitarily related as follows. Let us consider the ansatz
\begin{align}
  \dnsMat[\rMilne][0,\cstrVal]
    &= e^{i\goldOp[\rMilne]\relax[\cstrVal]}
      \,
      \dnsMat[\rMilne][0,0]
      \,
      e^{-i\goldOp[\rMilne]\relax[\cstrVal]}
      \;,
      \label{eq:dnsMatBlocksUnitaryRelation}
\end{align}
which is satisfies \cref{eq:dnsMatBlock}. Substituting this ansatz into
\cref{eq:entropyRenyiBlockDecomposition} we immediately recover
\cref{eq:entropyRenyiDecomposition}. Since the full set of R\'enyi entropies for
$n\in\naturals$ determines the spectrum of a density matrix, it conversely
follows that the ansatz \cref{eq:dnsMatBlocksUnitaryRelation} at least gives the
correct spectrum for $\dnsMat[\rMilne][0]$ and, because the ansatz satisfies
\cref{eq:dnsMatBlock}, also for the individual blocks
$\dnsMat[\rMilne][0,\cstrVal]$. We conclude that $\dnsMat[\rMilne][0,\cstrVal]$
for various $\cstrVal$ are unitarily related. In fact, let us further argue
that, \cref{eq:dnsMatBlocksUnitaryRelation} must also give the correct unitary
relation. To calculate expectation values in the state
$\dnsMat[\rMilne][0,\cstrVal]$, we should first project the vacuum density
matrix $\dnsMat[\rMilne][0]$ onto this block --- in terms of a path integral,
this means restriction to a sector with fixed electric flux near the entangling surface, precisely as in
\cref{eq:partFuncWithDelta}. As described below that
\namecref{eq:partFuncWithDelta}, the dependence on $\cstrVal$ merely enters as a
shift in the background by $\confPrimA[\rMilne]{\edge}\relax[\cstrVal]$. This
shift is precisely what the unitaries in \cref{eq:dnsMatBlocksUnitaryRelation}
achieve (as can be seen from \cref{eq:fieldShiftByStarAlgebra}).

Because the spectrum is the same for the blocks $\dnsMat[\rMilne][0,\cstrVal]$,
their R\'enyi entropies are all the same
\begin{align}
  \entropy[n]
  \left(
  \dnsMat[\rMilne][0,\cstrVal]
  \right)
  &=
    \entropy[n]
    \left(
    \dnsMat[\rMilne][0,0]
    \right)
    =
    \frac{
    n\log\left(
    \partFunc[\rMilne][][2\pi][0]
    \right)
    - \log\left(
    \partFunc[\rMilne][][2\pi n][0]
    \right)
    }{n-1}
\end{align}
which is why \cref{eq:entropyRenyiBlockDecomposition} decomposes into
\cref{eq:entropyRenyiDecomposition}. Similarly, the von Neumann entropies
\begin{align}
  \entropy[\vN](\dnsMat[\rMilne][0,\cstrVal])
  &= \entropy[\vN](\dnsMat[\rMilne][0,0])
    = (1-\invTemp\partial_\invTemp)
    \left.
    \partFunc[\rMilne][][\invTemp][0]
    \right|_{\invTemp=2\pi}
\end{align}
are all the same, and so the averaging in the bulk term of
\cref{eq:entropyvNBlockDecomposition} trivializes:
\begin{align}
  \entropy[\vN](\dnsMat[\rMilne][0])
  &= \entropy[\Sh][p]
    + \entropy[\vN](\dnsMat[\rMilne][0,0])
    \;.
    \label{eq:shannonvn}
\end{align}

\paragraph{Edge modes: their action, measure, and partition function.}

In the remainder of this section, we will focus on edge modes, evaluating their
on-shell action $\act[][\invTemp]\left[
  \confPrimA[\rMilne]{\edge}\relax[\cstrVal] \right]$ and subsequently their
partition function \cref{eq:partFuncRE} which determines their entropies
\labelcref{eq:entropyvNPartFunc,eq:entropyRenyiPartFunc}. These calculations
follow essentially from those of \cite{Donnelly:2015hxa,Donnelly:2014fua} which
consider entangling surfaces in interior of a spacetime. Our entangling surface,
on the other hand, lies on the spacetime boundary when considering either the
Milne and Minkowski perspectives. However, we may conformally embed these
spacetimes into the Einstein static universe, in which case the entangling
surface is merely the equator along a slice $S^3$ of constant Einstein static
time.
As we will demonstrate, it is convenient to carry out calculations more directly
in the Milne geometry which, as we see from \labelcref{eq:Milne-metric}, is
conformally static with $\hyp{3}$ time slices. Though the entangling surface lies
at the boundary $\partial\hyp{3}$, we shall recover results essentially
unchanged from \cite{Donnelly:2015hxa,Donnelly:2014fua} with the infinite volume
of $\partial\hyp{3}$ dropping out.

To begin, we must more precisely write down the edge modes
$\confPrimA[\rMilne]{\edge}$. Recall they were previously introduced in
\cref{eq:AcsAsLR,eq:AcsPrimeAsLR} and described in \cref{sec:milneSoft} as
modes localized near $\partial\cauchy[\rMilne]$ that carry the $\rMilne$ half
of the $\minkX^2=0$ shockwave in the Minkowski conformally soft wavefunctions $\confPrimA{\cs}$ (or $\confPrimA{\cs'}$)
--- see \cref{fig:finiteRegShockwave}.

It will be helpful to first study
 the $\confPrimA{\log,\pm}$ modes (introduced in \cref{eq:log}) from which $\confPrimA{\cs}$ is defined (in
\cref{eq:CS}):
\begin{align}
  \confPrimA{\log,\pm}
  &= -2 \milnet_\pm \, \diff_{\minkX} \confGauge{1,\pm}
  = 2 \left(
    - \diff_{\minkX}(\milnet_\pm \confGauge{1,\pm})
    + \diff\milnet_\pm \, \confGauge{1,\pm}
  \right)
    \;,
    \label{eq:AlogGaugeTransform}
  \\
  \confPrimF{\log,\pm}
  &= -2\, \diff\milnet_\pm \wedge \diff_{\minkX}\confGauge{1,\pm}
    \;.
\end{align}
Here, we have introduced the Milne time shifted according to the $i \epsilon$ prescription --- \cf{} \cref{eq:milnet}:
\begin{align}
    \milnet_\pm
  &= \frac{1}{2}\log(-\minkX_\pm^2)
    \;.
\end{align}
To better understand the function $\confGauge{1,\pm}$ given by
\cref{eq:cappuccino}, let us also rewrite the identity \eqref{eq:qX1} in Milne
coordinates\footnote{Note that it is not possible to choose the same branch for
  all the logarithms in \cref{eq:logqX,eq:hypK}. For concreteness, we may choose
  the branch for the logarithm in \cref{eq:hypK} to be over on the negative reals, as for the
  branch on the LHS \cref{eq:logqX} --- the latter was established in
  \cref{eq:argqX}. This forces the branch of $\log(\minkr/2)$ in \cref{eq:logqX}
  over $\minkr<0$ to depend on the sign $\pm$:
  \begin{align}
    \arg\minkr
    &= \mp \pi \stepFunc(-\minkr) \;.
  \end{align}
  However, this will not play any role in the celestial derivatives of
  \cref{eq:logqX}, in which we are primarily interested. \label{foot:argr}}:
\begin{align}
  -\nullUnit(\celw)\cdot \minkX_\pm
  &=
    \frac{\minkr}{2}\left(
    \fgr_\pm
    + \abs{\celwh-\minkzh}^2
    \right)\;,
    \label{eq:qX2}
    \\
      \log(-\nullUnit\cdot\minkX_{\pm})
  &= \log\frac{\minkr}{2} + 4\pi \hypK(\celw; \fgr_\pm, \minkz)\;,
    \label{eq:logqX}
\end{align}
where we have defined the shifted Fefferman-Graham coordinate $\fgr_\pm$ and the
function $\hypK$ over $\celw$, $\minkz$, and (not necessarily real) $\fgr$:
\begin{align}
  \fgr_\pm
  &= \frac{2(\minku \mp i\reg)}{\minkr}
    = \fgr \mp 2i \reg\, e^{-\milnet}\sqrt{\fgr}\;,
  \label{eq:shiftedfgr}
  \\
  \hypK(\celw; \fgr, \minkz)
  &= \frac{1}{4\pi}
    \log\left(  \fgr + \abs{\celwh-\minkzh}^2 \right)
    \;.
    \label{eq:hypK}
\end{align}
As we show in \cref{sec:asymptotics}, this function is a solution to the
massless scalar equation of motion on $\hyp{3}$ and therefore has the expected
asymptotics \labelcref{eq:hypKExpansion} of a massless bulk-to-boundary propagator on $\hyp{3}$ (\ie{} Euclidean
$\AdS_3$). Thus, combining
\cref{eq:cappuccino,eq:logqX,eq:hypKExpansion}, we find (in the small $\rho$ limit)
\begin{align}
  \pullback[\text{constant $\milnet,\fgr$}]
  \hodge \confPrimF{\log,\pm}
  &= -4\, \partial_\fgr \confGauge{1,\pm}\, \celVolForm
    = -4\pi \log(\fgr_\pm)\,
    \diff[\celw]\celDeltaFunc(\celw,\minkz) \,
    \celVolForm
    + \order{\fgr_\pm^0}
    \;.
    \label{eq:EFluxLog}
\end{align}


Note that, if $\reg$ is set to zero, the electric flux \labelcref{eq:EFluxLog}
diverges logarithmically at the $\hyp{3}$ boundary $\fgr=0$. One effect of
taking the difference \labelcref{eq:CS} to obtain $\confPrimA{\cs}$ is that the
electric flux of $\confPrimA{\cs}$ through the $\fgr=0$ surface becomes finite and
independent of $\reg$. Another effect is that $\confPrimA{\cs}$ dies off quickly
in the $\rMilne$ interior, with the rapidity of the falloff set by $\reg$. These
properties earn the restriction of $\confPrimA{\cs}$ to $\rMilne$ ---
\ie{} $\confPrimA[\rMilne]{\edge}$ --- the interpretation of an edge mode, as
described by \cite{Donnelly:2015hxa,Donnelly:2014fua}.

Unfortunately, the $\reg$ regulator inherited from conformal primary
wavefunctions --- see \cref{eq:frame} --- is not invariant under Milne time
translation and therefore does not easily analytically continue to thermal time
$i\milnet$. In contrast, \cite{Donnelly:2015hxa,Donnelly:2014fua} constructs
edge modes which are static. Searching for such solutions, one finds that the
electric flux at $\fgr=0$ inevitably diverges, as we saw above for
$\lim_{\reg\to 0} \confPrimA{\log,\pm}$. So our strategy is to introduce a
regulator surface at small $\fgr$  --- this is the same regulator surface \(\partial
  \mathcal{S}_2\) introduced around \cref{eq:partFuncSplit} --- and rescale the wavefunction as the regulator
is taken to be increasingly small. In this limit, the electric field lines of
the rescaled wavefunction will bunch up near the regulator surface which
approaches $\fgr=0$, similar to \cref{fig:milneEdgeE}.

In an attempt to make this Milne-static regulator scheme as close in spirit as
possible to the $\reg$ regulator inherited from the Minkowski conformal primary
wavefunctions, we shall also call the parameter controlling the regulator
surface $\reg$,
\begin{align}
\label{eq:rho-epsilon}
  \fgr|_{\partial\rMilne}
  &= \reg^2
    \;.
\end{align}
In both schemes, $\reg^2$ controls the $\fgr$-width or $\fgr$-position of the bunch of
electric field lines along a typical constant $\milnet$ surface. The
Milne-static edge modes with this cutoff surface are then given by
\begin{align}
    \confPrimA[\rMilne]{\edge}
    &= \frac{
      \confGauge{1}
    }{2\log(\reg^{-1})}
    \diff \milnet
      \;, 
  \quad {\rm with} ~~~
    \confGauge{1}(\celw;\fgr,\minkz)
  = -4\pi \diff[\celw] \hypK(\celw;\fgr,\minkz) \;,
    \;
    \label{eq:AREdgeStatic}
\end{align}
where we have discarded the old ``$\reg$'' inherited by the logarithmic modes
\labelcref{eq:AlogGaugeTransform} from the Minkowski wavefunctions, then
rescaled by the regulator surface's position at the new ``$\reg$'', so that
\cref{eq:cstrRBlkEdge} holds there. This can be seen from applying the same
rescaling to \cref{eq:EFluxLog}:
\begin{align}
\pullback[\partial\cauchy[\rMilne]]
\hodge \confPrimF[\rMilne]{\edge}
&=
2\pi\,
\diff[\celw]\celDeltaFunc(\celw,\minkz) \,
\celVolForm(\minkz)
  + \order{\frac{1}{\log(\reg^{-1})}}
  \;.
\end{align}
For the purposes of evaluating the action,
we are free to apply arbitrary pure gauge transformations and so we have discarded
the pure gauge piece of \cref{eq:AlogGaugeTransform} in favour of a static
electric potential \cite{Donnelly:2015hxa,Donnelly:2014fua}.
From \cref{eq:AREdgeStatic}, we can construct the $\cstrVal$-smeared modes \labelcref{eq:AERSmeared}:
\begin{align}
  \confPrimA[\rMilne]{\edge}\relax[\cstrVal]
    &= -\frac{
    1
    }{\log(\reg^{-1})}
    \left[
    \int \celVolForm(\celw)
    \cstrVal(\celw)
    \hypK(\celw)
    \right]
      \diff \milnet \;,
      \label{eq:AERSmearedStatic}
    \\
    \pullback[\partial\cauchy[\rMilne]]
    \hodge \confPrimF[\rMilne]{\edge}\relax[\cstrVal]
    &=
    \cstrVal(\minkz)\,
    \celVolForm(\minkz)
      + \order{\frac{1}{\log(\reg^{-1})}}
      \;,
      \label{eq:electricFluxERSmearedStatic}
\end{align}
whose action appears in the edge mode partition function \labelcref{eq:partFuncRE}.

The form of the static edge mode \labelcref{eq:AERSmearedStatic} is actually
quite universal and follows merely from the near-boundary asymptotics of
$\hyp{3}$ massless scalar solutions. Going to a different celestial conformal
frame corresponds to choosing a different set of Fefferman-Graham coordinates on
$\hyp{3}$. In any case, one may find static edge modes
$\confPrimA[\rMilne]{\edge}\relax[\cstrVal]$ of the form
\labelcref{eq:AERSmearedStatic} with electric flux
\labelcref{eq:electricFluxERSmearedStatic}, where $\reg^2$ is the cutoff in the
Fefferman-Graham bulk coordinate, $\celVolForm$ is the volume form on celestial
space, and $\hypK$ is the massless solution to the $\hyp{3}$ scalar equations of
motion that asymptotes to the Green's function $\celG$ on celestial space, which for the plane is given in \cref{eq:grape}. Thus,
our discussion in the remainder of this section is applicable to all celestial
conformal frames. For the remainder of the paper, we will focus on the spherical celestial conformal
frame and a corresponding spherical geometry for the entangling surface. {

In general, on-shell, the Euclidean action \labelcref{eq:actEuclideanPeriodic}
is given by
\begin{align}
  \act[][\invTemp]
  &\eqOnShell
    \frac{i}{2}\int_{\partial\rMilne,\, i\milnet\in[0,\invTemp]}
    A \wedge \hodge F
    \;.
    \label{eq:onShellAction}
\end{align}
Evaluating this in the configuration \labelcref{eq:AERSmearedStatic}, we find
\begin{align}
  \act[][\invTemp]
  [\confPrimA[\rMilne]{\edge}\relax[\cstrVal]]
  &= -\frac{\invTemp}{2\log(\reg^{-1})}
    \int \celVolForm(\celw)\, \cstrVal(\celw)
    \int \celVolForm(\minkz)\, \cstrVal(\minkz)
    \celG(\celw,\minkz)
    + \order{\frac{1}{\log^2(\reg^{-1})}}
\label{eq:actAER}
  \\
  &\eqReg 0
    \;,
    \label{eq:actAERVanishing}
\end{align}
which vanishes in the $\reg\to 0$ limit, as we have already deduced below
\cref{eq:actBetaDependence}.

In order to evaluate the edge mode partition function \labelcref{eq:partFuncRE},
we must specify the integration measure $\edgeVolForm[\cstrVal]$. Following
\cite{Donnelly:2015hxa,Donnelly:2014fua}, we can do so by drawing inspiration
from a discretized theory. In lattice gauge theory, one discretizes space into a
lattice and sums over variables $\link_i$ representing the electric flux through
each link $i$ on the lattice. Envisioning the entangling surface as punctured at
$N$ sites by links, a set of values for $\link_i$ corresponds roughly to a
configuration $\cstrVal$:
\begin{align}
  \frac{N}{\celVol} \sum_{i=1}^N \link_i\, \iota_i(\celw)
  \approx& \cstrVal(\celw)
           \;,
           &
           \iota_i(\celw)
  &= \begin{cases}
       1 & \text{if $\celw$ closest to site $i$}
           \\
           0 & \text{otherwise}
     \end{cases}
    \;.
\end{align}
With a compact $\U(1)$ gauge group, each link
variable $\link_i\in \elemCharge \integers$ is discretized in steps of the elementary
charge $\elemCharge$. Thus, there is a natural field space metric
\begin{align}
  \frac{1}{\elemCharge^2}
  \sum_{i=1}^N (\vary\link_i)^2
  &\approx
    \edgeMetricAll(\vary\cstrVal,\vary\cstrVal)
    = \frac{\celVol}{\elemCharge^2 N}
    \int \celVolForm\, (\vary\cstrVal)^2
    \label{eq:latticeSum}
\end{align}
inspired by the lattice theory. In particular, the sum over all possible values
for $\link_i\in \elemCharge \integers$ is analogous to path integration in field
space with the volume form $\edgeVolFormAll$ associated to the metric
$\edgeMetricAll$. Explicitly, in any basis $\{Y_j: j\ge 0\}$ of modes
which is orthonormal with respect to the more standard metric
$\scalarFieldMetric$,
\begin{align}
  \cstrVal(\celw)
  &= \sum_{j\ge 0}
    \cstrVal_j Y_j(\celw)
    \;,
  &
    \scalarFieldMetric(Y_j,Y_{j'})
  &= \int \celVolForm\, Y_j Y_{j'}
  = \delta_{j,j'}\;,
\end{align}
the volume form $\edgeVolFormAll$ reads\footnote{ In \(d+2=4\) bulk
  spacetime dimensions, this integration measure of
  \cite{Donnelly:2014fua,Donnelly:2015hxa} almost matches that of
  \cite{Ball:2024tba}, upon identifying \cref{eq:zetaRenScale} as the
  renormalization scale for \(\zeta\)-function renormalization --- see the
  surrounding discussion in \cref{sec:discussion}. The difference is that
  \cite{Ball:2024tba} lacks the factor of \(1/\elemCharge\) for each mode.

  In other spacetime dimensions \(d+2\), naively following
  \cref{eq:latticeSum} still gives \cref{eq:edgeVolFormAll} where \(\celVol\) is
  replaced by \(\mathrm{vol}^{(d)}\), while \cite{Ball:2024tba} continues to
  give \(\edgeVolFormAll[\cstrVal] =\bigFieldWedge_{j\ge 0} \frac{\fieldDiff
    \cstrVal_j}{2\pi \mu} \). Further comparisons
  with the measure of \cite{Ball:2024tba} will be made in
  \cref{foot:measureComparisonZeroMode} in relation to the zero mode.}
\begin{align}
  \edgeVolFormAll[\cstrVal]
  &=\bigFieldWedge_{j\ge 0}
    \frac{1}{\elemCharge}
    \sqrt{\frac{\celVol}{N}}
    \fieldDiff \cstrVal_j
    \;.
    \label{eq:edgeVolFormAll}
\end{align}
Here $\bigFieldWedge_{j\ge 0}$ denotes the wedge product over (discretized) field space.

Some explanation is required in the case of a noncompact celestial space, \eg{}
$\celPlane$. Here, $\celVol$ diverges and, to associate a finite area to each
link $1\le i \le N$, $N$ must diverge proportionately also. Moreover, one must
treat modes $Y_j$ where the index $j$ takes values in a continuum. To give
meaning to the path integral over functions $\cstrVal$ over celestial space, one
may introduce a celestial IR regulator. In the example of $\celPlane$, for
example, we may periodically identify $\celwh \sim \celwh+\celIRReg\, \naturals +
i \celIRReg\, \naturals$ where $\celIRReg$ is a large constant; we will see below that the edge mode partition function
\labelcref{eq:partFuncRE} is independent of $\celIRReg$.

There remains a subtlety to do with the constant mode when $\celVolForm$ diverges. For a compact celestial
space, the mean $\int \celVolForm \cstrVal$ of $\cstrVal$ is required to vanish
by Gauss's law. In the case of a noncompact celestial space, electric field
lines may enter $\hyp{3}$ through point(s) ``at $\infty$'' of the celestial
space and have a correspondingly non-vanishing net flux through all finite
points of celestial space. However, as we described below
\cref{eq:asymChargeEigenstate} in the $\celPlane$ case, such a solution is
rather singular. More generally, when $\cstrVal$ has non-vanishing mean in the
interior of celestial space, the on-shell action \labelcref{eq:onShellAction}
diverges due to pieces of $\partial\rMilne$ surrounding the ``$\infty$'' of
celestial space.\footnote{This can be argued in a variety of celestial conformal
  frames using Gauss's law and the static form $A=\gaugeParam\, \diff\milnet$ of
  the gauge potential. For
  concreteness, let us consider celestial conformal frames where surfaces of
  constant $\fgr$ are topologically similar to those of the $\celPlane$ case. By
  Gauss's law,
  \begin{align}
    \int_{\text{constant $\milnet,\fgr$}}\hodge F
    &= -2\,
  \partial_\fgr \int \celVolForm \gaugeParam
  \end{align}
  is constant in $\fgr$; when $\cstrVal$ has non-vanishing mean in the interior
  of celestial space, this constant is nonzero (as can be seen from evaluation
  at $\fgr=\reg^2$). Thus, $\int \celVolForm \gaugeParam$ grows linearly in
  $\fgr$. The contribution to the on-shell action \labelcref{eq:onShellAction}
  from a piece at constant $\fgr$ is proportional to $\partial_\fgr \int
  \celVolForm \gaugeParam^2$, which diverges as $\fgr\to \infty$.} (Note that
this is a separate issue from the previous paragraph; the action of the constant
mode is still divergent even if, for example, the celestial plane has been
compactified to a torus.)

In lattice gauge theory, the cure to the analogous problem is to restrict the
net $\link_i$ piercing the boundary $\partial\cauchy[\rMilne]$ to be vanishing
--- this can be implemented by the insertion of a Kronecker
$\kronDelta_{\sum_{i=1}^N \link_i,0}$:
\begin{align}
  \kronDelta_{\sum_{i=1}^N \link_i,0}
  &\approx \deltaFunc\left(
    \frac{1}{\elemCharge}
    \int \celVolForm \cstrVal
    \right)\;.
\end{align}
In the field theory, we should therefore insert the $\delta$-function of the RHS
into the path integral with measure \labelcref{eq:edgeVolFormAll}. For
concreteness, let us take $Y_0$ to be the constant mode over celestial space.
Then, the resulting integration measure over nonzero modes of $\cstrVal$ is
given by
\begin{align}
  \edgeVolForm[\cstrVal]
  &= \frac{1}{\sqrt{N}}\,
    \fieldPullback[\int \celVolForm \cstrVal=0]
    \edgeVolFormAll[\cstrVal]
    \label{eq:edgeVolForm}
    \\
    &= \frac{1}{\sqrt{N}}
      \bigFieldWedge_{j\ge 1}
    \frac{1}{\elemCharge}
    \sqrt{\frac{\celVol}{N}}
      \fieldDiff \cstrVal_j
      \;,
    &
      (\text{$Y_0$ constant})
      \label{eq:edgeVolFormExplicit}
\end{align}
where, as introduced below \cref{eq:sympConfSpace}, $\fieldPullback$ is the
field space pullback. In \cite{Donnelly:2015hxa,Donnelly:2014fua}, zero modes
are separately integrated over because their action does not take the same form
\labelcref{eq:actAER} as for the nonzero modes. However, as we have described in
our case, zero modes correspond either to no solutions (which must satisfy
Gauss's law) or singular solutions with infinite action (even at finite $\reg$
and after IR-regulating the celestial space). So, we will not integrate over
zero modes at all.\footnote{ The zero mode is also excluded from the
  edge mode path integral by \cite{Ball:2024tba}. Comparing
  \cref{eq:edgeVolFormAll,eq:edgeVolFormExplicit}, we see that our
  implementation of this exclusion comes with a factor of \(1/\sqrt{N}\). In
  \cite{Ball:2024tba}, the analogous factor is a ratio
  \(\abs{\mathcal{G}'}/\abs{\mathcal{G}} = \frac{\elemCharge}{2\pi\mu \sqrt{\mathrm{vol}^{(d)}}}\) between the sizes of the set
  \(\mathcal{G} = \{g:\partial\cauchy[\rMilne] \to U_{\elemCharge}(1)\}\) of all
  gauge-group-valued functions \(g(\minkz)\) on the entangling surface
  \(\partial\cauchy[\rMilne]\) and the set \(\mathcal{G}'\) of such nonconstant
  functions. Identifying the
  \(\zeta\)-function renormalization scale \labelcref{eq:zetaRenScale}, we see that in \(d+2=4\) dimensions
  \(\abs{\mathcal{G}'}/\abs{\mathcal{G}}=\elemCharge/\sqrt{N}\) has an extra
  factor of \(\elemCharge\). \label{foot:measureComparisonZeroMode}}

Finally, keeping the leading term in \cref{eq:actAER} and using
\cref{eq:edgeVolFormExplicit}, the edge mode partition function
\labelcref{eq:partFuncRE} can be evaluated as a Gaussian integral,
giving\footnote{Collecting the factors of \(\elemCharge\) in the difference in
  convention for the edge mode path integral measure used by
  \cite{Ball:2024tba}, the partition function
  \(\partFunc[\rMilne][\edge][\invTemp]\) obtained from \cite{Ball:2024tba}'s
  measure would have an extra factor of \(\elemCharge\) out front in
  \cref{eq:partFuncREExplicit} and lacks the factor of \(1/\elemCharge^2\) in the
  \(\det'\). In \cref{eq:edge-pf}, \(\elemCharge\) would appear as a factor out front and not in the parentheses. In \cref{eq:ee}, the \(\elemCharge\) dependence would instead be through a term \(\log\elemCharge\).}
\begin{align}
  \partFunc[\rMilne][\edge][\invTemp]
  &= \frac{1}{\sqrt{N}}
   \left( \det\nolimits'\left(
    \frac{2\pi\celVol \log(\reg^{-1})}{\elemCharge^2 N \invTemp}
    \,
    \celLap
    \right)\right)^{1/2}
    \;,
    \label{eq:partFuncREExplicit}
\end{align}
where $\det\nolimits'$ means the product of nonzero eigenvalues and we have used
\begin{align}
  \det\nolimits' \celG
  &= \frac{1}{\det\nolimits' \celLap}
    \;,
\end{align}
with $\celG$ acting as $\cstrVal \mapsto \int \celVolForm \celG \cstrVal$.

The formula for the edge mode partition function can be simplified  using the identity \cite{Donnelly:2015hxa}
\begin{align}
  \det\nolimits'(a\, \lap)
  &= a^{\zeta(0,\lap)} \det\nolimits'(\lap)
    \;,
  &
    \zeta(0,\lap)
  &= - \dim\ker\lap + \anomAct\;,
    \label{eq:zetaScalingIdentity}
\end{align}
where $\mathcal{A}$ is the anomaly coefficient, which in \(d=2\)
  dimensions is proportional to the Euler characteristic \(\chi\):
\begin{align}
  \mathcal{A}
  &= \frac{1}{6}\chi \;,
    &
    \chi
  &= \frac{1}{4\pi} \int \celVolForm \celRicci
    \;.
\end{align} 
With a single zero mode \(Y_0\), we have \(\dim\ker\celLap=1\).
  On the sphere, we of course also have
\begin{align}
  \mathcal{A}
  &= \frac{1}{3}
    \;,
  &
    \chi
  &= 2
    \;.
  &
    (\text{on \(S^2\)})
    \label{eq:anomOnS2}
\end{align}
We find that
\eqref{eq:partFuncREExplicit} simplifies to
\begin{equation}
\label{eq:edge-pf}
\partFunc[\rMilne][\edge][\invTemp]
= N^{-\mathcal{A}/2} \left(
  \frac{2\pi {\rm vol}^{(2)}\log(\epsilon^{-1})}{\elemCharge^2 \beta}
\right)^{\frac{\mathcal{A} - 1}{2}}
\left(\det\nolimits' (\Box^{(2)})\right)^{1/2} \;.
\end{equation} 
The edge mode contribution to the entanglement entropy is then 
\begin{equation}
\label{eq:ee}
\begin{split}
  S^\edge
  &= (1 - \beta \p_{\beta}) \log {}^R Z^E_{\beta}\left. \right|_{2\pi}\\
  &=   \frac{1}{2}
    \log \det\nolimits' (\Box^{(2)})
    + \frac{\mathcal{A}-1}{2}
    \log \frac{ {\rm vol}^{(2)}\log(\epsilon^{-1})}{\elemCharge^2}
    - \frac{\mathcal{A}}{2} \log N + \frac{\mathcal{A}-1}{2}.
\end{split}
\end{equation}

 We can also evaluate the R\'enyi entropy associated with the edge mode
  partition function. Substituting \eqref{eq:edge-pf} into
  \eqref{eq:entropyRenyiPartFunc} we easily evaluate the edge mode Renyi
  entropies
\begin{equation}
\label{eq:Renyi-ent}
S_n^{\edge} = S^{\edge} + \frac{\mathcal{A} - 1}{2}\left(\frac{\log n}{n - 1}-1\right).
\end{equation}
It is straightforward to see that in the limit $n \rightarrow 1$ this reduces to
the edge mode entropy \eqref{eq:ee}.
We discuss the various divergent terms appearing in these formulas below.


\section{Discussion}
\label{sec:discussion}


In this paper, 
we have employed the embedding of Minkowski space inside the
Einstein static universe, along with the Weyl invariance of four-dimensional free Maxwell theory to
identify partitions of $\mathscr{I}^+$ with Cauchy slices inside two Milne
regions of Minkowski geometries related by a conformal inversion. In this case,
conformal primary wavefunctions diagonalize the Milne
energy and hence form a basis for free Maxwell fields
confined to the Milne subregions. We found inversions to be closely related to
shadow transformations and used them to decompose the global conformal primary
wavefunctions in terms of linear combinations of conformal primary wavefunctions
with non-trivial support inside the two Milne regions. This allowed us to
identify the Minkowski vacuum state on $\mathscr{I}^+$ as a thermofield double
state with respect to the two subregions in the non-zero Milne mode sector. We
then analyzed the soft sector and showed that the constraint obeyed by physical
states of definite large gauge charge of the Minkowski theory can be understood
as due to sources for conformally soft fields that propagate in the inverted
Minkowski patch.

Along the way, we revisited the role of constraints and edge modes in free
Maxwell theory in four-dimensional Minkowski space. In particular, we have
established a precise relation between the logarithmic constituents of the
conformally soft modes \cite{Donnay:2018neh} and associated field strengths, and
the edge modes or electric field configurations confined to the entangling
surface of Donnelly and Wall \cite{Donnelly:2014fua, Donnelly:2015hxa}. Upon
identifying the regulator $\epsilon$ of the conformal primary wavefunctions with
the location of the ``brick wall'' near the entangling surface
\eqref{eq:rho-epsilon}, the ``radiative'' conformally soft gauge fields are
related to the static Donnelly-Wall gauge fields by a gauge transformation and
therefore give rise to identical field configurations and on-shell actions.
Consequently, the edge mode contribution to the vacuum entanglement entropy can
be reinterpreted as due to fluctuations in the large gauge charges of subregions
on $\mathscr{I}^+$.

Carrying our calculations to their conclusion, the net entanglement entropy
(including bulk modes) of the cut on $\mathscr{I}^+$ \eqref{eq:ee} takes the
usual form, \ie{} an area law divergence with a nonuniversal coefficient
\cite{Srednicki:1993im} and a subleading logarithmic contribution with a
universal coefficient \cite{Myers:2010tj,Casini:2011kv}. The soft (or edge) mode
contribution to the latter is readily evaluated from the Shannon entropy in
\cref{eq:shannonvn}.
For a spherical entangling surface, $S^\edge\simeq -\frac13\,\log(\zetaReg)$, where \(\zetaReg\) is a cutoff or
  renormalization energy scale \cite{Donnelly:2015hxa}. 
  As \cite{Ball:2024tba} points out, actually reading this off of \cref{eq:ee}
  is not as simple as \cite{Donnelly:2014fua,Donnelly:2015hxa} perhaps had
  intended, due to the exclusion of the zero mode in \(\det'\); we will
  elaborate on this below.
Ultimately, the edge mode contribution combines with
the von Neumann entropy of the hard modes to yield $S \simeq-31/45\,\log(\zetaReg)$, matching the expected result.

Similarly, the R\'enyi entropies \eqref{eq:entropyRenyiPartFunc}
  will have a similar form, with an area law divergence and a subleading
  logarithmic contribution. However, we note from eq.~\eqref{eq:Renyi-ent}, the
  edge mode contribution to these divergences in the R\'enyi entropies is
  precisely the same as their contribution to the von Neumann entropy. For example,
  with a spherical entangling surface, the logarithmic divergence is $S^\edge_n\simeq-\frac13\,\log(\zetaReg)$, identical to that above for the
  entanglement entropy. This is somewhat unusual because generally we expect the
  coefficients of these divergent terms to depend of the R\'enyi index $n$,
  \eg{} as was seen in two-dimensional CFTs
  \cite{Calabrese:2004eu,Calabrese:2005zw} and holographic CFTs
  \cite{Hung:2011nu}. However, in both of these examples, the full R\'enyi
  entropy is considered, whereas here we are only examining the contribution of
  the soft modes. 

Let us now discuss
the remaining contributions in the edge mode entanglement entropy
\eqref{eq:ee}. First, we consider the parameter $N$ in \eqref{eq:ee}, originally introduced as the
number of punctures of $\partial\cauchy[\rMilne]$ by lattice links. While it is
an obvious parameter of the lattice theory, it is not clear why, from the
field theory perspective, this parameter should enter through the path
integration measure \labelcref{eq:edgeVolFormExplicit}. Naively, one may be
tempted to drop $N$ from \cref{eq:edgeVolFormExplicit} and subsequently the
partition function \labelcref{eq:partFuncREExplicit}. In fact,
\cite{Donnelly:2015hxa,Donnelly:2014fua} offers another argument for dropping
$N$ from the latter, based on $\zeta$-function renormalization of the Laplacian
determinant.

Recall that the determinant of a Laplacian is UV divergent --- in particular,
its logarithm is the sum over the logarithms of arbitrarily large eigenvalues of
the Laplacian. Nonetheless, this can formally be resummed as the finite analytic
continuation of a $\zeta$-function and its derivative \cite{Hawking:1976ja}.
Using the identity \cref{eq:zetaScalingIdentity} from $\zeta$-function renormalization,
\cite{Donnelly:2015hxa,Donnelly:2014fua} pulls the coefficients of the Laplacian
out of the determinant just as we have done in \cref{eq:edge-pf}.
However, there are two points here to which we would like to draw
attention.

The first, as noted but brushed aside
by \cite{Donnelly:2015hxa,Donnelly:2014fua}, is the presence of the anomaly term in \cref{eq:zetaScalingIdentity}, which leads
  to an \(N\) dependence in \cref{eq:edge-pf}. 
A second observation is
that, though $\zeta$-function renormalization does not have a cutoff, a
renormalization scale is introduced for dimensional consistency as a coefficient
to the Laplacian within a determinant. Equivalently, the renormalization scale
must be introduced to ensure that the path integral measure is dimensionless.
Consideration of \cref{eq:edgeVolFormAll} or \cref{eq:edgeVolFormExplicit}
suggests that $\celVol/N$ plays the role of this renormalization scale. This
seems natural, given that $N$ was initially introduced to parametrize lattice
spacing, which in the lattice theory, also sets the renormalization scale.
This connection is noted also by \cite{Ball:2024tba} as they consider
  edge mode partition functions in \((d+2)\)-dimensional bulk spacetimes with
  \(\zeta\)-function renormalization. They identify the relation
\begin{align}
  \zetaReg &= \frac{1}{2\pi}\left( \frac{N}{\mathrm{vol}^{(d)}} \right)^{1/d}
        \label{eq:zetaRenScale}
\end{align}
between their \(\zeta\)-function renormalization energy scale \(\zetaReg\) and the
  lattice cutoff of \cite{Donnelly:2015hxa,Donnelly:2014fua}.

From this observation, the aforementioned
  universal term in entropy which is logarithmic in \(\zetaReg\) can be
  read off of \cref{eq:ee} --- it is precisely the term
\begin{align}
  S^\edge
  &\simeq
    - \frac{\mathcal{A}}{2} \log N
    \simeq -\mathcal{A}\log \zetaReg
    \;.
\end{align}
In particular, as written in \cref{eq:anomOnS2}, the celestial
  sphere has \(\mathcal{A}=1/3\). As \cite{Ball:2024tba} emphasizes, even though
  the UV-divergence tamed by \(\zeta\)-function renormalization arises from the
  \(\det\nolimits' (\Box^{(2)})\) appearing in
  \cref{eq:partFuncREExplicit,eq:edge-pf,eq:ee}, it is \emph{not} enough
  to consider only this term in \cref{eq:ee} when trying to deduce
  the universal logarithmic coefficient of entropy. Naively extracting the
  logarthmic dependence on the renormalization scale only from \(\det\nolimits'
  (\Box^{(2)})\) would be equivalent to only accounting for the \(\zetaReg\sim
  \sqrt{N/\celVol}\)-dependence appearing inside the \(\det'\) in
  \cref{eq:partFunc}. Due to the omission of the zero mode in \(\det'\), which
  precisely corresponds to the \(\dim\ker\lap\) term in
  \cref{eq:zetaScalingIdentity}, this leads to the incorrect answer \(S^\edge
  \stackrel{?}{\simeq} (1-\mathcal{A})\log\zetaReg\). We see that the extra
  $1/\sqrt{N}$ out front in \cref{eq:partFuncREExplicit} originating from
  killing the zero mode in the path integration measure
  \labelcref{eq:edgeVolForm} cancels precisely the $\dim\ker\celLap=1$ term in
  the identity \labelcref{eq:zetaScalingIdentity}. The result is that the dependence of
  the edge mode partition function \labelcref{eq:edge-pf} and entropy
  \labelcref{eq:ee} on $\zetaReg\sim \sqrt{N/\celVol}$ is purely due to \(\mathcal{A}\).

Taking a step further, we propose that $\zetaReg\sim \sqrt{N/\celVol}$ describes the
renormalization scale in the \emph{celestial CFT}. This
hypothesis is supported by the above observation that the scaling of the edge mode partition
function \labelcref{eq:edge-pf} with $\zetaReg$ is governed by $\anomAct$, in
a manner expected for a CFT --- this gives a literal meaning to $\anomAct$ as a
conformal anomaly in our case. Note that this is a nontrivial extension of the
usual derivation of the free scalar conformal anomaly under $\zeta$-function
renormalization \cite{Hawking:1976ja},\footnote{Originally, \cite{Hawking:1976ja}
  derives the anomalous trace of the stress tensor by varying the metric.
  However, as reviewed in \cite{Birrell:1982ix}, this is equivalent to varying
  the renormalization scale under $\zeta$-function renormalization.} which is
often given without explicit treatment of zero modes. 

There still remains in \cref{eq:partFuncREExplicit} and hence \cref{eq:ee} the
somewhat mysterious parameter $\reg$ setting the small $\sqrt{\fgr}$ regulator
required to define edge modes. Drawing intuition from AdS/CFT, one is led to
identify $\reg$ as a small-distance cutoff in the CFT
\cite{Susskind:1998dq,deHaro:2000vlm, Skenderis:2000in}. This would be
incongruous with the interpretation of \labelcref{eq:partFuncREExplicit} as a
renormalized CFT partition function. However, it is important to remember that
the \emph{full} Maxwell partition function \labelcref{eq:partFuncDecomposition}
also has a bulk contribution $\partFunc[\rMilne][][\invTemp][0]$. In fact, it is
shown in \cite{Donnelly:2015hxa,Donnelly:2014fua} that an opposite $\log\reg$
factor appears in the bulk contribution, so that the full partition function has
no such factor. A particular case where this cancellation must
  obviously happen is at \(\beta=2\pi\), where the conical singularity at the
  entangling surface disappears. The fact that the bulk and edge mode partition
  functions combine to agree with the full partition function on smooth
  Euclidean manifolds at \(\beta=2\pi\) has also been verified by
  \cite{Ball:2024tba}.

Superficially, the above cancellation is reminiscent of holographic renormalization
in AdS/CFT \cite{deHaro:2000vlm}; however, the `bare' bulk effective action
$\log \partFunc[\rMilne][][\invTemp][0]$ in our case has a peculiar
$\log\log\reg$ divergence. In fact, as pointed out in
\cite{Donnelly:2015hxa,Donnelly:2014fua}, the edge mode entropies
\labelcref{eq:entropyShPartFunc,eq:entropyClassicalRenyiPartFunc} also have
analogous $\log\log\reg$ divergences. We leave further study of these ideas for
future work.

As a final comment on the edge mode partition function
\labelcref{eq:partFuncDecomposition}, we note that it is invariant under
constant scalings of the celestial metric. (Here, we have in mind a fixed $N$,
which, according to our discussion in previous paragraphs, means proportionately
scaling the renormalization scale also.) In particular, the scalings of $\celVol$
and $\celLap$ cancel. For example, (at least to leading order in $\reg$) one may replace
$\celVol$ and $\celLap$ with the volume and Laplacian of the induced geometry on
$\partial\cauchy[\rMilne]$ (at $\fgr=\reg^2$). In fact, considering entangling surfaces
in the spacetime interior, refs.~\cite{Donnelly:2015hxa,Donnelly:2014fua} always works
with objects built from the induced geometry of the entangling surface.
Replacing $\celVol$ and $\celLap$ in our \cref{eq:partFuncDecomposition} with the
induced volume and Laplacian, we recover exactly the same expression for the
edge mode partition function as presented in 
\namecref{eq:partFuncDecomposition} (81) of \cite{Donnelly:2015hxa}.
Finally, in the case of a flat $\celPlane$ celestial space, we note that the
scaling symmetry of the edge mode partition function ensures that it is
independent of the period $\celIRReg$ used to regulate the celestial space, as
described below \cref{eq:edgeVolFormAll}.

We conclude here with a discussion of some open problems and future directions. \\

{\bf Bulk subregions and celestial CFT.} It would be interesting to develop a holographic interpretation of the
entanglement entropy \eqref{eq:ee} in the CCFT. As a starting point, one can construct celestial CFT operators
${}^\lMilne\shad{\mathcal{O}}^\Delta_a,{}^\rMilne{\mathcal{O}}^\Delta_a$ following the definition in
\cref{eq:confPrimOp}, using the Milne modes \eqref{eq:LR-dec} and inner products
\eqref{eq:limitMinkProdRRLL}.
In the $\eqIntegrated$ sense, ${}^\rMilne{\mathcal{O}}^\Delta_a$ is proportional
to annihilation and creation operators,
$\tensor[^\rMilne]{a}{}^{1-i\lambda}_a(\mathbf{w})$,
$\tensor[^\rMilne]{a}{}^{1-i\lambda}_{\bar{a}}(\mathbf{w})^\dagger$, for
$\Delta=1+i\lambda$ and $1-i\lambda$ (with $\lambda>0$), respectively. These
appear in a mode expansion of the field operator, multiplying the positive and
negative Milne frequency modes $({}^\rMilne{A}^{1+i\lambda})^a(\mathbf{w})$ and
$({}^\rMilne{A}^{1-i\lambda})^a(\mathbf{w})$, respectively.
Similarly, ${}^\lMilne\shad{\mathcal{O}}^\Delta_a$ is proportional to the
operators ${}^\lMilne{\shad{a}}^{1+i\lambda}_a$({\bf w}) and
${}^\lMilne{\shad{a}}^{1+i\lambda}_{\bar{a}}({\bf w})^\dagger$. Holographically,
${}^\lMilne\shad{\mathcal{O}}^\Delta_a,{}^\rMilne{\mathcal{O}}^\Delta_a$ may be
viewed as conformal primaries in
${}^{\lMilne}{\CFT}{},{}^{\rMilne}{\CFT}{}$ dual to the two bulk Milne
theories.

Our construction therefore divides the CCFT into two sectors. However, these
sectors are not independent. To see this, we define an entangling operator
\begin{align}
  \mathcal{K}^+
  = -\int_0^\infty d\lambda\,
    e^{-\pi\lambda}
    \int \celVolForm\,
    (\tensor[^\lMilne]{\shad{a}}{}^{1+i\lambda})^\dagger
    \cdot
    (\tensor[^\rMilne]{a}{^{1-i\lambda}})^\dagger
    \label{eq:K}
\end{align}
which relates the Minkowski and $\lMilne/\rMilne$ Milne vacua according to \cref{eq:TFD}. This can be rewritten in terms of the new $\lMilne,\rMilne$ conformal primaries as
\begin{equation}
  \mathcal{K}^+ 
  = -\int_0^\infty \frac{d\lambda\,
    e^{-\pi\lambda}}{(2\pi)^3}\,
  \frac{1+\lambda^2}{2\lambda}
  \!\int \celVolForm
  \tensor[^\lMilne]{\shad{\mathcal{O}}}{^{1- i\lambda}}
  \cdot
  \tensor[^\rMilne]{\mathcal{O}}{^{1+ i\lambda}},
  \label{eq:cftInteraction}
\end{equation}
which becomes an
interaction coupling the ${}^{\lMilne}{\CFT}{}$ and ${}^{\rMilne}{\CFT}{}$. For
example, amplitudes in the global vacuum state $\ket{0}$ are evaluated by CFT correlation
functions in the presence of this interaction:\footnote{The trivial scattering amplitudes of our free   theory can be equivalently described from the Minkowski or Milne perspectives,   so it suffices to consider the   ${}^\lMilne{\CFT}{},{}^\rMilne{\CFT}{}$ duals of the Milne   patches in \cref{eq:entanglementAsInteraction}. The nontrivial scattering of   interacting theories in Minkowski spacetime, however, would require a more   involved analysis sketched further below.}
\begin{align}
  \bra{0}
  \bullet
  \ket{0}
  &= \langle
    e^{-\mathcal{K}^+-(\mathcal{K}^+)^\dagger} \bullet\,
    \rangle_{{}^{\lMilne}{\CFT}{},{}^{\rMilne}{\CFT}{}}
    \;.
    \label{eq:entanglementAsInteraction}
\end{align}

With $\bullet$ in ${}^{\rMilne}{\CFT}{}$, the holographic dual of the
thermal expectation value in the state 
\begin{align}
  \tensor[^{\rMilne}]{\rho}{}[0]
  \equiv \tr_{\tensor[^L]{\mathcal{H}}{}}
  \ket{0}\bra{0}
  \propto e^{-2\pi \tensor[^\rMilne]{H}{}}
  \label{eq:thermalState}
\end{align}
arises by tracing out ${}^{\lMilne}{\CFT}{}$, leaving
\begin{align}
  &\qquad\qquad\langle   e^{-{}^\rMilne\mathcal{K}} \,\bullet\,
    \rangle_{{}^\rMilne{\CFT}{}}\qquad {\rm with}
   \qquad  {}^\rMilne\mathcal{K} 
    = -\int_0^\infty \frac{d\lambda\,
    e^{-2 \pi\lambda}}{(2\pi)^3}\,
    \frac{1+\lambda^2}{2\lambda}
    \!\int \celVolForm
    \tensor[^\rMilne]{\mathcal{O}}{^{1- i\lambda}}
    \cdot
    \tensor[^\rMilne]{\mathcal{O}}{^{1+ i\lambda}}\,.
    \label{eq:nice}
\end{align}
Expanding the exponential in \cref{eq:nice} yields a series of correlation
functions weighted by $(e^{-2\pi\lambda})^n$ and the entanglement entropy
corresponds to the von Neumann entropy of
this distribution. This procedure is somewhat analogous to
\cite{Mollabashi:2014qfa}, which examines entanglement between two interacting
CFTs. However, their framework allows for a standard evaluation of entanglement
entropy of the resulting mixed state, which is not the case for the CCFT.

We emphasize that the analysis herein relied on the theory being both Weyl invariant and free. In interacting theories, the unitary relation \eqref{eq:dnsMatBlocksUnitaryRelation} fails to
hold leading to the mixing of hard and soft
contributions to the entanglement entropy (see also below). If Weyl invariance
is also lost, \eg{} by introducing massive particles, we would still decompose the asymptotic states in terms of modes with support to the future and past of
the cut on $\mathscr{I}^+$. The Minkowski vacuum $\ket{0}$ would be some
entangled state of these modes. Further, these modes would again yield a
division of the CCFT into two sectors which interact through an entangling
operator. We leave an exploration of these important physically relevant generalizations to future work.\\

{\bf Relation to soft effective actions. }It was shown in \cite{Weinberg:1965nx,Kapec:2017tkm,Kapec:2021eug} that the exponentiation of IR divergences in QED implies that celestial amplitudes factorize according to 
\begin{equation}
\langle  \mathcal{O}_1 \cdots \mathcal{O}_n \rangle_{\mu} = e^{-\Gamma[\Lambda, \mu]}
 \langle  \mathcal{O}_1 \cdots \mathcal{O}_n \rangle_{\Lambda} ,
\end{equation}
where $\mu$ is an IR cutoff and $\Lambda$ is a soft-hard factorization scale. The real part of $\Gamma$ is due to exchanges of soft gravitons that exponentiate and the soft S-matrix is obtained from it by exponentiation \cite{Weinberg:1965nx}. It can be expressed in terms of the soft photon operators
\begin{equation}
\label{eq:soft-current}
J_a(x) \equiv \sum_i Q_i \p_a \log[-\hat{p}_i \cdot \hat{q}] = \sum_i J_a^i(x)
\end{equation}
and takes the form
\begin{equation}
\label{eq:QED}
{\rm Re} \Gamma = \alpha_{{\rm QED}} \int \frac{d^d x}{(2\pi)^d} \,[J_a(x)]^2.
\end{equation}
Here
\begin{equation}
\label{aqed}
\alpha = \frac{e^2}{8\pi} \int_{\mu}^{\Lambda} d\omega \,\omega^{d - 3}
\end{equation}
with $d$ the dimension of the celestial plane and
$(J_a)^ 2 \equiv J_a J^a$ with a sum over the polarization indices $a$ implicit. Note that $\alpha$ is divergent for $d = 2$, \ie, for a four-dimensional bulk, in the limit that the IR cutoff $\mu \rightarrow 0$.  

We now establish a concrete relationship between the soft S-matrix and the on-shell action of edge modes \eqref{eq:actAER}. To this end, we set $d = 2$ and  substitute \cref{eq:soft-current} into \cref{eq:QED} to find
\begin{equation}
\label{soft}
\begin{split}
\int \frac{d^2 x}{(2\pi)^2} [J_a(x)]^2 &= -\int\frac{d^2 x}{(2\pi)^2} \sum_{i, j} Q_i Q_j \p^a \p_a\left( \log[-\hat{p}_i \cdot \hat{q}]  \right) \log[-\hat{p}_j \cdot \hat{q}]\\
&=-\int\frac{d^2 x}{(2\pi)^2} \sum_{i, j} Q_i Q_j  \delta^{(2)}(z_i - x) (\Box^{(2)})^{-1} \delta^{(2)}(z_j - x)\\
&= -\frac{1}{e^2}\int\frac{d^2 x}{(2\pi)^2} q_H(x) (\Box^{(2)})^{-1} q_H(x).
\end{split}
\end{equation}
In the first line, we integrated by parts while in the second line, we used the Green's function for the 2D Laplacian
\begin{equation} 
\p^a \p_a \log[-\hat{p}_i \cdot \hat{q}] \equiv \Box^{(2)} \log[-\hat{p}_i \cdot \hat{q}] = \delta^{(2)}(z_i - x), \quad  \log[-\hat{p}_i \cdot \hat{q}] = (\Box^{(2)})^{-1}\delta^{(2)}(z_i - x)\,.
\end{equation}
In the final expression, we defined the charge distribution
\begin{equation}
q_H(x) \equiv \sum_{i} e\, Q_i\, \delta^{(2)}(z_i - x).
\end{equation}
Note that charge conservation implies that
\begin{equation}
\int d^2x\, q_H(x) = \sum_i Q_i = 0\,.
\end{equation}
It is important to notice that \cref{soft} is only non vanishing in the presence of hard charges (otherwise $q_H(x) = 0$). 

We can compare and contrast $\Gamma$ with the on-shell action \eqref{eq:actAER} with $\beta = 2\pi$
\begin{equation} 
\label{on-shell}
I_{\rm on-shell}[q] = \frac{\pi}{\log \epsilon} \int \frac{d^2 x}{(2\pi)^2} q(x) (\Box^{(2)})^{-1} q(x),
\end{equation}
where $q(x)$ is instead the soft charge associated with the Milne patch. This becomes proportional to ${\rm Re}\Gamma$ subject to the identification
\begin{equation} 
\label{eq:soft-hard}
q(x) =  q_H(x)
\end{equation}
with 
\begin{equation}
\epsilon^{-1} = \frac{\Lambda}{\mu}.
\end{equation}
It is important
to note that the $q_H$ appearing in the soft S-matrix represent physical charges
that exchange soft photons, while $q(x)$ appearing in the on-shell action are
associated with conformally soft field configurations, which as discussed in
section \ref{sec:entanglement}, are non-vanishing even in the free theory. The
identification \eqref{eq:soft-hard} may be understood via our embedding inside
the Einstein static universe, where $q(x)$ may be regarded as turned on by
 charged particles inside the inverted Minkowski patch, as illustrated in figure \eqref{fig:bun}. Note that these sources are infinitely boosted from the perspective of the inverted patch, but accelerating from the perspective of the Einstein cylinder where they follow trajectories with sharp cusps. Straightening out these trajectories, the charged particles $q_H(x) = q(x)$ may then be regarded as the continuation of the trajectories into the original Minkowski patch. Note, however, that in the limit as $\epsilon \rightarrow 0$, the on-shell action vanishes, while the exponent of the soft S-matrix diverges. It would be interesting to understand this relation better. 

 The soft effective actions are closely related to the dressings leading to finite scattering amplitudes in the limit as the infrared cutoff is removed \cite{Kulish:1970ut}. These dressings were shown in \cite{Arkani-Hamed:2020gyp} to be given by inner products of gauge fields with conformally soft wavefunctions. As remarked in section \ref{sec:CS-decompositions}, the matching condition implies the presence of a pure-gauge, boundary term in the large-$r$ expansion of conformally soft wavefunctions, which has been ignored in previous works. It would be important to check whether this term affects the dressing, for example by evaluating its inner product with plane waves. If it does, it would be interesting to understand whether it can also be extracted from the momentum space scattering amplitude, and study its implications for the loop corrected stress tensor \cite{He:2017fsb,Pasterski:2022djr,Donnay:2022hkf}.\\
 
{\bf Gravity and spacetime fluctuations.} The logarithm of the soft S-matrix
\eqref{eq:QED} admits a straightforward generalization to gravity
\cite{Himwich:2020rro, Arkani-Hamed:2020gyp, Kapec:2021eug}. This suggests that
our analysis could apply to theories of gravity at least in the soft sector. One
naive obstacle is that Einstein gravity in  four dimensions is not a conformal
theory. Nevertheless, for gravitons in the Minkowski vacuum, the traces of the Riemann tensor vanish and
hence all effects are captured by the Weyl tensor. Of course, the latter transforms
covariantly under Weyl rescalings \cite{PhysRevLett.10.66}. Since the
gravitational soft (as well as hard) charges are extracted precisely from
components of the Weyl tensor and the associated constraints, it is plausible
that the analysis herein could generalize to gravity at least in the
infrared/vacuum sector. This would be particularly relevant for a better
understanding of the vacuum spacetime fluctuations proposed in
\cite{Verlinde:2019xfb,Zurek:2020ukz,Verlinde:2022hhs} as observational
signatures of quantum gravity. Relatedly, it would be useful to better
understand the bulk meaning of shadow transforms in non-conformally invariant
theories and their role in characterizing subregions of null infinity. We leave
a complete understanding of the relation between soft effective actions and
boundary actions in asymptotically flat spacetimes, as well as its implications
for the future.
\\

{\bf Implications for $\mathcal{N} = 4$ SYM and holography.} One may expect that the methods developed in this paper could be used to study edge modes in $\mathcal{N} = 4$ SYM and their relation to IR divergences. The exponentiation of IR divergences is in this case complicated by the non-abelian nature of the gauge fields, which by analogy with QED is suggestive of a rich structure of edge mode entanglement entropy in this context. On the one hand, Wilson loop techniques that proved very useful in the study of infrared and collinear divergences \cite{Caron-Huot:2013fea,Caron-Huot:2016tzz} may also be applicable to entanglement computations. On the other hand, one may hope that asymptotic symmetries will provide a complementary perspective and potentially new tools to tackle IR divergences.  It would, for example, be very interesting to explore the relation between memory effects associated with the edge/conformally soft modes and Wilson loop computations in $\mathcal{N} = 4$ SYM, as well as their AdS$_5$ holographic duals. \\

{\bf Cutoffs and mixing of soft and hard modes.} Needless to say, it would be
important to better understand the mixing between soft and hard modes at finite
cutoff. As remarked in \cite{Donnelly:2014fua, Donnay:2015abr}, the
double-logarithmic terms in the edge mode entanglement entropy \eqref{eq:ee}
 should cancel against similar terms in
the entanglement entropy of bulk modes. It would be interesting to understand
this cancellation by explicitly computing the divergent contributions to the
entanglement of hard modes extracted from \cref{eq:TFD}, as well as by
generalizing the analysis to the case of Maxwell fields coupled to matter. In this case \eqref{eq:TFD} should acquire cutoff-dependent corrections which will be interesting to work out. Moreover, our relation between conformally
soft mode field configurations and sources in the inverted patch could provide a
connection between the Donnelly-Wall edge mode entropy and the complementary
perspective on the logarithmic coefficient of entanglement of Maxwell fields due
to charge fluctuations in the UV completed theory developed in
\cite{Casini:2019nmu}. 

Relatedly, it would be interesting to
understand the covariance properties of the edge mode entropy under
transformations of the cutoff. Being an entanglement entropy across a cut of
$\mathscr{I}^+$, \eqref{eq:ee} may potentially be regarded as a contribution to entanglement in
a Carrollian field theory and may therefore transform covariantly under Carrollian symmetries (see \cite{Grumiller:2023rzn} for a recent discussion of 3-dimensional asymptotically flat bulk theories).  Note however that in our case, the entangling surface is not an interval in the Carrollian field theory, but  rather a subregion on a null surface, or a Cauchy surface in the ambient Minkowski space. The entanglement entropy will then be measuring correlations among fields in the future and past of a ``time'' slice in the Carrollian field theory. We leave a better understanding of any connections to Carrollian quantities to future work.


\section*{Acknowledgements}
We would like to thank Laurent Freidel, Prahar Mitra and especially Sabrina
Pasterski for helpful discussions, as well as Temple He and Kathryn Zurek for
collaboration on related topics. We would also like to thank
  Sabrina Pasterski for her comments on a nearly final draft, as well as Adam
  Ball, Albert Law, and Gabriel Wong for useful discussions and  sharing a draft of their
  paper \cite{Ball:2024tba} with us. Research at Perimeter Institute is supported in part
by the Government of Canada through the Department of Innovation, Science, and
Economic Development Canada and by the Province of Ontario through the Ministry
of Colleges and Universities. HZC was supported by the Natural Sciences and
Engineering Research Council of Canada through a Doctoral Canadian Graduate
Scholarship. Additionally, HZC is supported by a Fundamental Physics Fellowship
through the University of California, Santa Barbara. RCM is supported in part by
a Discovery Grant from the Natural Sciences and Engineering Research Council of
Canada and by funding from the BMO Financial Group. The authors were also
supported by the Simons Foundation through the ``It from Qubit'' collaboration.
Over the past three years while this project has been ongoing, AMR was further supported by a Stephen Hawking fellowship at Perimeter Institute, by the Heising Simons Foundation collaboration on ``Observational Signatures of Quantum Gravity'', as well as by the European Commission through a Marie Sklodowska-Curie fellowship, grant agreement 101063234.
\vskip 1cm


\appendix
\section{Preliminaries and conventions}
\label{sec:conventions}

\subsection{Fock space construction}
\label{app:Fock}

Let us review the standard Fock space
construction for the non-soft modes. By studying $\symp{\phSpace}$, the idea is
to identify appropriate creation $(a^M)^\dagger$ and annihilation $a^M$
operators. In particular, the $a^M$ are taken to annihilate the associated
vacuum state $\ket{0}$:
\begin{align}
  a^M \ket{0}
  &= 0
\end{align}
The Fock space is then generated by the creation operators $(a^M)^\dagger$
acting on $\ket{0}$.

More precisely, these creation and annihilation operators are holomorphic and
antiholomorphic phase space coordinates which parametrize the non-soft part of
\cref{eq:generalModeExpansion}. Considering real fields for simplicity, we write
\begin{align}
  \allFields
  &= \allFields^\soft + a^M U_M + a^{\bar{M}} U_{\bar{M}} \;,
  &
    a^{\bar{M}}
  &= (a^M)^\dagger \;,
  &
    U_{\bar{M}}
  &= (U_M)^* \;,
\end{align}
where $U_M$ and $U_{\bar{M}}$ are non-soft solutions of the free equations of motion. In formal language
\cite{Ashtekar:1975zn}, the choice of complex structure $J$,
\begin{align}
  J\, U_M
  &= i U_M \;,
  &
    J\, U_{\bar{M}}  &= -i U_{\bar{M}} \;,
    \label{eq:complexStructure}
\end{align}
distinguishes creation operators from annihilation operators, which in turn
selects the vacuum state $\ket{0}$. Note that having the expected vanishing
commutators
\begin{align}
  [a^M, a^N]
  &= [a^{\bar{M}}, a^{\bar{N}}]
    = 0
\end{align}
is equivalent to the complex structure $J$ preserving $\symp{\phSpace}^{-1}$ and
hence $\symp{\phSpace}$, \ie{}
\begin{align}
  \symp{\phSpace}(U, U')
  &= \symp{\phSpace}(J U, J U')
    \;.
\end{align}
Moreover, note that the Hilbert space inner product between one-particle states
\begin{align}
  \fieldHermMetric(U, U')
  &= \bra{0}
    \symp{\phSpace}(U',\allFields)
    \symp{\phSpace}(U,\allFields)
    \ket{0}
    \;,
  &
    \fieldHermMetric(U_M, U_N)
  &=
    \fieldHermMetric(U_{\bar{M}}, U_{\bar{N}})
    =
    \fieldHermMetric(U_{\bar{M}}, U_N)
    = 0
\end{align}
doubles as a Hermitian metric\footnote{Note that we have flipped the ordering of
  $U$ and $U'$ in the first equality to ensure the standard convention of
  $h_{\bar{M} N}=0$ for the Hermitian metric.} on the complex manifold of the
non-soft phase space. Indeed, from \cref{eq:starAlgebra}, we can derive the
K\"ahler relation
\begin{align}
  2 \fieldHermMetric(U,U') &= \fieldMetric(U,U') + i \symp{\phSpace}(U,U')
\end{align}
between the Hermitian metric $\fieldHermMetric$, symplectic form $\symp{\phSpace}$, and
Riemannian metric
\begin{align}
  \fieldMetric(U,U')
  &=
    \bra{0}
    \symp{\phSpace}(U,\allFields)
    \symp{\phSpace}(U',\allFields)
    +
    \symp{\phSpace}(U',\allFields)
    \symp{\phSpace}(U,\allFields)
    \ket{0}
    = \symp{\phSpace}(J U,U') \;,
    \label{eq:kahlerRiemanMetric}
\end{align}
whose positive-definitess is another constraint on $J$. Clearly, $\fieldMetric$ and $\fieldHermMetric$ and
are also invariant under $J$:
\begin{align}
  \fieldMetric(U,U')
  &= \fieldMetric(JU, JU')
    \;,
  &
    \fieldHermMetric(U,U')
  &= \fieldHermMetric(JU, JU')
    \;.
    \label{eq:hgJInvariance}
\end{align}
To summarize, the non-soft phase space is a K\"ahler manifold on which the
choice of complex structure $J$ determines the vacuum $\ket{0}$ on which the
Fock space is constructed \cite{Ashtekar:1975zn}.

In general, the choice of $J$ and hence $\ket{0}$ is not unique.\footnote{See
  section 4.4 of \cite{Wald:1995yp} for a discussion of the unitary equivalence
  between Fock spaces built from different choices of vacua.} However, on a
stationary spacetime with time-like Killing vector $\xi$, a natural and standard
choice is \cite{Ashtekar:1975zn}
\begin{align}
  J &= -(-\lie{\xi}^2)^{-1/2} \lie{\xi} \;,
\end{align}
where $\lie{\xi}$ is the spacetime Lie derivative. This complex structure
distinguishes, in the sense of \cref{eq:complexStructure}, solutions $U_M$ and
$U_{\bar{M}}$ of positive and negative frequencies\footnote{Note that our
  definitions of positive and negative frequencies is standard, but opposite to
  that of \cite{Ashtekar:1975zn}. Specifically, if $\xi=\partial_t$ then modes
  $U_M\sim e^{-i\omega t}$ and $U_{\bar{M}} \sim e^{i\omega t}$ with $\omega>0$
  are said to have positive and negative frequencies respectively.} respectively
under the $\xi$ flow.\footnote{For conformal field theories (like the
  four-dimensional Maxwell theory), it is also natural to consider the complex
  structure $J$ associated to a time-like conformal Killing vector $\xi$. In
  \cref{sec:tfd}, we will apply this approach in the Milne patch of Minkowski
  space, where the flow orthogonal to the hyperbolic spatial slices is a
  conformal Killing vector. Further, we will find that the Milne vacuum defined
  in this way differs from the Minkowski vacuum associated to the usual Killing
  flow along the Cartesian time direction.}

We conclude this section by discussing some properties of the symplectic product \eqref{eq:sympProd}.  From \cref{eq:complexStructure,eq:kahlerRiemanMetric,eq:hgJInvariance}, we see
that
\begin{align}
  \solnProd{U_M}{U_N}
  &= \fieldMetric(U_M, U_{\bar{N}}) \;,
  &
    \solnProd{U_{\bar{M}}}{U_{\bar{N}}}
  &= -\fieldMetric(U_{\bar{M}}, U_N) \;,
  &
    \solnProd{U_M}{U_{\bar{N}}}
  &= \solnProd{U_{\bar{M}}}{U_N}
    = 0 \;.
    \label{eq:solnProdModeProperties}
\end{align}
Conversely, the positive-definiteness of the first product,
negative-definiteness of the second, and the vanishing of the last two are
sufficient indications that the division of modes between $U_M$ and
$U_{\bar{M}}$ gives a valid choice of complex structure $J$.

\subsection{Minkowski spacetime, celestial space, and null momenta}
\label{sec:minkConventions}
The Minkowski metric in Cartesian coordinates takes the form
\begin{align}
  \diff s_{\mink}^2
  &= \minkMet_{\mu\nu} \, \diff \minkX^\mu \diff \minkX^\nu \;,
  &
    \minkMet_{\mu\nu} &= \diag(-1,1,1,1) \;.
\end{align}
The spacetime covariant derivative denoted $\covD$, as in \cref{eq:vacuum-ME},
and reduces to a partial derivative in Cartesian coordinates:
\begin{align}
  \covD_\mu
  &= \partial_\mu
\end{align}
Spacetime tensor indices shall be lowered and raised by $\minkMet$ as needed. We
will sometimes use $\cdot$ to mean a contraction of tensors and $\bullet^2$ to
mean the contraction of tensor with itself. Additionally, we will sometimes
write $\minkX$ to mean a vector or dual vector field, whose Cartesian components
are given by $\minkX^\mu$ or $\minkX_\mu$. To illustrate these conventions, we
may write
\begin{align}
  \lap
  &\equiv \covD^2
    =\covD \cdot \covD
    = \covD^\mu \covD_\mu
    = \minkMet^{\mu\nu} \covD_\mu \covD_\nu
    = -(\partial_0)^2 + (\partial_1)^2 + (\partial_2)^2+(\partial_3)^2 \;,
  \\
  \minkX^2
  &= \minkX\cdot \minkX
    = \minkX^\mu \minkX_\mu
    = \minkMet_{\mu\nu} \minkX^\mu \minkX^\nu
    = -(\minkX^0)^2 + (\minkX^1)^2 + (\minkX^2)^2+(\minkX^3)^2 \;.
\end{align}

Points in the celestial space will be denoted
by $\celw$ or $\minkz$. To initially define certain objects and discuss
conformal transformations, it will be useful to refer to complex coordinates
here $(\celwh,\celwa)$ or $(\minkzh,\minkza)$; otherwise, we will try to be as
covariant as possible, employing general coordinate indices
$\alpha,\beta,\ldots$ or abstract tensor indices $a,b,\ldots$ (which if present,
merely help indicate tensor structure and how tensors are contracted --- see
\eg{} \cite{Wald:1984rg}). The metric for the celestial plane is given by
\begin{align}
  \diff s_{\celPlane}^2
  &= \celMet_{\alpha \beta} \diff \celw[^\alpha] \diff \celw[^\beta] \;,
  &
    \celMet_{\celwh \celwh}
  &= \celMet_{\celwa \celwa} = 0 \;,
  &
    \celMet_{\celwh \celwa}
  &= \celMet_{\celwa \celwh}
    = \frac{1}{2} \;.
    \label{eq:yogurt}
\end{align}
The celestial covariant derivative $\celCovD$ reduces to partial derivatives in
complex coordinates:
\begin{align}
  \celCovD_{\celwh}
  &= \partial_{\celwh} \;,
  &
    \celCovD_{\celwa}
  &= \partial_{\celwa} \;.
\end{align}
We shall also use $\cdot$ and $\bullet^2$ to mean contraction and squaring of
tensors in celestial space, where indices are lowered and raised by the metric
$\celMet$ and its inverse. For instance, we may illustrate this by defining the
celestial Laplacian
\begin{align}
  \celLap
  &\equiv \celCovD^2
    = \celCovD\cdot\celCovD
    = \celCovD^a \celCovD_a
    = \celMet^{a b} \celCovD_a \celCovD_b
    = 4 \partial_{\celwh} \partial_{\celwa} \;.
\end{align}
We shall use $\celVolForm$ and $\celDeltaFunc(\celw,\celw')$ to denote the
celestial volume form and associated $\deltaFunc$-function distribution, which
satisfy
\begin{align}
  \int \celVolForm(\celw') \celDeltaFunc(\celw,\celw') f(\celw')
  &= f(\celw) \;.
    \label{eq:celDeltaFunc}
\end{align}
In complex coordinates, they read 
\begin{align}
  \celVolForm
  &= \frac{i}{2}
    \diff \celwh \wedge \diff \celwa \;,
  &
    \celDeltaFunc(\celw,\celw')
  &= -2i\, \deltaFunc[2](\celwh-\celwh',\celwa-\celwa') \;.
\end{align}
An object which will appear often in this paper is the celestial Green's
function $\celG$ which satisfies
\begin{align}
  \celLap[\celw] \celG(\celw,\celw')
  &= \celLap[\celw'] \celG(\celw,\celw')
    = \celDeltaFunc(\celw,\celw') \;,
    \label{eq:grape}
\end{align}
and, in complex coordinates, is given by
\begin{align}
  \celG(\celw,\celw')
  &= \frac{1}{4\pi}\log[(\celwh-\celwh')(\celwa-\celwa')] \;.
    \label{eq:celG}
\end{align}
As exemplified by $\celLap[\celw]$ and $\celLap[\celw']$, wherever it is
ambiguous, we shall subscript differential operators with the spacetime or
celestial point on which they act. For the celestial sphere, $\gamma_{w\bar{w}}$ in \eqref{eq:yogurt} is replaced by $\gamma_{w\bar{w}} = \frac{2}{(1+w\bar{w})^2}$ and the derivatives are replaced by covariant derivatives on the sphere.

One nice property of the null vector $\nullUnit$ in \eqref{eq:null-vector} is that it manifests the
correspondence
\begin{align} \SO(1,3) &= \SL(2,\complexes)
\end{align}
between the Lorentz group of four-dimensional Minkowski spacetime and the global
conformal group on the two-dimensional celestial space. In particular, for any
Lorentz transformation $\lorentz \in \SO(1,3)$ and corresponding
$\SL(2,\complexes)$ transformation
\begin{align} \confTrans: (\celwh, \celwa) &\mapsto \left( \frac{a \celwh +b}{c
      \celwh +d} , \frac{\bar{a} \celwa+\bar{b}}{\bar{c} \celwa+\bar{d}} \right)
  & (ad - bc = 1)
    \label{eq:confTrans}
\end{align}
we have
\begin{align} \nullUnit( \confTrans(\celw)) &= (c \celwh+d)^{-1} (\bar{c}\celwa
  + \bar{d})^{-1} \lorentz \cdot\, \nullUnit(\celw) \;.
\end{align}

The null vector $\nullUnit$ is also useful for parametrizing the directions in
which null momenta point:
\begin{align} \minkMom^\mu &= (\minkMom^0, \spacMom) =\celFreq
  \nullUnit^\mu(\celw) \;,
\end{align}
With this, the natural integration measure and $\deltaFunc$-distribution over
null momenta are given by
\begin{align} \frac{\diff^3\spacMom}{\minkMom^0} &= \celFreq\, \diff\celFreq
  \wedge \celVolForm \;, & \minkMom^0 \, \deltaFunc[3](\spacMom - \spacMom') &=
  \frac{\deltaFunc(\celFreq - \celFreq')}{\celFreq} \celDeltaFunc(\celw,\celw').
                                                                               \label{eq:pizza}
\end{align}


\section{Near boundary asymptotics}
\label{sec:asymptotics}

In this appendix, we carefully work out the leading and subleading terms in the
large-$r$ expansions of the logarithmic building blocks of conformally soft
wavefunctions. These expansions contain logarithmic terms in $r$, whose
coefficients are distributions on the celestial space needed for consistency
with the free equations of motion on the hyperbolic slices of the Milne patch.
These logarithmic contributions appear to have been missed in previous works
such as \cite{Donnay:2018neh}.

\subsection{Asymptotics of $\log(-\nullUnit\cdot\minkX_\pm)$}
\label{sec:logqXAsymptotics}
Let us begin by considering the large $\minkr$ expansion of the scalar function
$\log(-\nullUnit\cdot\minkX_\pm)$ appearing in the definition \cref{eq:cappuccino} of the gauge parameter
$\confGauge{1,\pm}$ of asymptotic symmetries.

Our starting point is \cref{eq:logqX}, which relates our function of interest $\log(-\nullUnit\cdot\minkX_\pm)$ to the function $\hypK(\celw;\fgr,\minkz)$ defined in \cref{eq:hypK}. Thus, the
near-$\nullInfty^{\pm}$ (\ie{} large $\minkr$) asymptotics of
$\log(-\nullUnit\cdot\minkX_\pm)$, apart from a term $\log(\minkr/2)$, is
identical to its small $\minku\mp i\reg$ asymptotics and also to the small
$\fgr_\pm$ asymptotics of $4\pi\hypK(\celw;\fgr_\pm,\minkz)$. Here, we are not necessarily restricting to a Milne patch, but if one does
so, then in the $\reg\to 0$ limit, $\fgr_\pm$ reduces to the real positive
Fefferman-Graham coordinate \labelcref{eq:fgr} along hyperbolic slices.

Obviously, at leading order, $\hypK(\celw;\fgr,\minkz)$ at $\fgr=0$ reduces to
the celestial Green's function \labelcref{eq:celG}:
\begin{align}
  \hypK(\celw;0,\minkz)
  &= \celG(\celw,\minkz)
    \label{eq:hypKLeading}
\end{align}
Also useful for this paper will be its next order term near $\fgr=0$, which we
derive in the following sections using a direct method, as well as a
more elegant analysis of scalar equations of motion on $\hyp{3}$.

\subsubsection{Direct derivation of the next order term}
Naively, one might expect
\begin{align}
  \hypK(\celw;\fgr,\minkz)
  -\celG(\celw,\minkz)
  &= \frac{1}{4\pi} \log\left(
    \frac{\fgr}{|\celwh-\minkzh|^2} + 1
    \right)
    \label{eq:hypKSubtracted}
\end{align}
to be $\order{\fgr}$ but in fact there is a lower order term as we now show.
Note that terms of noninteger order in $\fgr$ such as this must be
$\celw=\minkz$ contact terms, otherwise the naive Taylor expansions of
\cref{eq:hypK,eq:hypKSubtracted} in $\fgr$ would be valid.

We have already noted the zeroth order contribution \labelcref{eq:hypKLeading} to
$\hypK(\celw;\fgr,\minkz)$ and we now want to find all contact terms below
$\order{\fgr}$. For this purpose, we shall evaluate the derivative
$\partial_\fgr\hypK(\celw;\fgr,\minkz)$ and isolate any terms of divergent order
in $\fgr$. Since we are searching for contact terms which are best formulated as
distributions, let us consider integrating
$\partial_\fgr\hypK(\celw;\fgr,\minkz)$ against an arbitrary smooth function
$f(\minkz)$ over celestial space (which falls off sufficiently quickly at large
$\minkzh$):
\begin{align}
  \int \celVolForm(\minkz) \,
  \partial_\fgr \hypK(\celw;\fgr,\minkz)
  f(\minkz)
  &= \frac{1}{4} \int_0^\infty \diff\abs{\celwh-\minkzh}^2\,
    \frac{1}{\fgr + \abs{\celwh-\minkzh}^2} g(\abs{\celwh-\minkzh}^2)
    \label{eq:smearedPartialhypK1}
  \\
  &= \frac{1}{4} \int_{\fgr+\posReals} \diff\celIntVar\,
    \frac{g(\celIntVar-\fgr)}{\celIntVar}
    \;,
    \qquad
    (\celIntVar=\fgr+\abs{\celwh-\minkzh}^2).
    \label{eq:smearedPartialhypK2}
\end{align}
Here, we have defined the mean
\begin{align}
  g(\abs{\minkzh-\celwh}^2)
  &= \frac{1}{2\pi} \int \diff\celpsi\,
    f(\minkz)
  &
    (\minkzh-\celwh = \abs{\minkzh-\celwh}\, e^{i\celpsi})
    \label{eq:celAngMean}
\end{align}
of $f(\minkz)$ in the angular direction around $\celw$. Because $f(\minkz)$ is
smooth, $g(\abs{\minkzh-\celwh}^2)$ is a smooth function in
$\abs{\minkzh-\celwh}^2$, even at $\celw=\minkz$. The reason the integral
\eqref{eq:smearedPartialhypK2} diverges in the $\fgr\to 0$ limit is that the
endpoint of the contour of integration approaches $\celIntVar=0$, where the
integrand blows up like $ \celIntVar^{-1}$. Evaluating the divergence of the
integral, we find
\begin{align}
  \int \celVolForm(\minkz) \,
  \partial_\fgr \hypK(\celw;\fgr,\minkz)
  f(\minkz)
  &=  -\frac{\log \fgr}{4} g(0) + \order{\fgr^0} \;.
    \label{eq:smearedPartialhypKExpansion}
\end{align}
Identifying $g(0)=f(\celw)$, we therefore conclude that
\begin{align}
  \hypK(\celw;\fgr,\minkz)
  &= \celG(\celw,\minkz) -\frac{\fgr\log\fgr}{4} \celDeltaFunc(\celw,\minkz)
    + \order{\fgr} \;.
    \label{eq:hypKExpansion}
\end{align}

It is interesting to note that the $\fgr\log\fgr$ term, in essence, captures the
divergence of the `bare' linear order term obtained from a Taylor expansion of
$\hypK$ in $\fgr$:
\begin{align}
  \left[ \partial_\fgr \hypK(\celw;\fgr,\minkz)  \right]_{\fgr=0}
  &= \partial_{\abs{\celwh-\minkzh}^2} \celG(\celw,\minkz)
    = \frac{1}{4\pi \abs{\celwh-\minkzh}^2}
    \label{eq:hypKBareLinear}
\end{align}
Note that this bare term is precisely what one finds in the integrand of
\cref{eq:smearedPartialhypK1} if $\fgr$ is immediately set to zero there. When
smeared against any continuous function that is nonvanishing at $\minkz=\celw$,
it diverges; however, what our analysis has demonstrated is that this divergence
should really be thought of as a $\celDeltaFunc$-function occurring at a
marginally lower order in $\fgr$. We will see another manifestation of this in
\cref{sec:alpha1SquaredExpansion}.

\subsubsection{Asymptotics of scalar equations of motion on $\hyp{3}$}
\label{sec:hypK}

The expansion \labelcref{eq:hypKExpansion} can be more easily deduced by considering
the scalar equation of motion on $H_3$. Working in the coordinates introduced
in \cref{sec:milnegeo} and restricting to a hyperbolic slice inside the
Milne patch, this takes the form
\begin{equation}
  \label{eq:H3eom}
  \rho\, \partial_{\rho}^2 \varphi +
  \frac{1}{4}\celLap \varphi = 0.
\end{equation}
It is straightforward to verify that ${\rm K}({\bf w; \rho, z})$ defined in
\cref{eq:hypK} satisfies this equation. Its small $\rho$ expansion can be
derived by
requiring that the ansatz
\begin{equation}
  {\rm K}({\bf w}; \rho, z)
  = \celG({\bf w}, {\bf z})
  + \rho \log \rho {\rm K}^{\rm log,1}({\bf w},{\bf z})
  + \mathcal{O}(\rho)
\end{equation}
satisfies \cref{eq:H3eom} order by order in the small $\rho$ expansion. At the
leading order, \cref{eq:H3eom} then implies
\begin{equation}
  \frac{1}{4}\celLap \celG({\bf w}, {\bf z}) + {\rm K}^{\rm log, 1}({\bf w, z})
  = 0
\end{equation}
from which we immediately deduce
\begin{equation} {\rm K}^{\rm log, 1}({\bf w, z}) = -\frac{1}{4} \delta({\bf w,
    z}).
\end{equation}
This agrees with \eqref{eq:hypKExpansion} as promised.

\subsection{Near boundary expansions of other functions}
\label{sec:quadFuncsAsym}
While the asymptotics of field configurations such as $\confGauge{1,\pm}$ (and
therefore $\confPrimA{1,\pm}$) are readily obtained by differentiating
\cref{eq:logqX,eq:hypKExpansion} in the celestial directions, some other
functions whose pullback to $\nullInfty^{\pm}$ are considered in this paper are
quadratic in $\log(-\nullUnit\cdot\minkX_{\pm})$. Here, we evaluate the
asymptotic expansions of such functions also.

\subsubsection{Asymptotics of $\log^2(-\nullUnit\cdot\minkX_\pm)$}
We start by squaring the relation \eqref{eq:logqX} between
$\log\left(-\hat{q}\cdot X_{\pm} \right)$ and $\hypK$
\begin{align}
  \log^2(-\nullUnit\cdot\minkX_\pm)
  &= \log^2 \frac{\minkr}{2}
    + 8\pi \log\left(
    \frac{\minkr}{2}
    \right)
    \hypK(\celw;\fgr_\pm,\minkz)
    + (4\pi)^2 \hypK^2(\celw;\fgr_\pm,\minkz).
    \label{eq:logsq}
\end{align}
The small $\rho$ expansion of $\hypK$ was already discussed in
\cref{sec:logqXAsymptotics}, so it remains to analyze the last term in
\cref{eq:logsq}. Just as before, we are looking for contact terms, so we follow
the same approach of integrating
\begin{align}
  \partial_\fgr \hypK^2(\celw;\fgr,\minkz)
  &= \frac{1}{8\pi^2}
    \frac{\log(\fgr+\abs{\celwh-\minkzh}^2)}{\fgr + \abs{\celwh-\minkzh}^2}
\end{align}
against an arbitrary smooth function $f({\bf z})$ over celestial
  space. We find
\begin{align}
  \int \celVolForm(\minkz) \,
  \partial_\fgr \hypK^2(\celw;\fgr,\minkz)
  f(\minkz)
  &= \frac{1}{8\pi} \int_{\fgr+\posReals} \diff\celIntVar\,
    \frac{\log\celIntVar}{\celIntVar} g(\celIntVar-\fgr)
    \;
  \\
  &= -\frac{1}{16\pi} \log^2(\fgr) g(0)
    + \order{\fgr^0}, \qquad  (\celIntVar=\fgr+\abs{\celwh-\minkzh}^2),
    \label{eq:K2-int}
\end{align}
where $g$ is defined in \cref{eq:celAngMean}. In the last line
  we used that $g$ is smooth together with the (second of the) identities
\begin{align}
  \int^\celIntVar \diff\celIntVar'\left[
  (\celIntVar')^{n-1}\log\celIntVar'
  \right]
  &= \celIntVar^n \left[
    \frac{\log\celIntVar}{n}
    - \frac{1}{n^2}
    \right]
    + \text{constant} \;,
  & (n\ne 0)
    \label{eq:logIntegraln}
  \\
  \int^\celIntVar \diff \celIntVar'\left[
  (\celIntVar')^{-1}\log\celIntVar'
  \right]
  &= \frac{\log^2\celIntVar}{2}
    + \text{constant} \;.
    \label{eq:logIntegral0}
\end{align}
We conclude from \cref{eq:K2-int}, upon integrating over $\rho$, that
\begin{align}
  \hypK^2(\celw;\fgr,\minkz)
  &= \celG^2(\celw,\minkz)
    +\frac{\fgr(-\log^2\fgr + 2\log\fgr)}{16\pi} \celDeltaFunc(\celw,\minkz)
    + \order{\fgr} \;.
    \label{eq:hypKSquaredExpansion}
\end{align}
In particular, note that this could not be deduced by simply
  squaring \eqref{eq:hypKExpansion}, due to $\celG$ being singular in the limit
  ${\bf w} \rightarrow {\bf z}$.

\subsubsection{Asymptotics of $\log(-\nullUnit\cdot\minkX_{\pm}) \celLap[\celw]\log(-\nullUnit\cdot\minkX_{\pm})$}
\label{app:KboxK}
Next, we consider the function
\begin{align}
  \log(-\nullUnit\cdot\minkX_{\pm}) \celLap[\celw]\log(-\nullUnit\cdot\minkX_{\pm})
  &= 4\pi \left[ \log\frac{\minkr}{2} + 4\pi \hypK(\celw; \fgr_\pm, \minkz) \right]
    \celLap[\celw] \hypK(\celw; \fgr_\pm, \minkz) \;.
    \label{eq:logqXlaplogqX}
\end{align}
It may be tempting to simply plug in \cref{eq:hypKExpansion} to the RHS. Indeed,
this can be done for the $\log(\minkr/2)\celLap[\celw]\hypK$ term, because
$\log(\minkr/2)$ is constant in the celestial directions. However, note that,
according to eqs. \eqref{eq:hypKLeading} and \eqref{eq:grape},
\begin{align}
  \celLap[\celw] \hypK(\celw; 0, \minkz)
  &= \celDeltaFunc(\celw,\minkz)\;,
\end{align}
just as $\hypK(\celw;0,\minkz)$ diverges at $\celw=\minkz$. Thus, we must be
more careful with the term in \cref{eq:logqXlaplogqX}; our strategy will be
similar to our derivation of \cref{eq:hypKExpansion}, starting with a retreat to
finite $\fgr$ (which is again taken to be complex) where
\begin{align}
  \hypK(\celw; \fgr, \minkz)\,
  \celLap[\celw]
  \hypK(\celw; \fgr, \minkz)
  &= \frac{\fgr}{(2\pi)^2}
    \frac{
    \log\left(  \fgr + \abs{\celwh-\minkzh}^2 \right)
    }{
    \left( \fgr + \abs{\celwh-\minkzh}^2 \right)^2
    }
\end{align}
is smooth. We wish to evaluate the $\fgr\to 0$ expansion of the above in a
distributional sense, similar to what was performed in \cref{eq:smearedPartialhypKExpansion}.

For an arbitrary smooth celestial function $f(\minkz)$ (that does not grow too quickly at
large $\minkzh$), let us therefore consider
\begin{align}
  \int \celVolForm(\minkz)
  \hypK(\celw; \fgr, \minkz)\,
  \celLap[\celw] \hypK(\celw; \fgr, \minkz)
  f(\minkz)
  &=
    \frac{\fgr}{4\pi}
    \int_{\fgr+\posReals} \diff\celIntVar \,
    \frac{\log \celIntVar}{\celIntVar^2}
    g(\celIntVar-\fgr)
    \;,
    \quad
    (\celIntVar=\fgr+\abs{\celwh-\minkzh}^2),
    \label{eq:smearedKlapK}
\end{align}
with $g(\abs{\minkzh-\celwh}^2)$ again denoting the angular mean
\eqref{eq:celAngMean} of $f(\minkz)$ around $\minkz=\celw$. Let us aim to
evaluate \cref{eq:smearedKlapK} up to $\order{\fgr}$ corrections, so that we need
only keep the $\fgr\to 0$ divergences of the integral. Again, the divergence of
the integral in the $\fgr\to 0$ limit arises from the $\nu=\fgr$ end point. So,
considering the indefinite integrals \eqref{eq:logIntegraln} and \eqref{eq:logIntegral0},
we can pick out the divergences of the integral in \cref{eq:smearedKlapK} in the
$\fgr\to 0$ limit, giving
\begin{align}
  \begin{split}
    \MoveEqLeft[3]
    \int \celVolForm(\minkz)
    \hypK(\celw; \fgr, \minkz)\,
    \celLap[\celw] \hypK(\celw; \fgr, \minkz)
    f(\minkz)
    \\
    &=
      \frac{1}{4\pi}
      \left\{
      \left[ \log(\fgr) + 1 \right]g(0)
      - \fgr\log(\fgr)\left( 1 + \frac{\log\fgr}{2} \right) g'(0)
      \right\}
      + \order{\fgr}
      \;.
  \end{split}
\end{align}
So, observing
\begin{align}
    g'(0) &= \frac{\celLap}{4} f(\celw) \;,
            \label{eq:celAngMeanAtZero}
\end{align}
we conclude that
\begin{align}
  \begin{split}
    \MoveEqLeft[3] \hypK(\celw; \fgr, \minkz)\,
    \celLap[\celw] \hypK(\celw; \fgr, \minkz)
    \\
    &=
      \frac{1}{4\pi}
      \left\{
      \left[ \log(\fgr) + 1 \right]
      \celDeltaFunc(\celw,\minkz)
      - \frac{\fgr\log\fgr}{4}
      \left( 1 + \frac{\log\fgr}{2} \right)
      \celLap \celDeltaFunc(\celw,\minkz)
      \right\}
      + \order{\fgr} \;.
  \end{split}
\end{align}
Finally, combining this result with \cref{eq:hypKExpansion,eq:logqXlaplogqX},
\begin{align}
  \begin{split}
    \MoveEqLeft[1]
    \log(-\nullUnit\cdot\minkX_{\pm}) \celLap[\celw]\log(-\nullUnit\cdot\minkX_{\pm})
    \\
    &=
      4\pi \left\{
      \left[ \log(\minku \mp i\reg) + 1 \right]
      \celDeltaFunc(\celw,\minkz)
      - \frac{\fgr_\pm \log\fgr_\pm}{4}
      \left( 1 + \log\frac{\minkr}{2} + \frac{\log\fgr_\pm}{2} \right)
      \celLap \celDeltaFunc(\celw,\minkz)
      \right\}
      \\
    &\phantom{{}={}}
      {+ \log(\minkr)\order{\fgr_\pm}}
      + \order{\fgr_\pm} \;.
  \end{split}
  \label{eq:logqXlaplogqXAsym}
\end{align}

\subsubsection{Asymptotics of $(\confGauge{1,\pm})^2$}
\label{sec:alpha1SquaredExpansion}
Next, we consider the function
\begin{align}
  (\confGauge{1,\pm})^2
  &= \left[\celCovD\log(-\nullUnit\cdot\minkX_{\pm})\right]^2
    = (4\pi)^2
    \left[\celCovD \hypK(\celw;\fgr_\pm, \minkz)\right]^2
\end{align}
for which, again, a naive substitution of \cref{eq:hypKExpansion} fails to give
a well-defined answer. Let us more carefully evaluate the small $\fgr$ expansion
of
\begin{align}
  \left[\celCovD \hypK(\celw;\fgr, \minkz)\right]^2
  &= \frac{1}{(2\pi)^2}\frac{\abs{\celwh-\minkzh}^2}{\left(\fgr + \abs{\celwh-\minkzh}^2\right)^2} \;.
\end{align}

Again, we consider integration against smooth functions $f(\minkz)$ (with
sufficient falloff at large $\minkzh$). As before, denoting by $g$ the angular mean
\cref{eq:celAngMean} of $f$, we have
\begin{align}
  \begin{split}
    \MoveEqLeft[3]
    \int \celVolForm(\minkz) f(\minkz)
    \left[\celCovD \hypK(\celw;\fgr, \minkz)\right]^2
    \\
    &=
      \frac{1}{4\pi}
      \int_{\fgr+\posReals} \diff\celIntVar \,
      \frac{\celIntVar-\fgr}{\celIntVar^2}
      g(\celIntVar-\fgr),
      \qquad
      (\celIntVar=\fgr+\abs{\celwh-\minkzh}^2)
  \end{split}
  \\
    &=-\frac{1}{4\pi}\left\{
      \log(\fgr) g(0)
      +
      \int_{\fgr+\posReals} \diff\celIntVar \,
      \left[
      \log(\celIntVar)g'(\celIntVar-\fgr)
      +
      \frac{\fgr}{\celIntVar^2}
      g(\celIntVar-\fgr)
      \right]
      \right\}
      \;,
      \label{eq:smeareddKSquared1}
\end{align}
where we have integrated by parts for the second equality. Now, the integral of
the last integrand's first term is simply a way of writing $g'(\abs{\celw-\minkz}^2)$
integrated in $\minkz$ against $4\hypK(\celw;\fgr,\minkz)$:
\begin{align}
  \begin{split}
    \int \celVolForm(\minkz)
    \left[\celCovD \hypK(\celw;\fgr, \minkz)\right]^2
    f(\minkz)
    =&\, -\frac{1}{4\pi}
\Bigg[
       \log(\fgr)
      g(0)
      +
      4\int \celVolForm(\minkz)
      \hypK(\celw;\fgr,\minkz)
      g'(\abs{\celw-\minkz}^2)
      \\
      &\, + \fgr
      \int_{\fgr+\posReals} \frac{\diff\celIntVar}{\celIntVar^2} \,
        g(\celIntVar-\fgr)
        \Bigg]
      \;.
  \end{split}
\end{align}
Meanwhile, the last integral multiplies $\fgr$ and so we may reduce attention to
its divergences, as we did for \cref{eq:smearedKlapK}, which arise again from
the endpoint $\celIntVar=\fgr\to 0$. Altogether, using
\cref{eq:hypKExpansion,eq:celAngMean,eq:celAngMeanAtZero}, and picking out the
endpoint divergences of the last integral, we find:
\begin{align}
  \begin{split}
    \int \celVolForm(\minkz)
    \left[\celCovD \hypK(\celw;\fgr, \minkz)\right]^2
    f(\minkz)
    =&\, -\frac{1}{4\pi}
\Bigg[
      \log(\fgr)
      f(\celw)
      +
      4\int \celVolForm(\minkz)
      \celG(\celw,\minkz)
       \partial_{\abs{\celwh-\minkzh}^2}f(\minkz)
    \\ &\,
         + f(\celw)
         - \frac{\fgr \log\fgr}{4} \celLap f(\celw)
        \Bigg]
        + \order{\fgr}
      \;,
  \end{split}
  \label{eq:smeareddKSquared2}
\end{align}
where the partial derivative $\partial_{\abs{\celwh-\minkzh}^2}$ is in the polar
coordinate system around
$\celw=\minkz$.

While it is tempting to move this derivative off of $f$ through
integration by parts (so that the RHS is explicitly displayed as a function
integrated against $f$), note that this will result in what we previously
referred to as the bare distribution in \cref{eq:hypKBareLinear}. Moreover,
looking back at \cref{eq:smeareddKSquared1}, we realize that all we would
achieve is to undo the integration by parts performed there --- this gave rise
to the $\log\fgr$ term. The interpretation of the first two terms on the RHS
of \cref{eq:smearedKlapK} is therefore of a `renormalized' way to express the
bare distribution \cref{eq:hypKBareLinear} acting on the function $f$: the
divergences of the smeared bare distribution have been extracted as a
`counterterm' (the first term on the RHS) at marginally lower order
in $\fgr$, leaving behind a `renormalized' contribution (the second term on the
RHS).

At any rate, we will leave \cref{eq:smeareddKSquared2} in this form as it
honestly associates a well-defined distribution with each order in $\fgr$. Note,
however that when considering the difference between $(\confGauge{1,+})^2$ and
$(\confGauge{1,-})^2$ (which is the context in which the main text of this paper
requires the asymptotics of $(\confGauge{1,\pm})^2$), the second term on the RHS
of \cref{eq:smeareddKSquared2} cancels. So, we may compactly write
\begin{align}
  \begin{split}
  \frac{(\confGauge{1,+})^2 - (\confGauge{1,-})^2}{2\pi i}
  =&\, 4\pi[\regStepFunc{\reg}(-\minku)-\stepFunc(-\minkr)] \celDeltaFunc(\celw,\minkz)
    \\
    &- \frac{i}{2} (\fgr_+\log\fgr_+ - \fgr_-\log\fgr_-) \celLap \celDeltaFunc(\celw,\minkz)
      + {\eval{\order{\fgr}}_{\fgr_-}^{\fgr^+}}
  \end{split}
  \label{eq:differOfconfGauge2Asym}
\end{align}
where we have used\footnote{Recall footnote \ref{foot:argr}, which explains the
  presence of the $\stepFunc(-\minkr)$.}
\begin{align}
  \frac{\log\fgr_+ - \log\fgr_-}{2\pi i}
  &= -\regStepFunc{\reg}(-\minku)+\stepFunc(-\minkr) \;,
\end{align}
with the regulated step function defined in \cref{eq:regStepFunc}.

\subsubsection{Asymptotics of
  $\log(-\nullUnit\cdot\minkX_\pm)\diff[\minkz]\confGauge{1,\pm}$}
Using \cref{eq:cappuccino,eq:logqX},
\begin{align}
  \log(-\nullUnit\cdot\minkX_\pm)\,
  \diff[\minkz]\confGauge{1,\pm}
  &= -4\pi \left[ \log\frac{\minkr}{2} + 4\pi \hypK(\celw; \fgr_\pm, \minkz) \right]
    \diff[\celw]\diff[\minkz] \hypK(\celw; \fgr_\pm, \minkz).
\end{align}
There are two cases to be considered namely
\begin{align}
  \hypK(\celw; \fgr, \minkz)
  \partial_{\celwh}\partial_{\minkza}
  \hypK(\celw; \fgr, \minkz)
    &= -\frac{1}{4}
      \hypK(\celw; \fgr, \minkz)\celLap[\celw]\hypK(\celw; \fgr, \minkz)
\end{align}
and
\begin{align}
\hypK(\celw; \fgr, \minkz)
  \partial_{\celwh}\partial_{\minkzh}
  \hypK(\celw; \fgr, \minkz)
  &= \frac{1}{(4\pi)^2}\log\left( \fgr + \abs{\celwh-\minkzh}^2  \right)\left(
    \frac{\celwa-\minkza}{ \fgr + \abs{\celwh-\minkzh}^2 }
    \right)^2.
\end{align}
The former has already been evaluated in section \ref{app:KboxK}, so we are left
with the latter. Integrating it against a smooth function on the sphere with
sufficiently fast fall-off at infinity,
\begin{align}
\label{eq:KppK}
  \begin{split}
    \MoveEqLeft[4]\int \celVolForm(\minkz)
    \hypK(\celw; \fgr, \minkz)
    \partial_{\celwh}\partial_{\minkzh}
    \hypK(\celw; \fgr, \minkz)
    f(\minkz)
    \\
    &=
      \frac{1}{16\pi}
      \int_{\fgr+\posReals} \diff\celIntVar \,
      \frac{(\celIntVar-\fgr)\log \celIntVar}{\celIntVar^2}
      h(\celIntVar-\fgr)
      \;,
  \end{split}
  \qquad
  (\celIntVar=\fgr+\abs{\celwh-\minkzh}^2),
\end{align}
where we defined
\begin{align}
  h(\abs{\minkzh-\celwh}^2)
  &= \frac{1}{2\pi} \int \diff\celpsi\,
    e^{-2i\celpsi} f(\minkz)
  &
    (\minkzh-\celwh = \abs{\minkzh-\celwh} e^{i\celpsi}).
\end{align}
Considering the Taylor expansion of $f(z)$ about $z=w$, one
  finds that
\begin{align}
  h(0)
  &= 0 \;,
  &
    h'(0)
  &= \frac{\partial_{\celwh}^2}{2} f(\celw) \;.
    \label{eq:h}
\end{align}

Integrating the first term of \cref{eq:KppK} by parts and using \cref{eq:h}, we then find
\begin{align}
  \begin{split}
    \MoveEqLeft[1]\int \celVolForm(\minkz)
    \hypK(\celw; \fgr, \minkz)
    \partial_{\celwh}\partial_{\minkzh}
    \hypK(\celw; \fgr, \minkz)
    f(\minkz)
    \\
    &=
      -\frac{1}{16\pi}
      \int_{\fgr+\posReals} \diff\celIntVar \left[
     \frac{1}{2} \log^2(\celIntVar)\,
      h'(\celIntVar-\fgr)
      + \frac{\fgr\log \celIntVar}{\celIntVar^2}
      h(\celIntVar-\fgr)
      \right]
  \end{split}
  \\
    &=
      -  \frac{1}{2} \int \celVolForm(\minkz) \,
      \hypK^2(\celw;\fgr,\minkz) \,
      h'(\abs{\celwh-\minkzh}^2)
      -
      \frac{\fgr}{16\pi} \int_{\fgr+\posReals} \diff\celIntVar
      \frac{\log \celIntVar}{\celIntVar^2}
      h(\celIntVar-\fgr)
  \\
    &=
      -  \frac{1}{2} \int \celVolForm(\minkz) \,
      \celG^2(\celw;\minkz) \,
      \frac{\celwa-\minkza}{\celwh-\minkzh}
      \partial_{\abs{\celwh-\minkzh}^2}f(\minkz)
      +\frac{\fgr\log^2\fgr}{32\pi}
      \partial_{\celwh}^2 f(\celw)
      + \order{\fgr}\;,
      \label{eq:hypKddhypKExpansion}
\end{align}
where we have used \cref{eq:hypKSquaredExpansion} in the last equality. Putting everything together, we find

  
\begin{align}
  \begin{split}
    \MoveEqLeft[0]
    \frac{
    \log(-\nullUnit\cdot\minkX_+)\,
    \partial_{\minkzh} \confGauge{1,+}[\celwh]
    -\log(-\nullUnit\cdot\minkX_-)\,
    \partial_{\minkzh} \confGauge{1,-}[\celwh]
    }{
    2\pi i
    }
    \\
    &=
      4\pi \stepFunc(-\minkr)
      \partial_{\minkzh} \partial_{\celwh} \celG(\celw,\minkz)
      + \left[
      \frac{\fgr \log\fgr}{2i}
      \left(\log\frac{\minkr}{2} + \frac{\log\fgr}{2} \right)
      \partial_{\minkzh} \partial_{\celwh} \celDeltaFunc(\celw,\minkz)
      {+ \log(\minkr)\order{\fgr}}
      + \order{\fgr}
      \right]^{\fgr_+}_{\fgr_-}
      \;.
  \end{split}
\end{align}
Combining with \cref{eq:logqXlaplogqXAsym}, we obtain

\begin{align}
  \begin{split}
    \MoveEqLeft
    \frac{
    \log(-\nullUnit\cdot\minkX_+)\,
    \diff[\minkz] \confGauge{1,+}
    -\log(-\nullUnit\cdot\minkX_-)\,
    \diff[\minkz] \confGauge{1,-}
    }{
    2\pi i
    }
    \\
  =&\,
     {2}\pi [\stepFunc(-\minkr)-\regStepFunc{\reg}(-\minku)]
     \celMet \celDeltaFunc(\celw,\minkz)
     +
     4\pi \stepFunc(-\minkr)
     \diff[\minkz] \diff[\celw]\celG(\celw,\minkz)
    \\
    &\, + \left\{
    \frac{\fgr \log\fgr}{2 i}
      \left[
      \celMet \frac{\celLap}{2} +
      \left( \log\frac{\minkr}{2} + \frac{\log\fgr}{2} \right)
      \diff[\minkz] \diff[\celw]
      \right]
      \celDeltaFunc(\celw,\minkz)
      {+ \log(\minkr)\order{\fgr}}
      + \order{\fgr}
      \right\}^{\fgr_+}_{\fgr_-}
      \;.
  \end{split}
  \label{eq:logqXpzalphaDifferAsym}
\end{align}

\subsubsection{Asymptotics of
  $\log(-\nullUnit\cdot\minkX_\pm)\partial_\minku \confGauge{1,\pm}$}
Using \eqref{eq:logqX} and \eqref{eq:shiftedfgr}, we find
\begin{align}
  \log(-\nullUnit\cdot\minkX_\pm)\,
  \partial_\minku \confGauge{1,\pm}
  &= -\frac{8\pi}{\minkr} \left[ \log\frac{\minkr}{2} + 4\pi \hypK(\celw; \fgr_\pm, \minkz) \right]
    \diff[\celw] \partial_{\fgr_\pm} \hypK(\celw; \fgr_\pm, \minkz)
\end{align}
It is straightforward to evaluate
\begin{align}
  \hypK(\celw; \fgr, \minkz)
  \partial_{\celwh}\partial_{\fgr}
  \hypK(\celw; \fgr, \minkz)
  &= \frac{\minkza-\celwa}{(4\pi)^2}
    \,
    \frac{\log\left( \fgr + \abs{\celwh-\minkzh}^2  \right)}{ \left(\fgr + \abs{\celwh-\minkzh}^2  \right)^2}.
\end{align}
As before we integrate this against a smooth function on the sphere to find
\begin{align}
    \int \celVolForm(\minkz)
    \hypK(\celw; \fgr, \minkz)
    \partial_{\celwh}\partial_{\fgr}
    \hypK(\celw; \fgr, \minkz)
    f(\minkz)
    &=
      \frac{1}{16\pi}
      \int_{\fgr+\posReals} \diff\celIntVar \,
      \frac{\log \celIntVar}{\celIntVar^2}
      j(\celIntVar-\fgr)
      \;,
  \quad
  (\celIntVar=\fgr+\abs{\celwh-\minkzh}^2).
\end{align}
Here we defined
\begin{align}
  j(\abs{\minkzh-\celwh}^2)
  &= \frac{1}{2\pi} \int \diff\celpsi\,
    (\minkza-\celwa)f(\minkz),
  &
    (\minkzh-\celwh = \abs{\minkzh-\celwh} e^{i\celpsi}),
\end{align}
which can be shown to satisfy
\begin{align}
  j(0) &= 0 \;,
         &
  j'(0)
  &= \partial_{\celwh} f(\celw).
\end{align}
Using \eqref{eq:logIntegraln} and \eqref{eq:logIntegral0} we find the divergent part of the integral to be
\begin{align}
    \int \celVolForm(\minkz)
    \hypK(\celw; \fgr, \minkz)
    \partial_{\celwh}\partial_{\fgr}
    \hypK(\celw; \fgr, \minkz)
    f(\minkz)
    &=
      -\frac{\log\fgr}{16\pi}
       \left( 1 + \frac{\log\fgr}{2} \right) j'(0)
      + \order{\fgr^0}
      \;.
\end{align}
It follows that
\begin{align}
  \hypK(\celw; \fgr, \minkz)
  \partial_{\rho} \diff[\celw]
  \hypK(\celw; \fgr, \minkz)
  &=
    -\frac{\log\fgr}{16\pi}
    \left( 1 + \frac{\log\fgr}{2} \right)
    \diff[\celw] \celDeltaFunc(\celw,\minkz)
    + \order{\fgr^0}.
\end{align}
Putting everything together, we find
\begin{align}
  \begin{split}
  \log(-\nullUnit\cdot\minkX_\pm)\,
  \partial_\minku \confGauge{1,\pm}
  &=
    \frac{2\pi\log\fgr_\pm}{\minkr}
    \left( 1 + \log\frac{\minkr}{2} + \frac{\log\fgr_\pm}{2} \right)
    \diff[\celw] \celDeltaFunc(\celw,\minkz)
    \\
    &\phantom{{}={}}+ \minkr^{-1} \left[
    \log(\minkr)\order{\fgr_{\pm}^0}
    +\order{\fgr_{\pm}^0}
    \right]
    \;.
  \end{split}
  \label{eq:logqXpualphaAsym}
\end{align}



\bibliographystyle{bibstyle.bst}
\bibliography{biblio.bib}

\end{document}